%% file: mismeasure.tex
\documentclass[11pt]{article}

\usepackage{geometry}
\usepackage{fullpage}
\usepackage{graphicx}
\usepackage{pdflscape}
\usepackage{multirow}
\usepackage{ragged2e}

\usepackage{dcolumn}
\usepackage{booktabs,calc}
\usepackage{longtable}
\usepackage{array}
\usepackage{paralist}
\usepackage{verbatim}
\usepackage{subfig}
\usepackage{setspace}
\usepackage{amsmath}
\usepackage{amssymb}
\usepackage{amsthm}
\usepackage{natbib}
\usepackage[linktocpage=true, colorlinks=true, linkcolor=blue, citecolor=black]{hyperref}
\usepackage{enumerate}
\usepackage{authblk}
\usepackage{placeins}
\usepackage{multicol}
\usepackage{array}
\usepackage[usenames, dvipsnames]{xcolor}
\usepackage{url}
\usepackage{listings}

\usepackage{epigraph}
\usepackage{etoolbox}

\makeatletter
\patchcmd{\epigraph}{\@epitext{#1}}{\itshape\@epitext{#1}}{}{}
\makeatother

\lstdefinestyle{myListingStyle} 
    {
        basicstyle = \small\ttfamily,
        breaklines = true,
    }

\usepackage{threeparttable}

\usepackage{siunitx} 
    \defcitealias{ETH1}{CSA, 2014}
    \defcitealias{ETH2}{CSA, 2015}
    \defcitealias{ETH3}{CSA, 2017}
    \defcitealias{ETH4}{CSA, 2020}
    \defcitealias{ETH5}{CSA, 2024}
    
    \defcitealias{MWI1}{NSO, 2012}
    \defcitealias{MWI2}{NSO, 2015}
    \defcitealias{MWI3}{NSO, 2020a}
    \defcitealias{MWI4}{NSO, 2017}
    \defcitealias{MWI5}{NSO, 2020b}
    
    \defcitealias{NGR1}{NIS, 2014}
    \defcitealias{NGR2}{NIS, 2016}
    
    \defcitealias{NGA1}{NBS, 2012}
    \defcitealias{NGA2}{NBS, 2014}
    \defcitealias{NGA3}{NBS, 2016}
    \defcitealias{NGA4}{NBS, 2019}
    
    \defcitealias{TZA1}{TNBS, 2011}
    \defcitealias{TZA2}{TNBS, 2012}
    \defcitealias{TZA3}{TNBS, 2015}
    \defcitealias{TZA4}{TNBS, 2017}
    \defcitealias{TZA5}{TNBS, 2021}
    \defcitealias{TZA6}{TNBS, 2023}
    
    \defcitealias{UGA1}{UBOS, 2012}
    \defcitealias{UGA2}{UBOS, 2014a}
    \defcitealias{UGA3}{UBOS, 2014b}
    \defcitealias{UGA4}{UBOS, 2016}
    \defcitealias{UGA5}{UBOS, 2019}
    \defcitealias{UGA6}{UBOS, 2020}
    \defcitealias{UGA7}{UBOS, 2021}

\begin{document}
	
\title{The Mismeasure of Weather: Using Remotely Sensed Earth Observation Data in Economic Contexts\thanks{A pre-analysis plan for this research has been filed with Open Science Framework (OSF): \href{https://osf.io/8hnz5/}{https://osf.io/8hnz5/}. We gratefully acknowledge funding from the World Bank Living Standards Measurement Study (LSMS) and the Knowledge for Change Program (KCP). This paper has been shaped by conversations with Marc Bellemare, Leah Bevis, as well as seminar participants at the Methods and Measurement Conference 2021, the AAEA annual meetings in Chicago and Atlanta, the $31^{st}$ triennial ICAE conference, and participants in presentations at Arizona State University, the University of Minnesota, the World Bank, and Virginia Tech. We are especially grateful to Reece Branham, Alison Conley, Kieran Douglas, Rodrigo Guerra-Su, Emil Kee-Tui, Brian McGreal, and Jacob Taylor for their diligent work as research assistants and to Oscar Barriga Cabanillas and Aleksandr Michuda for early help in developing the Stata \texttt{wxsum} package. Thanks also to Laura Josephson for logistical support. We are solely responsible for any errors or misunderstandings.}}

	\author[1]{Anna Josephson}
	\author[1]{Jeffrey D. Michler}
	\author[2]{Talip Kilic}
	\author[2]{Siobhan Murray}
	\affil[1]{\small \emph{Department of Agricultural and Resource Economics, University of Arizona}}
	\affil[2]{\small \emph{Development Data Group, World Bank}}

\date{August 2024}
\maketitle

\thispagestyle{empty}

\begin{center}\begin{abstract}
\noindent The availability of weather data from remotely sensed Earth observation (EO) data has reduced the cost of including weather variables in econometric models. Weather variables are common instrumental variables used to predict economic outcomes and serve as an input into modelling crop yields for rainfed agriculture. The use of EO data in econometric applications has only recently been met with a critical assessment of the suitability and quality of this data in economics. We quantify the significance and magnitude of the effect of measurement error in EO data in the context of smallholder agricultural productivity. We find that different measurement methods from different EO sources: findings are not robust to the choice of EO dataset and outcomes are not simply affine transformations of one another. This begs caution on the part of researchers using these data and suggests that robustness checks should include testing alternative sources of EO data.   
	\end{abstract}\end{center}


	{\small \noindent\emph{JEL Classification}: C38, C81, D83, O13, Q12 }

\newpage
\onehalfspacing


\epigraph{Science must be understood as a social phenomenon... not the work of robots programmed to collect pure information. Scientific findings should not be elevated to the status of immutable truths. }{--- \textup{Stephen Jay Gould}, The Mismeasure of Man}

\section{Introduction}

There are 37 weather stations in the World Meteorological Organization's (WMO) database on the entire continent of Africa. These stations provide weather data for 1.1 billion persons living on the continent \citep{TzachorEtAl23}. This figure stands in contrast to the 636 weather stations across the European Union and the United States, which provide weather data for 1.2 billion persons in those two regions \citep{TzachorEtAl23}. Further, because of their uneven dispersion across the African continent, the weather stations which do exist cover only about 40 percent of the African population. Weather stations on the continent are often so far apart that the data collected are of limited use, a condition exacerbated by under-investment in their maintenance, which results in a deterioration of the frequency and quality of data reporting. Only one in five weather stations in Africa met the WMO's reporting standards as of 2019. 

The lack of station data across Africa means that, for much of the continent, the truth of the weather is unmeasured. The emergence of Earth Observation (EO) and improvements in weather modeling have enabled the development of remotely sensed weather datasets to fill this gap. These products provide an estimate of the on-the-ground conditions of weather, albeit detected from a distance. These EO products can provide myriad measures of weather phenomenon, like precipitation, temperature, wind speed, or humidity. Each EO product uses a different combination of weather sensors and methods for interpolating and interpreting the data from those sensors. And so, while each EO product endeavors to measure the objective truth of weather conditions, each produces its own ``truth.'' In theory, these ``truths'' should all be the same. In practice, they are not. Figures~\ref{fig:rain_res} and \ref{fig:temp_res} show these different ``truths'' across six remotely sensed EO precipitation products and three temperature products. One precipitation product reports rainfall of less than five mm while a different product reports rainfall of more than 47 mm for the same location on the same day. Temperature also varies by EO product. One product reports a maximum temperature of 23$^{\circ}$ C while another reports the maximum temperature that day as 27$^{\circ}$ C.

Where does this disparity leave researchers who want to incorporate EO weather data into economic contexts? If using ground-based weather station data, the gold standard for weather measurement, is not possible, then we are left to use EO sources. But, their mismeasurement of weather can introduce two issues. First, if economists are unaware of these measurement issues, we may be using the ``wrong'' data, or a poor proxy, to measure the truth of what weather occurred where and when. This can introduce either classical or non-classical measurement error into our estimates and predictions. Second, if economists are aware that different EO products report different numbers, then we can engage in $p$-hacking, HARKing, and other unethical behavior by choosing the EO product that gives us the results we want. 

To determine the magnitude of this disparity and its impact on the use of EO data in economic contexts, we integrate nine EO geospatial weather datasets with georeferenced longitudinal household survey data. The survey data comes from six Sub-Saharan African countries and was collected under the World Bank's Living Standards Measurement Study - Integrated Surveys on Agriculture (LSMS-ISA) initiative. We assess the degree to which mismeasurement matters by modeling the relationship between weather and smallholder agricultural productivity, as captured in the LSMS-ISA survey data. We use a simple heuristic to compare coefficients from regressions of yield on weather as measured by one EO product to coefficients when weather is measured by a different EO product. Our goal is to provide guidance to researchers looking to integrate EO weather data with socioeconomic survey data regarding the degree to which their results may be dependent on their choice of remote sensing EO product.

We find that the different measurement methods by EO products affect not only the cardinality of coefficients across the products but also coefficient ordinality. The first result is unsurprising. Obviously, continuous variables measured differently by different sources will result in different effect sizes. In applied economics, we are typically not too concerned about differences in relative size (cardinality) across specifications, as we understand that different choices in data cleaning and modelling will affect the exact size of a coefficient. What we do care about is getting the order correct (ordinality). That using different EO products affects the ordinality of coefficients means that EO products are not simply affine transformations of one another. One can completely reverse the ordering of coefficient size across EO products by simply changing the specification, country, or dependent variable. A researcher can essentially get whatever result she or he wants through the judicious choice of which EO weather product to use. 

That different EO sources produce a different ordering of outcomes suggests that many results in economics that rely on EO weather data, or any model-derived EO product, may not be robust to the choice of EO product. As \cite{Jain20} has pointed out, this is a concern for the large body of economic research that has relied on EO weather data for identification of causal effects. This literature includes important contributions to understanding human capital formation \citep{MacciniYang09, ShahSteinberg17, GargEtAl20}, labor markets \citep{Jayachandran06, ChenEtAl17, Kaur19, Morten19}, conflict and institutions \citep{BruckerCiccone11, Sarsons15, KonigEtAl18}, agricultural production and economic growth \citep{MiguelEtAl04, DescheneGreenstone07, BarriosEtAl10, DellEtAl12, YehEtAl20}, intra-household bargaining power \citep{CornoEtAl20}, technology adoption \citep{Taraz18, JagnaniEtAl21, AragonEtAl21, TesfayeEtAl21}, and extreme weather impacts \citep{WinemanEtAl17, MichlerEtAl19, McCarthyEtAl21a}. Our findings suggest that economists should exercise caution when seeking to combine EO data with socioeconomic survey data, as the selection of different EO sources may produce very different results, thus contributing to this literature on the use of EO data in economic applications.

Our paper also contributes to the nascent, but growing, literature on the economic implications of measurement error. Much of the research in this stream has focused on mismeasurement in the context of agricultural production and focuses on mismeasurement in self-reported data \citep{Carlettoetal17, AbayEtAl19, GourleyEtAl19, KosmowskiEtAl19, Lobelletal19, Abay20, AbayEtAl21, GollinUdry21, KilicEtAl21, MichlerEtAl22}. A related literature focuses on mismeasurement in everything from prices \citep{Fitzpatrick23}, to consumption \citep{AbateEtAl23}, to time use \citep{FieldEtAl23}, to willingness to pay \citep{JackEtAl22}. A smaller literature has begun to document, for economists, the presences of measurement error in EO data \citep{Gibson21, GibsonEtAl21, AlixGarciaandMillimet23}, an issue long known and discussed in GIS, remote sensing, and geophysical science \citep{Wilkinson96}. We add to this literature as the first attempt, to our knowledge, at understanding the impact of mismeasurement in EO products in estimating policy relevant relationships between climate and development outcomes in low- and middle-income countries.

Understanding the differences in results that arise when a researcher chooses one EO product over another also has practical consequences, due to the importance of smallholder agriculture for rural livelihoods in low- and middle-income countries. Agriculture is both as a source of income and employment, and as such understanding how weather affects agricultural productivity on smallholder farms is of policy interest. Thus, these results are important for informing policies to develop and advise on improved agricultural technologies that can mitigate the risks posed by climate variability and extreme weather events, and to provide social protection measures to smallholder farmer that are exposed to climate shocks.


\section{Measuring the Truth: Gauge Station Data and Earth Observation Data in Africa}

The goal of weather data products is to measure and report on the objective fact that is the volume of precipitation and the temperature in degrees in a given location at a given time. In theory, this goal is easily accomplished using technology that has existed for centuries. One simply needs a rain gauge to collect precipitation, a mercury thermometer to measure temperature, and someone to record this data with pencil on paper. What could be simpler than measuring the weather? Humans have been reporting on the weather since the invention of writing and have been using scientific instruments to record precipitation and temperature since the 1600s \citep{LundstadEtAl23}.

The problem with measuring the truth of the weather is not the technology but the personnel. It is insufficient to simply have the instruments to capture the data. One requires a person to record and report the data. In our setting, where we use LSMS-ISA data that provides tens of thousands of observations across 13 years and six countries, one would need to provide each household with the appropriate instruments and sufficient compensation to record the daily record of the weather. Additionally, one would need to employ individuals to collect the information from around the country and do quality checks on the data. The cost of a rain gauge and a thermometer might be trivial. The cost of converting that data into a geospatially explicit time series spanning across decades and countries is where measuring the truth becomes a problem.

Given the prohibitively high costs of recording and reporting the weather for every household in a nationally representative survey, two alternatives exist. The first alternative is to rely on gauge data from nationally run meteorological stations. This is considered the gold standard of weather data, as the quality control process in capturing, recording, and reporting the weather should be pristine. However, not everyone lives near a weather station. And in low-income countries, weather stations can be sparsely distributed \citep{TzachorEtAl23}. Figure~\ref{fig:stations} maps the locations of LSMS-ISA households with locations of Global Historical Climatology Network (GHCN) weather stations.\footnote{GHCN is a database which addresses the critical need for historical daily weather records. It is a composite of climate records from numerous sources, merged together and assessed for quality. At its best, the data includes more than 40 meteorological elements, including temperature daily maximum and minimum, temperature at observation time, precipitation, snowfall, snow depth, evaporation, wind movement and maximums, soil temperature, and more.} These maps show the limited density of these stations in Africa. Consider the case of Uganda, in which there are no in-country stations currently reporting to the GHCN. This means that the closest weather station for LSMS-ISA households in Uganda are one station in Kenya and one in Tanzania. And so, the second alternative (and the one most widely employed in economic research at this time) is to rely on remotely sensed EO data. Using EO data provides global-scale data on the weather at various spatial and temporal resolutions, overcoming the sparseness of station data.

Spatial datasets of weather variables, like precipitation and temperature, that are produced using remotely sensed EO data, are not direct measurements of the variable of interest. Satellite sensors provide spatially continuous observation of reflectance from the Earth's surface in different parts of the magnetic spectrum. These values are used to estimate related phenomena, such as cloud presence, cloud top temperature or earth surface temperature. The continuous datasets are then used in combination with directly observed, but often sparsely distributed, gauge data to produce weather variables. Some inputs are common across products, but there are differences in other inputs as well as modeling techniques. 

The type of analysis matters in assessing weather datasets for use in economic contexts. Is the goal to understand climate trends; to capture characteristics of a particular agricultural season; or to identify extreme weather events occurring in near real-time? This can help determine the relative importance of different dataset characteristics, such as spatial detail, temporal frequency and length of record, with respect to the intended analysis. 

In an ideal setup, we would compare objective truth of rainfall and data measured on the ground, with remote sensing EO sources, and simply determine which EO product best matches with the measured truth. However, as we allude to, it is not possible to measure the truth, through direct measurement or through gauge station data. This leaves us with our EO weather data. We discuss the particular EO sources which we use and their various interpolation methods in Section~\ref{sec:weatherdata}. It is this second-best and its outcomes for the measurement and potential mismeasurement of agricultural production and productivity that we explore in this paper. 

With this said, we acknowledge that EO data are not intended for field-level analyses - that is, they are not explicitly designed to fix the problems identified. And so, we are pushing these data to the limits of precision. While this is justified due to lack of availability of higher resolution alternatives, we appreciate that spatial aggregation likely itself creates some messiness within the data.


\section{Data} \label{sec:data}

To understand the possible sources of measurement error associated with integrating spatial, remotely sensed EO weather data with socioeconomic data, we combine publicly available EO weather data products with publicly available unit-record survey data that have been generated as part of the World Bank LSMS-ISA initiative and that are made available through the World Bank Microdata Library.

The publicly available data contains obfuscated GPS coordinates to protect the privacy of survey participants. Specifically, spatial anonymization in the LSMS-ISA is achieved by only releasing GPS coordinates for enumeration area (EA) clusters, not households \citep{Murray21}. Further, following a method pioneered by the Demographic and Health Surveys (DHS), the EA cluster coordinates are displaced. Urban clusters are displaced in any direction within a two kilometers radius while rural clusters are displaced in any direction within a five kilometers radius. Additionally, one percent of rural clusters are displaced up to ten kilometers.

In trying to assess measurement error in EO weather data, the displacement of coordinates coming from spatial anonymization obviously can have a confounding effect, since one would be unable to disentangle mismeasurement in the EO product from mismeasurement introduced by the mismatch between the true household location and the displaced EA cluster coordinates. However, for this analysis, members of the authorship group were granted access to the true household coordinates - not just the un-displaced EA cluster coordinates - and matched the EO data to the true household location data. Thus, the temperature and precipitation reported by each EO product is measured at the exact location of the household. While this gives our analysis an added degree of accuracy in that we can precisely match household point location to EO grid cell, \cite{MichlerEtAl22} show that, in similar analyses, using the obfuscated coordinates has no meaningful impact on results.

In the remainder of this section, we describe the sources of weather data and household data. We then discuss descriptive statistics on the combined weather-household datasets.


\subsection{Earth Observation Weather Data} \label{sec:weatherdata}

In this study, the most important data are our EO weather data. To align with the use of these data by economists, while also representing a wide range of the types of available data in this domain, we use several public weather data sources that represent different modeling types, input sources, and spatial resolutions. To ensure consistency and to ensure the creation of common metrics across the analysis, we imposed two inclusion criteria. The source had to have (1) high temporal resolution (daily), and (2) a minimum 30-year length of record at least up to 2021.\footnote{This criteria resulted in the exclusion of some EO products that have been frequently used by economists, including the monthly \emph{Terrestrial Air Temperature and Precipitation} from the Center for Climatic Research at the University of Delaware.} Table~\ref{tab:weather} describes each data source, including the length of record, spatial and temporal resolution, and the type of data recorded.

We classify EO weather data products by the method through which they generate precipitation and temperature estimates. The first type of product we use merges gauge data with data from meteorological satellites, which provide indirect information at full coverage. EO products of this type include the African Rainfall Climatology version 2 (ARC2), the the Climate Hazards group InfraRed Precipitation with Station Data (CHIRPS), and Tropical Applications of Meteorology using SATellite data and ground-based observations (TAMSAT) \citep{ARC2, CHIRPS, TAMSAT}.

ARC2 is a merged gauge data and EO product that provides daily rainfall outputs for the African continent. The dataset, produced by the National Oceanic and Atmospheric Administration (NOAA) Climate Prediction Center (CPC) provides improvements over ARC1 and a longer length of record compared to the rainfall estimate (RFE), the operational dataset of USAID's Famine Early Warning Systems Network (FEWSNET) program. Inputs are Global Telecommunications System (GTS) rain gauge data over Africa, geostationary Meteosat infrared (IR) imagery, and polar-orbiting microwave Special Sensor Microwave Image (SSM/I) and Advanced Microwave Sounding Unit (AMSU-B).

Validation efforts by \cite{ARC2} found that low reporting rates for some GTS stations degrades model performance in those regions. Other findings are a general tendency to underestimate rainfall, which is enhanced in areas of high relief or complex topography.\footnote{Data and technical documentation are available for download from \url{https://www.cpc.ncep.noaa.gov/products/international/data.shtml}.}

Like ARC2, the CHIRPS rainfall dataset builds on established techniques for merging gauge and EO data. Produced by the Climate Hazards Group at University of California, Santa Barbara this dataset is designed for monitoring of drought and environmental change at a global level. To minimize latency, there are two products, a preliminary version with two day lag, and final output available at three weeks. Outputs are available at time-steps from six hours to three months. As inputs, CHIRPS makes use of a monthly climatology CHPclim, Tropical Rainfall Measuring Mission Multi-satellite Precipitation Analysis version 7 (TMPA 3B42 v7) and global Thermal Infrared Cold Cloud Duration (TIR CCD) from two NOAA archives. The EO data are then merged with gauge data from five public archives, including the Global Historical Climatology Network (GHCN) and GTS, several private sources, and meteorological agencies. While targeted gauge data collection efforts resulted in a greater number of input stations for years prior to 2010, the number of stations going forward is more limited, particularly in Sub-Saharan Africa. Detailed metadata by country is available and may be a useful reference to determine if coverage for a region of interest is sufficient for the analysis.

Validation for select countries found that the climatology input CHPclim outperformed other climatology datasets in data sparse regions and complex terrain \citep{CHIRPS}. Further, in an evaluation of wet season statistics CHIRPS showed less bias than other rainfall sources and good correspondence with Global Precipitation Climatology Centre (GPCC) estimates.\footnote{Data and technical documentation are available for download from \url{https://data.chc.ucsb.edu/products/CHIRPS-2.0/}.}

The TAMSAT rainfall dataset is the highest spatial resolution gridded dataset used in this analysis. Inputs are similar to other merged gauge and EO products: Meteosat TIR imagery, purposefully collected archival (1983-2010) rain gauge data from meteorological agencies and other sources and GTS gauge data. Rainfall estimation is based on cold cloud duration (CCD) inferred from TIR and calibrated using gauge data within discrete calibration zones.

Validation of TAMSAT found a mean underestimation of rainfall of approximately four mm per dekad, though the bias was not always negative \citep{TAMSAT}. Due to differences in methodology from CHIRPS and ARC2 precipitation products, TAMSAT is not affected by inconsistency in gauge data inputs. This makes it suitable for placing rainfall variability in the context of a long-term climatology and thus detecting unusually wet or dry conditions.\footnote{Data and technical documentation are available for download from
\url{http://www.tamsat.org.uk/data/}.}

The second type of product uses assimilation models to combine a large number of observations from different sources (e.g., satellites, weather stations, ships, aircraft) to produce a model of the global climate system or a particular atmospheric phenomenon. Outputs are inferred or predicted based on the system state and understanding of interactions between model variables. EO products of this type include the European Centre for Medium-Range Weather Forecasts ERA5 and the NASA Modern-Era Retrospective analysis for Research and Applications (MERRA-2) \citep{ERA5, MERRA2}.

ERA5, based on the European forecasting model ECMWF, is one of two assimilation model datasets used in this paper. The inputs are far too numerous to mention but include a range of satellite inputs as well as gauge datasets. There are a wide range of outputs as well, including 2-meter air temperature and rainfall, available at sub-daily intervals and differentiated vertically. ERA5 is coarser spatial resolution than the global and regional merged rainfall datasets, but more detailed than MERRA-2. 

The sheer number and complexity of outputs can be a deterrent to the use of weather variables from assimilation models. Uncertainty or lack of understanding about inaccuracies associated with individual output variables of assimilation models, compared to other types of models, is another reason to carefully consider their suitability for particular research \citep{Parker16}. Nevertheless, reanalysis datasets are used in a broad range of applications and even outperform other gridded climate datasets in some settings \citep{ZandlerEtAl20}.\footnote{Data and technical documentation are available for download from \url{https://cds.climate.copernicus.eu}.}

The second reanalysis dataset used in this analysis is MERRA-2, a product of NASA's Goddard Earth Observing System, version 5 (GEOS-5) assimilation model. Specifically we make use of the variables T2MMEAN from the statD daily statistics collection, and PRECTOTLAND from the Land Surface Diagnostics collection.\footnote{Data and technical documentation are available for download from \url{https://disc.gsfc.nasa.gov/}.}

Last, we consider an EO data product produced primarily from gauge data, using only spatial interpolation techniques to produce a continuous surface from observed measurements. The NOAA Climate Prediction Center (CPC) Unified Gauge-Based Analysis of Daily Precipitation and Temperature datasets were created using all information sources available at CPC and undergoes extensive pre-processing and cleaning, including comparison with contemporaneous data from satellite and other sources \citep{CPC}.

NOAA's Climate Prediction Center (CPC) Unified Gauge-based (CPC-U) datasets for daily temperature and precipitation do not incorporate EO data in the estimation of weather variables. Instead, an optimal interpolation (OI) technique is used on gauge data for precipitation, and Shepard's algorithm for temperature. CPC-U provides systematic global datasets for validation and climate monitoring. GTS is a primary input data source, with some national collections, but density is most sparse over Africa.

As to be expected, even though the OI interpolation performs better than other techniques, a cross-validation exercise shows performance to degrade significantly with increasing distance to nearest station \citep{CPC}. As a result, this dataset may not be suitable for analysis in some parts of Africa, with high spatial variation and low density of stations.\footnote{Data and technical documentation are available for download from \url{https://psl.noaa.gov/data/gridded/data.cpc.globalprecip.html} for precipitation and
\url{https://psl.noaa.gov/data/gridded/data.cpc.globaltemp.html} for temperature.}


\subsection{Household Survey Data} \label{sec:householddata}

The World Bank Living Standards Measurement Study - Integrated Surveys on Agriculture (LSMS-ISA) is a household survey program that provides financial and technical assistance to national statistical offices in Sub-Saharan Africa for the design and implementation of national, multi-topic longitudinal household surveys with a focus on agriculture. Our analysis uses data from multiple rounds of panel household surveys from Ethiopia, Malawi, Niger, Nigeria, Uganda and Tanzania. Table~\ref{tab:lsms} provides a summary of the countries, years, and observations used in the analysis.

In Ethiopia, we use data from the 2011/12, 2013/14, 2015/16, 2018/19, and 2021/22 rounds of the Ethiopia Socioeconomic Survey (ESS), which has been conducted by the Central Statistical Agency of Ethiopia \citepalias{ETH1, ETH2, ETH3, ETH4, ETH5}. The Wave 1 data is representative at the regional level for the most populous regions in the country while Waves 2 and 3 expanded to include 1,500 households in urban areas. The 2018/19 round represented a total refresh of the panel, making it nationally representative as well as representative of each of the 11 regions in Ethiopia. The 2020/21 round followed up with the households from 2018/19, creating a new panel distinct from the three-round panel that ran from 2011 to 2016. Because of COVID and civil conflict in Ethiopia during the 2021/22 data collection effort, this fifth wave contains no households from Tigray and is thus not nationally representative, though it remains representative of ten of the regions. After data cleaning to remove urban and non-agricultural rural households, we are left with 10,674 observations from 5,333 distinct household across five survey waves.

In Malawi, the LSMS-ISA data includes two separate surveys that form three different data sets: the cross-sectional Integrated Household Survey (IHS), and the longitudinal Integrated Household Panel Survey (IHPS) \citepalias{MWI1, MWI2, MWI4, MWI3, MWI5}. These two surveys form three data sets: three cross sections (2010/11, 2016/17, 2019/20), two waves of panel data known as the short-term panel (2010 and 2013), and five waves of panel data known as the long-panel (2010, 2013, 2016, and 2019). This analysis relies on the short-term and long-term data sets from the IHPS, which are representative at the national-, urban/rural-, and regional-level. The long-term panel tracks households that move or split off, meaning the sample size grows from round-to-round. After data cleaning to remove tracked and non-agricultural households, we are left with 8,897 observations from 3,833 distinct households across four survey years.

In Niger, we use two waves, the first from 2011 and the second from 2014 \citepalias{NGR1, NGR2}. Unlike the other countries from the LSMS-ISA used in this paper, there has been no follow-up on data collection in Niger since 2014. The sample is representative at the national and urban/rural-level. Data cleaning and removal of non-agricultural households gives us 3,913 observations from 2,320 distinct households across two survey waves.

In Nigeria, we use the data from the 2010/11, 2012/13, 2015/16, and 2018/19 rounds of the General Household Survey - Panel, which is representative at the national and urban/rural-level \citepalias{NGA1, NGA2, NGA3, NGA4}. Similar to in Ethiopia, there was a panel refresh in 2018/19 that started a new panel. Unlike in Ethiopia, in Nigeria the refresh was partial, with the survey continuing to track 1,425 of the original households. Because there is not yet a second round of data for the refreshed households, we exclude them from the analysis Data cleaning and removal of non-agricultural households yields 9,145 observations from 3,412 distinct households across four survey waves.

In Tanzania, the data come from the 2008/09, 2010/11, 2012/13, 2014/15, 2019/20, and 2020/21 rounds of the Tanzania National Panel Survey (TZNPS) \citepalias{TZA1, TZA2, TZA3, TZA4, TZA5, TZA6}. The sample is representative for the nation, and provides estimates of key socioeconomic variables for mainland rural areas, Dar es Salaam, other mainland urban areas, and Zanzibar. As in Nigeria, thr fourth wave (2014/15) was a partial refresh, with data being collected from a sub-sample of the original panel households (known as the extended panel) plus the addition of completely new households (known as the refresh panel). The extended panel households were re-interviewed in 2019/20 in what is known as the Sex-Disaggregated Data, which added improved individual-level data. The refresh panel was followed up with in 2020/21. The 2020/21 survey round also included a booster sample, that will continue to be followed in subsequent rounds. However, since at this time there is only one round of observations for the booster sample, we exclude them from the analysis. Focusing on rural, crop producing households we have 9,916 observations from 4,804 distinct households across five survey waves.

In Uganda, we use the data from the 2009/10, 2010/11, 2011/12, 2013/14, 2015/16, and 2019/20 rounds of the Uganda National Panel Survey (UNPS) \citepalias{UGA1, UGA2, UGA3, UGA4, UGA5, UGA7}. The original households were followed for three waves and then starting in 2013/14, a subset of households were cycled out and replaced with new households. This continuous, partial refresh process means that households remain in the sample for different lengths of time, though by 2019/2020 all of the original households have been replaced. The original sample, as well as the refresh, was designed to be representative at the national-, urban/rural- and regional-level. Focusing on rural, crop producing households we have 11,692 observations from 4,003 distinct households across six survey waves.

For the analysis, we combine data from the six countries and all waves to generate a single cross-country panel dataset which includes 54,237 observations from 23,705 distinct households. For estimation, we include two measures of agricultural productivity: yield (kg/ha) of the primary cereal crop and the value (2015 USD/ha) of all seasonal crop productivity on the farm. Summary statistics for these outputs as well as measured input are in Table~\ref{tab:sumstattab} in the Appendix.


\subsection{Descriptive Statistics} \label{sec:summarystats}

Our pre-analysis plan specifies that we will examine 22 different ways to measure precipitation and temperature. These variables and definitions are presented in Table~\ref{tab:Wvar} in the Appendix Section~\ref{sec:app_varweather}. For parsimony, we focus all of our analysis in the paper on only four of these 22 variables: (1) total seasonal rainfall, (2) number of days without rain, (3) mean seasonal temperature, and (4) growing degree days (GDD). These four variables are indicative of a number of different ways to measure precipitation and temperature (volume/degree v. count). We spend what may be more space than is typical in an economics paper discussing the descriptive statistics around these measures, but this is essential for understanding they way in which each EO product differs in its measurement of rainfall and temperature. In summarizing the descriptive statistics, we highlight a trends or stylized facts about differences in measurement (a) between EO products within a country and (b) within an EO product between countries.

Figure~\ref{fig:density_rf} presents the distribution of total rainfall (measured in mm) during the growing season, by country and EO product.\footnote{We discuss our calculation of growing season in the Appendix Section~\ref{sec:appRS_gs}.} Not surprisingly, there are substantial differences in the distribution of rainfall across countries. Ethiopia and Uganda average around 650mm of rainfall a season while Malawi, Nigeria, and Tanzania average around 900mm of rainfall a season. By contrast, Niger averages only around 350mm of rainfall in a season. These level differences in the quantity of precipitation in a country are clearly due to geographic characteristics. When we consider all countries together, some general trends emerge with respect to EO sources. A first trend (T1) is that ERA5 tends to have long tails, indicating this reanalysis data product reports dramatically more rainfall than other EO source.\footnote{These long tails are known by \cite{ERA5}. A few times per year, the rainfall can be extremely large in small areas, in an event known as a ``rain bomb''. This can create the long tails we observe on the high rainfall end.} For example, in Ethiopia and Tanzania, ERA5 reports some households receiving 6,000-7,000mm of rainfall in a season - six to seven meters - while no other EO product reports even half that amount. A second trend (T2) is that CPC tends to measure less rainfall relative to other EO products. For example, in Ethiopia and Malawi, CPC reports only half as much rainfall as the other products.

However, there are important exceptions to these trends. In Ethiopia, Malawi, and Uganda, both T1 and T2 hold but in Niger, neither holds. In fact, nearly the opposite is seen: ERA5 measures lower rainfall, with a distribution below those of all other EO products. CPC's distribution is nearly identical to the distribution of the non-ERA5 products. Next, in Nigeria, T2 holds (CPC reports less rainfall than the other products) but T1 does not hold. The largest volume of rainfall reported by ERA5 and is almost the exact same amount as reported by CHIRPS, MERRA-2, and TAMSAT. Finally, in Tanzania, T1 holds (ERA5 reports values more than double any other product) but T2 does not. Here CHIRPS and TAMSAT report less rainfall than the other products and both report zero rainfall for some households, something that does not happen for any other products in any country but Niger. The differences in rainfall reported by each EO product are likely due to a combination of the country's geography (topography, hydrology) and the method each product uses to convert EO signals to estimates of the volume of precipitation.

Figure~\ref{fig:norain_rf} further explores these differences by estimating the mean number of days without rain reported by each EO product for each country. Mean estimates are generated using a fractional-polynomial and graphs include $95\%$ confidence intervals on the mean estimates. We again make note of several trends. One trend (T3) is that ARC2, CHIRPS, and CPC form a cluster, measuring more days without rain, relative to the other EO sources. Another trend (T4) is that the two reanalysis data sets, ERA5 and MERRA-2 form a different cluster, measuring fewer days without rain, relative to other EO sources. A final trend (T5) is that TAMSAT reports substantially fewer days without rain than the first group but substantially more than the second, ending up somewhere in the middle.

Again, there are exceptions to these trends. In Ethiopia, Nigeria, and Tanzania, T3, T4, and T5 hold. While MERRA-2 reports the fewest days without rain in all four countries, the EO product that reports the most days without rain varies from country to country. Additionally, the size of the gap between the products reporting drier conditions and TAMSAT varies, with TAMSAT reporting nearly as little rain in Tanzania as the others while it reports nearly half the number of days without rain relative to the other products in Ethiopia and Nigeria. Turning to Malawi, T4 and T5 (TAMSAT in the middle) hold, though T3 does not. Instead, CPC measures about 20 percent more days without rain than ARC2 or CHIRPS. Next, looking at Niger, T3 and T5 hold. Unlike in other countries, in Niger ERA5 does not agree with MERRA-2 but rather reports very few days without rain, similar to ARC2, CHIRPS, and CPC. Finally, in Uganda, only T4 (ERA5 and MERRA-2 cluster) holds. TAMSAT and CPC report the fewest days without rain while ARC2 and CHIRPS fall in between this high reporting cluster and the low reporting re-analysis cluster.

In Figure~\ref{fig:density_tp} we present the distribution of mean daily seasonal temperature (measured in $^{\circ}$C), by country and EO product. Compared to the distribution of mean daily rainfall, the figures show much tighter distributions around mean temperatures. All EO sources in all countries produce essentially the same results for temperatures. This results is our first temperature trend (T6): there is relatively low variation across EO sources within a country. This is largely due to the relatively low temperature variation in these tropical and sub-tropical countries. While there is some variation country-to-country in terms of which EO product reports the hottest average temperature, these differences are not agronomically meaningful.

Figure~\ref{fig:gdd_tp} estimates the mean GDDs in a year using a fractional-polynomial and includes $95\%$ confidence intervals on the mean estimates. As with number of days without rain, GDD represents a relative coarsening of the data by converting measured temperature into a count variable for the number of days in which temperature fell within a given range. Compared to mean daily temperature, there is less overlap in the values each EO product produces, though relative to the rainfall measures there is again substantially less variation across temperature products. This gives us our final trend (T7): there is relatively low variation in GDD across EO sources. 

Examining each country individually, we see T7 most clearly in Ethiopia and Uganda where there is overlap in the confidence intervals for all three products across almost all years. Nigeria also adheres to T7 in that several products overlap and all have downward trajectories in the number of GDD per year. By contrast, Malawi, Niger, and Tanzania all deviate from T7, though each deviates in its own way. In Malawi, there is effectively no overlapping in the confidence intervals of the three products and while CPC and ERA5 have flat slopes, MERRA-2 produces a strongly divergent downward trajectory. Niger is similar to Malawi in that there is almost no overlap in confidence intervals and MERRA-2 has a strong downward trajectory. However, in Niger, CPC and ERA5 have strong upward trajectories. Further, in Malawi MERRA-2 produces the fewest GDDs while in Niger it produces the most. Finally, in Tanzania, there is almost perfect agreement between ERA5 and MERRA-2, both of which show no change in GDD over time. By contrast, CPC is in almost complete disagreement, producing fewer and fewer GDDs over time.

Summarizing the descriptive evidence: there can be substantial differences in the weather reported by each EO product. Again, this is not surprising since each product uses different data and different methods to produce gridded weather data. If all that these descriptive figures showed was one EO product consistently reporting higher levels of rainfall and temperature than other products, researchers could easily adjust their methods to account for this over reporting. However, Figures~\ref{fig:density_rf} through~\ref{fig:gdd_tp} document not just differences in cardinality (one product reports a few more mm of rain) but in ordinality. In most countries ERA5 reports the most rain but in Niger it reports the least. In most countries TAMSAT falls in the middle in terms of how many days are without rain, but in Uganda it reports the driest conditions. And while compared to rainfall, there is much more agreement among the EO temperature products, there are still significant deviations, particularly in the number of GDD produced by the products. In Malawi, MERRA-2 produces the fewest GDD but in Niger it produces the most. In most countries, CPC is in agreement with ERA5 but in Tanzania CPC disagrees with both ERA5 and MERRA-2. Based on these changes in the ordering of EO products across countries, we conclude that researchers' selection of EO product cannot be country-agnostic.


\section{Analysis Plan}  \label{sec:analysisplan}

A common feature of research on measurement error is that the researchers have a measure of objective truth. They can trace the boundaries of a plot of land \citep{AbayEtAl19}, they can weigh the quantity of harvest \citep{Lobelletal19}, they can determine the exact location of a household \citep{MichlerEtAl22}. In our setting, we lack an objective measure of the truth regarding what the weather was like for a given household in a given year. And, given the paucity of gauge station data on the Continent, the typical gold standard for weather data does not exist. This lack of an objective measure for quantifying mismeasure in EO weather data informs how we implemented our research design. First, we developed a pre-analysis plan and registered it at Open Science Framework \citep{PAP}. While pre-analysis plans have become common in experimental economics, they are still relatively uncommon for binding researchers' hands when using observational data \citep{JanzenMichler21}. The use of a pre-analysis plan allowed us to pre-define the sources of data for inclusion in the study, what metrics would be tested using what functional forms, and how we would compare results across models in the absence of formal statistical tests. Second, we adopted a blinding strategy to help ensure objectivity in the implementation of the pre-analysis plan. As such, the authors were divided into two groups: the Data Generating Group and the Data Analysis Group. Authors Kilic and Murray were in the Data Generating Group and had full responsibility for extracting the EO data and matching it to the household records in the household survey data to create a number of different paired EO-survey datasets. In these datasets, the source of the EO data was anonymized prior to sharing with the Data Analysis Group. Authors Josephson and Michler made up the Data Analysis Group and had full responsibility for cleaning the agricultural productivity data, running the regressions, and conducting and writing the analysis. The pre-specified analysis was carried out on the blinded datasets and these results were posted to \href{arXiv.org}{arXiv.org} prior to unblinding \citep{MichlerEtAl20}. The generation of datasets in this manner preserves the objectivity of any findings regarding differences in outcomes between different EO products. 

The following analysis, and the associated results, were all pre-specified. If methods, approaches, or inference criteria differ from our plan, we highlight these differences. Results arising from these deviations in our plan should be interpreted as exploratory.


\subsection{Estimation}\label{sec:eststart}

We follow the basic model of \cite{DescheneGreenstone07}:

\begin{equation}
Y_{ht} = \alpha_{h} + \gamma_{t} + \sum_{j}^{J} \beta_{j} f_{j} \left( W_{jht} \right) + u_{ht}
\end{equation}

\noindent where $Y_{ht}$ is our outcome variables from the LSMS-ISA-supported household surveys, described above, for household $h$ in year $t$, log transformed using the inverse hyperbolic sine. We control for year fixed-effects $(\gamma_{t})$ and include household fixed-effects $(\alpha_{h})$ in some specifications. The function $f_{j} \left( W_{jht} \right)$ represents our weather variables of interest where $j$ represents a particular measurement of weather. Finally, $u_{ht}$ is an idiosyncratic error term clustered at the household-level.

From this general set-up, we estimate three versions of the model:\footnote{In our pre-analysis plan we defined three additional models that include weather and its square to allow for nonlinearities in the way weather is related to crop productivity. Our results do not change when we use the quadratic specifications, so in this paper we focus solely on the linear version of the model. All results are available in the populated pre-analysis plan through the World Bank \citep{MichlerEtAl21WB}.}

\begin{subequations}
\begin{align}
Y_{ht} &= \alpha + \beta_{1} W_{ht} + u_{ht} \label{eq:linear} \\
Y_{ht} &= \alpha_{h} + \gamma_{t} +  \beta_{1} W_{ht} + u_{ht} \label{eq:linearFE} \\
Y_{ht} &= \alpha_{h} + \gamma_{t} +  X_{ht} \pi + \beta_{1} W_{ht} + u_{ht}  \label{eq:linearCOV}
\end{align}
\end{subequations}

\noindent  Model~\eqref{eq:linear} examines the simple correlation between a weather metric $(W_{ht})$ and crop productivity $(Y_{ht})$, measured as either yield of the main crop or the total value of on-farm production per hectare. This model, which lacks any control variables, implicitly treats the data as if it was a pooled cross section. We refer to this model as the ``weather only'' model in our discussion. Model~\eqref{eq:linearFE} accounts for the fact that we have multiple observations from a household over time by including household fixed effects $(\alpha_{h})$. This model also includes year fixed effects $(\gamma_{t})$ that control for country-wide changes, such as civil conflict, the pandemic, or changes in government, that vary over time. We refer to this model as the ``weather with fixed effects'' model. Model~\eqref{eq:linearCOV} adds measured inputs as covariates $(X_{ht})$. These include quantity of fertilizer per hectare, the amount of labor days per hectare, and indicators for if the household applied pesticide, herbicide, or irrigation. We refer to this model as the ``weather with fixed effects and inputs'' model. 

All of the regression models are estimated for each permutation of the data. This is a substantial number of regressions, given the number of models (three), countries (six), variables defined (14 rainfall, eight temperature variables), EO products (six rainfall, three temperature), and the number of outcomes (two). This gives us a total of 3,888 regressions: for each of our three models and two outcomes on the 648 different versions of the data. By varying specifications and data we define a robust set of outcomes regarding mismeasurement in EO data absent objective measure of rainfall and temperature.


\subsection{Inference}

In the majority of economics research, empirical results are presented in tables, with coefficient estimates and statistics for inference. In our case, because of the large number of regressions estimated, standard modes of inference and traditional presentations of results are not appropriate. Instead, we rely on a series of heuristic methods and criteria to make inference, evaluate the results, and present our findings.

In discussing our method for inference and interpretation of results, it is useful to reflect on the characteristics necessary for our purposes. First, a heuristic should be agnostic about the sign of a coefficient, as some weather metrics that we test are likely to be positively correlated with outcomes (e.g., mean rainfall) while others are likely to be negatively correlated (e.g., days without rain). While we are agnostic about sign generally, we are interested in the ordering of coefficients and so, do not completely disregard this dimension in our heuristics. Second, a heuristic should be able to determine whether or not a weather metric is significantly correlated with outcomes. Our prior is that weather is significantly correlated with outcomes, regardless of direction. This maintained assumption is based on the frequency with which weather is used in the economics literature to predict all sorts of outcomes, from crop production to migration to economic growth. Thus, identification of statistical significance is an important dimension to be included in our heuristics. Finally, a heuristic should be able to measure the amount of unexplained variance in the model after controlling for weather. This is also line with our prior, we expect weather to reduce the amount of unexplained variance in a model, all else being equal. 

The full set of heuristics developed in our pre-analysis plan and corresponding results are presented in \cite{MichlerEtAl21WB}.\footnote{The three metrics are (1) mean log likelihood values, (2) share of coefficient $p$-values significant at standard levels ($0.01$, $0.05$, and $0.10$), and (3) coefficient size with $95\%$ confidence intervals.} Because of our interest in the order of coefficients, the most relevant heuristics for this analysis is the first, related to directionality and ordering. To this end, we use specification charts to examine the actual regression coefficients and estimated confidence intervals for a subset of regressions.

An important caveat for consideration with respect to our results: the significance of a point estimate does not necessarily imply that the model is correctly specified, that the point estimate is agronomically meaningful, or, in particular, that the point estimate has the correct sign. These results and the associated figures simply allow us to visualize the variability in the number of significant coefficients, their magnitude, and ordering across specifications. Variability within these domains may suggest that measurement error exists.


\section{Results} \label{sec:ext}

As with the descriptive statistics, we focus our results on four variables, which are representative of the overall trends across variables. They are (1) total seasonal rainfall, (2) days without rain, (3) mean seasonal temperature, and (4) growing degree days. These four variables present a series of different ways to measure precipitation and temperature (e.g., volume/degree v. count) and allow us to focus our discussion on the differences in measurement (a) between EO products within a country and (b) within an EO product between countries.

\subsection{Coefficient Size}

Following our described heuristics, we present our results in specification charts. We include visual representations of both point estimates and $95\%$ confidence intervals for single regressions, with 36 regressions per rainfall metric per country and 18 regressions per temperature metric per country. The markers in the space below the coefficients indicate which specific version of a regression the coefficient comes from. The gray circles at the bottom indicate the econometric specification (weather only, weather with FE, and weather with fixed effects and inputs). The gray squares indicate the dependent variable (either total farm value or the quantity of maize harvest). The diamonds indicate which EO product the weather data comes from as well as if the sign and significance of the coefficient. Gray diamonds indicate not significant, blue diamonds indicate positive and significant, and red diamonds indicate negative and significant. Each chart is divided into three regions which correspond to the econometric specification that the results come from. This is because we expect the magnitude of the coefficients to differ depending on the inclusion or exclusion of controls. Within specification, results are sorted by coefficient size.

In understanding and interpreting the specification charts, one wants to look for patterns in the data. If differences in the values of rainfall and temperature that different products reported were simply affine transformations of each other, then we should see a consistent pattern in the location of the diamonds (though not in their color). Said another way, if all that the differences in measurement produced was a change in cardinality, then the size and significance of coefficients would change (their color) but not the ordering of coefficients (the location of the diamonds). If instead the differences in EO products are such that it leads to changes in ordinality, then we should see no consistent pattern in the location of the diamonds. At one extreme, in which results are wholly robust to the choice of EO product, then the pattern in the location of the diamonds should repeat in each section of the chart and across charts. At the other extreme, in which results lack any robustness to the choice of EO product, then the diamonds should look like a cloud of noise.

Figures~\ref{fig:pval_v5} and \ref{fig:pval_v10} present specification charts for mean daily rainfall and number of days without rain, by country. It is immediately obvious that the location of the diamonds looks more like a cloud of noise than a repeated pattern both across specifications within a country and across countries. Consider, as an example, Ethiopia in Figure~\ref{fig:pval_v5}: in the specification in which only weather variables are included, all coefficients are positive and significant with ARC2 producing small but positive and significant results and MERRA-2 reporting large, positive and significant results. But, when fixed effects are included (with and without inputs) the pattern changes. It is not just that all coefficients get larger or smaller. Rather the ordering changes. ARC2 now produces the largest coefficients, and the only that remain positive and significant, while MERRA-2 produces some of the smallest, which tend to be negative and significant. Similar re-orderings occur in each country and for days without rain, though not all are as stark as mean daily rainfall in Ethiopia.

Next, we turn to temperature in Figures~\ref{fig:pval_v15} and \ref{fig:pval_v19}. As with rainfall, results differ by EO product. And again, as with rainfall, coefficient ordering shifts across models. This is more evident with the smaller number of EO sources used in the temperature regressions. Most frequently, countries follow a trend of one set of ordering for the weather only model specification, which shifts to a different ordering for the models which include fixed effects. Consider, as an example, Niger in Figure~\ref{fig:pval_v15}: in the first panel ERA5 produces large negative and significant coefficients while MERRA-2 produces small coefficients that are either negative and significant or not significant. In the models with FE, MERRA-2 produces the largest negative coefficient while ERA5 produces the largest coefficient, now positive and significant. While we see similar reorderings do appear in other countries for other temperature metrics, the most striking element of the temperature results are that we see strong patterns in the specification charts for a couple countries, specifically Nigeria and Uganda. As we saw in the descriptive statistics, there is less variation in the temperatures that the different EO products report relative to the temperature products and thus less changes to the ordinality of coefficients.

Before moving on the a closer examination of changes in ordinality in regression results, two other regularities in the specification charts bear commenting upon. First, for both rainfall and temperature, regardless of the EO product or the weather metric, weather becomes much less important in predicting agricultural productivity once household FEs are included. Over and over again, regressing output on just rainfall or temperature shows a significant relationship between the two. But once household FEs are included to control for time-invariant unobservables, the relationship goes away. Weather is a time varying phenomenon, often considered exogenous or as good as randomly determined. It is not immediately obvious why the inclusion of a time-invariant control destroys the relationship between weather and agricultural production. A likely reason, which we leave unexplored, is that weather in a given location does not vary that much over time. Thus, including household FEs, assuming a household does not move, is essentially a location FE. Once location is controlled for, weather varies very little year-to-year and small deviations from the time average in a location has little to no impact on agricultural production in the year. This raises questions about the use of weather as an instrument (as it is correlated with time-invariant household heterogeneity, specifically location) as well as the ability of economists to use weather to predict the effects of climate change (by definition, a slow process) on economic indicators.

A second regularity is that the sign on the weather coefficients can vary substantially. This is related to, but distinct from the ordinality issue. Within a single econometric specifications, rainfall from one EO product will be positively correlated with agricultural production while in rainfall from another product will be negatively correlated. Shifts in significance across specification can be the result of changes in relative size (cardinality). But was should raise concerns among applied economists is that in every country a research could produce whatever relationship they wanted between weather and the dependent variable by judiciously choosing the EO product, the weather metric, and the econometric specification. Essentially, a researcher using EO weather data has almost unlimited degrees of freedom to $p$-hack their way to his or her desired results.

\subsection{Coefficient Ordering}\label{sec:ordering}

While the specification charts summarize a large amount of information in a compact form, it can be difficult to detect the extent to which the diamonds indicating EO product form a pattern or are patternless. To dig further into coefficient ordering, we present a series of figures which represent the order (ranking) of coefficients within a country, across models, by EO source using bumpline functions \citep{bumpline}. These are presented for total seasonal rainfall in Figure~\ref{fig:bump_train} and mean daily temperature in Figure~\ref{fig:bump_meant}. Results for days without rain and GDD are functionally similar and so we present figures for those variables in the Appendix. Within each country panel, the x-axis (each column) presents a different outcome variable (value or quantity) and a different econometric specification (weather only, weather with FE, and weather with fixed effects and inputs). The y-axis (each row) represents the ranking of the coefficient size for each EO source within each regression. They are ordered from one to six, with one representing the largest coefficient and six representing the smallest coefficient. These results give greater insight to the changes in ordinality across specifications and across countries.

We first consider the findings from total seasonal rainfall (see Figure~\ref{fig:bump_train}). Going country by country: in Ethiopia, we see that in the weather only specification, CPC and TAMSAT have large coefficients while ERA5 has the smallest. This ordering changes once FE are included, with ERA5 producing the largest coefficients and TAMSAT and CPC producing the smallest. In Malawi, the same trend does not hold. Rather, there appear to be trends based on the dependent variable, with ARC2 and TAMSAT producing large coefficients when the value of total farm production is the dependent variable and small coefficients when the dependent variable is quantity of maize yield. Thus, what drives heterogeneity in results in Ethiopia differs in Malawi. Results in Niger appear almost chaotic, with order of EO products jumping around with any change in specification of dependent variable. It is difficult to identify any trends other than a lack of trends. Next, we consider Nigeria, which does not demonstrate a specification- or dependent variable-based pattern, but generally has ARC2, CPC, and ERA5 ordered in the last three places with MERRA-2, CHIRPS, and TAMSAT in the top three places. Compared to the previous countries, the ordering of EO products in Nigeria appears robust, suggesting one would get similar results regardless of which EO product one chose. This can be confirmed by referencing the specification charts in Figure~\ref{fig:pval_v5}. A similar result holds in Tanzania, where we observe that ARC2 generally falls into the fourth position and TAMSAT into the sixth. Finally, we consider Uganda which somewhat resembles Ethiopia. In the weather only specifications MERRA-2 and ERA5 produce the largest coefficients while ARC2 and TAMSAT produce the smallest. This order reverses when FEs are included. Overall, based on the findings for total seasonal rainfall, we can draw some general trends about ordering or re-ordering of coefficients based on EO products within countries, but we cannot draw trends across countries. Which EO products produce the largest or smallest coefficients varies from country to country suggesting the how well an EO product captures the truth about weather depends on the geography and climate of where it is looking. 

Moving to mean daily temperature (see Figure~\ref{fig:bump_meant}), the figures are presented with the same column and row structure, but instead of having six ranking places, there are only three. This somewhat simplifies the view. Beginning with Ethiopia, in the weather only specification, ERA5 produces the largest coefficients while CPC and MERRA-2 which places for the smallest. When we include FEs, CPC and MERRA continue to switch places based on the dependant variable but ERA5 now always produces the smallest coefficients. Next, in Malawi, we see a clear and consistent ordering of outcomes: MERRA-2 always produces the largest coefficients while CPC and ERA5 switch regarding which one produces the smallest based on the dependant variable. Like with Nigeria and rainfall, one get similar results in Malawi regardless of what EO product was selected. Turning to Niger, MERRA-2 produces largest coefficients in the weather only specifications while ERA5 produces the smallest. This order changes once FEs are added to the specification. We next consider Nigeria, which does not present a consistent pattern in the weather only specification but produces a consistent ordering of EO products once FEs are added. For Tanzania, a similar but less consistent pattern exists. There is very little stability in ordering of coefficient size in the weather only specification but a fair amount of consistency in the specifications that include FEs, with MERRA-2 always producing the smallest coefficients. Finally, in Uganda, there exists a pattern very similar to Malawi. CPC always produces the largest coefficients while MERRA-2 and ERA5 swap places for the smallest coefficient. Overall, while there is must fewer changes in ordinality for EO temperature data compared to rainfall data within a country, there again is little consistency across country. In Ethiopia and Uganda, CPC tends to produce the largest coefficients while in Malawi and Nigeria it is MERRA-2 and in Niger and Tanzania it is ERA5. But ERA5 produces the smallest coefficients in Ethiopia, Malawi, and Uganda, while MERRA-2 produces the smallest coefficients Niger and Tanzania. The EO product that produces the largest or smallest coefficients varies from country-to-country and again suggests that the accuracy of these products is a function of geography and climate. Not all products perform equally well across space. This means what results a researcher gets will depend on their choice of EO product as well as their country of study.

Taken together, the preponderance of evidence from our regressions leads us to conclude that our results are not robust to the choice of remote sensing EO weather product. The correlation between rainfall or temperature and agricultural productivity varies not only across countries but within a country based on the econometric specification or how productivity is quantified. Our findings show that different measurement methods from different EO sources result in different ordering of coefficients and not just a change in the relative size of the coefficients.


\section{Conclusions and Recommendations for Researchers} \label{sec:best}

The use of remotely sensed EO weather data in economic contexts has become commonplace. Researchers use a variety of EO products to produce a host of different weather metrics to predict numerous different economic outcomes. If each EO product reported the same values for rainfall and temperature for a given location for a given time on a given day, then the choice of EO product would be immaterial. Unfortunately, this is not the case. Each product, while attempting to measure the objective truth of the weather, produces its own ``truth.'' The findings produced are not immutable truths. 

Using a LSMS-ISA data from six African countries in combination with nine EO weather data products, we demonstrate the lack of consistency in how EO weather products report on the weather. Importantly, the differences across products does not simply affect the cardinality, or relative size, of coefficients from different EO products. The shifts we see cannot be ascribed to an affine transformation of one EO product by another. Rather, what we document are changes in ordinality, the ranking of coefficients by size. In some countries one EO product may produce large, positive and significant coefficients while another EO product produces negative and significant coefficients. This order can swap if one moves to another country or if, within a give country, one changes the econometric specification or how the agricultural productivity is quantified. The unwary researcher may end of with results and draw conclusions that are a function of their choice of EO product and could have ended up with completely different results and conclusions if they had chosen a different EO product. Knowing this, the unscrupulous researcher can select the EO product that measures the ``truth'' in the way that produces the results and conclusion they want. 

Given the lack of gold standard gauge station data in many countries, where does this disparity in how EO products measure the weather leave researchers who want to incorporate this type of data into economic contexts? As we cannot know the truth of what weather occurred where, we are left in a second-best world where more work may be required to demonstrate the accuracy of our scientific findings. Based on the results in this paper, we recommend that researchers carefully choose which remotely sensed EO source to use in their analysis. More justification needs to be given in papers for the choice of one EO data set over another. In EO product documentation there is often a list of known issues, such as over-estimation of rainfall events or difficulty in capturing precipitation in rugged terrain. In the same way economists already document and justify choices in terms of variable selection, we also need to document and justify our selection of EO product. Additionally, researchers may need to demonstrate that key results are not a function of their choice of EO product. Economists already verify the robustness of results to changes in data and specification. Checks for robustness to alternative EO weather products may also be called for, particularly if weather is a key component of the story or plays a role in causal identification.

Potentially the most important recommendation resulting from our findings is the need to recognize, as Stephen Jay Gould wrote, that science is a social phenomenon, not the collection and production of pure information. And, the results of scientific inquiry should never be interpreted as immutable facts. For years the economics profession has used remotely sensed EO data, weather or otherwise, as if it perfectly reported on the objective facts. We have done this ourselves, having chosen one EO product over another simply because it was easy to download or work with. By keeping Gould's caution about science and scientific findings in the forefront of our minds, we reduce the probability of blindly accepting the truth or accuracy of the data we use. As researchers, we should cultivate skepticism, but we should also cultivate empathy for all of us engaged in this social phenomenon called science.


\newpage
\singlespacing
\bibliographystyle{chicago}
\bibliography{LSMSref}


\newpage 
\FloatBarrier

\begin{landscape}
\begin{table}[htbp]	\centering
    \caption{Sources of Weather Data}  \label{tab:weather}
	\scalebox{0.9}
	{ \setlength{\linewidth}{.1cm}\newcommand{\contents}
		{\begin{tabular}{p{0.65\linewidth} lllll}
            \\[-1.8ex]\hline 
			\hline \\[-1.8ex]
            dataset & Length of record & Resolution & Time step & Data & Units \\
            \midrule
            \multicolumn{6}{l}{\textbf{Precipitation}} \\
            -Africa Rainfall Climatology version 2 (ARC2) & 1983-current & 0.1 deg & daily & total precip & mm \\
            -Climate Hazards group InfraRed Precipitation with Station data (CHIRPS) & 1981-current & 0.05 deg & daily & total precip & mm \\
            -CPC Global Unified Gauge-Based Analysis of Daily Precipitation & 1979-current & 0.5 deg & daily & total precip & mm \\
            -European Centre for Medium-Range Weather Forecasts (ECMWF) ERA5 & 1979-current & 0.28 deg & hourly & total precip & m \\
            -Modern-Era Retrospective analysis for Research and Applications, version 2 (MERRA-2) Surface Flux Diagnostics & 1980-current & 0.625x0.5 deg & hourly & rain rate & kg m$^2$ s$^1$ \\
            -Tropical Applications of Meteorology using SATellite data  and ground-based observations (TAMSAT) & 1983-current & 0.0375 deg & daily & total precip & mm \\
            \midrule
            \multicolumn{6}{l}{\textbf{Temperature}} \\
            -CPC Global Unified Gauge-Based Analysis of Daily Temperature & 1979-current & 0.5 deg & daily & min, max temp & C \\
            -European Centre for Medium-Range Weather Forecasts (ECMWF) ERA5 & 1979-current & 0.28 deg & hourly & mean temp & K \\
            -Modern-Era Retrospective analysis for Research and Applications, version 2 (MERRA-2) statD & 1980-current & 0.625x0.5 deg & daily & mean temp & K \\
			\\[-1.8ex]\hline 
			\hline \\[-1.8ex]
    		\multicolumn{6}{l}{\footnotesize \textit{Note}: The table summarizes the Earth observation sources and related details for precipitation and temperature data.}
    	\end{tabular}}
	\setbox0=\hbox{\input{tables/summary_stats}}
    \setlength{\linewidth}{\wd0-2\tabcolsep-.25em}
    \input{tables/summary_stats}}
\end{table}
\end{landscape}


\begin{table}[htbp]	\centering
    \caption{Sources of Household Data}  \label{tab:lsms}
	\scalebox{0.9}
	{ \setlength{\linewidth}{.1cm}\newcommand{\contents}
		{\begin{tabular}{llrrr}
            \\[-1.8ex]\hline 
			\hline \\[-1.8ex]
            Country & Survey Name & Years & Original $n$ & Final $n$  \\
            \midrule
            Ethiopia & Ethiopia Socioeconomic Survey (ESS) & 2011/2012 & 3,969 & 1,689 \\
                        & & 2013/2014 & 5,262 & 2,865 \\
                        & & 2015/2016 & 4,954 & 2,718 \\
                        & & 2018/2019 & 7,527 & 1,996 \\
                        & & 2021/2022 & 4,999 & 1,406 \\
            Malawi & Integrated Household Panel Survey (IHPS)  & 2010/2011 & 3,247 & 2,250 \\
                        & & 2013 & 4,000 & 2,472 \\
                        & & 2016 & 2,508 & 1,845 \\
                        & & 2019 & 3,178 & 2,330 \\
            Niger & Enqu\^{e}te Nationale sur les Conditions de Vie des & 2011 & 3,968 & 2,223 \\
                        & $\:$ M\'{e}nages et l'Agriculture (ECVMA) & 2014 & 3,617 & 1,690 \\
            Nigeria & General Household Survey (GHS) & 2010/2011 & 4,916 & 2,674 \\
                        & & 2012/2013 & 4,716 & 2,768 \\
                        & & 2015/2016 & 4,581 & 2,783 \\
                        & & 2018/2019 & 4,976 & 920 \\
            Tanzania & Tanzania National Panel Survey (TZNPS) & 2008/2009 & 3,280 & 2,001 \\
                        & & 2010/2011 & 3,924 & 2,013 \\
                        & & 2012/2013 & 5,015 & 1,889 \\
                        & & 2014/2015 & 3,352 & 2,127 \\
                        & & 2019/2020 & 1,184 & 312 \\
                        & & 2020/2021 & 4,709 & 1,564 \\
            Uganda & Uganda National Panel Survey (UNPS) & 2009/2010 & 2,975 & 1,883 \\
                        & & 2010/2011 & 2,716 & 1,886 \\
                        & & 2011/2012 & 2,850 & 2,020 \\
                        & & 2013/2014 & 3,119 & 2,190 \\
                        & & 2015/2016 & 3,305 & 1,868 \\
                        & & 2019/2020 & 3,098 & 1,845 \\
            \midrule
            Total & 6 countries & 27 waves & 105,945 & 54,237 \\
			\\[-1.8ex]\hline 
			\hline \\[-1.8ex]
    		\multicolumn{5}{l}{\footnotesize \textit{Note}: The table summarizes the household data details for each country, per LSMS Basic Information Documents.}
    	\end{tabular}}
	\setbox0=\hbox{\input{tables/summary_stats}}
    \setlength{\linewidth}{\wd0-2\tabcolsep-.25em}
    \input{tables/summary_stats}}
\end{table}


\newpage 
\FloatBarrier 

\begin{figure}[!htbp]
	\begin{minipage}{\linewidth}
		
		\caption{Varying Resolution of Rainfall Measurement}
    	\label{fig:rain_res}
		\begin{center}
			\includegraphics[width=.9\linewidth,keepaspectratio]{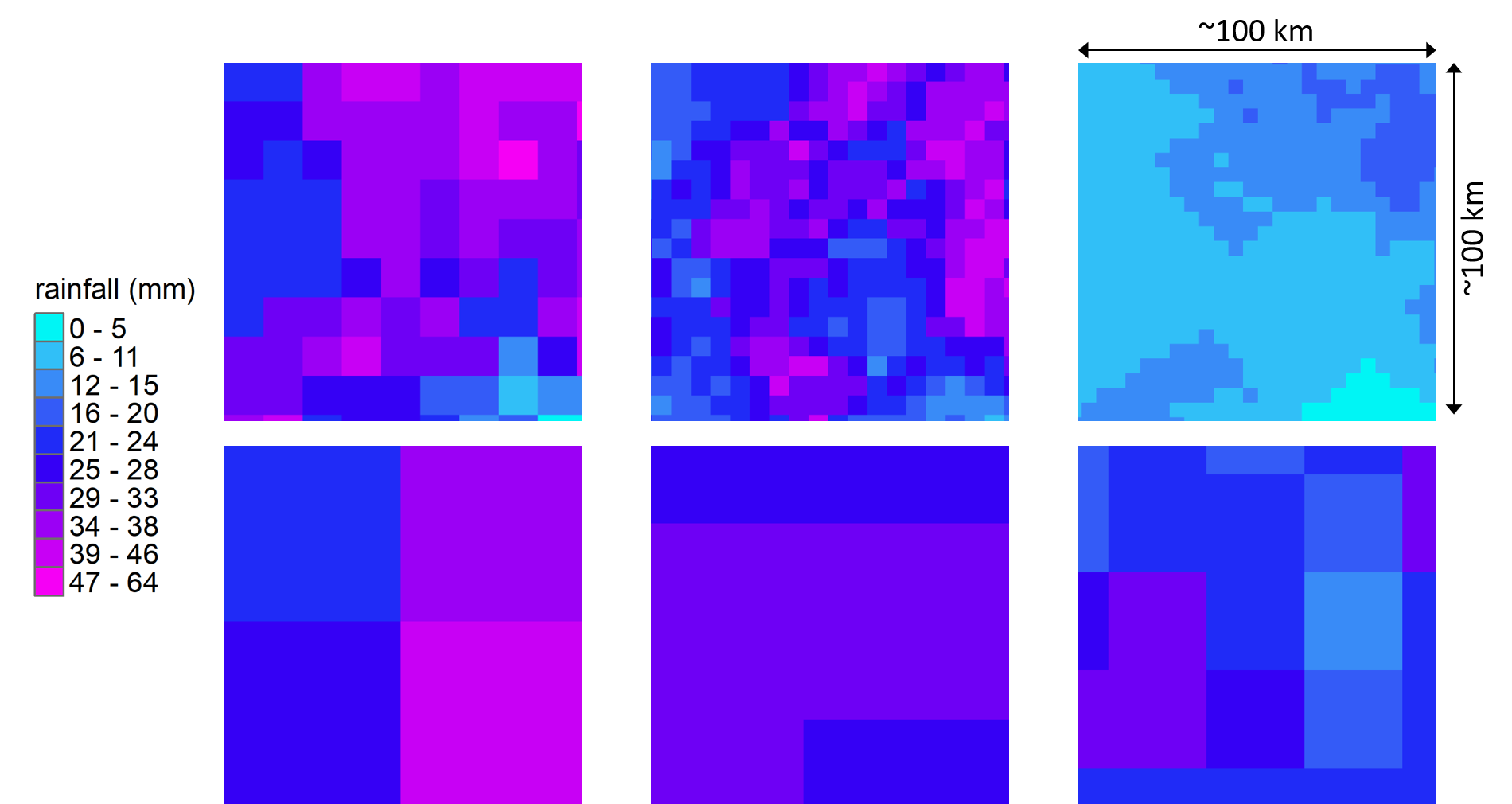}
		\end{center}
		\footnotesize  \textit{Note}: The figure captures rainfall as measured by all six precipitation products for the same 100km x 100km area on a single day (7 January 2010).

	    \vspace{2cm}
	
		\caption{Varying Resolution of Temperature Measurement}
    	\label{fig:temp_res}
		\begin{center}
			\includegraphics[width=.9\linewidth,keepaspectratio]{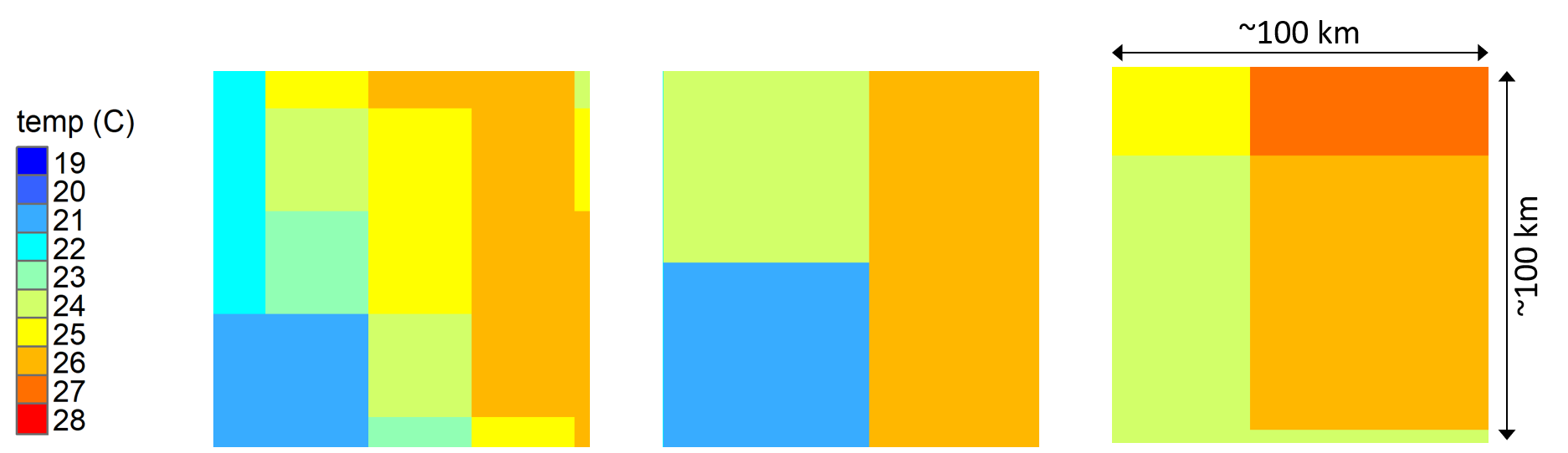}
		\end{center}
		\footnotesize  \textit{Note}: The figure captures temperature as measured by all three temperature products for the same 100km x 100km area on a single day (7 January 2010).
	\end{minipage}
\end{figure}


\newpage 
\FloatBarrier 

\begin{figure}[!htbp]
	\begin{minipage}{\linewidth}	
		\caption{In-Country Weather Stations}
    	\label{fig:stations}
		\begin{center}
			\includegraphics[width=.8\linewidth,keepaspectratio]{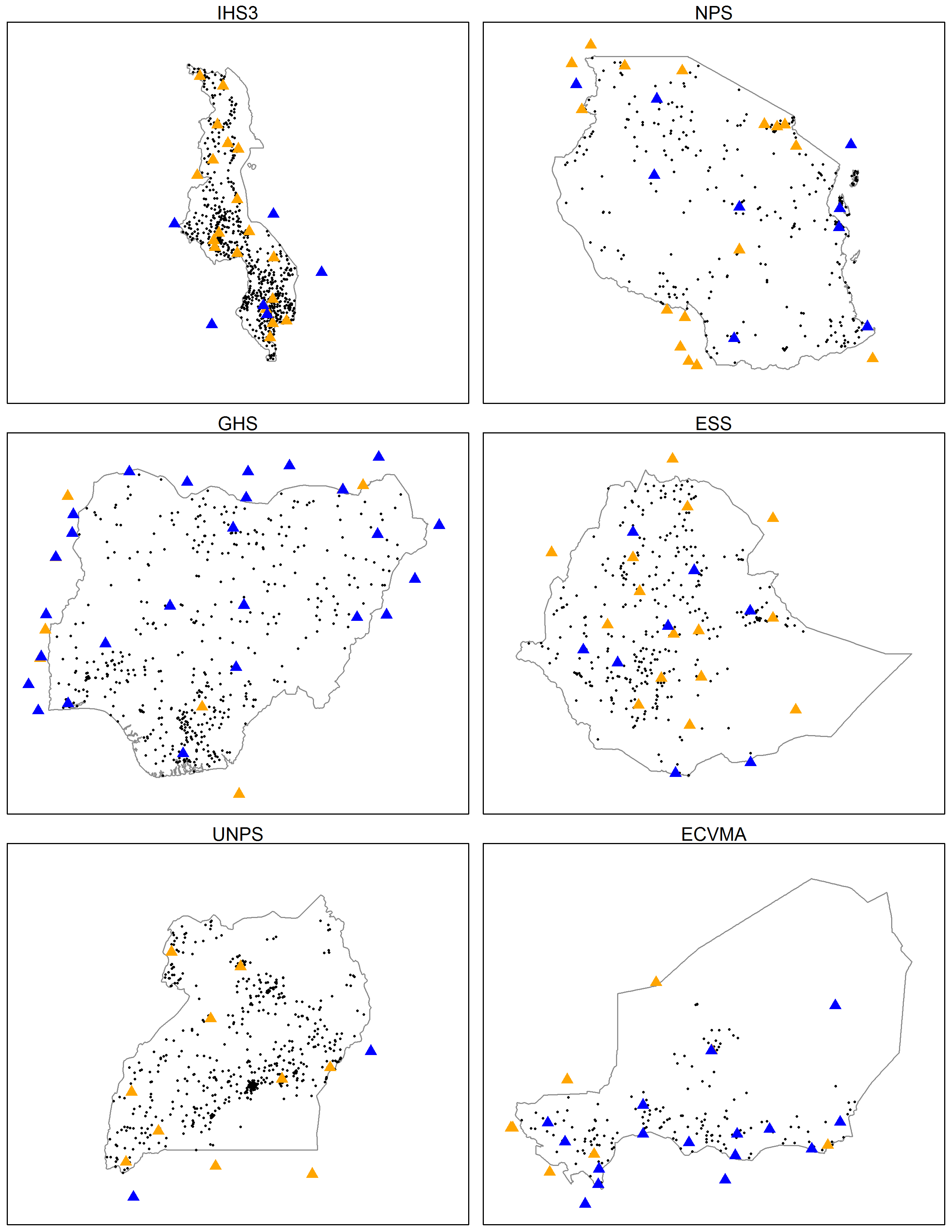}
		\end{center}
		\footnotesize  \textit{Note}: Maps display LSMS-ISA household locations along with locations of  Global Historical Climatology Network (GHCN) weather stations. Small black dots represent the approximate location of all surveyed communities (anonymized public coordinates). Blue and organge triages represent the GHCN stations located within or near the country border, where blue symbols are stations with at lesat some rainfall or temperature data reported during the time period of interest. Note that for some households the closest station is outside the national boarders. Blue dots with green halos are the stations that continue to report data to the GHCN network. In the case of Uganda, no in-country stations currently report to the network, meaning the closest weather station data for LSMS-ISA households in Uganda are one station in Kenya and one in Tanzania.
	\end{minipage}
\end{figure}


\newpage 
\begin{landscape}
\begin{center}
\begin{figure}[!htbp]
	\begin{minipage}{\linewidth}
		\caption{Distribution of Total Seasonal Rainfall}
		\label{fig:density_rf}
		\begin{center}
			\includegraphics[width=\linewidth,keepaspectratio]{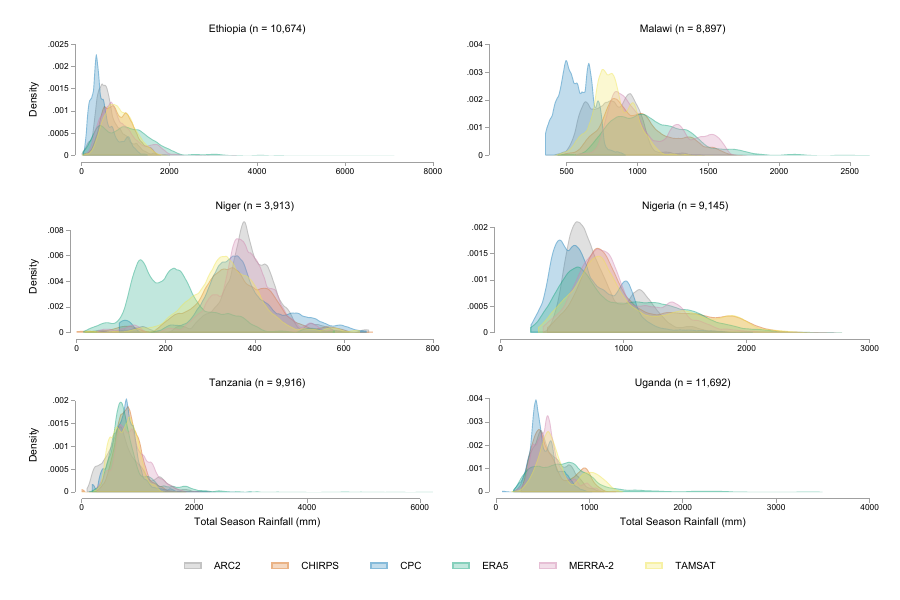}
		\end{center}
		\footnotesize  \textit{Note}: The figure presents the distribution of total seasonal rainfall pooled across all years, disaggregated by Earth observation source. Each distribution (Earth observation product) in each panel is constructed using all household-year observations $(n)$. Variation in distributions do not come from variation in the household data that is paired with the Earth observation data. Rather, variation in distributions within a panel is solely due to differences in the value of precipitation reported by the Earth observation source.
	\end{minipage}	
\end{figure}
\end{center}    
\end{landscape}

\newpage 
\begin{landscape}
\begin{center}
\begin{figure}[!htbp]
	\begin{minipage}{\linewidth}
		\caption{Average Number of Days Without Rain}
		\label{fig:norain_rf}
		\begin{center}
			\includegraphics[width=\linewidth,keepaspectratio]{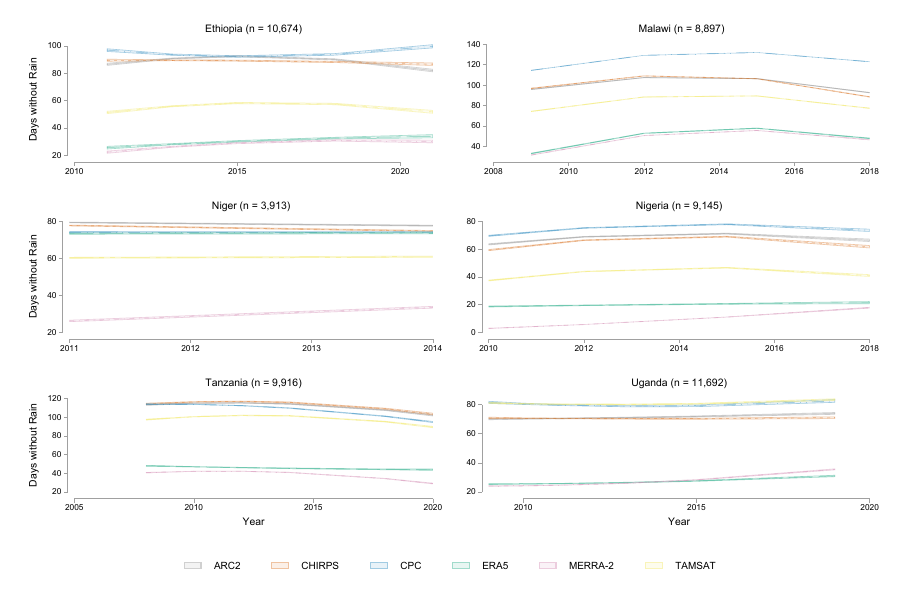}
		\end{center}
		\footnotesize  \textit{Note}: The figure presents the mean number of days without rain $(< 1mm)$ in a year, disaggregated by Earth observation source. Means are calculated by fitting a fractional-polynomial line to the data, with $95\%$ confidence interval represented by shaded area. Each line (Earth observation product) in each panel is constructed using all household-year observations $(n)$. Variation in lines do not come from variation in the household data that is paired with the Earth observation data. Rather, variation in lines within a panel is solely due to differences in the number of days without rain reported by the Earth observation source.
	\end{minipage}	
\end{figure}
\end{center}    
\end{landscape}

\newpage 
\begin{landscape}
\begin{center}
\begin{figure}[!htbp]
	\begin{minipage}{\linewidth}
		\caption{Distribution of Mean Daily Temperature}
		\label{fig:density_tp}
		\begin{center}
			\includegraphics[width=\linewidth,keepaspectratio]{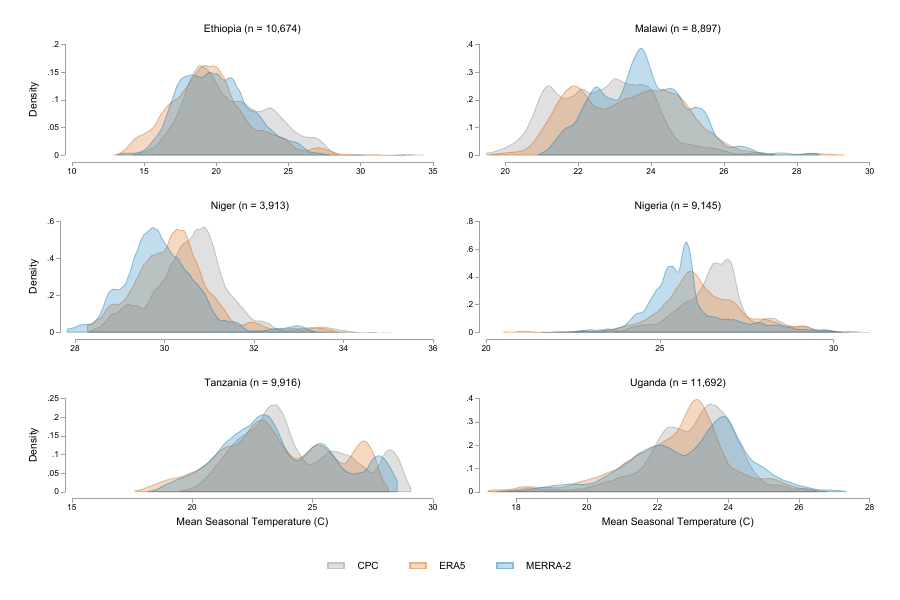}
		\end{center}
		\footnotesize  \textit{Note}: The figure presents the distribution of mean daily temperature pooled across all years, disaggregated by Earth observation source. Each distribution (Earth observation product) in each panel is constructed using all household-year observations $(n)$. Variation in distributions do not come from variation in the household data that is paired with the Earth observation data. Rather, variation in distributions within a panel is solely due to differences in the temperature reported by the Earth observation source.
	\end{minipage}	
\end{figure}
\end{center}    
\end{landscape}

\newpage 
\begin{landscape}
\begin{center}
\begin{figure}[!htbp]
	\begin{minipage}{\linewidth}
		\caption{Average Growing Degree Days (GDD)}
		\label{fig:gdd_tp}
		\begin{center}
			\includegraphics[width=\linewidth,keepaspectratio]{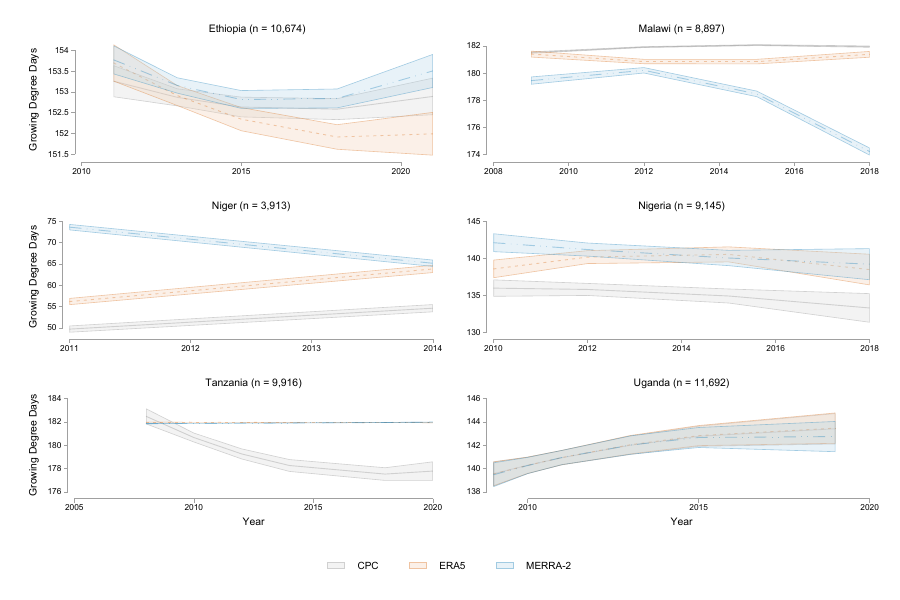}
		\end{center}
		\footnotesize  \textit{Note}: The figure presents the mean growing degree days in a year, disaggregated by Earth observation source. Means are calculated by fitting a fractional-polynomial line to the data, with $95\%$ confidence interval represented by shaded area. Each line (Earth observation product) in each panel is constructed using all household-year observations $(n)$. Variation in lines do not come from variation in the household data that is paired with the Earth observation data. Rather, variation in lines within a panel is solely due to differences in the daily temperatures reported by the Earth observation source.
	\end{minipage}	
\end{figure}
\end{center}    
\end{landscape}



\begin{center}
\begin{figure}[!htbp]
	\begin{minipage}{\linewidth}
		\caption{Specification Charts for Total Seasonal Rainfall}
		\label{fig:pval_v5}
		\begin{center}
			\includegraphics[width=.49\linewidth,keepaspectratio]{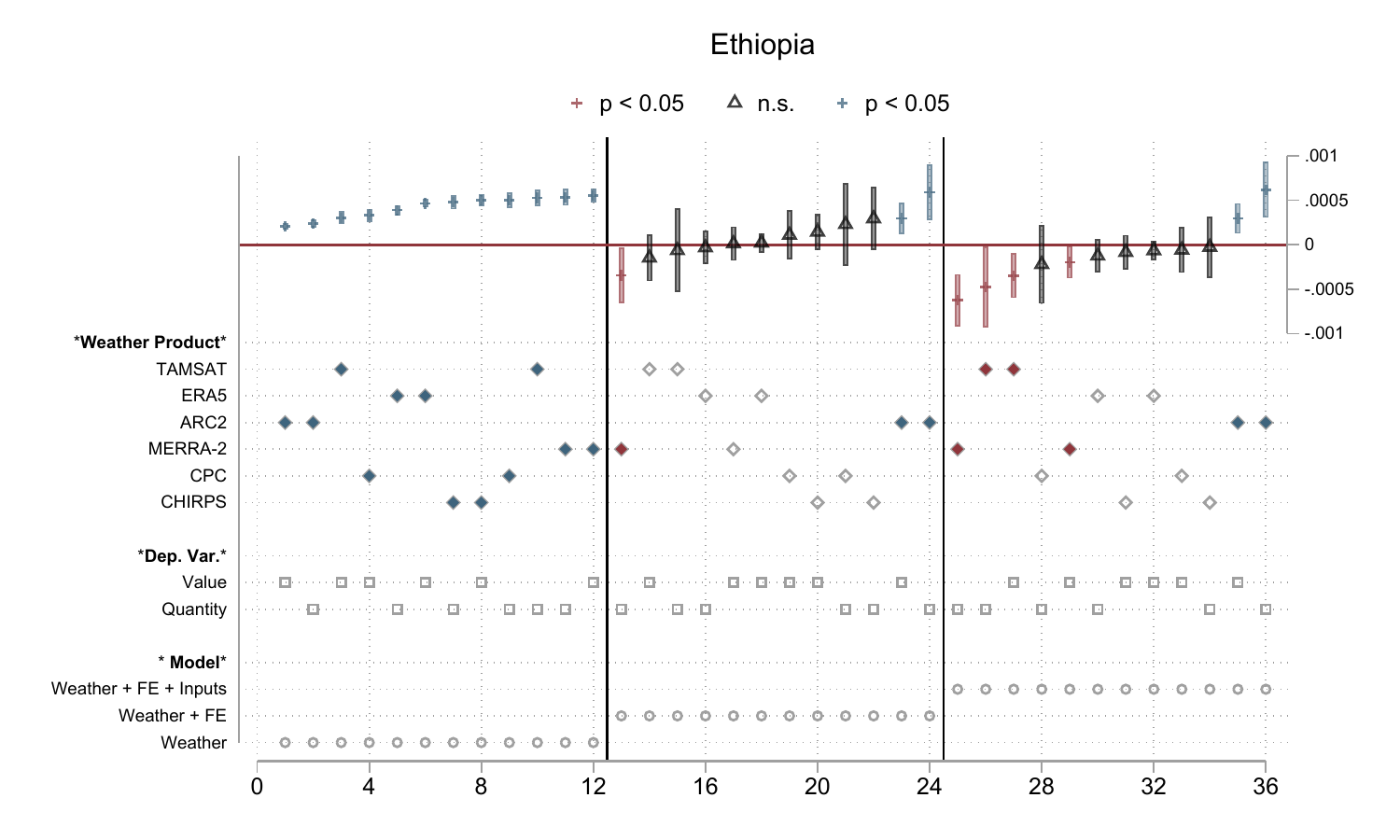}
			\includegraphics[width=.49\linewidth,keepaspectratio]{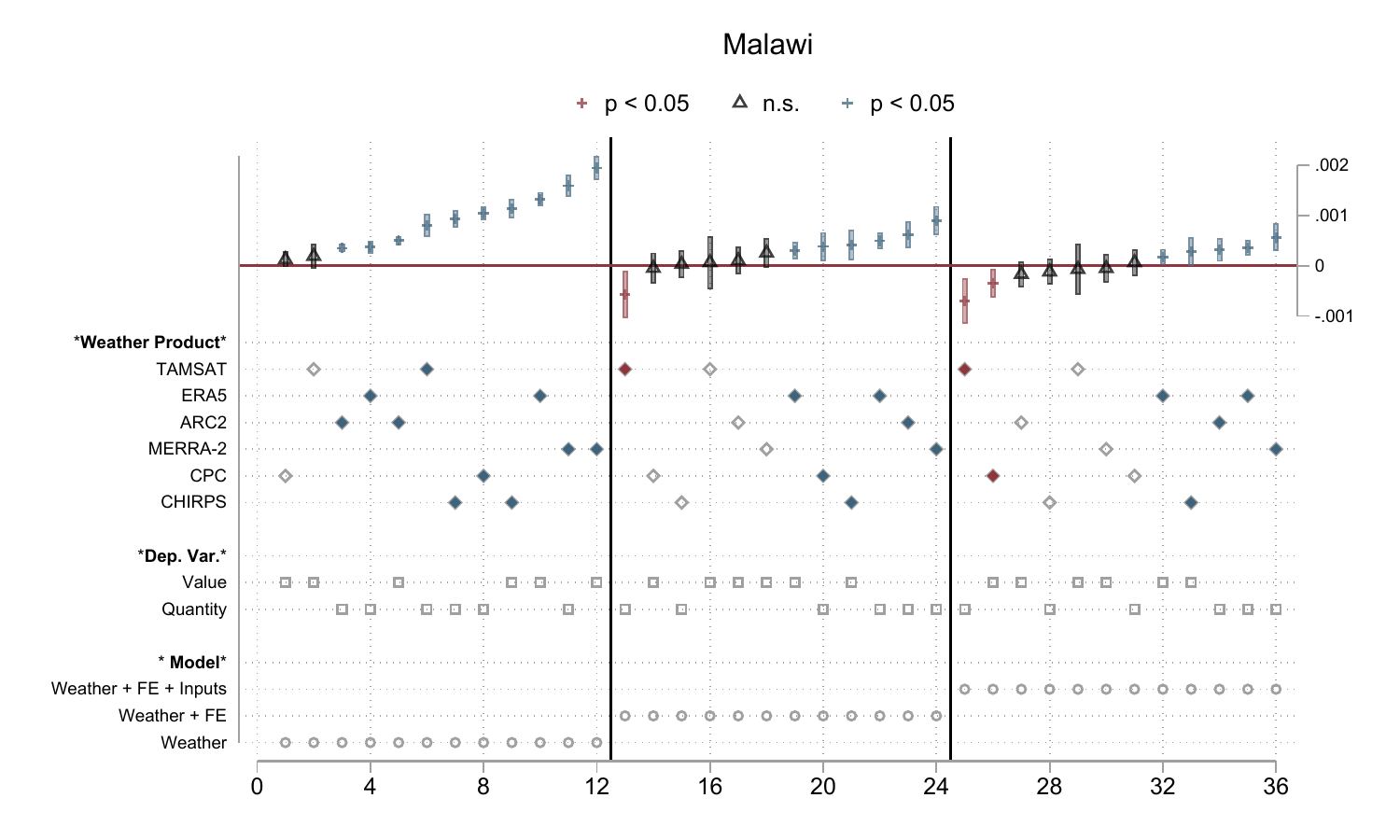}
			\includegraphics[width=.49\linewidth,keepaspectratio]{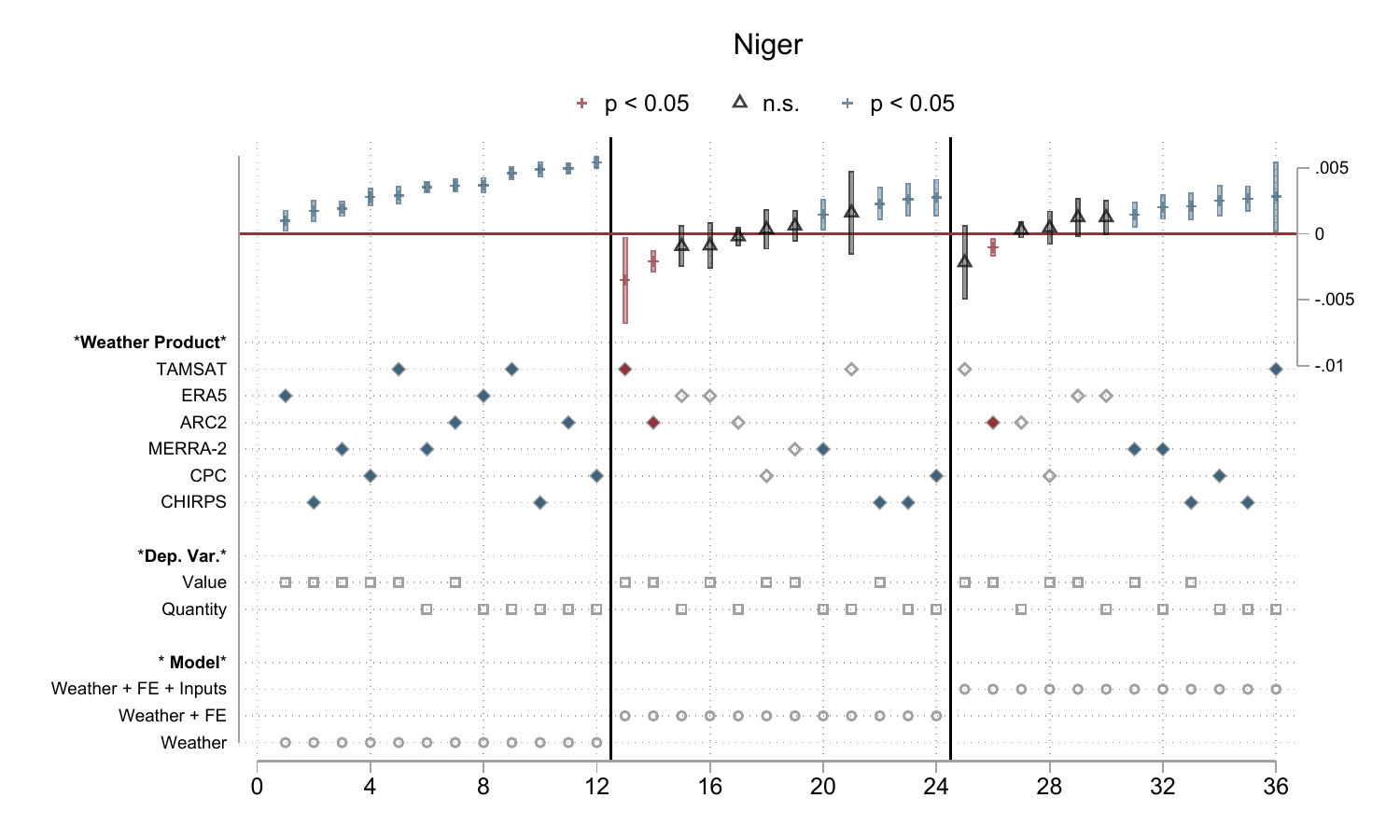}
			\includegraphics[width=.49\linewidth,keepaspectratio]{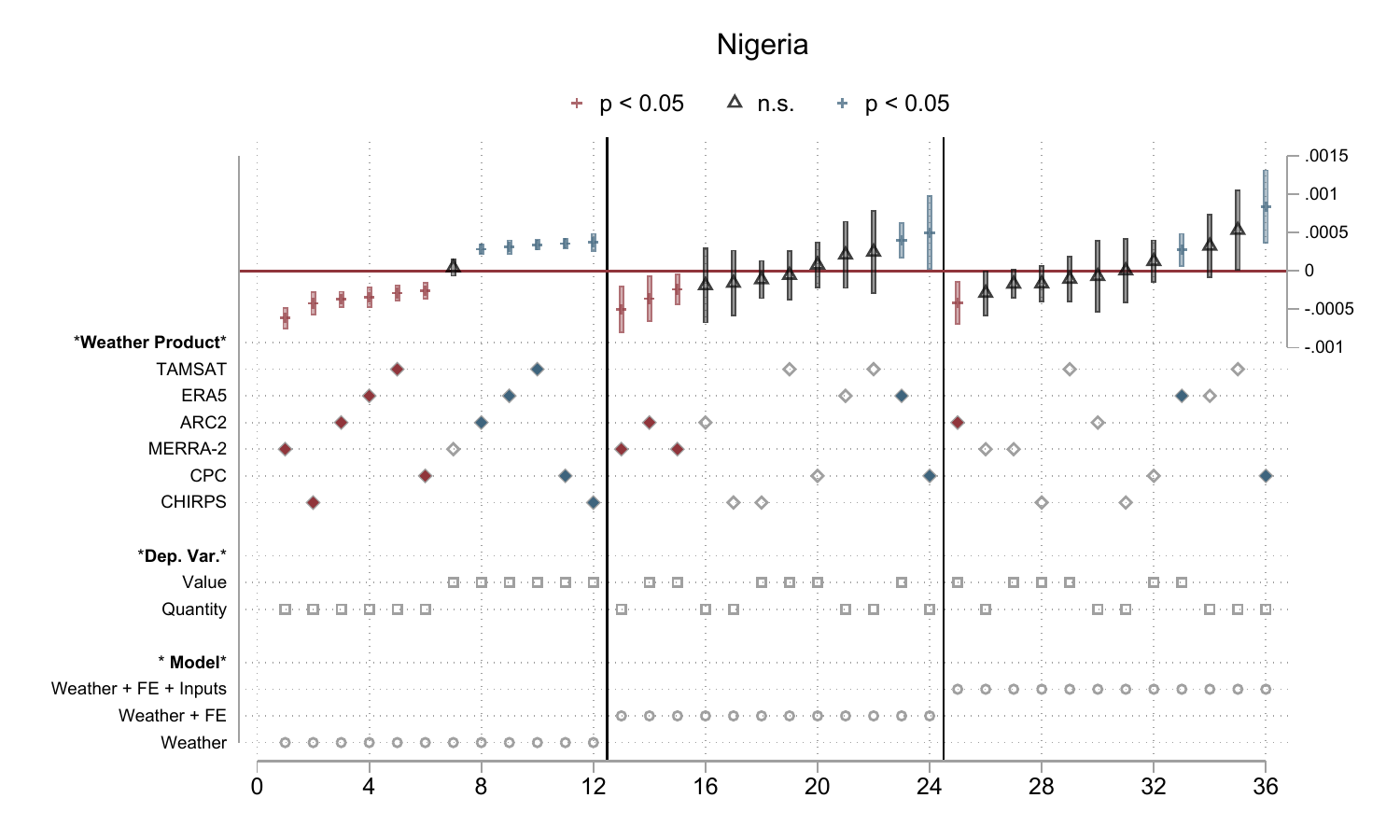}
			\includegraphics[width=.49\linewidth,keepaspectratio]{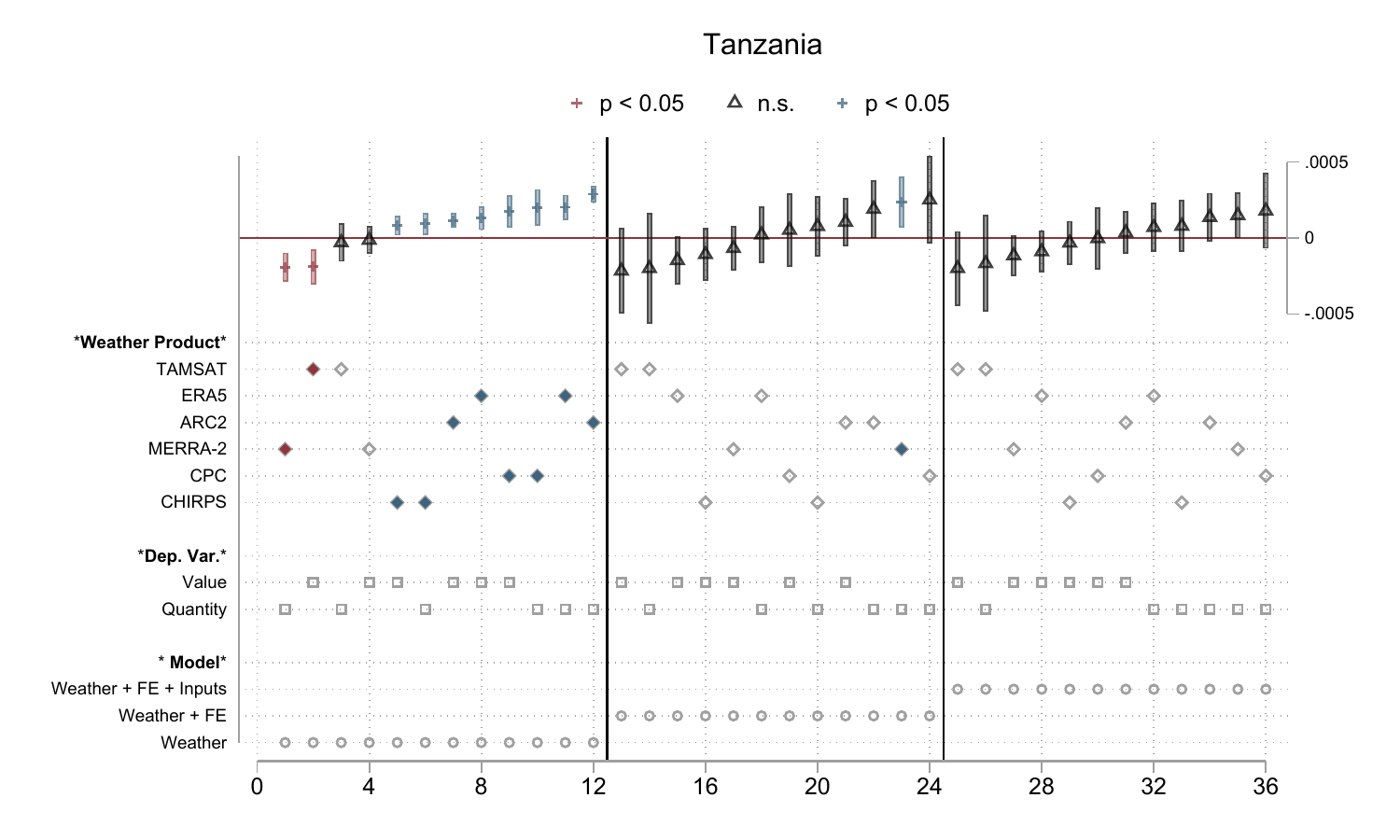}
			\includegraphics[width=.49\linewidth,keepaspectratio]{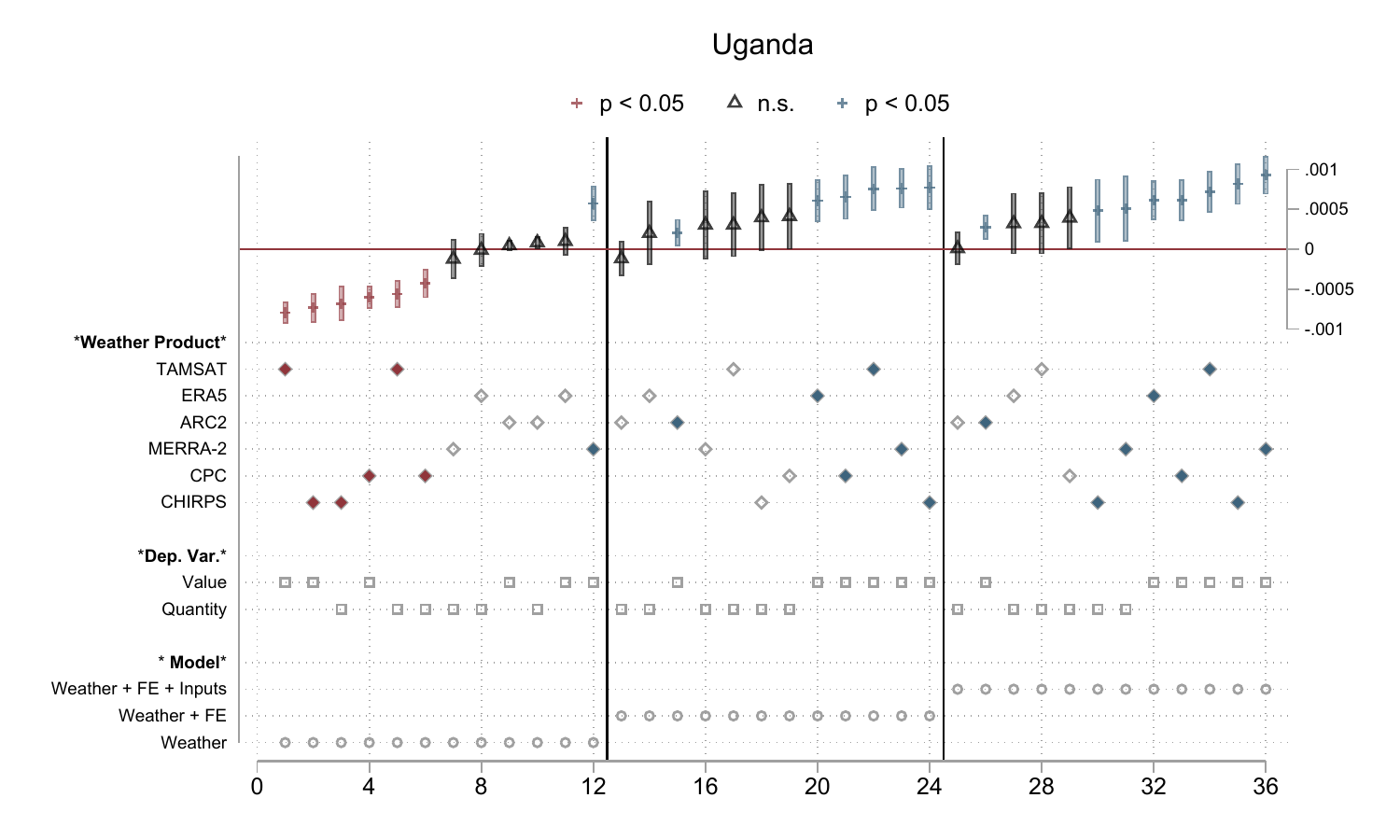}
		\end{center}
		\footnotesize  \textit{Note}: The figure presents specification curves, where each panel represents a different country, with three different models presented within each panel. Each panel includes 36 regressions, where each column represents a single regression. Significant and non-significant coefficients are designated at the top of the figure. For each Earth observation product, we also designate the significance and sign of the coefficient with color: red represents coefficients which are negative and significant; white represents insignificant coefficients, regardless of sign; and blue represents coefficients which are positive and significant.  
	\end{minipage}	
\end{figure}
\end{center}

\begin{center}
\begin{figure}[!htbp]
	\begin{minipage}{\linewidth}
		\caption{Specification Charts for Number of Days With No Rain}
		\label{fig:pval_v10}
		\begin{center}
			\includegraphics[width=.49\linewidth,keepaspectratio]{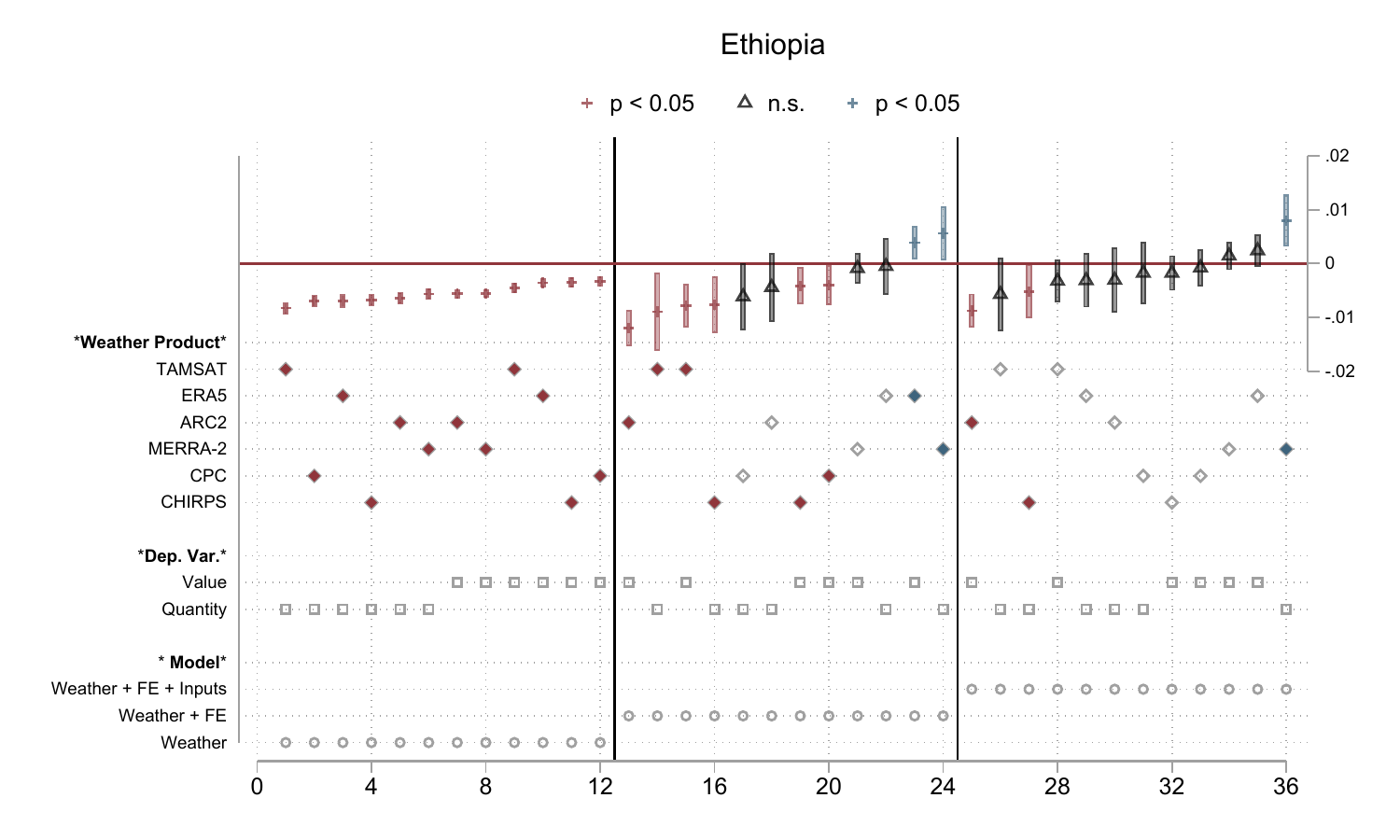}
			\includegraphics[width=.49\linewidth,keepaspectratio]{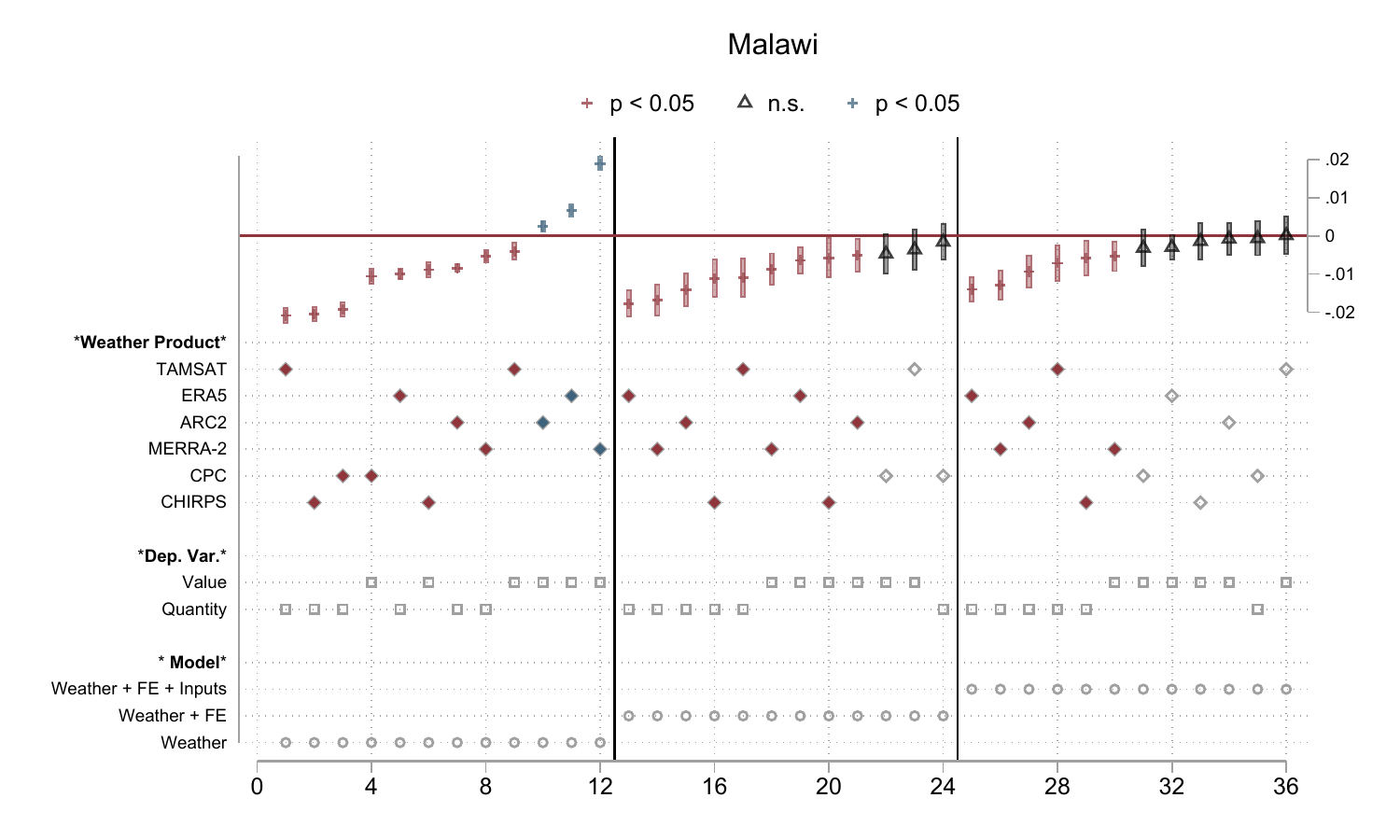}
			\includegraphics[width=.49\linewidth,keepaspectratio]{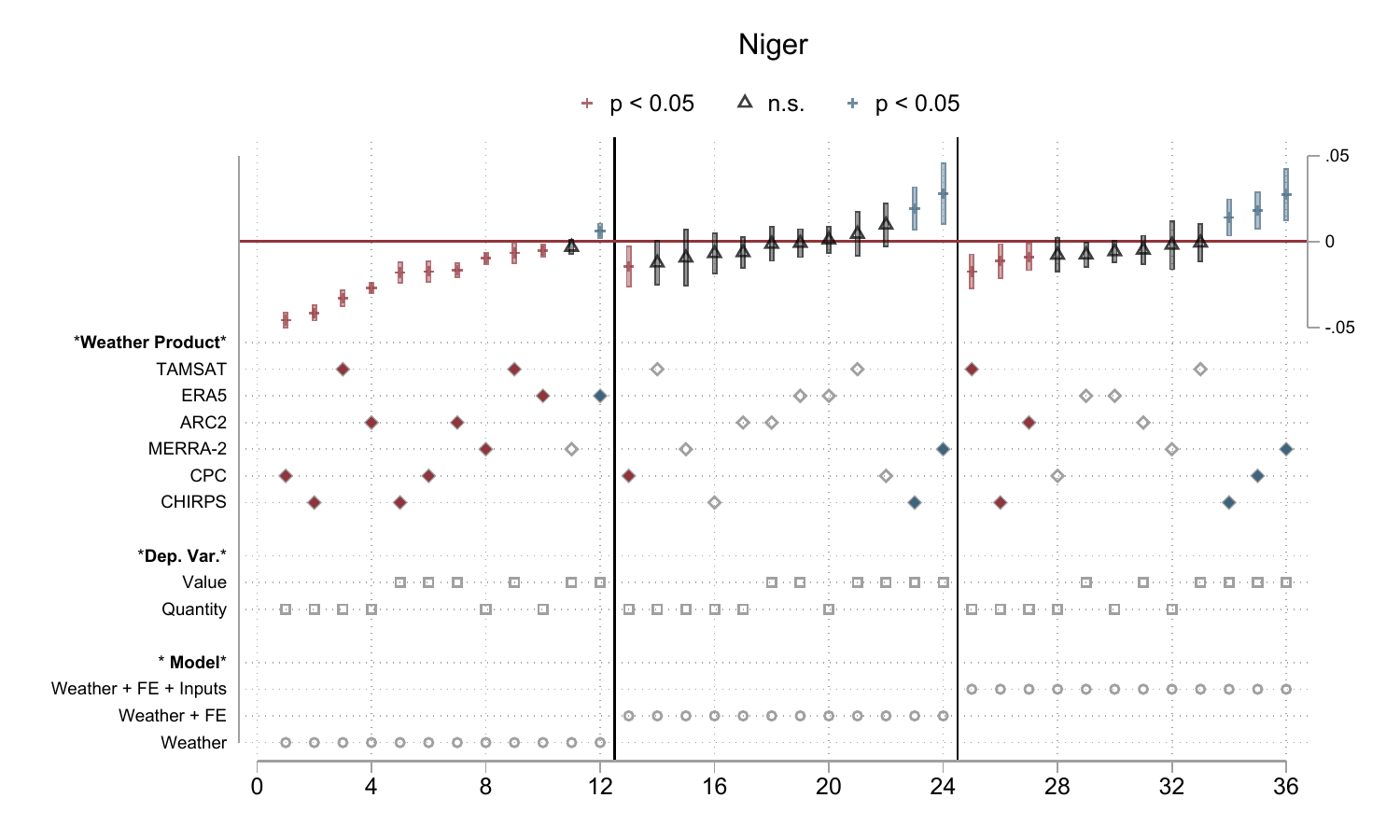}
			\includegraphics[width=.49\linewidth,keepaspectratio]{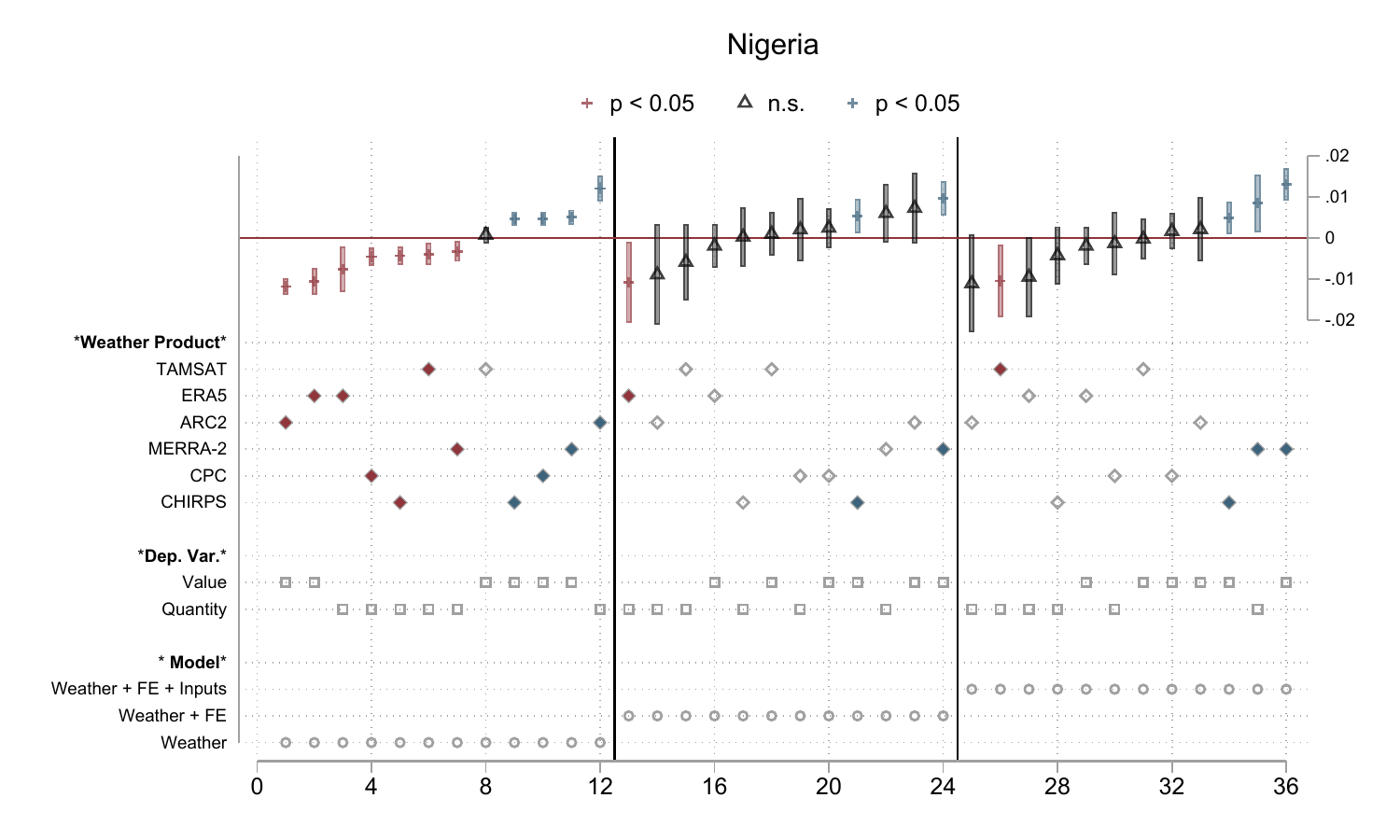}
			\includegraphics[width=.49\linewidth,keepaspectratio]{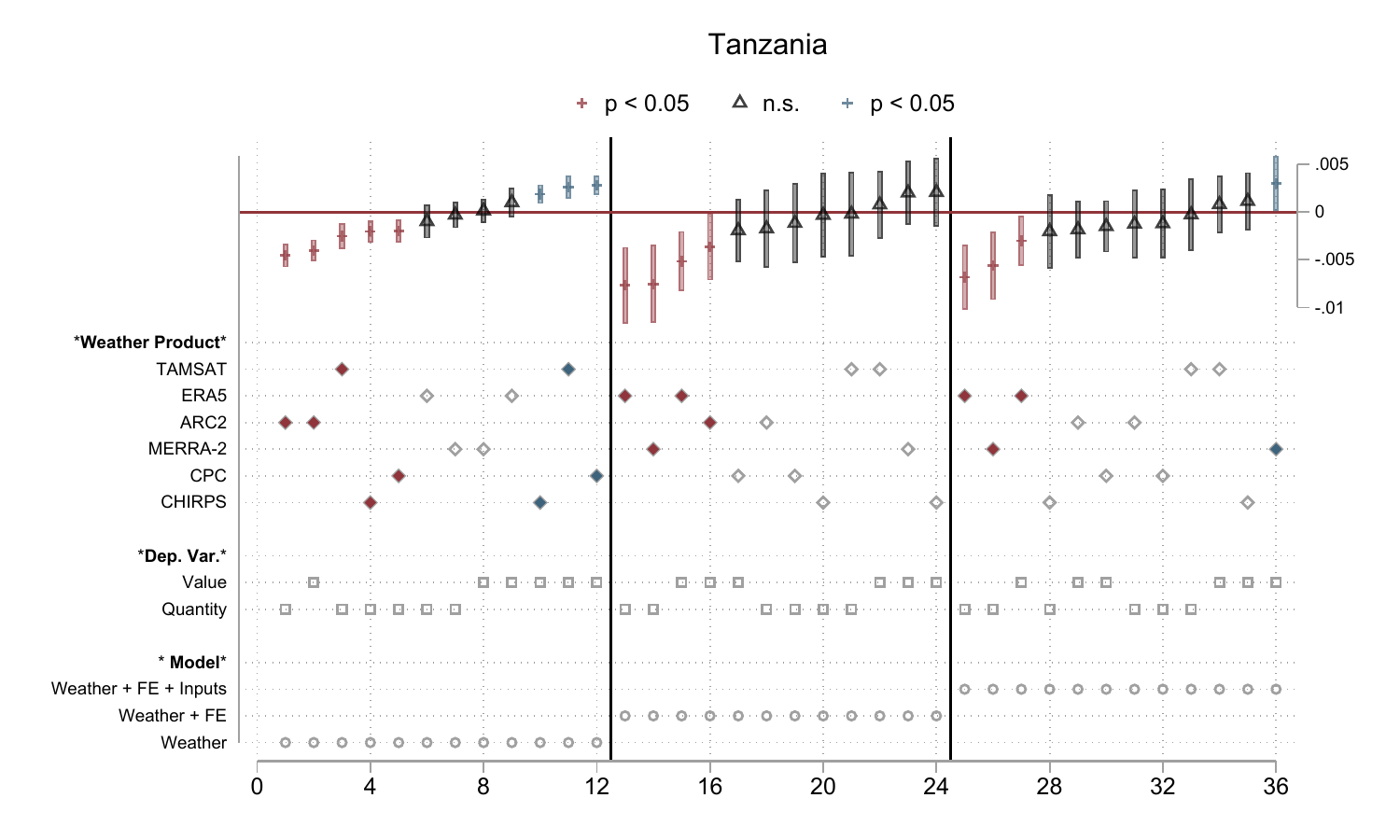}
			\includegraphics[width=.49\linewidth,keepaspectratio]{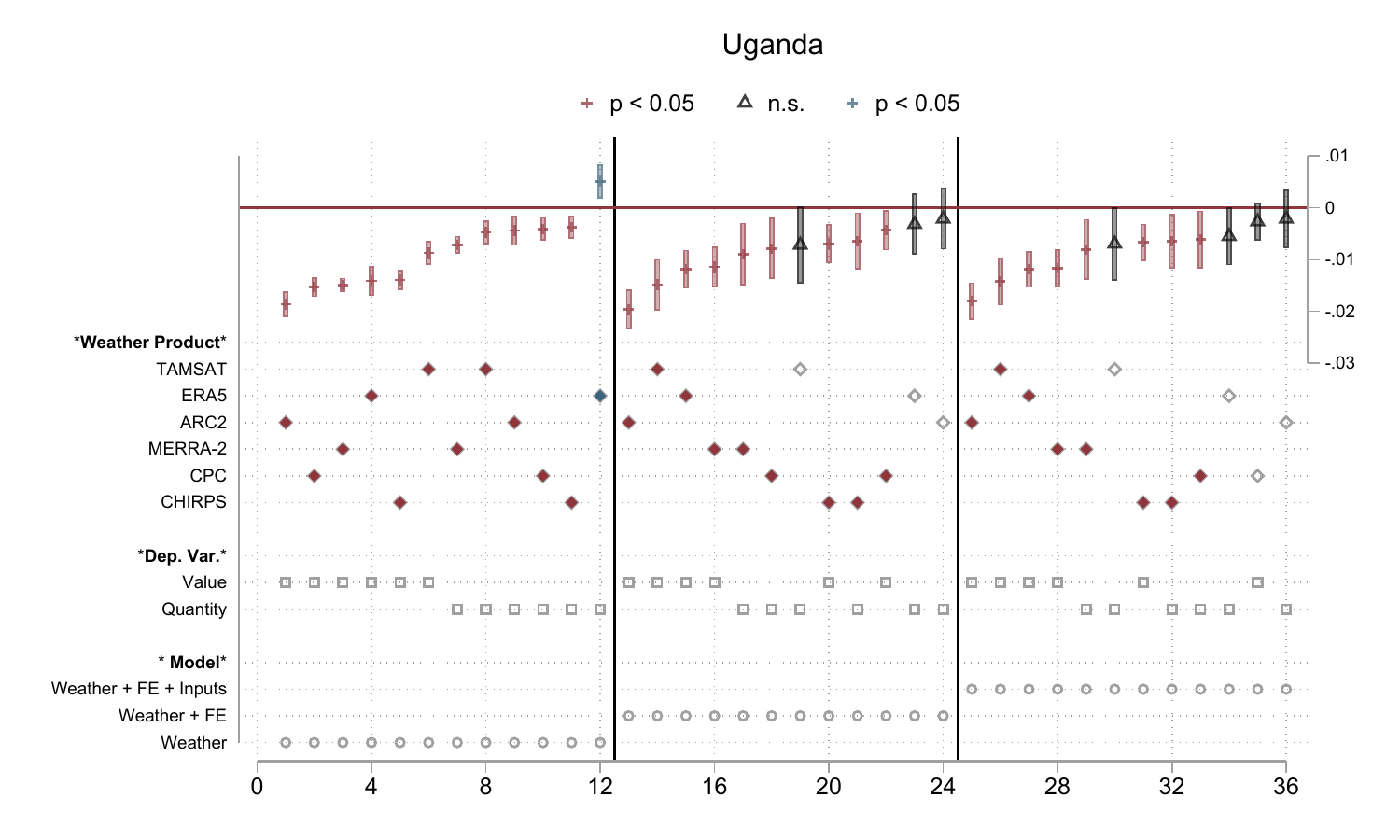}
		\end{center}
		\footnotesize  \textit{Note}: The figure presents specification curves, where each panel represents a different country, with three different models presented within each panel. Each panel includes 36 regressions, where each column represents a single regression. Significant and non-significant coefficients are designated at the top of the figure. For each Earth observation product, we also designate the significance and sign of the coefficient with color: red represents coefficients which are negative and significant; white represents insignificant coefficients, regardless of sign; and blue represents coefficients which are positive and significant.  
	\end{minipage}	
\end{figure}
\end{center}


\begin{center}
\begin{figure}[!htbp]
	\begin{minipage}{\linewidth}
		\caption{Specification Charts for Mean Daily Temperature}
		\label{fig:pval_v15}
		\begin{center}
			\includegraphics[width=.49\linewidth,keepaspectratio]{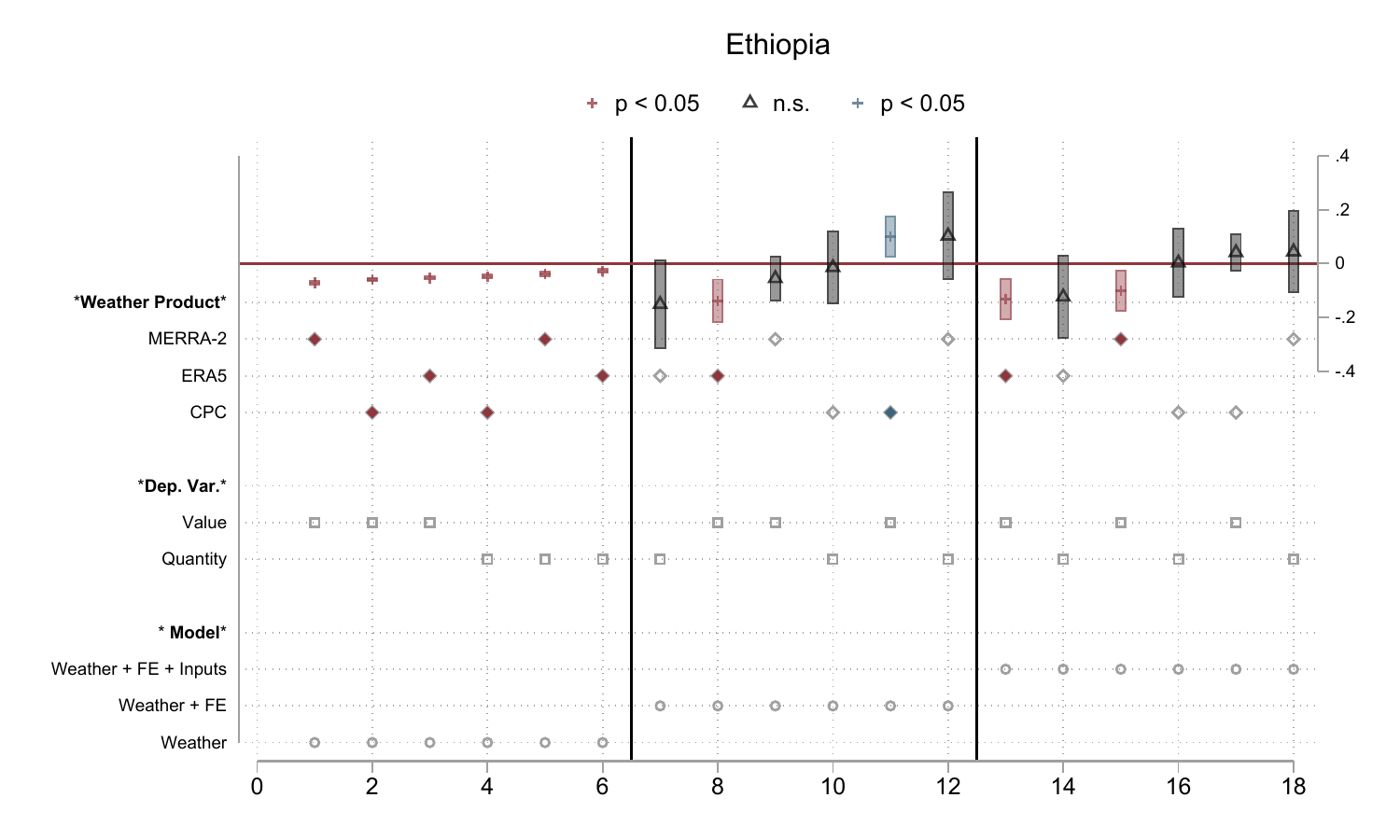}
			\includegraphics[width=.49\linewidth,keepaspectratio]{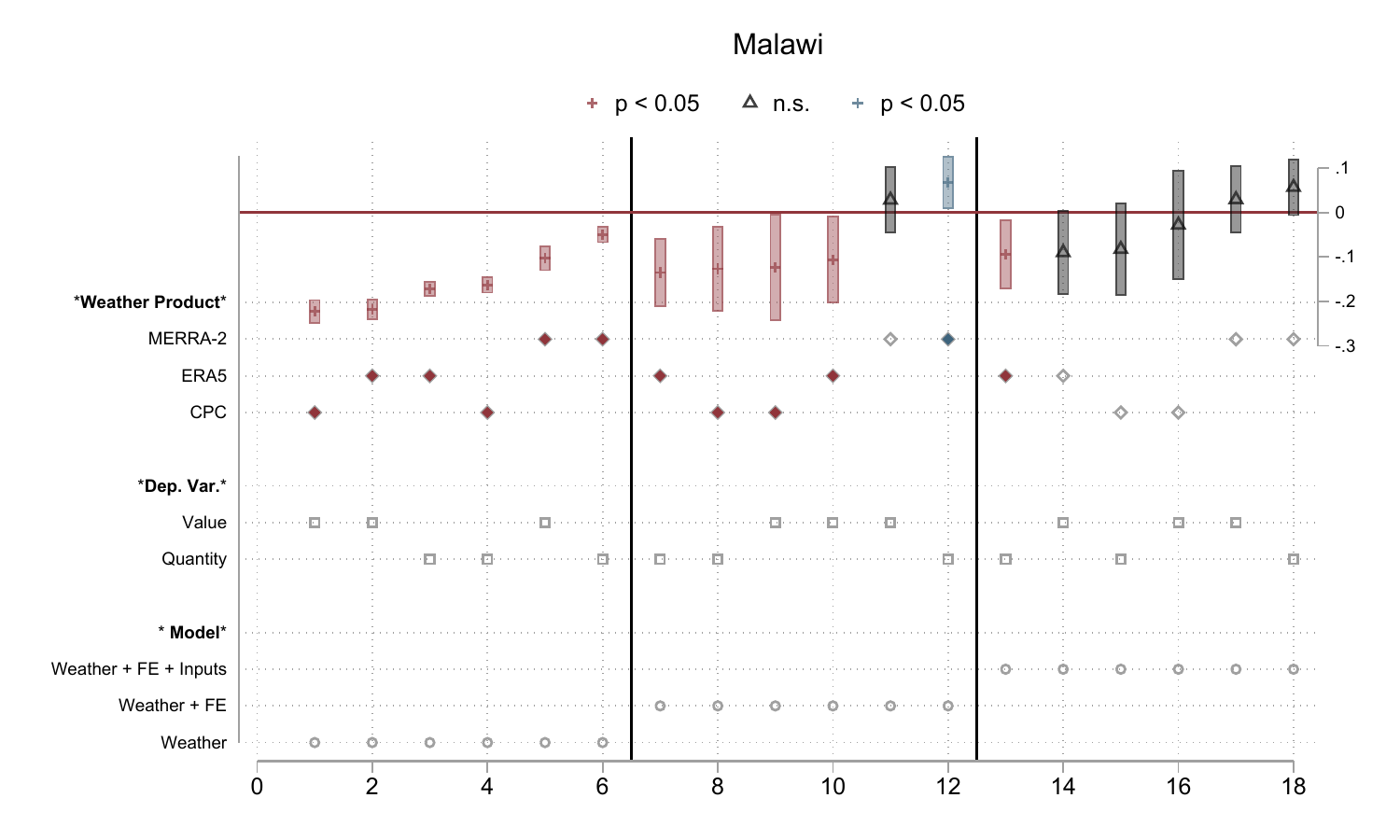}
			\includegraphics[width=.49\linewidth,keepaspectratio]{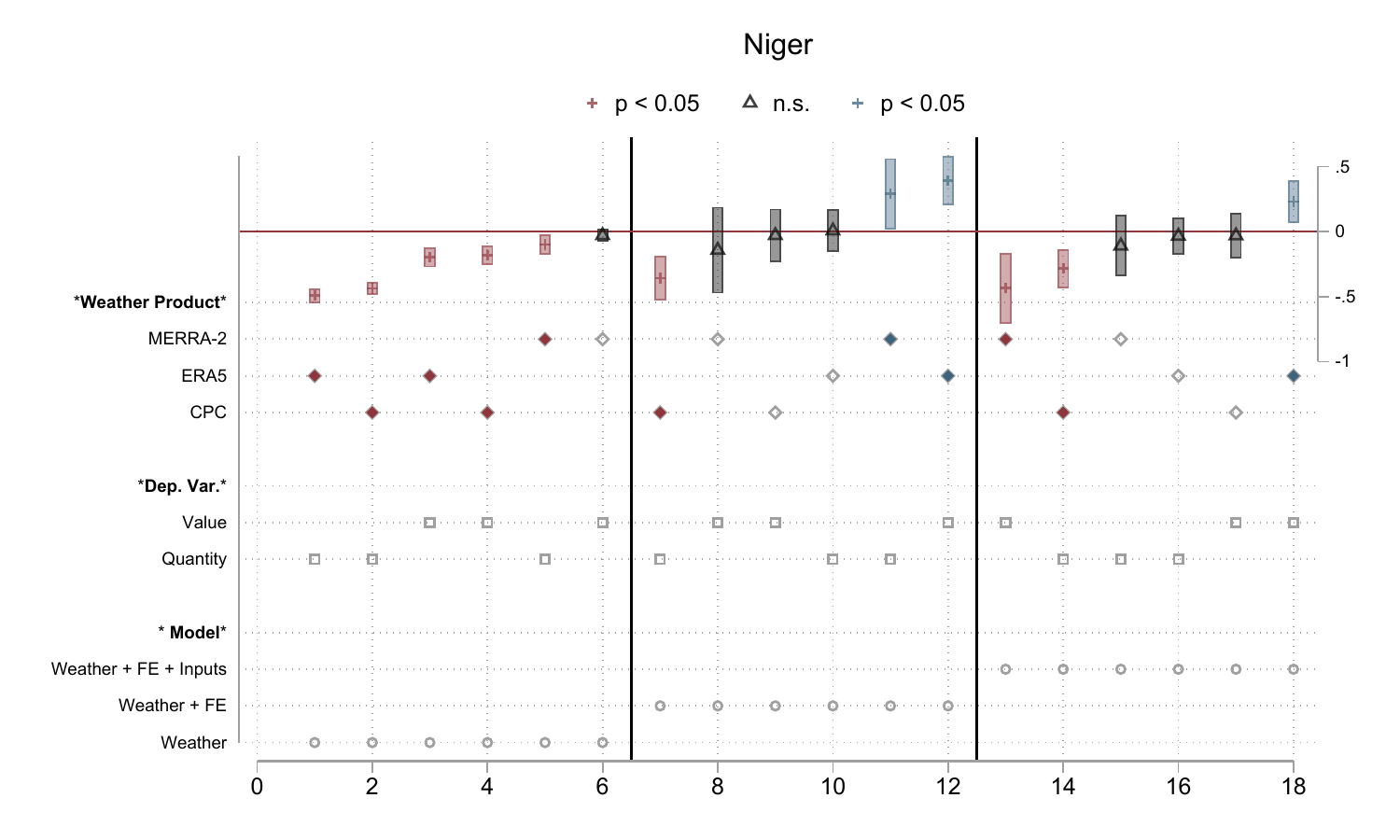}
			\includegraphics[width=.49\linewidth,keepaspectratio]{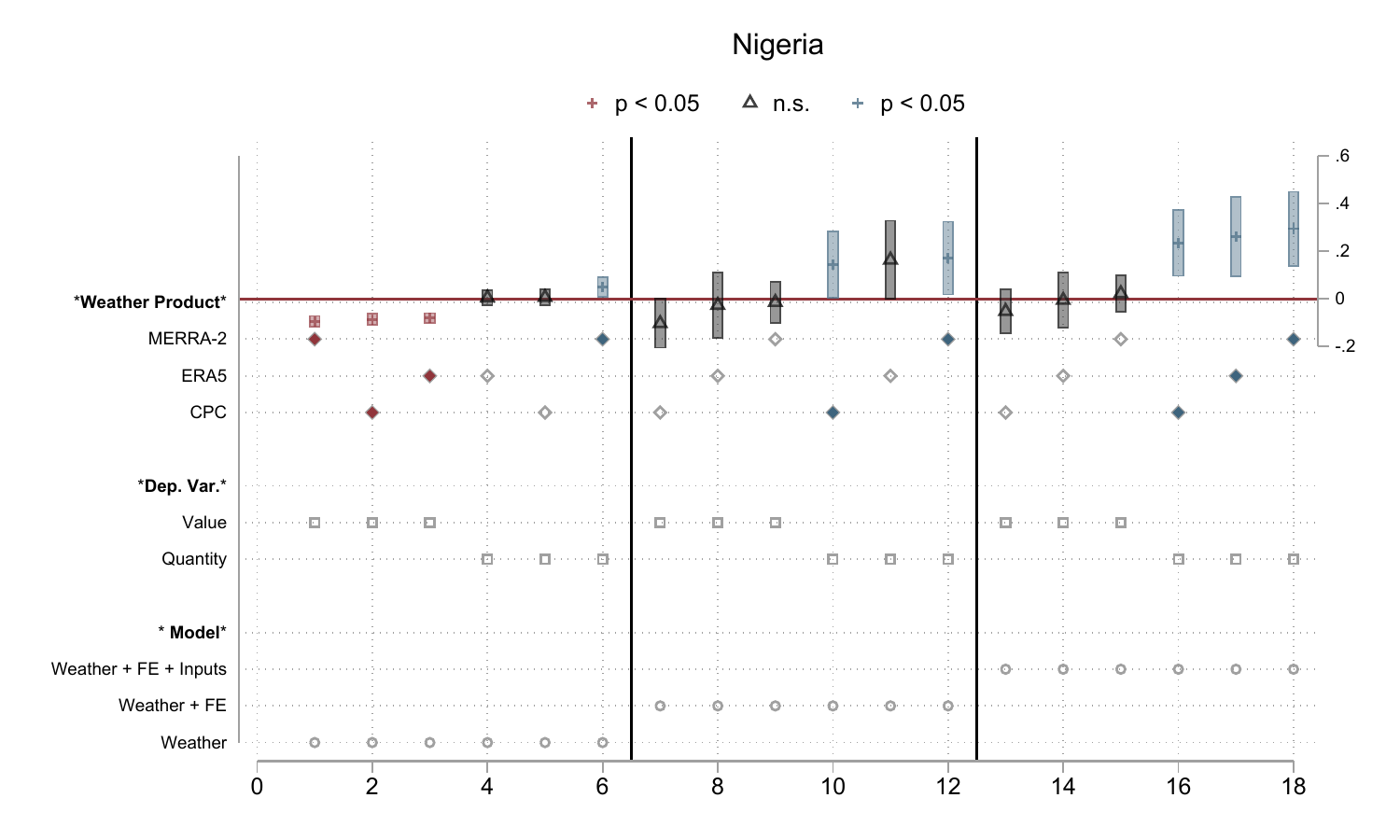}
			\includegraphics[width=.49\linewidth,keepaspectratio]{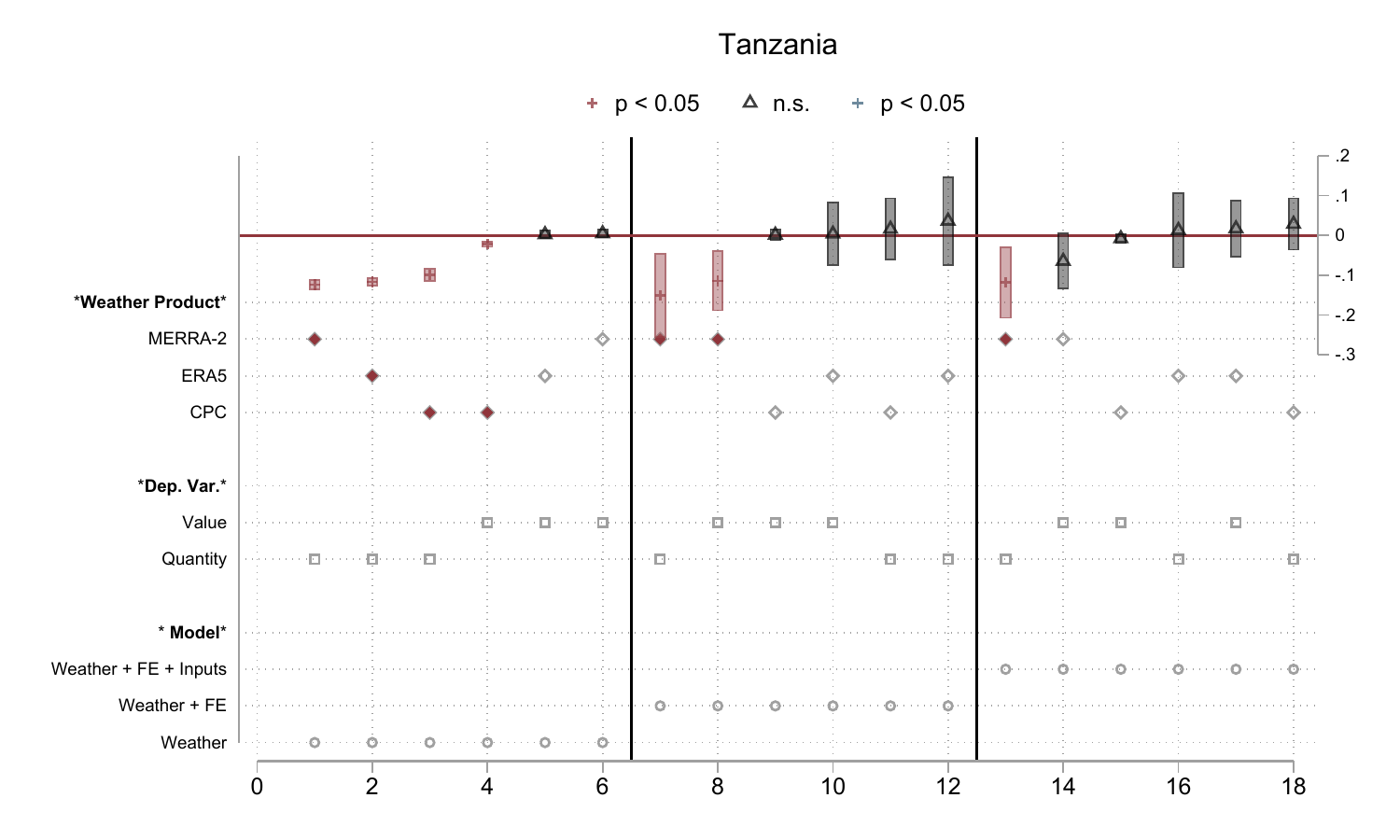}
			\includegraphics[width=.49\linewidth,keepaspectratio]{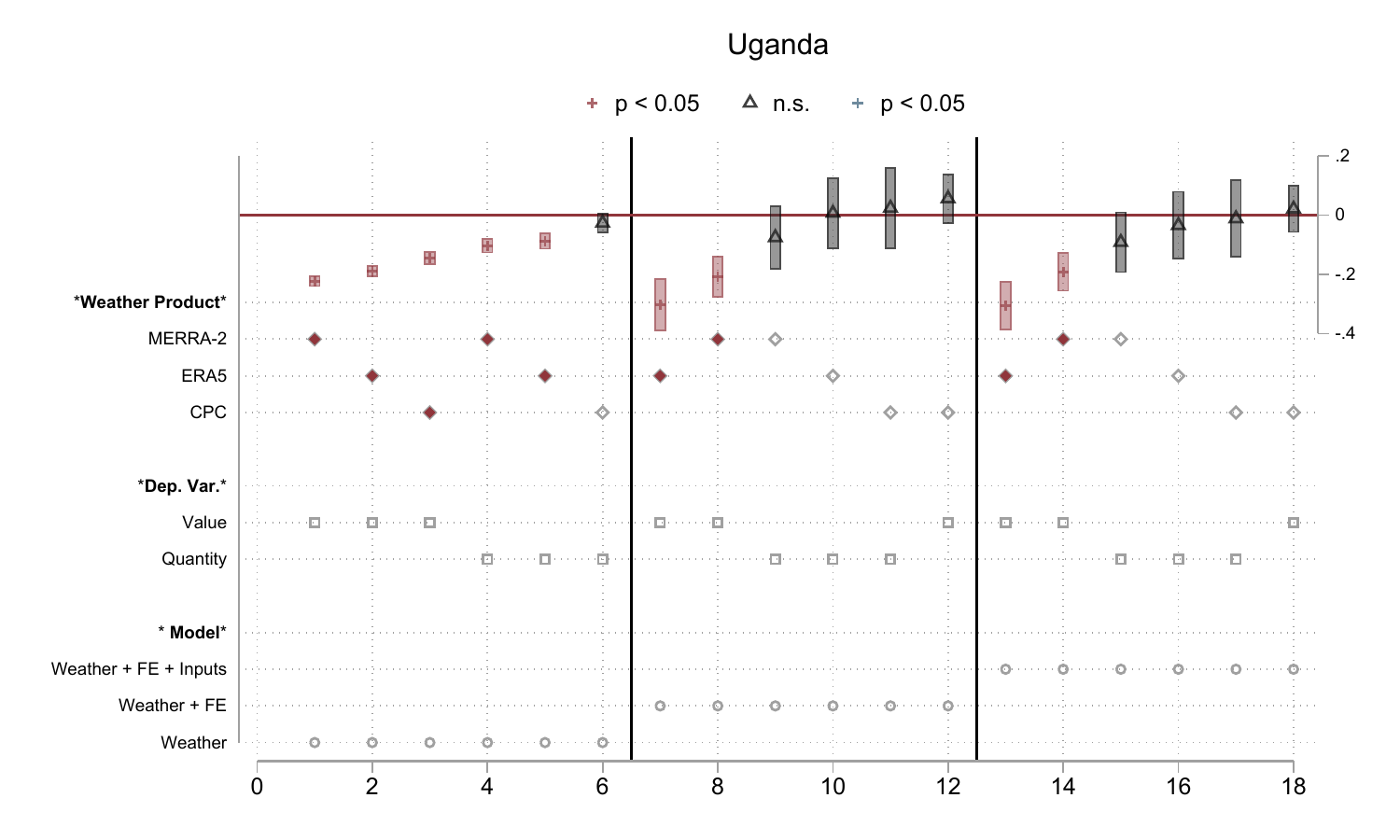}
		\end{center}
		\footnotesize  \textit{Note}: The figure presents specification curves, where each panel represents a different country, with three different models presented within each panel. Each panel includes 18 regressions, where each column represents a single regression. Significant and non-significant coefficients are designated at the top of the figure. For each Earth observation product, we also designate the significance and sign of the coefficient with color: red represents coefficients which are negative and significant; white represents insignificant coefficients, regardless of sign; and blue represents coefficients which are positive and significant.  
	\end{minipage}	
\end{figure}
\end{center}

\begin{center}
\begin{figure}[!htbp]
	\begin{minipage}{\linewidth}
		\caption{Specification Charts for Growing Degree Days (GDD)}
		\label{fig:pval_v19}
		\begin{center}
			\includegraphics[width=.49\linewidth,keepaspectratio]{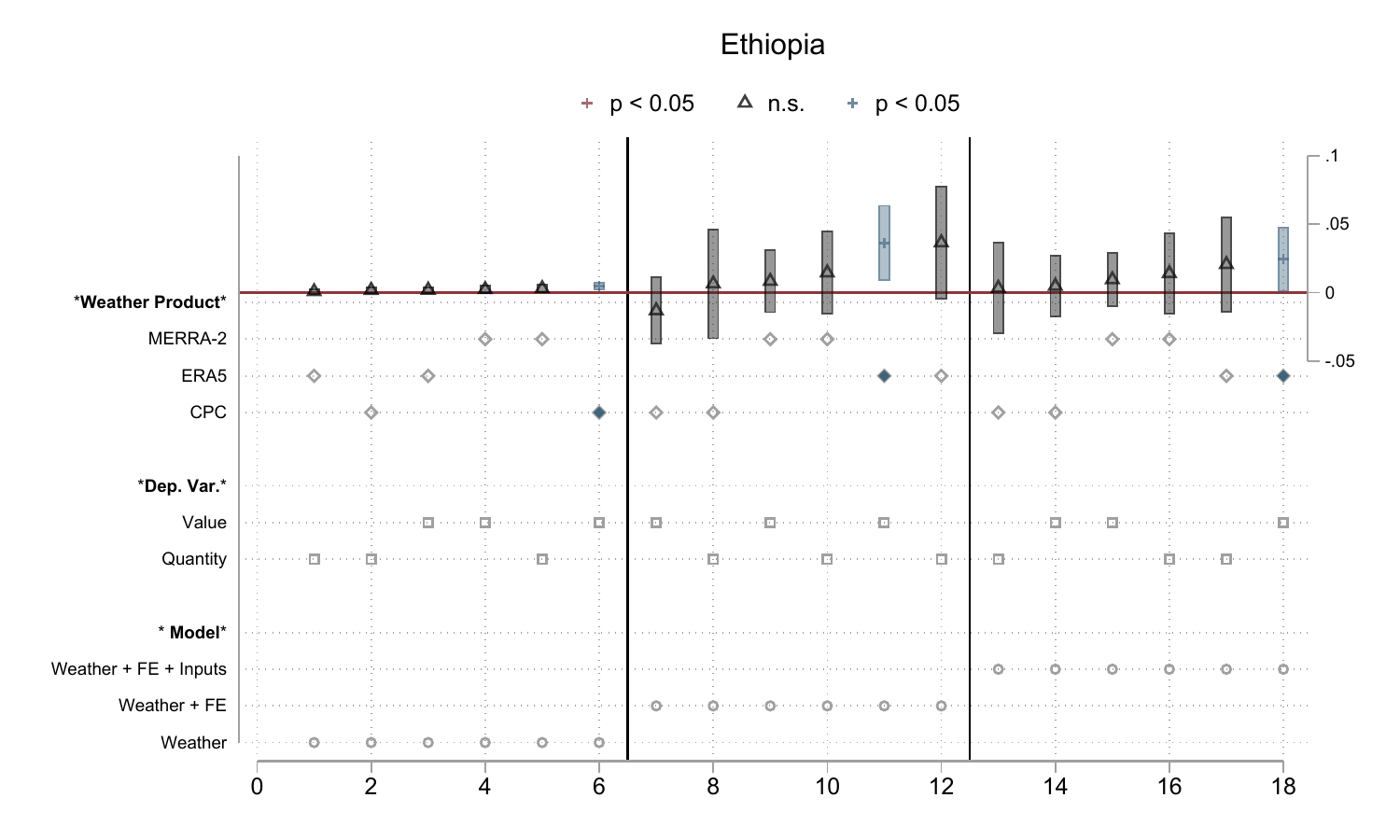}
			\includegraphics[width=.49\linewidth,keepaspectratio]{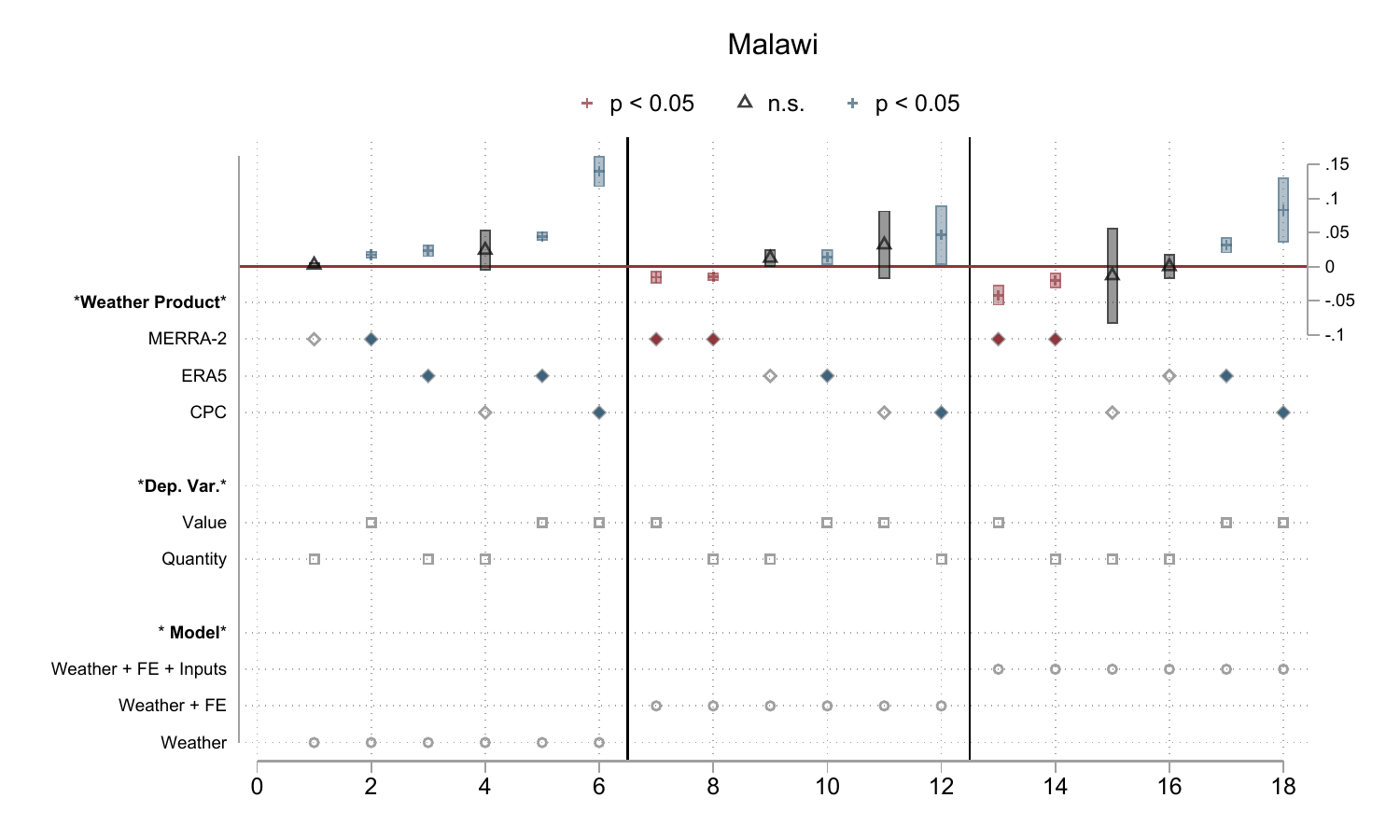}
			\includegraphics[width=.49\linewidth,keepaspectratio]{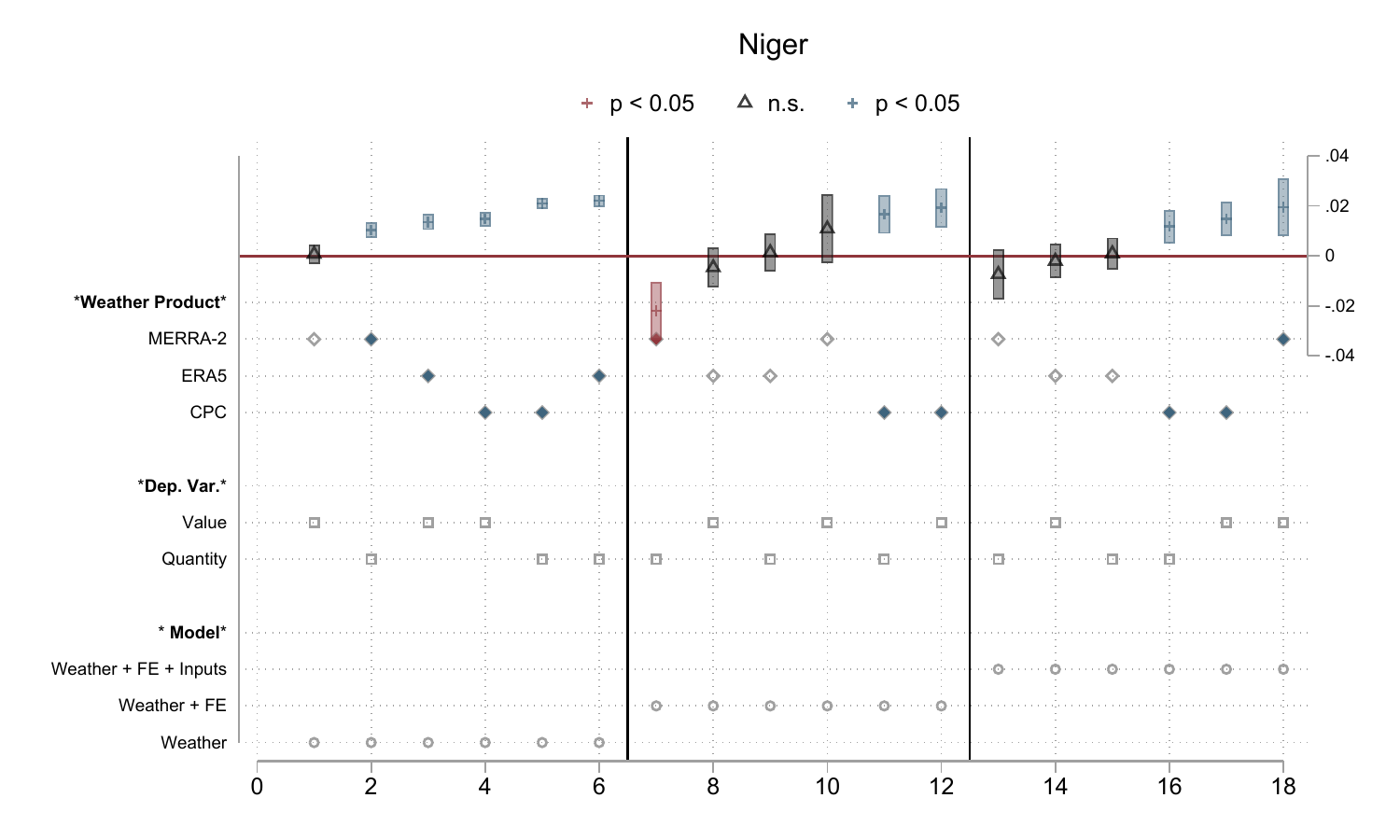}
			\includegraphics[width=.49\linewidth,keepaspectratio]{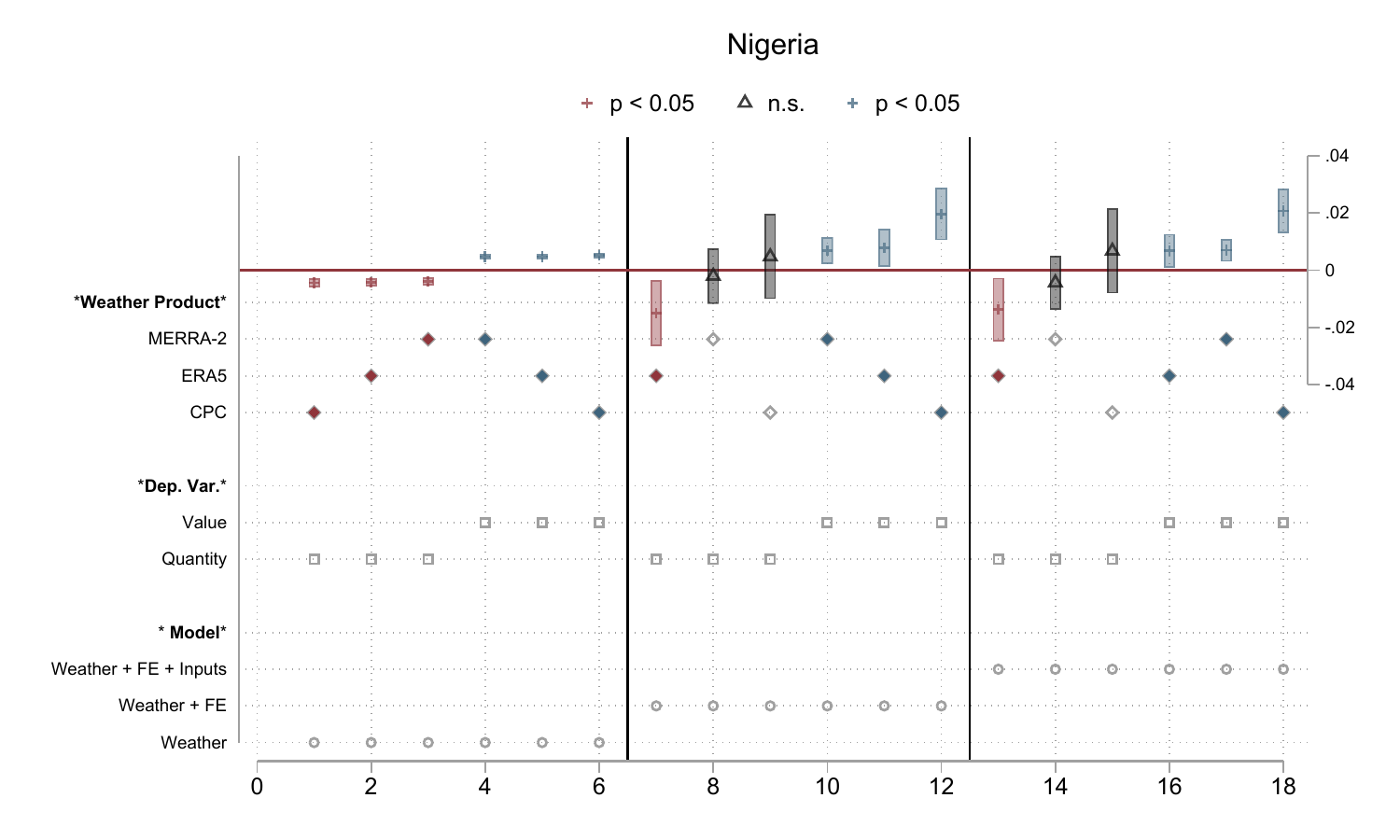}
			\includegraphics[width=.49\linewidth,keepaspectratio]{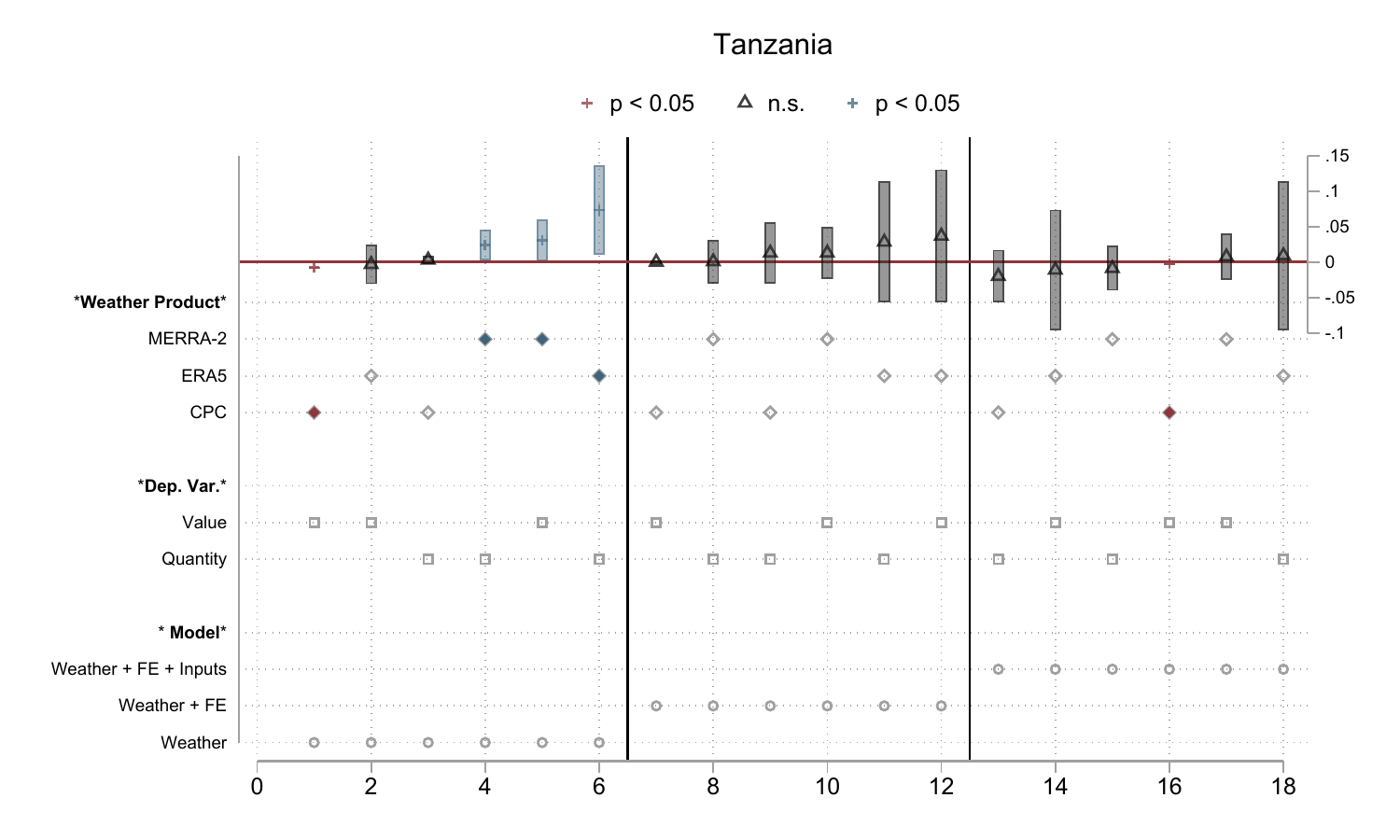}
			\includegraphics[width=.49\linewidth,keepaspectratio]{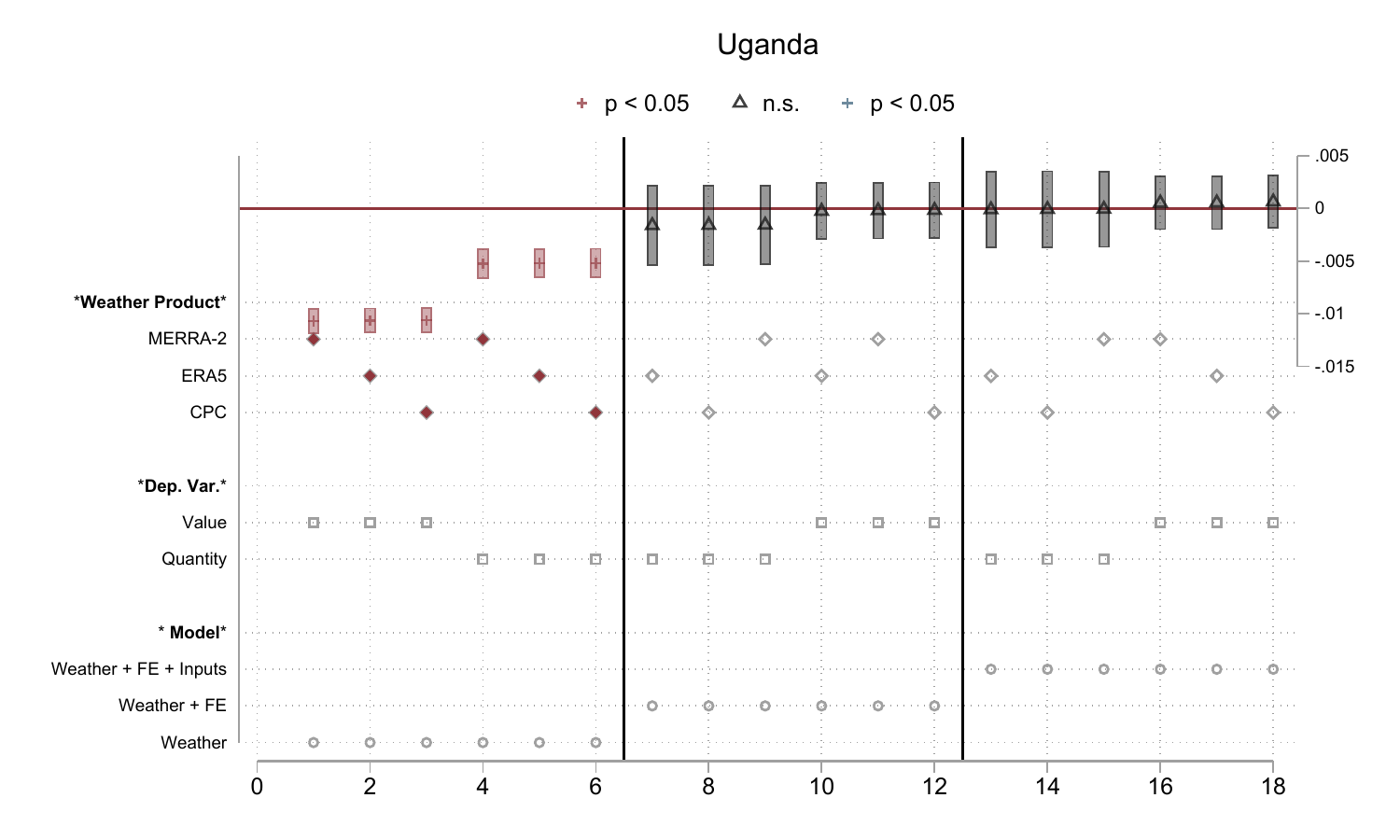}
		\end{center}
		\footnotesize  \textit{Note}: The figure presents specification curves, where each panel represents a different country, with three different models presented within each panel. Each panel includes 18 regressions, where each column represents a single regression. Significant and non-significant coefficients are designated at the top of the figure. For each Earth observation product, we also designate the significance and sign of the coefficient with color: red represents coefficients which are negative and significant; white represents insignificant coefficients, regardless of sign; and blue represents coefficients which are positive and significant.  
	\end{minipage}	
\end{figure}
\end{center}



\newpage 
\begin{landscape}
\begin{center}
\begin{figure}[!htbp]
	\begin{minipage}{\linewidth}
		\caption{Bumpline: Total Rainfall Growing Season}
		\label{fig:bump_train}
		\begin{center}
			\includegraphics[width=\linewidth,keepaspectratio]{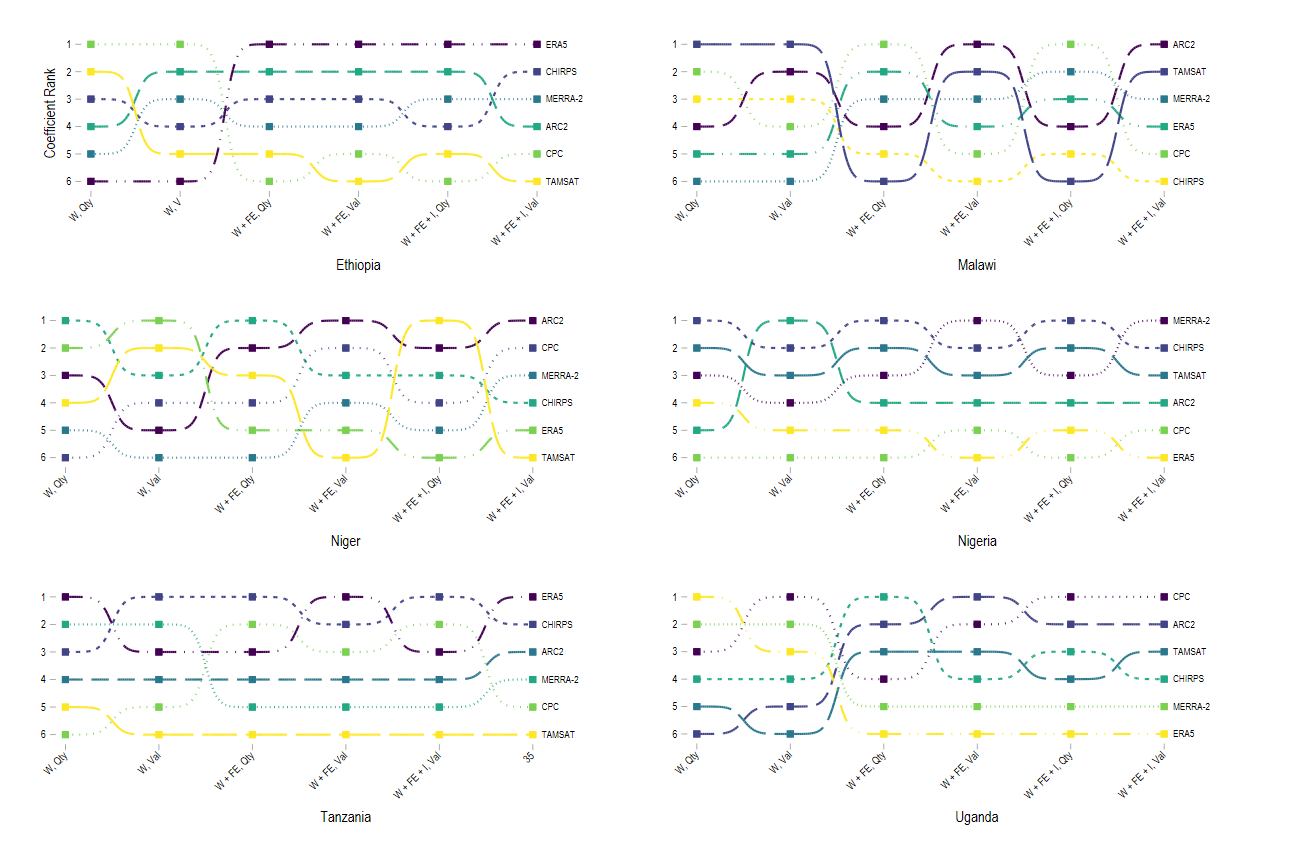}
		\end{center}
		\footnotesize  \textit{Note}: The figure presents a bumpline plot, where each panel represents a different country, where each column represents a different models and a different outcome variable for each model; each row represents the rank of each coefficient (one to six, with one representing the largest number and six representing the smallest number) within those regressions. Each panel includes 36 regressions. The bumpline package was developed by \cite{bumpline}. 

	\end{minipage}	
\end{figure}
\end{center}    
\end{landscape}


\newpage 
\begin{landscape}
\begin{center}
\begin{figure}[!htbp]
	\begin{minipage}{\linewidth}
		\caption{Bumpline: Mean Temperature}
		\label{fig:bump_meant}
		\begin{center}
			\includegraphics[width=\linewidth,keepaspectratio]{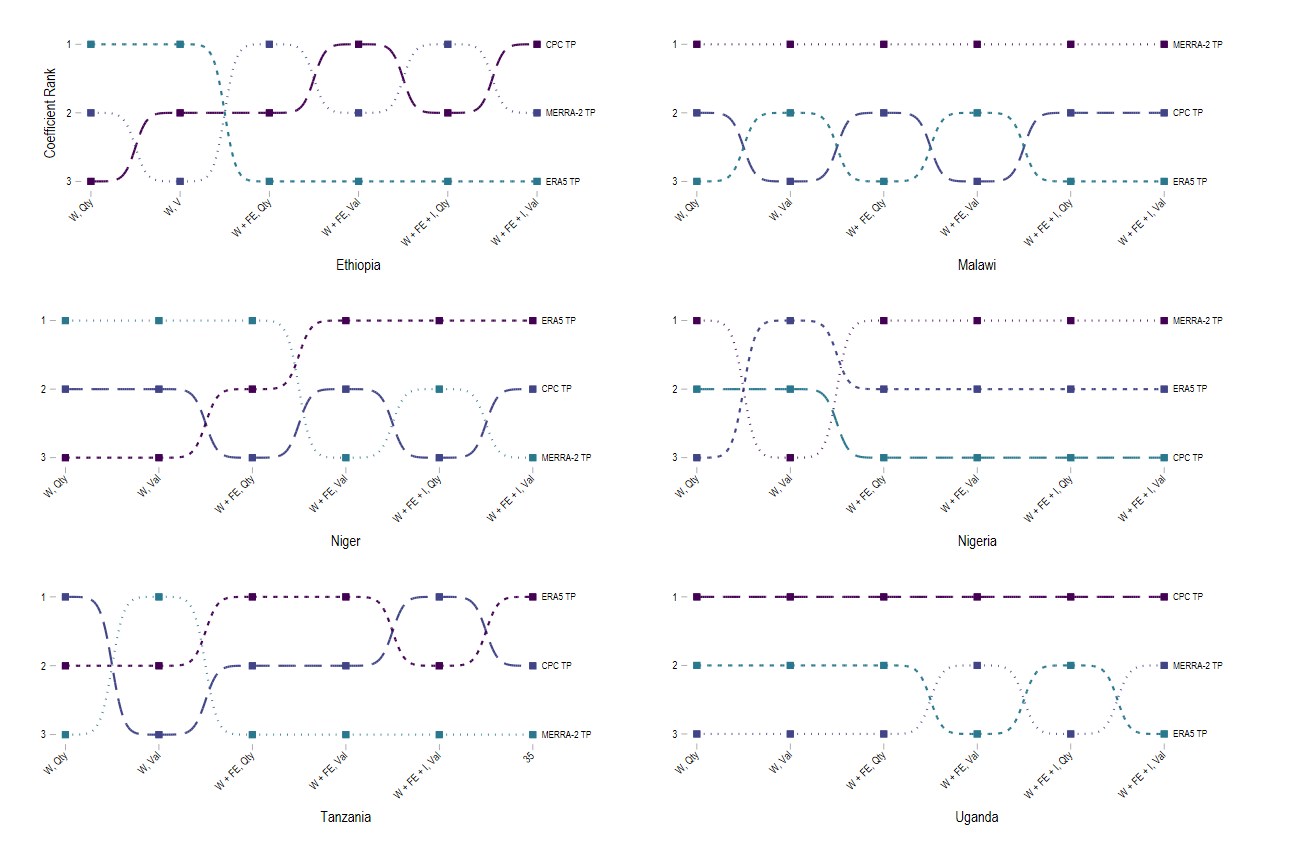}
		\end{center}
		\footnotesize  \textit{Note}: The figure presents a bumpline plot, where each panel represents a different country, where each column represents a different models and a different outcome variable for each model; each row represents the rank of each coefficient (one to six, with one representing the largest number and six representing the smallest number) within those regressions. Each panel includes 18 regressions. The bumpline package was developed by \cite{bumpline}. 
	\end{minipage}	
\end{figure}
\end{center}    
\end{landscape}


\clearpage
\newpage
\appendix
\onehalfspacing

\begin{center}
	\section*{Appendix to ``The Mismeasure of Weather: Using Remotely Sensed Weather in Economic Contexts''} \label{sec:app}
\end{center}


\subsection*{Defining Growing Season} \label{sec:appRS_gs}

\setcounter{table}{0}
\renewcommand{\thetable}{A\arabic{table}}
\setcounter{figure}{0}
\renewcommand{\thefigure}{A\arabic{figure}}

We define growing season following the FAO \href{http://www.fao.org/agriculture/seed/cropcalendar/welcome.do}{crop calendar} for each country. Table~\ref{tab:growseason} presents details for each country on the growing season used, as well as whether that season spans years and whether it is unimodal or bimodal. Earth observation data used in our analysis follows the defined growing season in each respective country.

It is important to note that both Malawi and Tanzania span calendar years. This means that the growing season begins in one year and extends into the year that follows. Consider, Malawi. The growing season begins on 1 October and ends on 30 April. This means that it would begin 1 October 2024 and would end 30 April 2025. 

Similarly, both Nigeria and Uganda have bimodal seasons. Season modality designates whether different regions within the countries have different growing seasons. In both Nigeria and Uganda, the northern part of the country has a different growing season from the southern part of the country. In these cases we designate the modality of the season, and also provide the growing season dates for both regions. 


\subsection*{Weather Variables}\label{sec:app_varweather}

\begin{table}[htbp]	\centering
	\caption{Weather Variables \& Transformations} \label{tab:Wvar}
	\scalebox{0.9}
	{ \setlength{\linewidth}{.1cm}\newcommand{\contents}
		{\begin{tabular}{ll}
			\\[-1.8ex]\hline 
			\hline \\[-1.8ex]
			\multicolumn{2}{l}{\emph{\textbf{Panel A}: Rainfall}} \\
			\multicolumn{1}{l}{Daily rainfall} & \multicolumn{1}{l}{In mm} \\
			\multicolumn{1}{l}{Mean} & \multicolumn{1}{p{11cm}}{The first moment of the daily rainfall distribution for the growing season$^\dagger$} \\
			\multicolumn{1}{l}{Median} & \multicolumn{1}{p{11cm}}{The median daily rainfall for the growing season$^\dagger$} \\
			\multicolumn{1}{l}{Variance} & \multicolumn{1}{p{11cm}}{The second moment of the daily rainfall distribution for the growing season$^\dagger$} \\
			\multicolumn{1}{l}{Skew} & \multicolumn{1}{p{11cm}}{The third moment of the daily rainfall distribution for the growing season$^\dagger$} \\
			\multicolumn{1}{l}{Total} & \multicolumn{1}{p{11cm}}{Cumulative daily rainfall for the growing season$^\dagger$} \\
			\multicolumn{1}{l}{Deviations in total rainfall} & \multicolumn{1}{p{11cm}}{The z-score for cumulative daily rainfall for the growing season$^\dagger$} \\
			\multicolumn{1}{l}{Scaled deviations in total rainfall} & \multicolumn{1}{p{11cm}}{The z-score for cumulative daily rainfall for the growing season$^\dagger$} \\
			\multicolumn{1}{l}{Rainfall days} & \multicolumn{1}{p{11cm}}{The number of days with at least 1 mm of rain for the growing season$^\dagger$} \\
			\multicolumn{1}{l}{Deviation in rainfall days} & \multicolumn{1}{p{11cm}}{The number of days with rain for the growing season minus the long run average$^*$} \\
			\multicolumn{1}{l}{No rain days} & \multicolumn{1}{p{11cm}}{The number of days with less than 1 mm of rain for the growing season$^\dagger$} \\
			\multicolumn{1}{l}{Deviation in no rain days} & \multicolumn{1}{p{11cm}}{The number of days without rain for the growing season minus the long run average$^*$} \\
			\multicolumn{1}{l}{Share of rainy days} & \multicolumn{1}{p{11cm}}{The percent of growing season days with rain$^\dagger$} \\
			\multicolumn{1}{l}{Deviation in share of rainy days} & \multicolumn{1}{p{11cm}}{The percent of growing season days with rain minus the long run average$^\dagger$ $^*$} \\
			\multicolumn{1}{l}{Intra-season dry spells} & \multicolumn{1}{p{11cm}}{The maximum length of time (measured in days) without rain during the growing season$^\dagger$} \\
			\midrule
			& \\
			\multicolumn{2}{l}{\emph{\textbf{Panel B}: Temperature}} \\
			\multicolumn{1}{l}{Daily average temperature} & \multicolumn{1}{l}{In $^{\circ}$Celsius} \\
			\multicolumn{1}{l}{Daily maximum temperature} & \multicolumn{1}{l}{In $^{\circ}$Celsius} \\	
			\multicolumn{1}{l}{Mean} & \multicolumn{1}{p{11cm}}{The first moment of the daily temperature distribution for the growing season$^\dagger$} \\
			\multicolumn{1}{l}{Median} & \multicolumn{1}{p{11cm}}{The median daily temperature for the growing season$^\dagger$} \\
			\multicolumn{1}{l}{Variance} & \multicolumn{1}{p{11cm}}{The second moment of the daily temperature distribution for the growing season$^\dagger$} \\
			\multicolumn{1}{l}{Skew} & \multicolumn{1}{p{11cm}}{The third moment of the daily temperature distribution for the growing season$^\dagger$} \\
			\multicolumn{1}{l}{Growing degree days (GDD)} & \multicolumn{1}{p{11cm}}{The number of days within bound temperature for the growing season, following \cite{RS1991}$^\dagger$} \\
			\multicolumn{1}{l}{Deviation in GDD} & \multicolumn{1}{p{11cm}}{GDD for the growing season minus the long run average$^\dagger$ $^*$} \\
			\multicolumn{1}{l}{Scaled deviation in GDD} & \multicolumn{1}{p{11cm}}{The z-score for GDD} \\
			\multicolumn{1}{l}{Maximum temperature} & \multicolumn{1}{p{11cm}}{The average maximum daily temperature} \\
			\\[-1.8ex]\hline 
			\hline \\[-1.8ex]
			\multicolumn{2}{p{\linewidth}}{\footnotesize  \textit{Note}: The table presents definitions for included weather variables and transformations from weather sources defined in Table~\ref{tab:weather}. $^\dagger$Growing season determined for each country following \href{http://www.fao.org/agriculture/seed/cropcalendar/welcome.do}{FAO crop calendar} (see Table~\ref{tab:growseason}). $^*$For variables when ``long run'' is referenced, long run is defined as the entire length of the weather dataset. While each weather source has a different start date, to ensure blinding all datasets were shortened to 1983, which is the latest start date of the data sources.} \\
		\end{tabular}}
	\setbox0=\hbox{\input{tables/summary_stats}}
    \setlength{\linewidth}{\wd0-2\tabcolsep-.25em}
    \input{tables/summary_stats}}
\end{table}


\newpage
\begin{table}[htbp]	\centering
	\caption{Growing Seasons} \label{tab:growseason}
	\scalebox{0.9}
	{ \setlength{\linewidth}{.1cm}\newcommand{\contents}
		{\begin{tabular}{llll}
			\\[-1.8ex]\hline 
			\hline \\[-1.8ex]
			& \multicolumn{1}{c}{Growing Season} & \multicolumn{1}{c}{Span Calendar Years} & \multicolumn{1}{c}{Season Modality}  \\
			\multicolumn{1}{l}{\href{http://www.fao.org/giews/countrybrief/country.jsp?code=ETH}{Ethiopia}} & \multicolumn{1}{l}{1 March - 30 November} & \multicolumn{1}{c}{no} & \multicolumn{1}{c}{unimodal}  \\
			\multicolumn{1}{l}{\href{http://www.fao.org/giews/countrybrief/country.jsp?code=MWI}{Malawi}} & \multicolumn{1}{l}{1 October - 30 April} & \multicolumn{1}{c}{yes} & \multicolumn{1}{c}{unimodal}  \\
			\multicolumn{1}{l}{\href{http://www.fao.org/giews/countrybrief/country.jsp?code=NER}{Niger}} & \multicolumn{1}{l}{1 June - 30 November} & \multicolumn{1}{c}{no} & \multicolumn{1}{c}{unimodal}  \\
			\multicolumn{1}{l}{\href{http://www.fao.org/giews/countrybrief/country.jsp?code=NGA}{Nigeria}} & \multicolumn{1}{l}{\emph{North}: 1 May - 30 September} & \multicolumn{1}{c}{no} & \multicolumn{1}{c}{bimodal}  \\
			 & \multicolumn{1}{l}{\emph{South}: 1 March - 31 August} &  &  \\
			\multicolumn{1}{l}{\href{http://www.fao.org/giews/countrybrief/country.jsp?code=TZA}{Tanzania}} & \multicolumn{1}{l}{1 November - 30 April} & \multicolumn{1}{c}{yes} & \multicolumn{1}{c}{unimodal}  \\
			\multicolumn{1}{l}{\href{http://www.fao.org/giews/countrybrief/country.jsp?code=UGA}{Uganda}} & \multicolumn{1}{l}{\emph{North}: 1 April - 30 September}  & \multicolumn{1}{c}{no} & \multicolumn{1}{c}{bimodal} \\
	    	& \multicolumn{1}{l}{\emph{South}: 1 February - 31 July} & &  \\
			\\[-1.8ex]\hline 
			\hline \\[-1.8ex]
			\multicolumn{4}{p{\linewidth}}{\footnotesize  \textit{Note}: \footnotesize The table presents the growing season ranges, as defined by following FAO \href{http://www.fao.org/agriculture/seed/cropcalendar/welcome.do}{crop calendar} for each country, respectively.} \\
		\end{tabular}}
	\setbox0=\hbox{\input{tables/summary_stats}}
    \setlength{\linewidth}{\wd0-2\tabcolsep-.25em}
    \input{tables/summary_stats}}
\end{table}

\subsection*{Household Data Summary Statistics}


\begin{table}[htbp]	\centering
    \caption{Household Data Summary Statistics \label{tab:sumstattab}}
	\scalebox{.9}
	{ \setlength{\linewidth}{.2cm}\newcommand{\contents}
		{\input{tables/summary_stats}}
	\setbox0=\hbox{\input{tables/summary_stats}}
    \setlength{\linewidth}{\wd0-2\tabcolsep-.25em}
    \input{tables/summary_stats}}
\end{table}

\subsection*{Specification and Bumpline Charts}



\begin{center}
\begin{figure}[!htbp]
	\begin{minipage}{\linewidth}
		\caption{Specification Charts for Mean Daily Rainfall}
		\label{fig:pval_v1}
		\begin{center}
			\includegraphics[width=.49\linewidth,keepaspectratio]{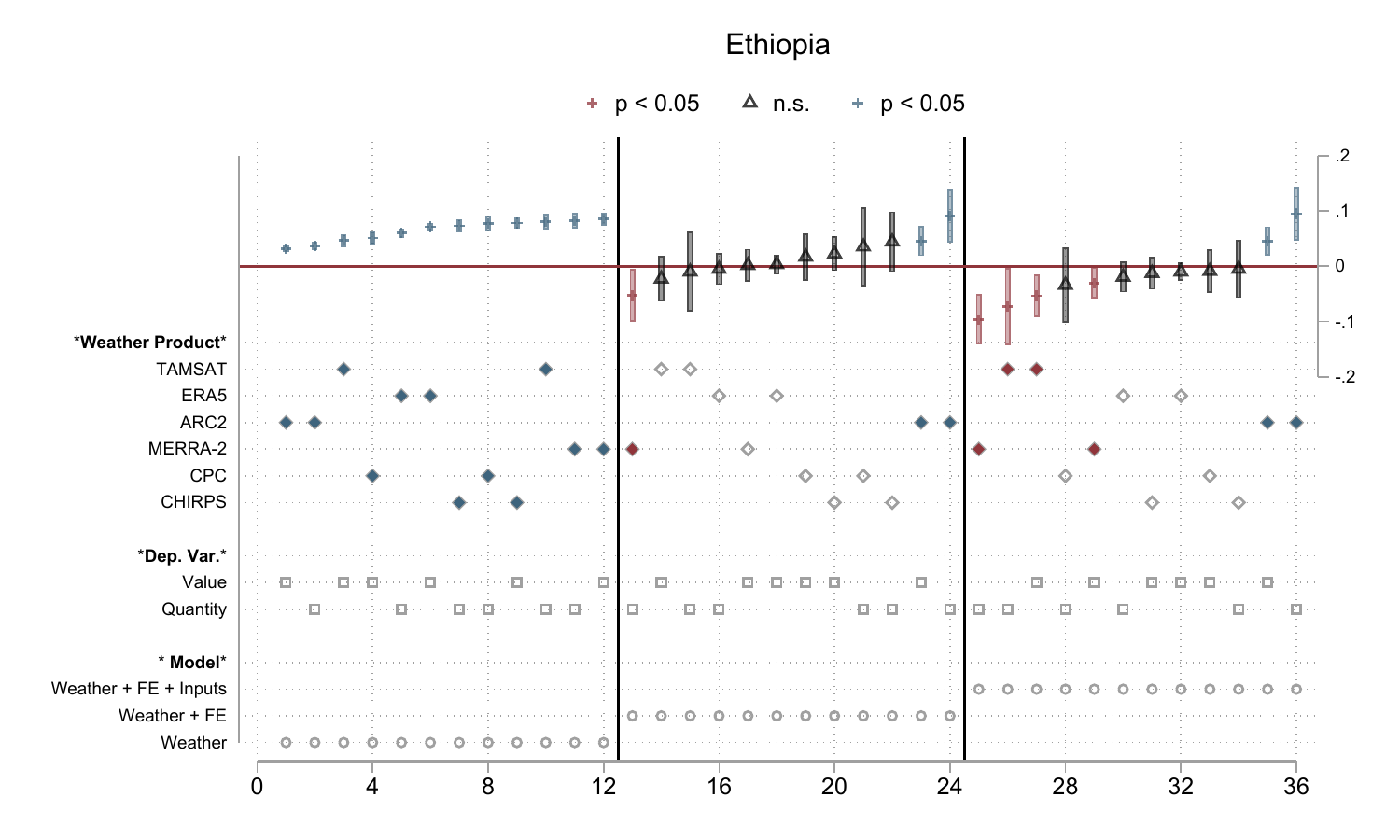}
			\includegraphics[width=.49\linewidth,keepaspectratio]{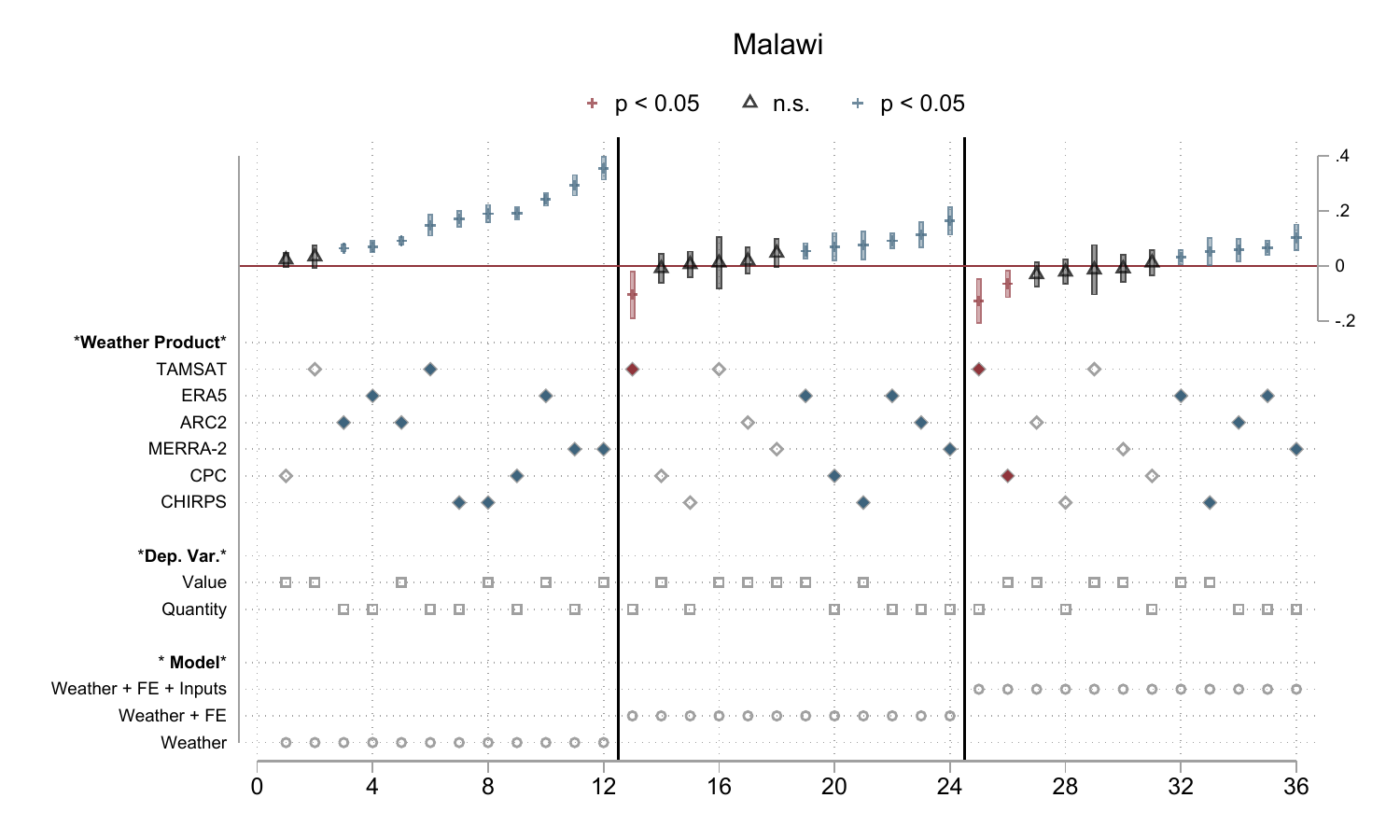}
			\includegraphics[width=.49\linewidth,keepaspectratio]{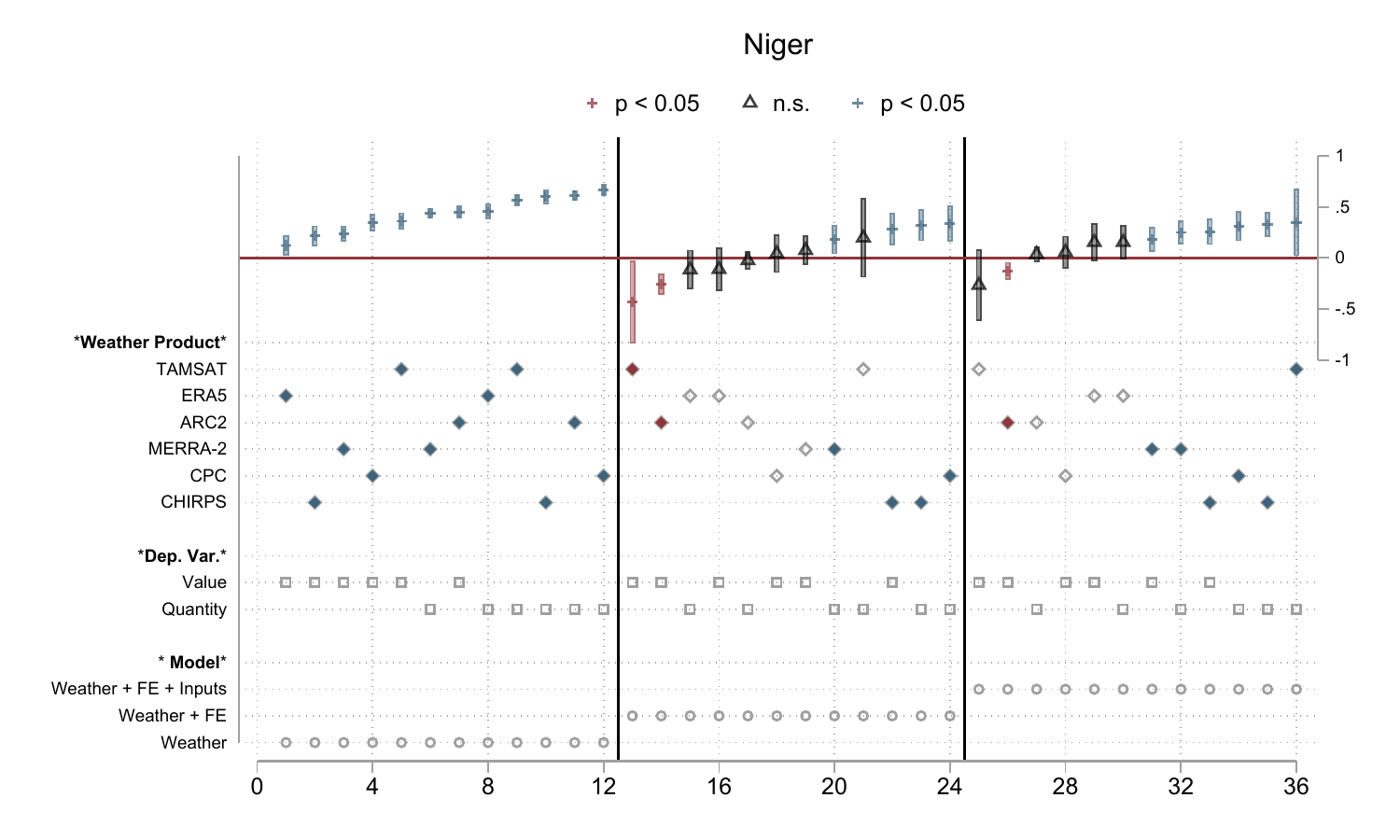}
			\includegraphics[width=.49\linewidth,keepaspectratio]{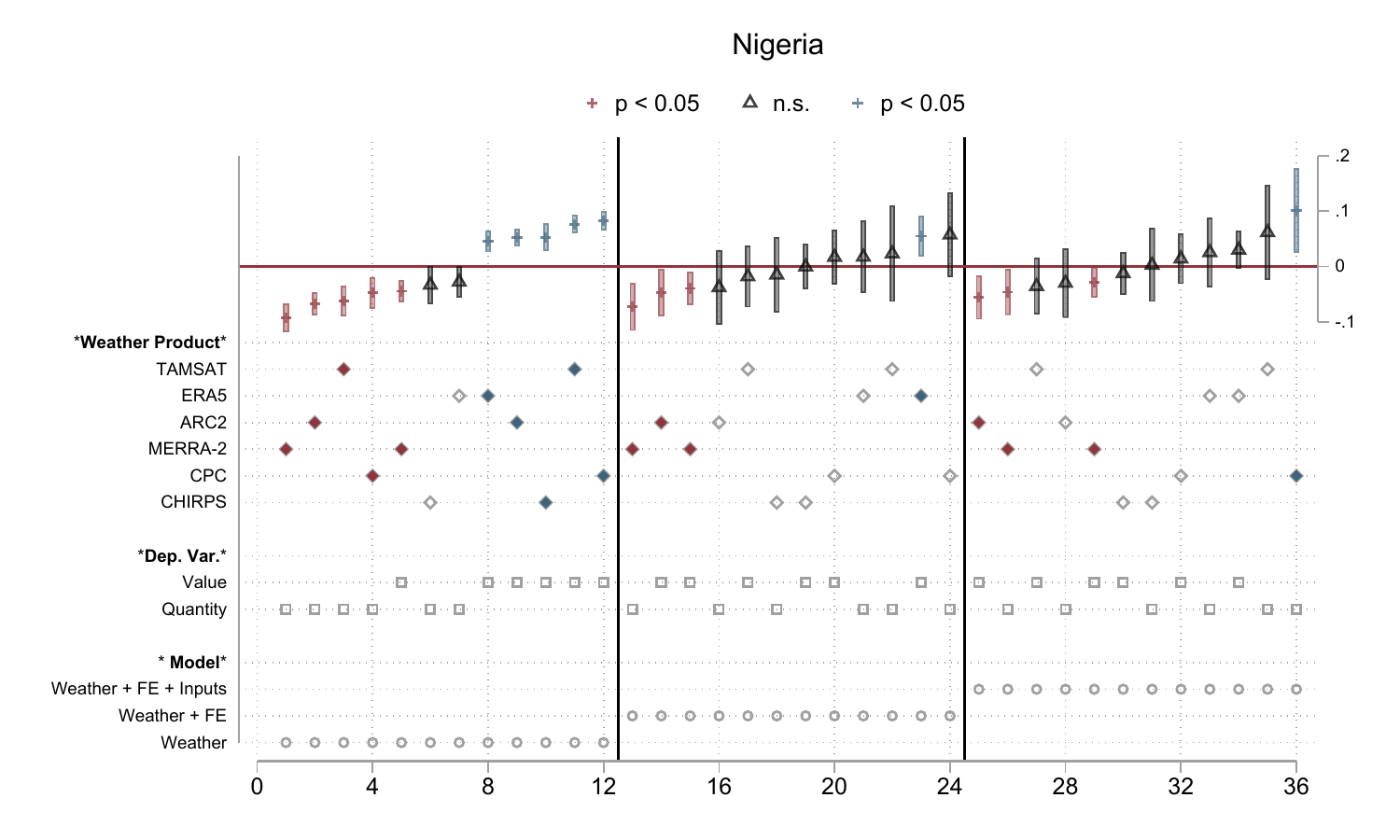}
			\includegraphics[width=.49\linewidth,keepaspectratio]{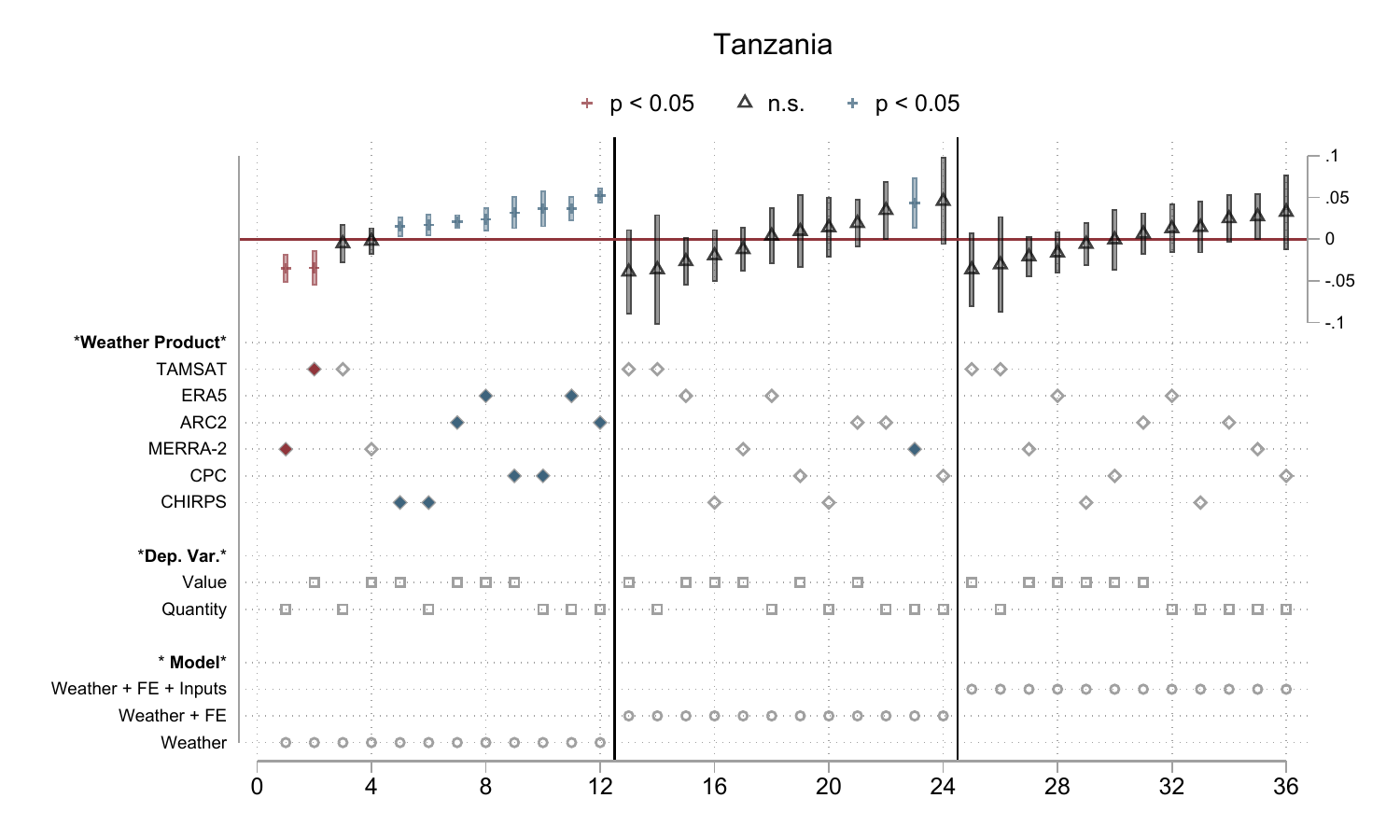}
			\includegraphics[width=.49\linewidth,keepaspectratio]{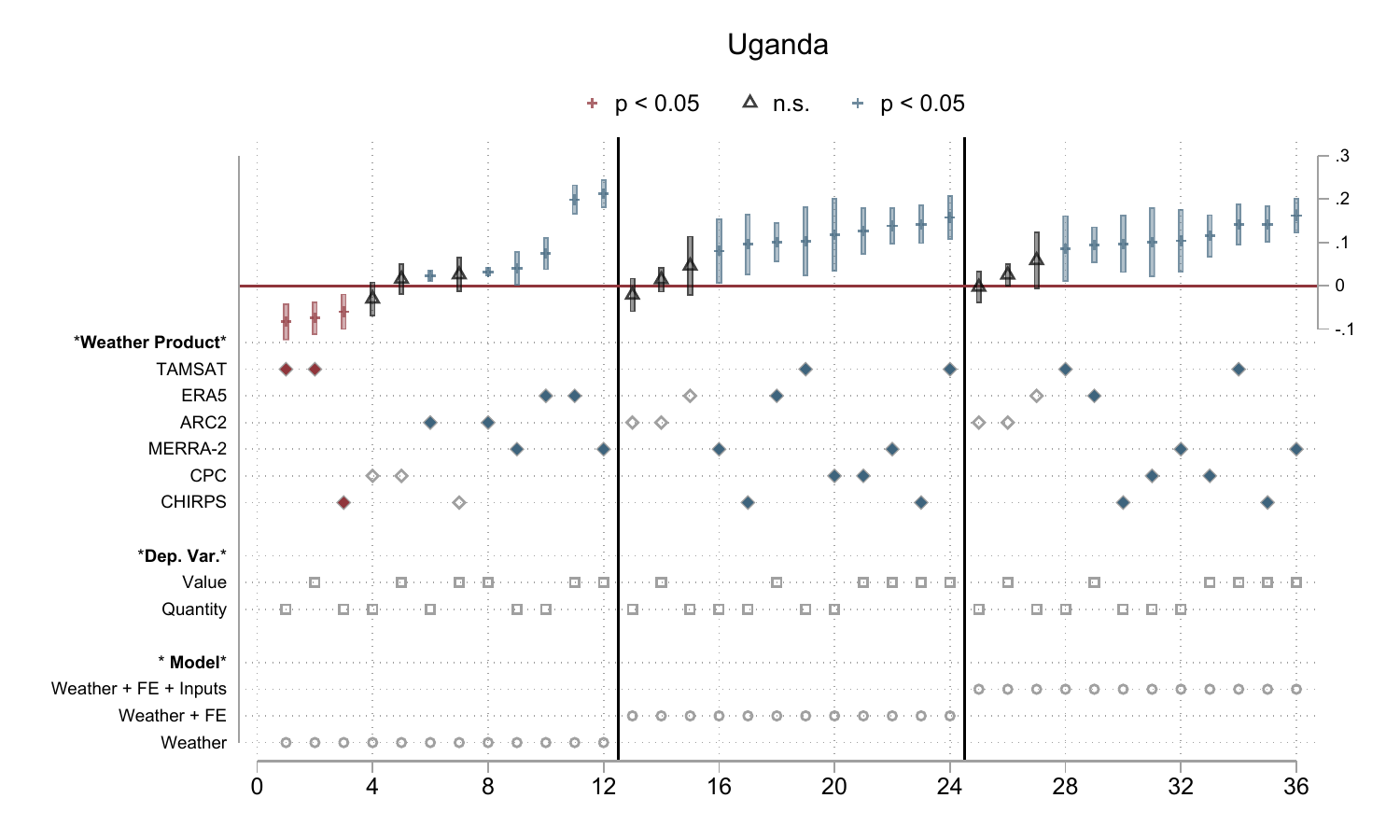}
		\end{center}
		\footnotesize  \textit{Note}: The figure presents specification curves, where each panel represents a different country, with three different models presented within each panel. Each panel includes 36 regressions, where each column represents a single regression. Significant and non-significant coefficients are designated at the top of the figure. For each Earth observation product, we also designate the significance and sign of the coefficient with color: red represents coefficients which are negative and significant; white represents insignificant coefficients, regardless of sign; and blue represents coefficients which are positive and significant.  
	\end{minipage}	
\end{figure}
\end{center}

\begin{center}
\begin{figure}[!htbp]
	\begin{minipage}{\linewidth}
		\caption{Specification Charts for Median Daily Rainfall}
		\label{fig:pval_v2}
		\begin{center}
			\includegraphics[width=.49\linewidth,keepaspectratio]{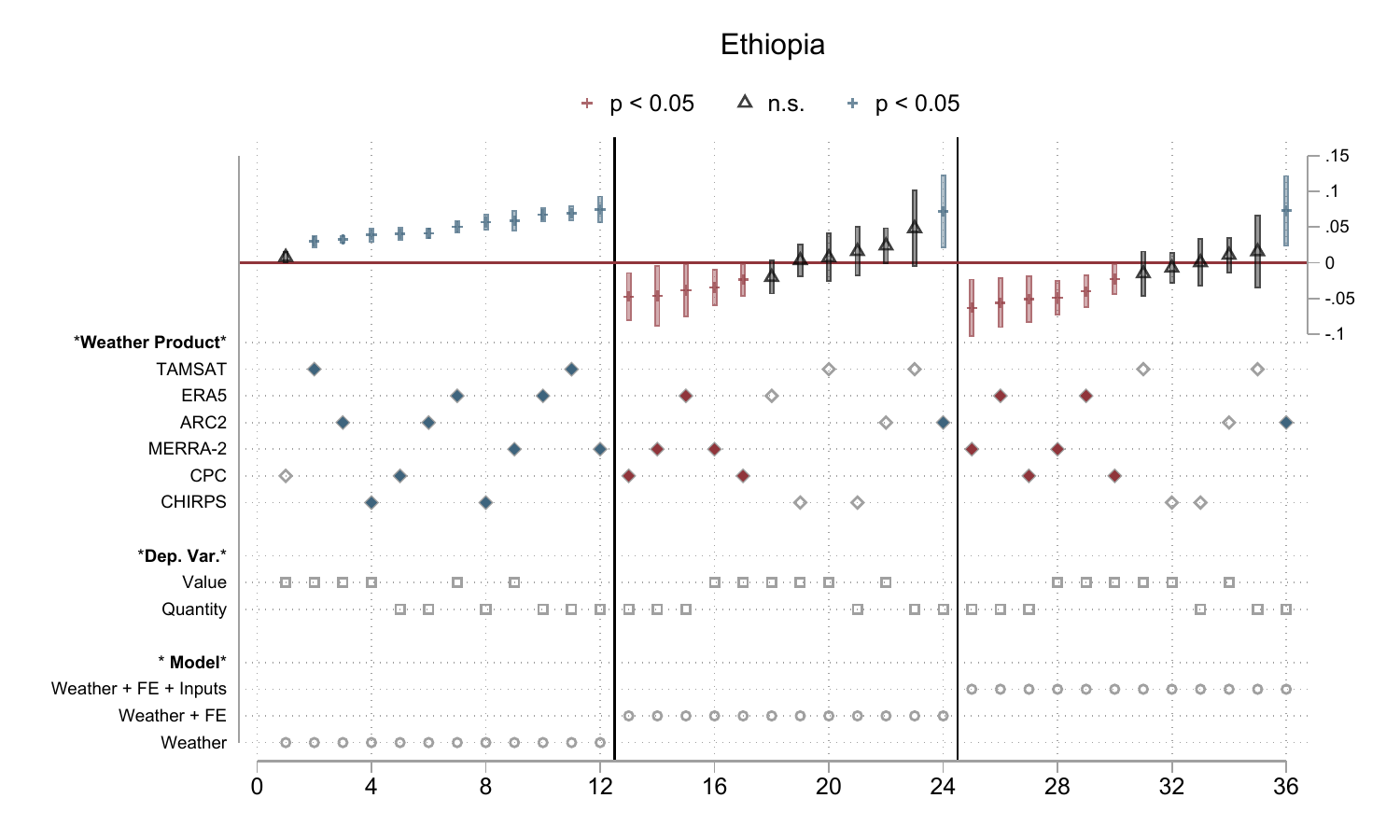}
			\includegraphics[width=.49\linewidth,keepaspectratio]{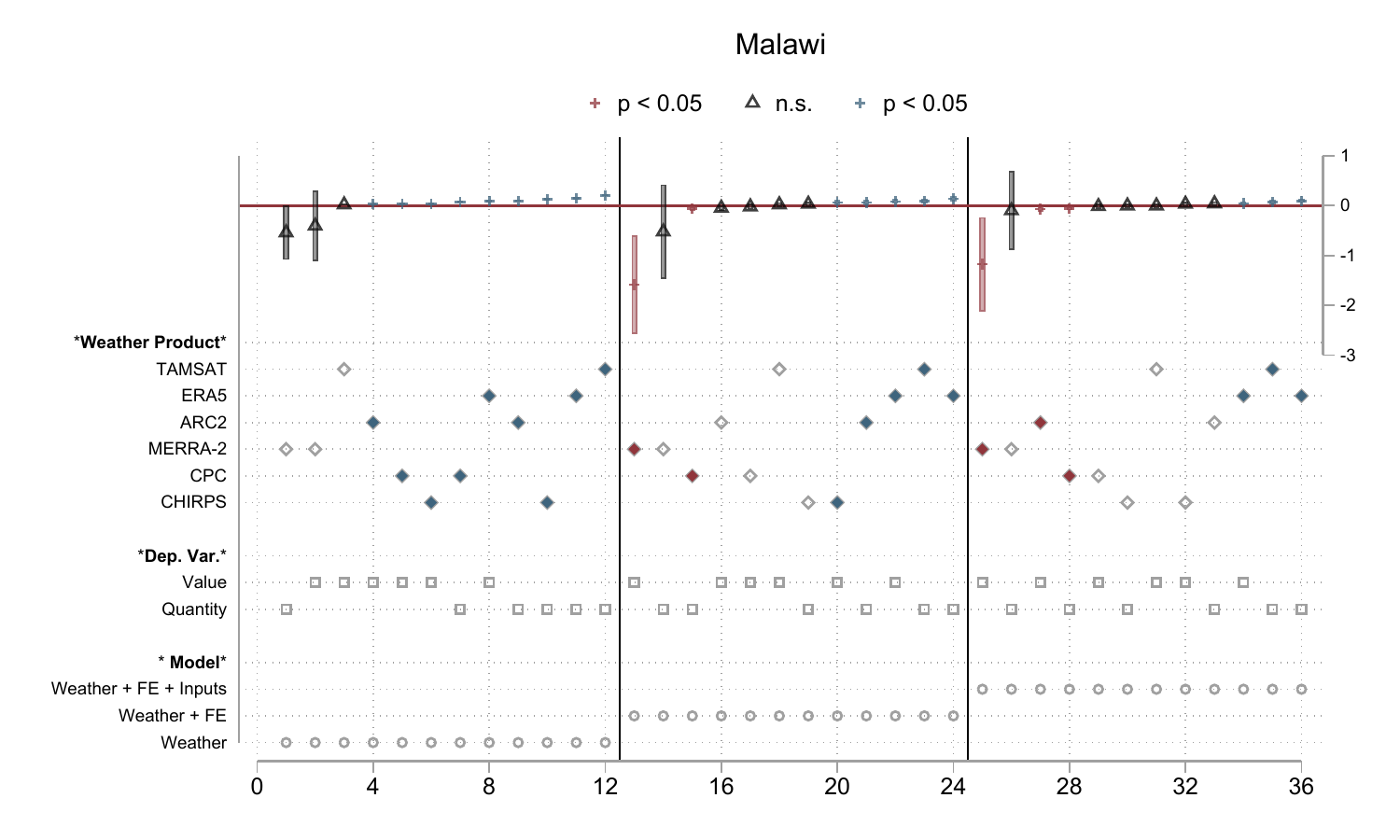}
			\includegraphics[width=.49\linewidth,keepaspectratio]{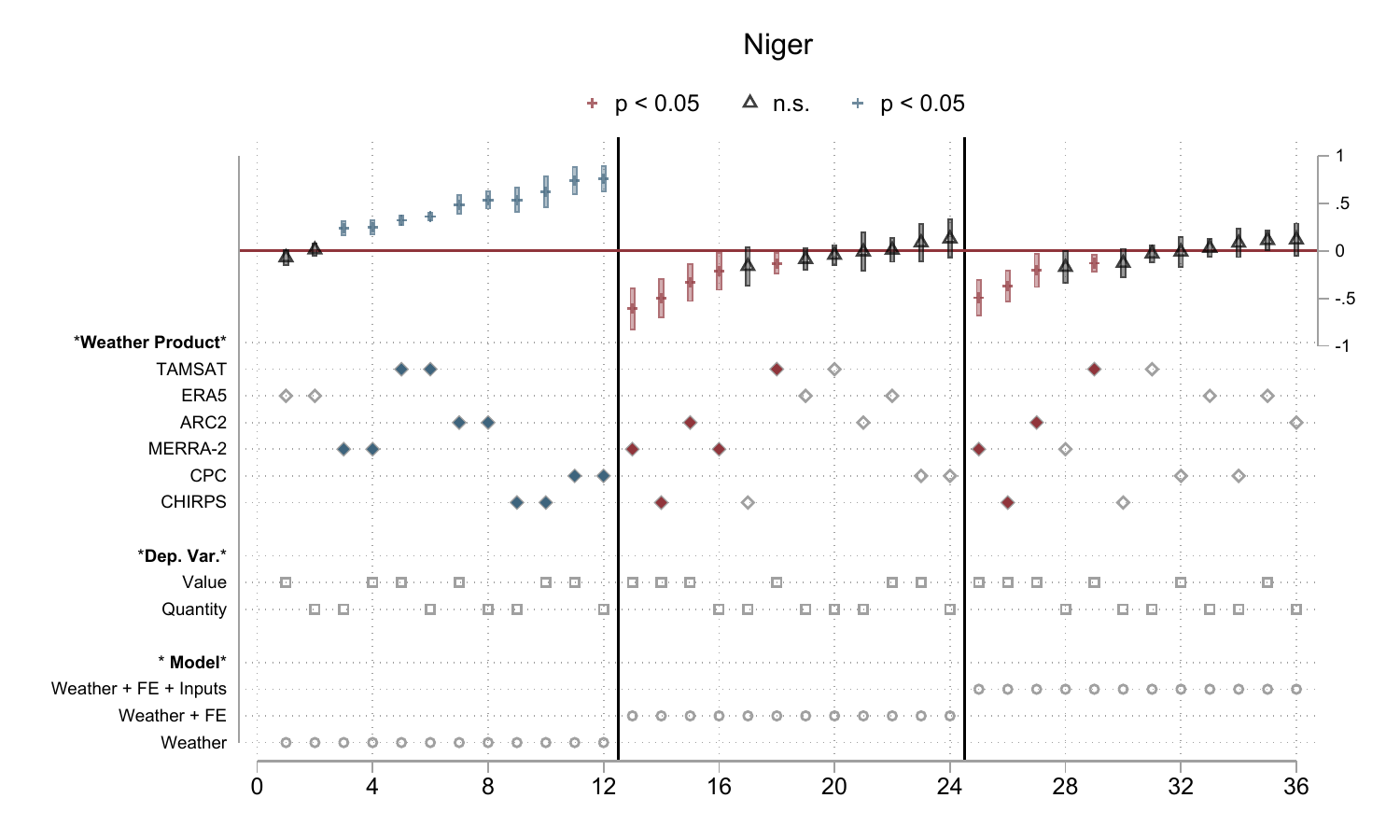}
			\includegraphics[width=.49\linewidth,keepaspectratio]{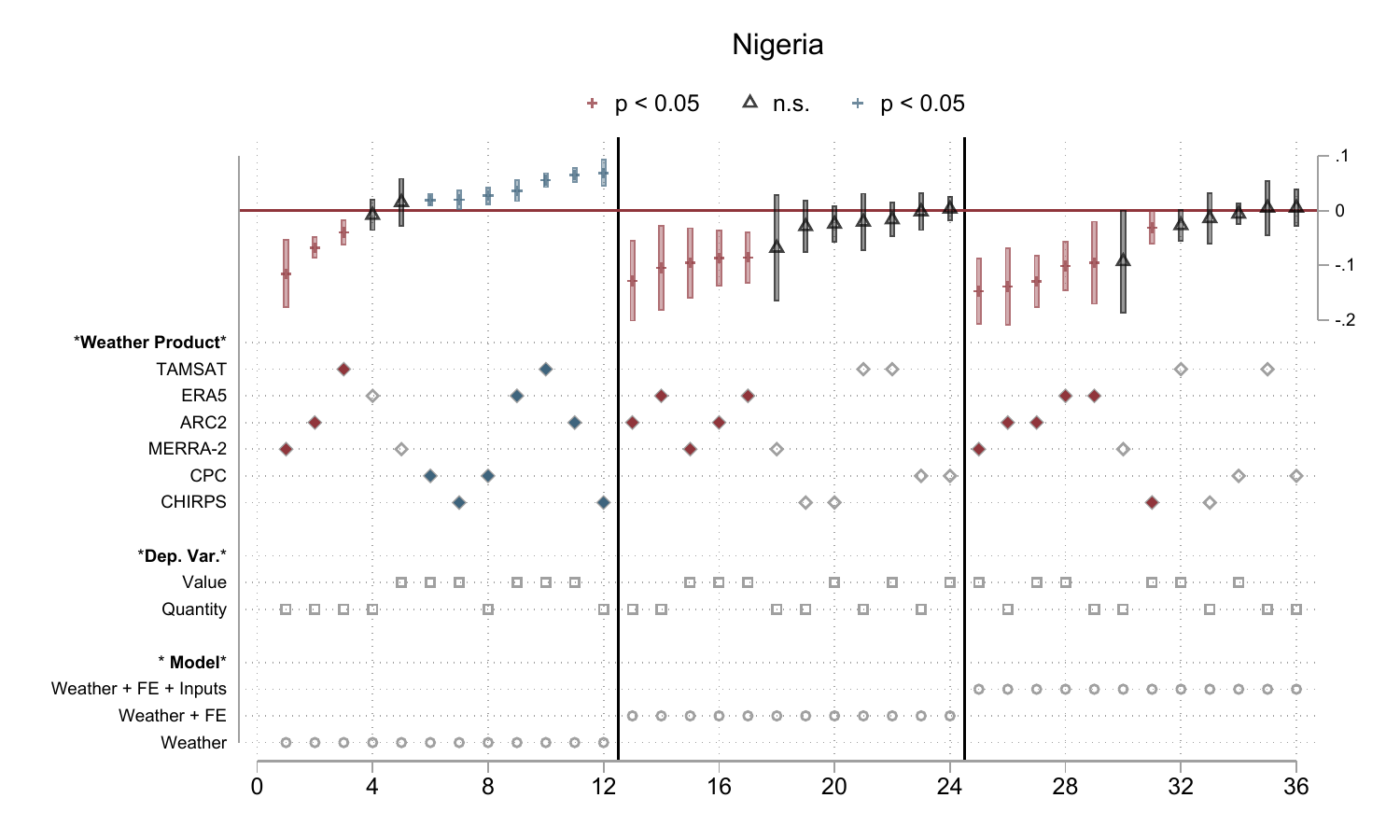}
			\includegraphics[width=.49\linewidth,keepaspectratio]{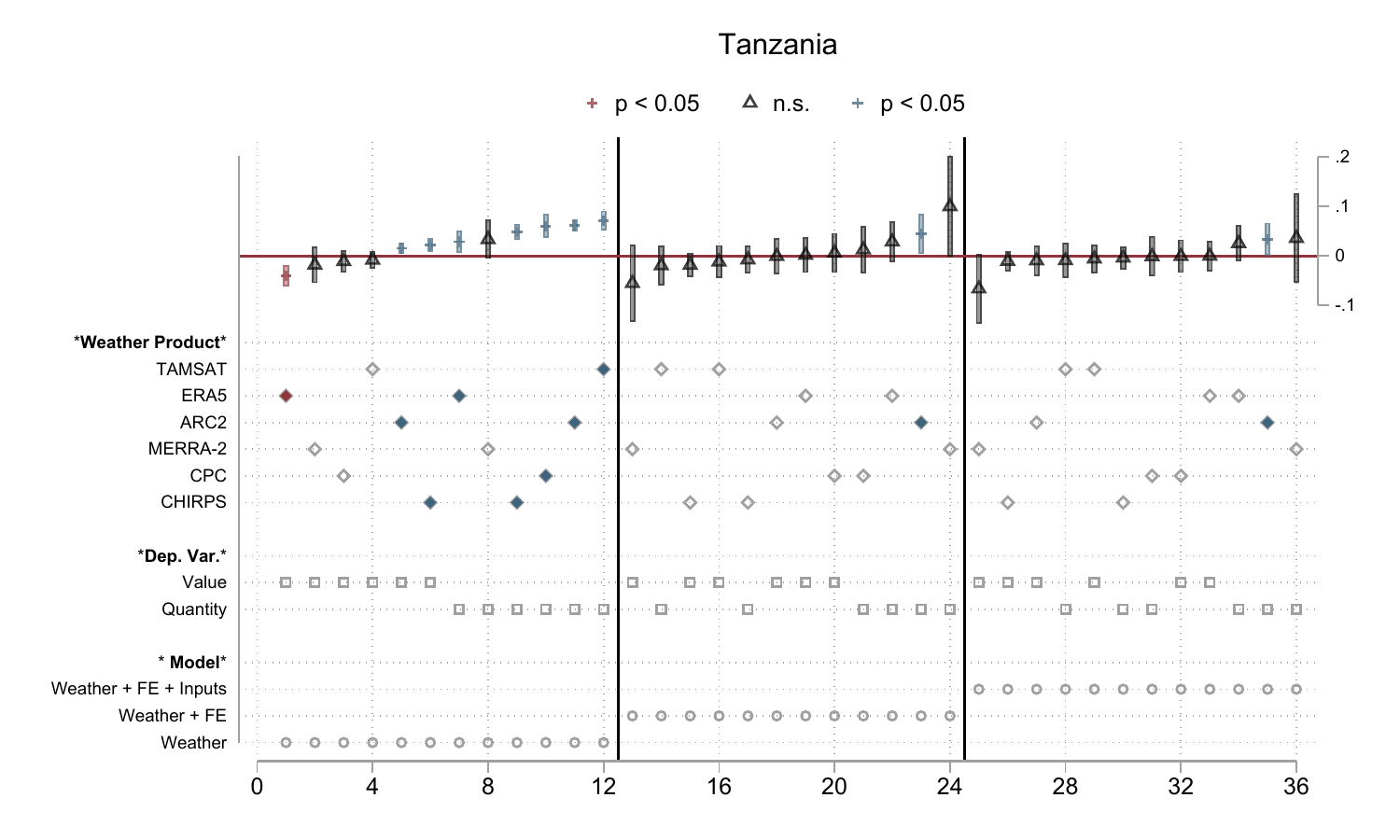}
			\includegraphics[width=.49\linewidth,keepaspectratio]{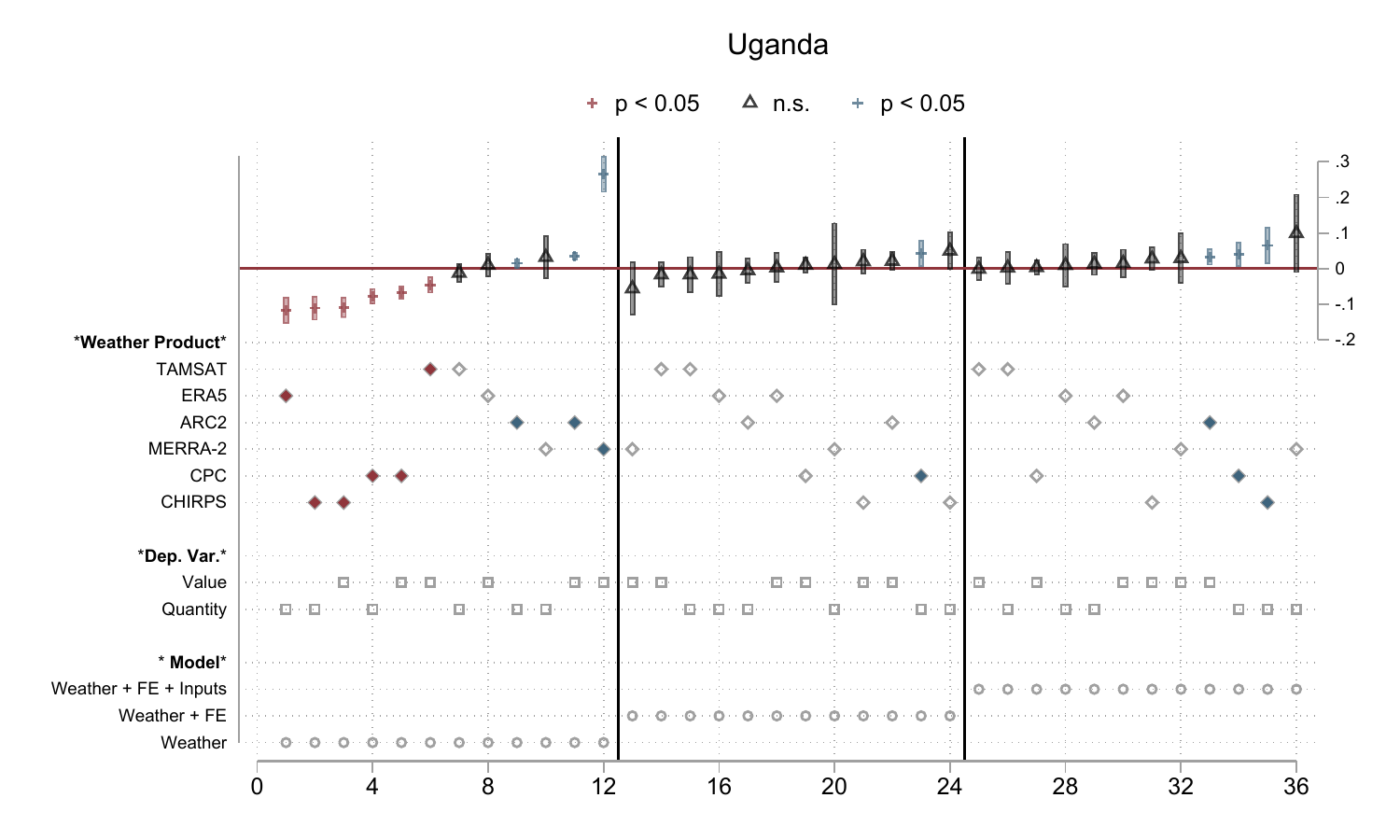}
		\end{center}
		\footnotesize  \textit{Note}: The figure presents specification curves, where each panel represents a different country, with three different models presented within each panel. Each panel includes 36 regressions, where each column represents a single regression. Significant and non-significant coefficients are designated at the top of the figure. For each Earth observation product, we also designate the significance and sign of the coefficient with color: red represents coefficients which are negative and significant; white represents insignificant coefficients, regardless of sign; and blue represents coefficients which are positive and significant.  
	\end{minipage}	
\end{figure}
\end{center}

\begin{center}
\begin{figure}[!htbp]
	\begin{minipage}{\linewidth}
		\caption{Specification Charts for Variance of Daily Rainfall}
		\label{fig:pval_v3}
		\begin{center}
			\includegraphics[width=.49\linewidth,keepaspectratio]{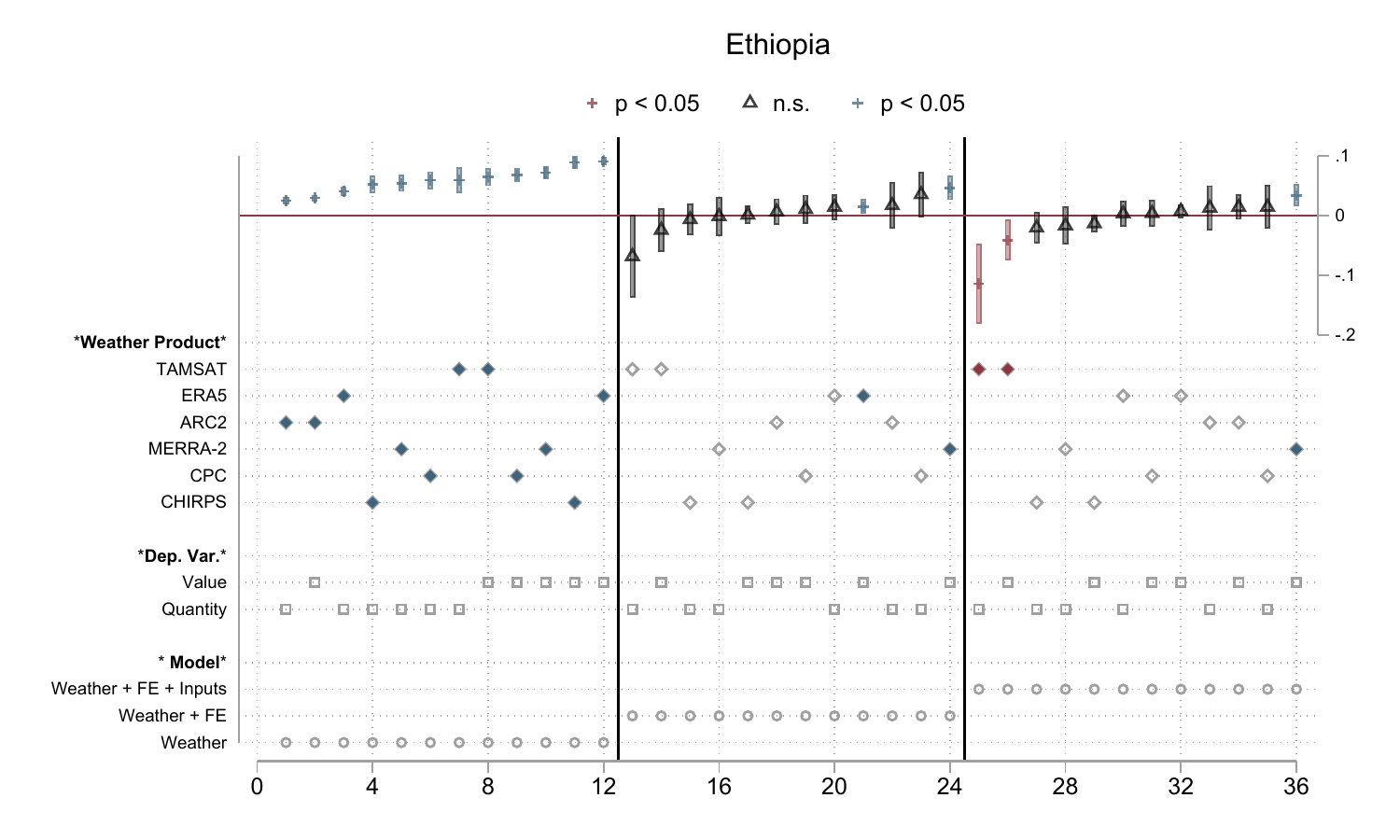}
			\includegraphics[width=.49\linewidth,keepaspectratio]{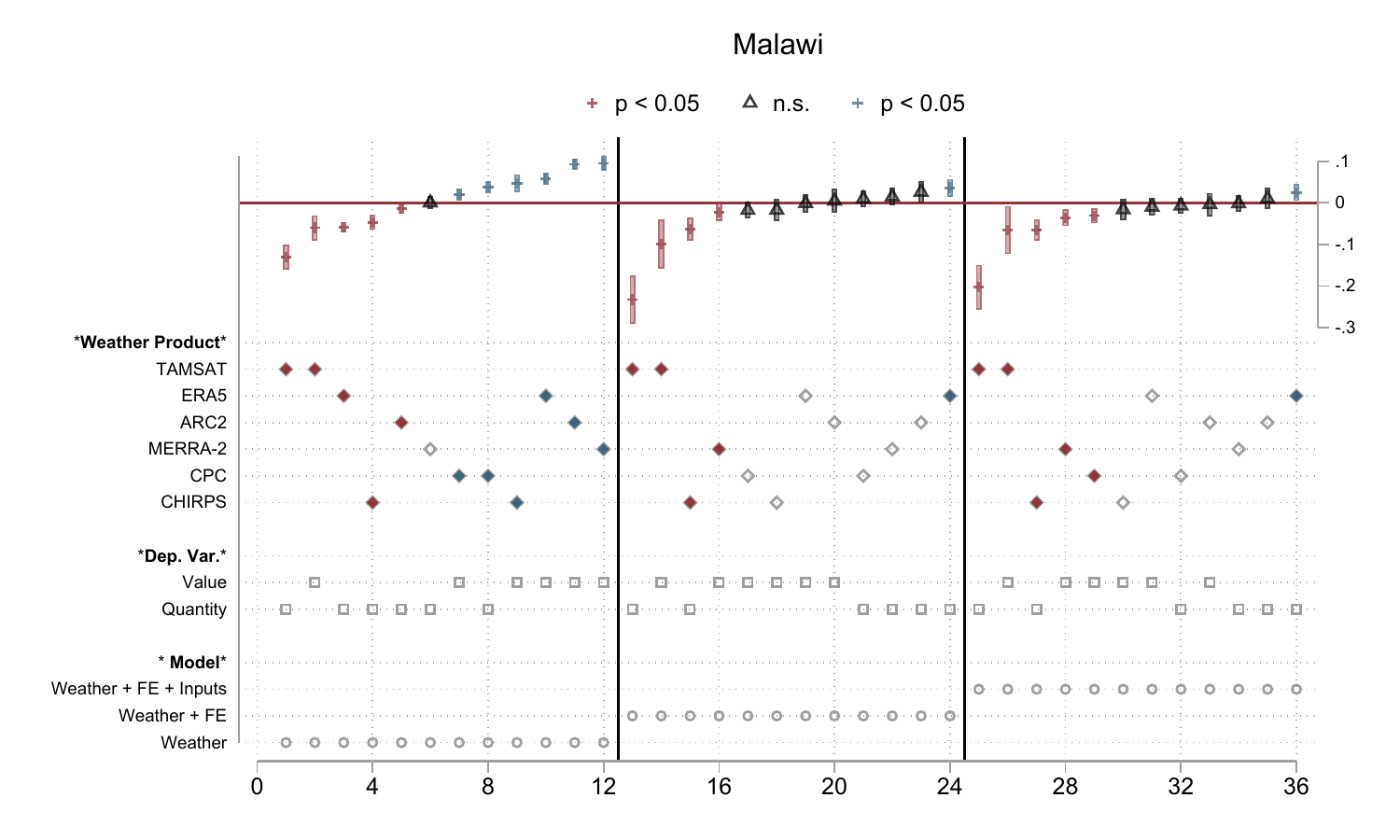}
			\includegraphics[width=.49\linewidth,keepaspectratio]{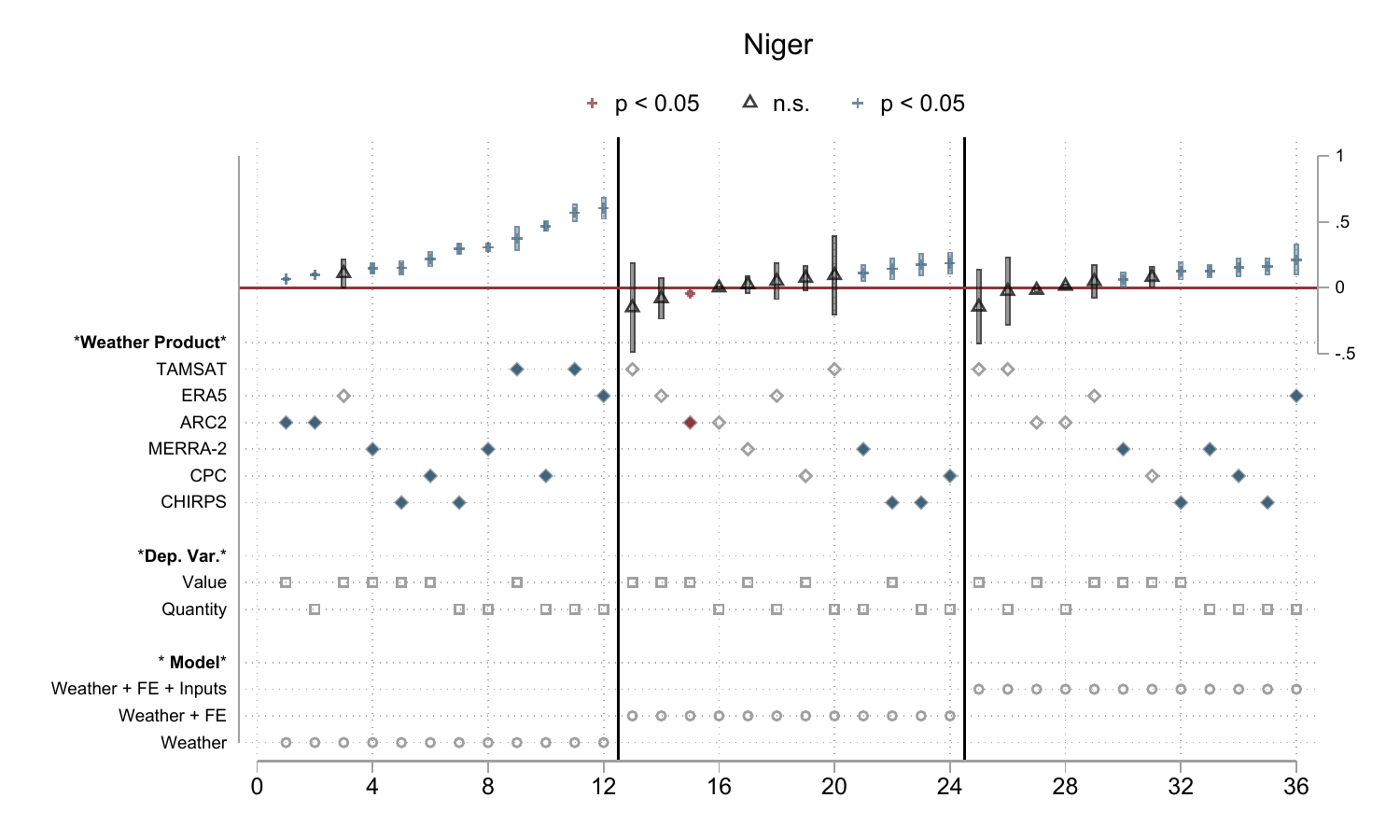}
			\includegraphics[width=.49\linewidth,keepaspectratio]{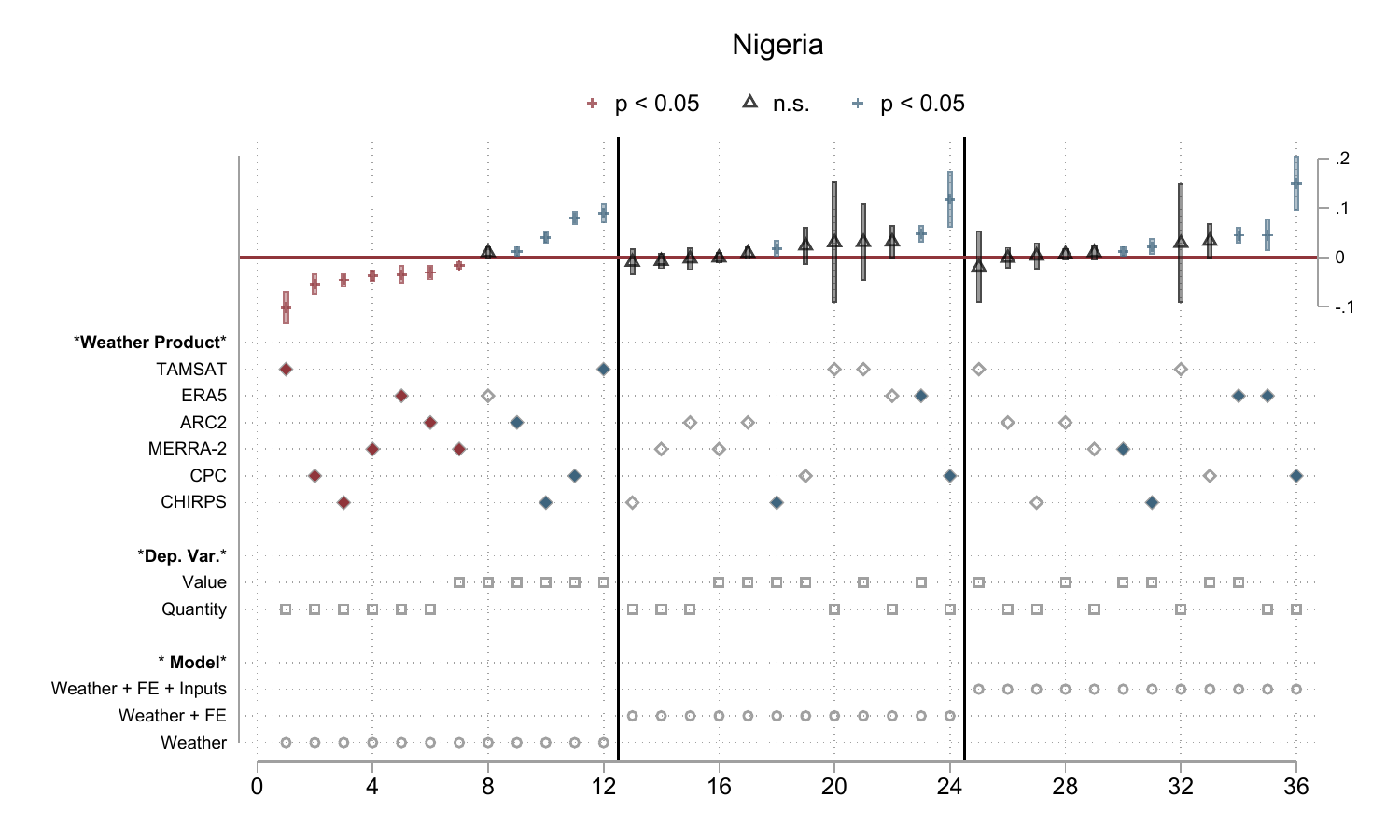}
			\includegraphics[width=.49\linewidth,keepaspectratio]{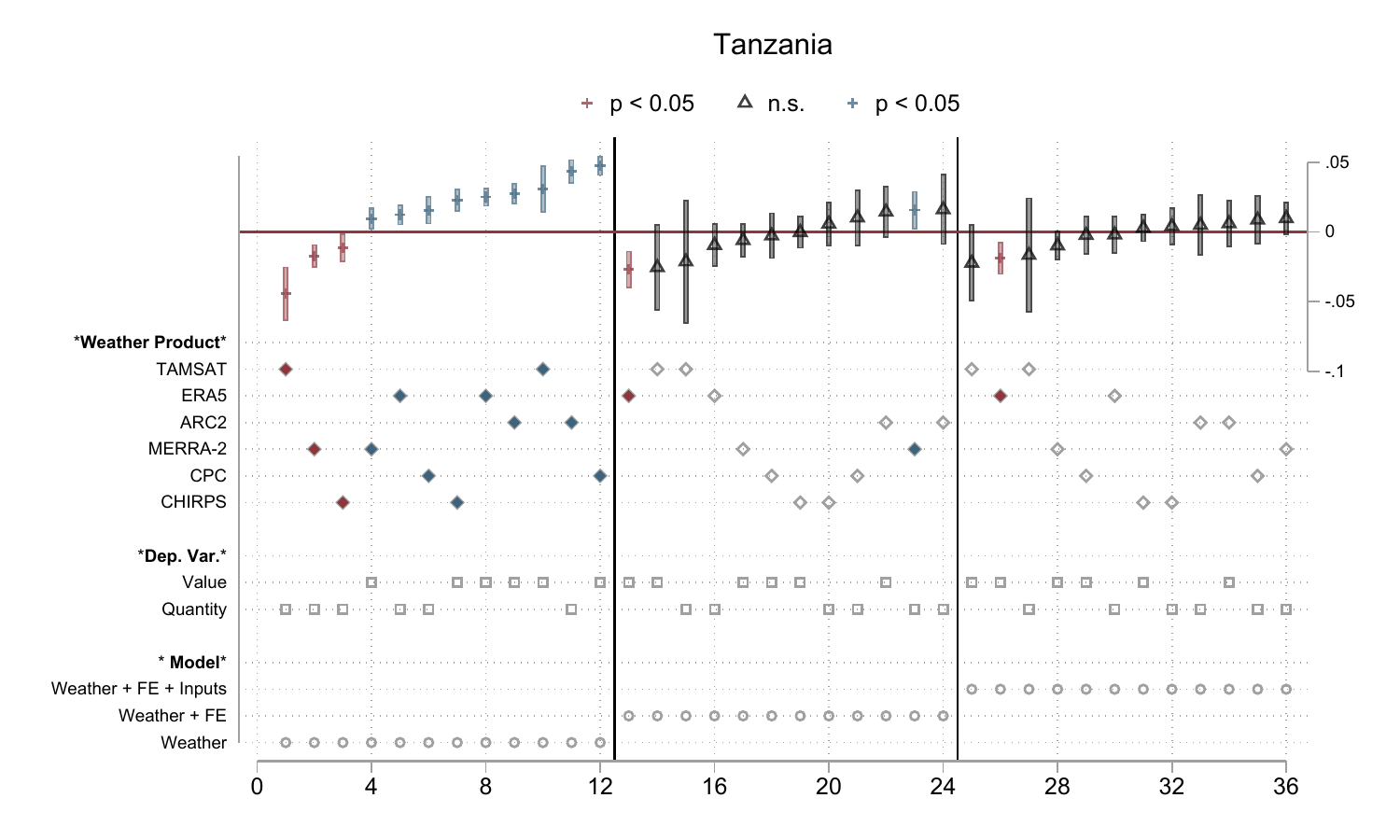}
			\includegraphics[width=.49\linewidth,keepaspectratio]{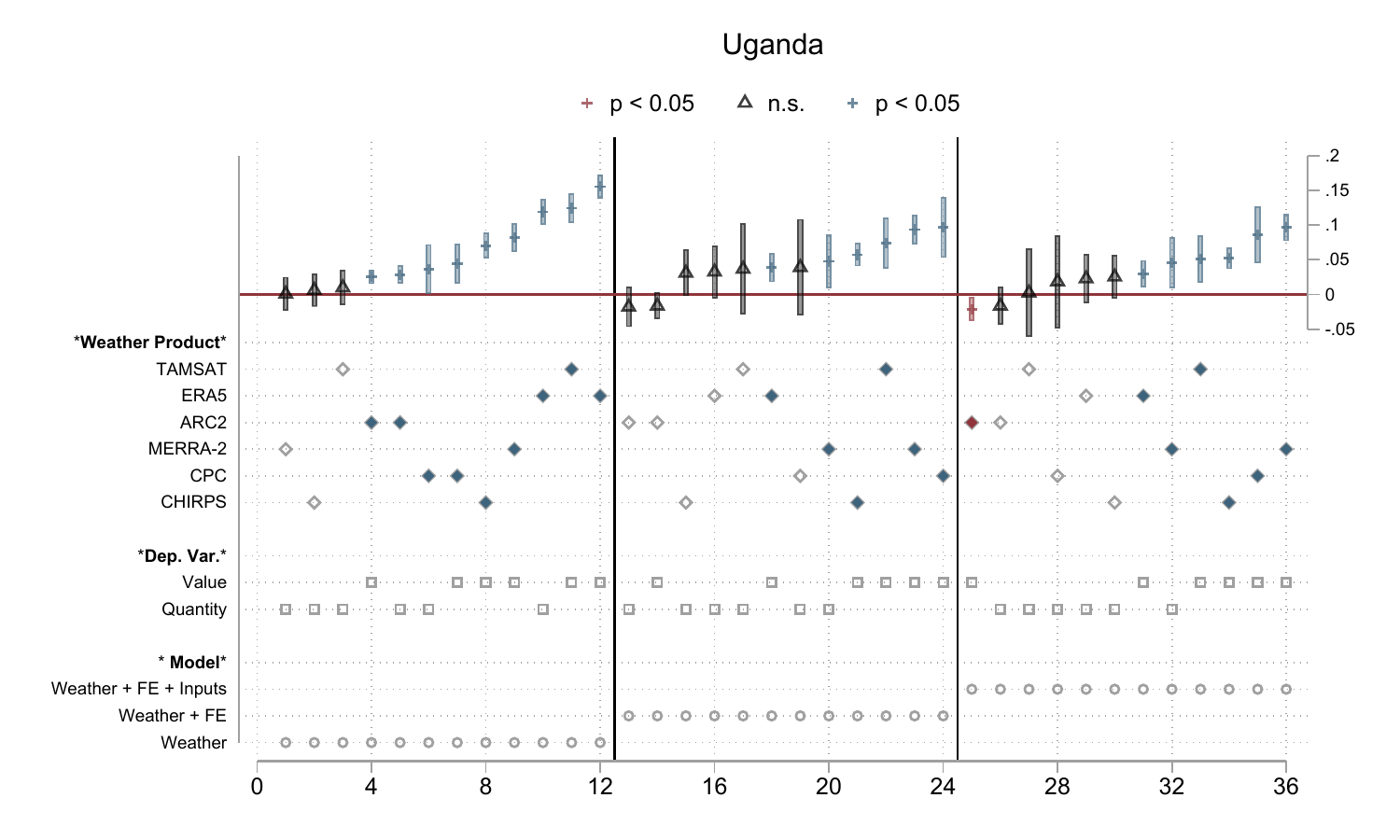}
		\end{center}
		\footnotesize  \textit{Note}: The figure presents specification curves, where each panel represents a different country, with three different models presented within each panel. Each panel includes 36 regressions, where each column represents a single regression. Significant and non-significant coefficients are designated at the top of the figure. For each Earth observation product, we also designate the significance and sign of the coefficient with color: red represents coefficients which are negative and significant; white represents insignificant coefficients, regardless of sign; and blue represents coefficients which are positive and significant.  
	\end{minipage}	
\end{figure}
\end{center}

\begin{center}
\begin{figure}[!htbp]
	\begin{minipage}{\linewidth}
		\caption{Specification Charts for Skew of Daily Rainfall}
		\label{fig:pval_v4}
		\begin{center}
			\includegraphics[width=.49\linewidth,keepaspectratio]{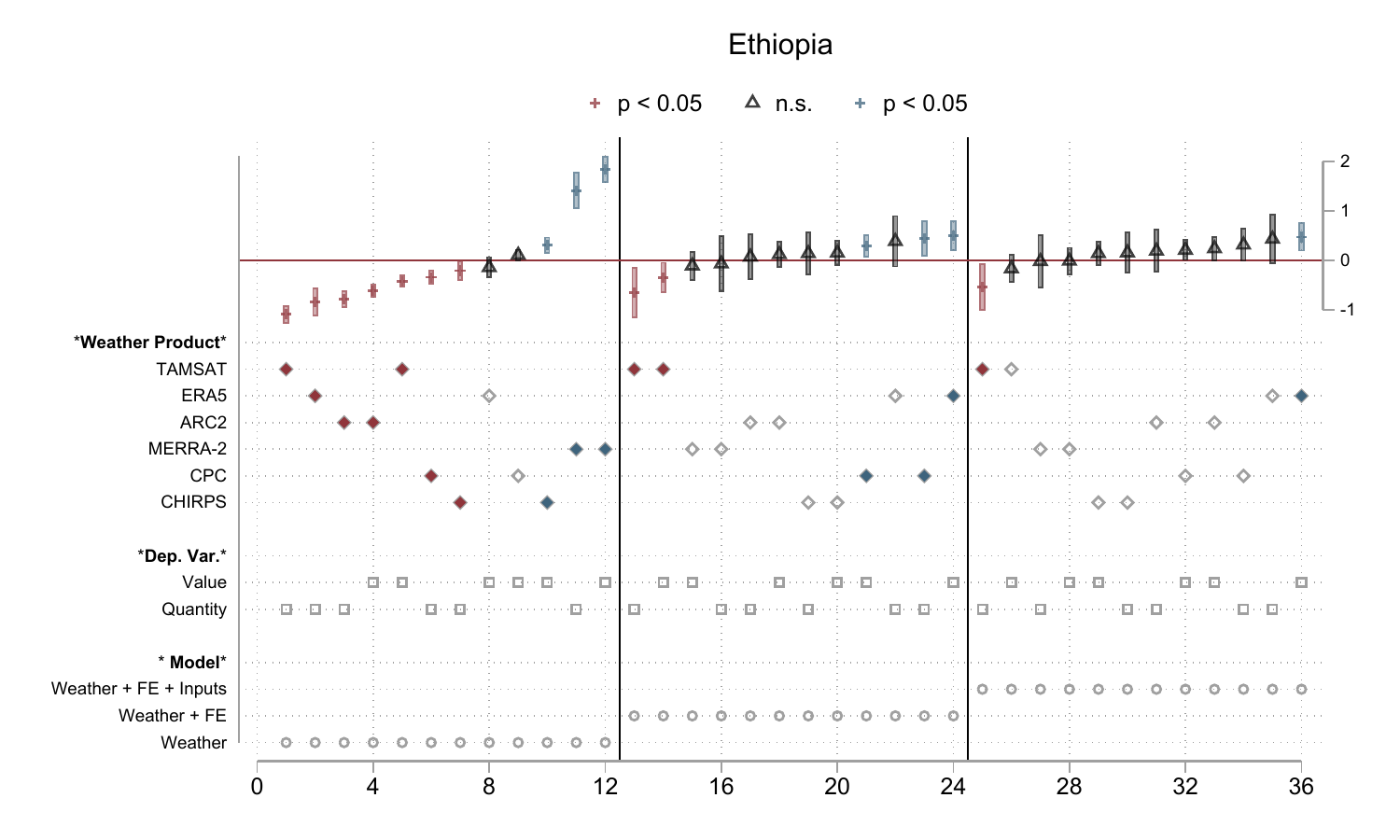}
			\includegraphics[width=.49\linewidth,keepaspectratio]{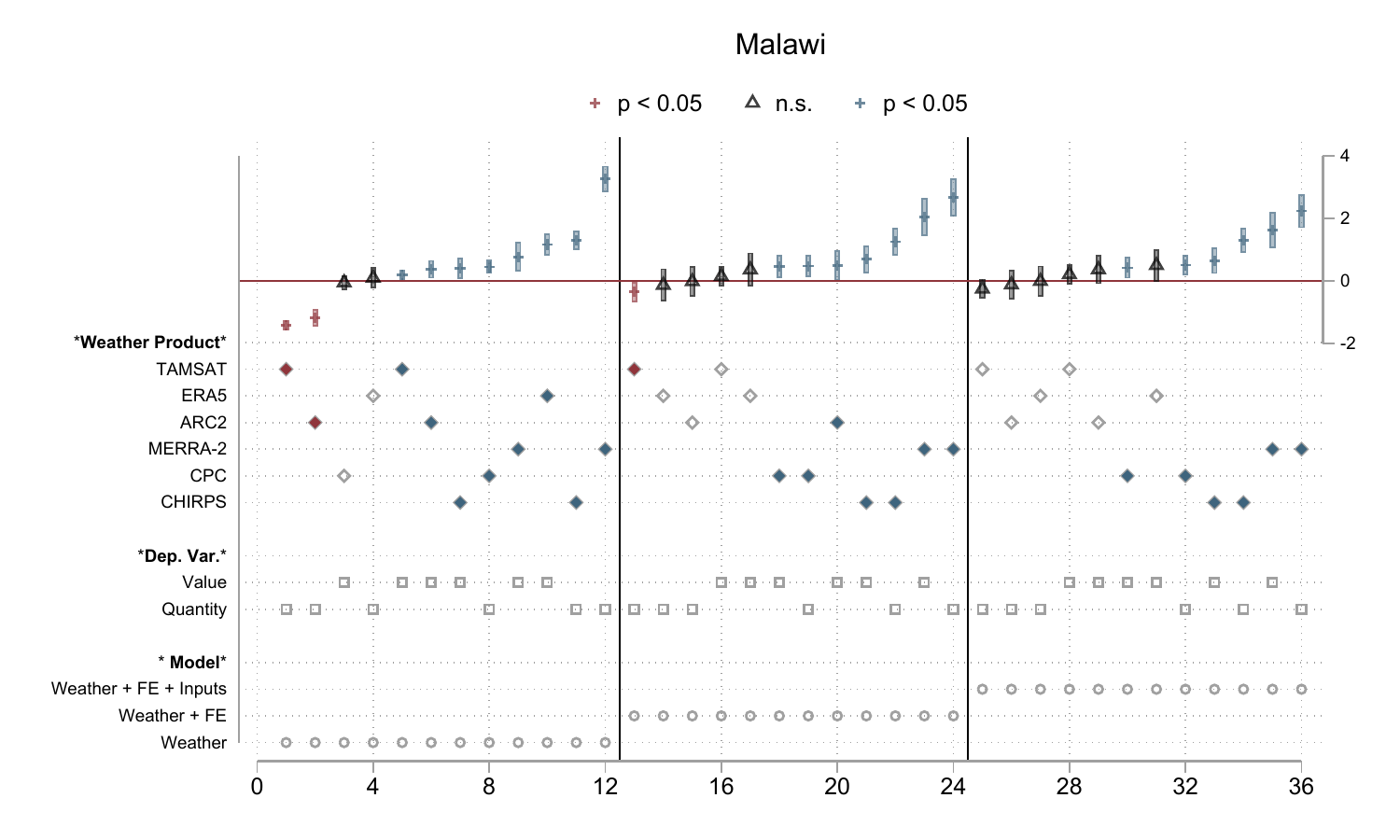}
			\includegraphics[width=.49\linewidth,keepaspectratio]{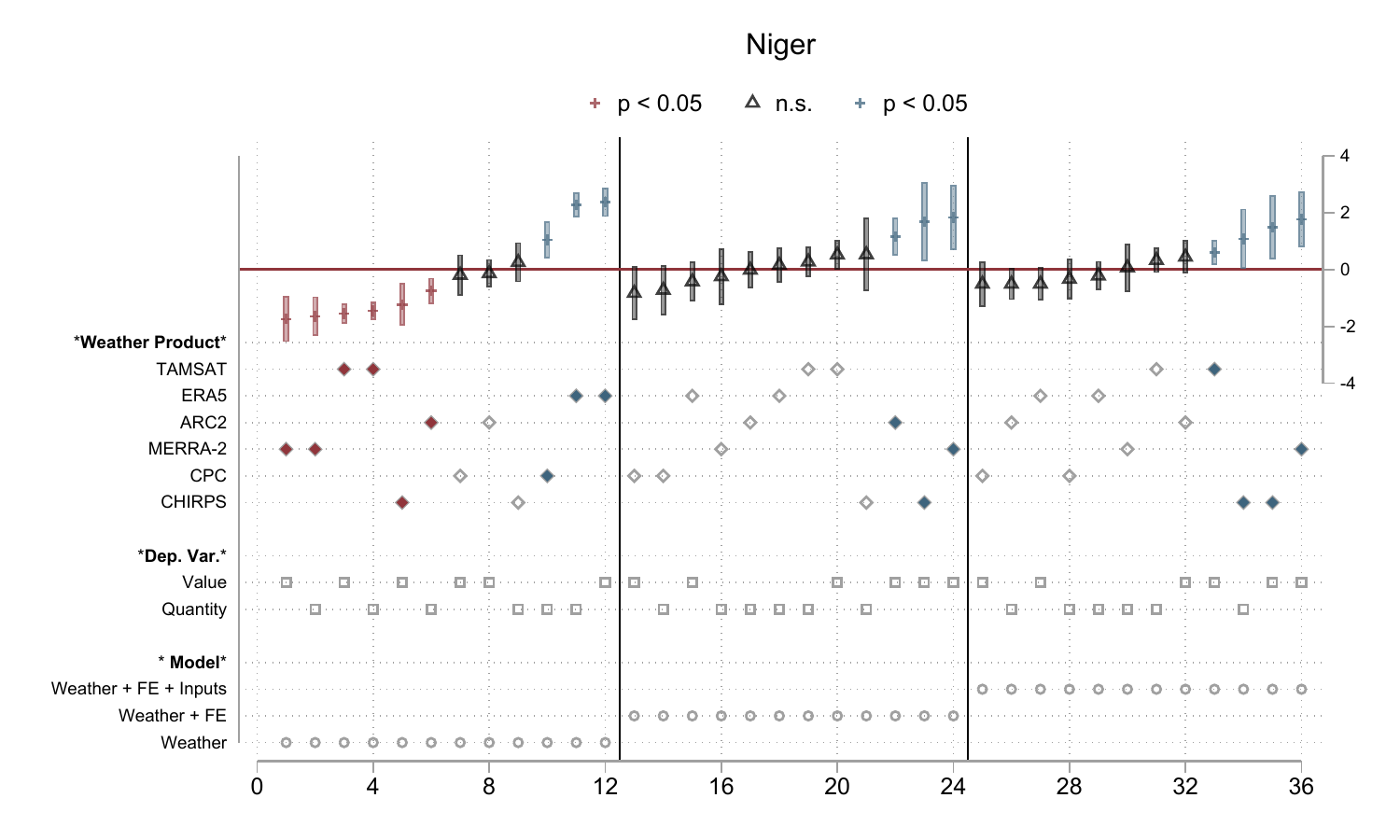}
			\includegraphics[width=.49\linewidth,keepaspectratio]{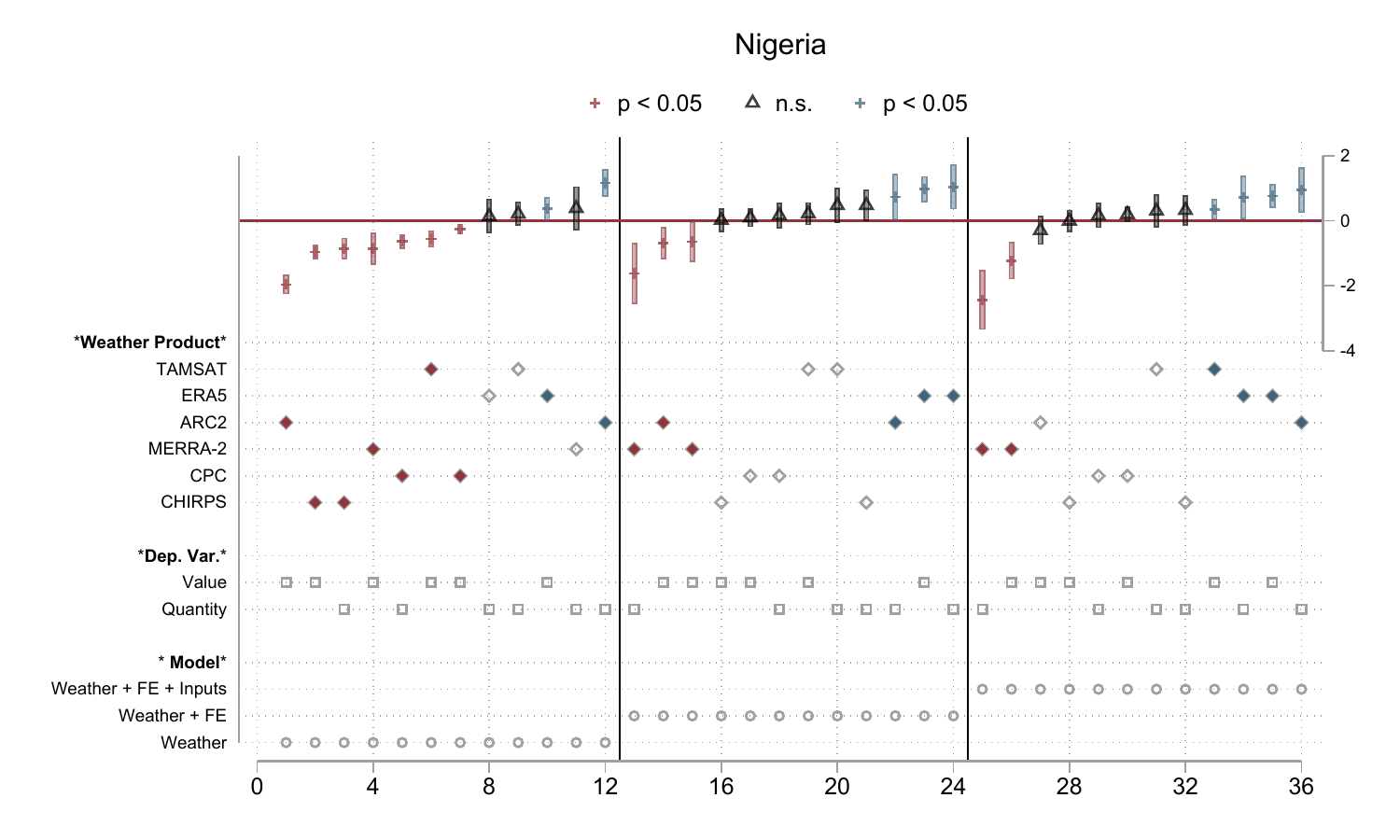}
			\includegraphics[width=.49\linewidth,keepaspectratio]{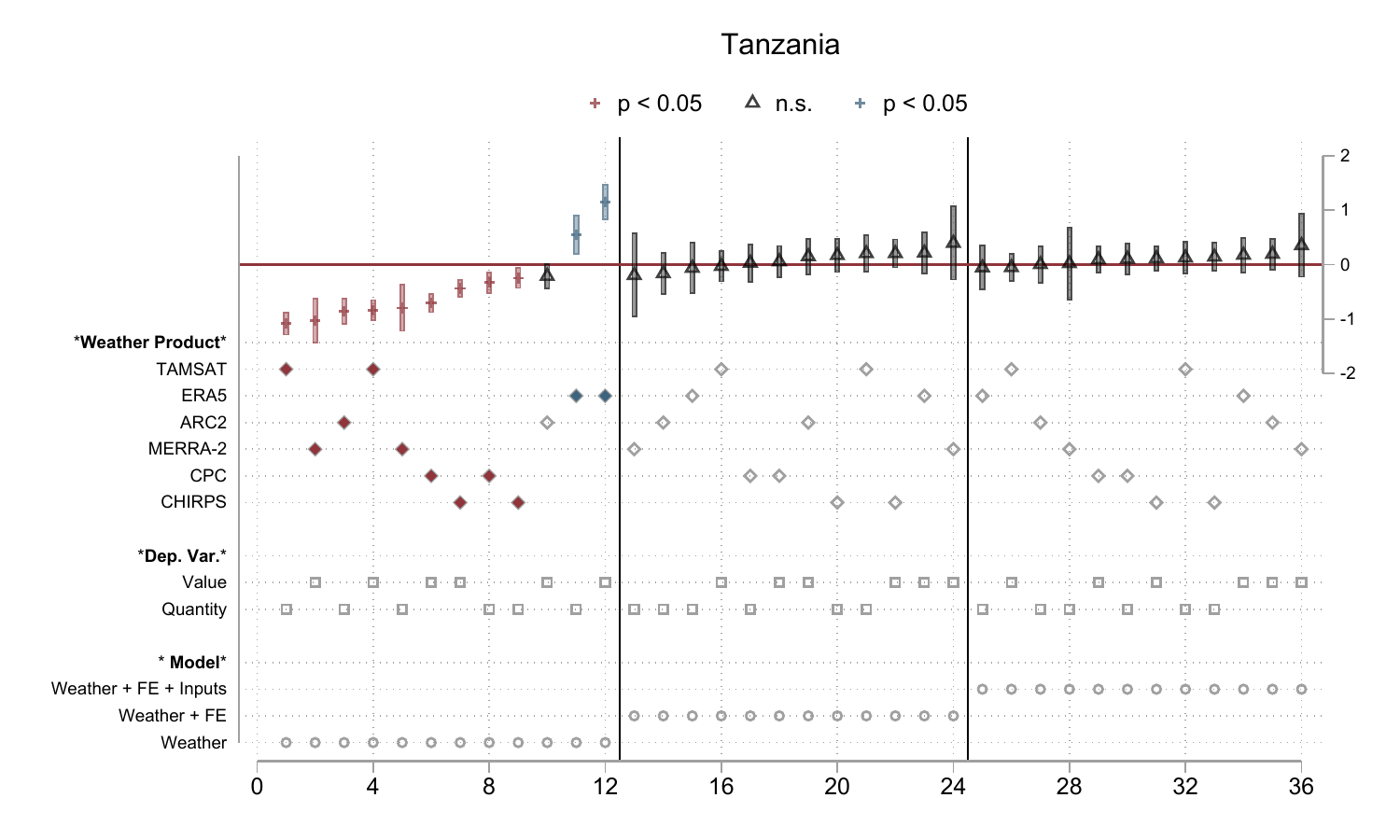}
			\includegraphics[width=.49\linewidth,keepaspectratio]{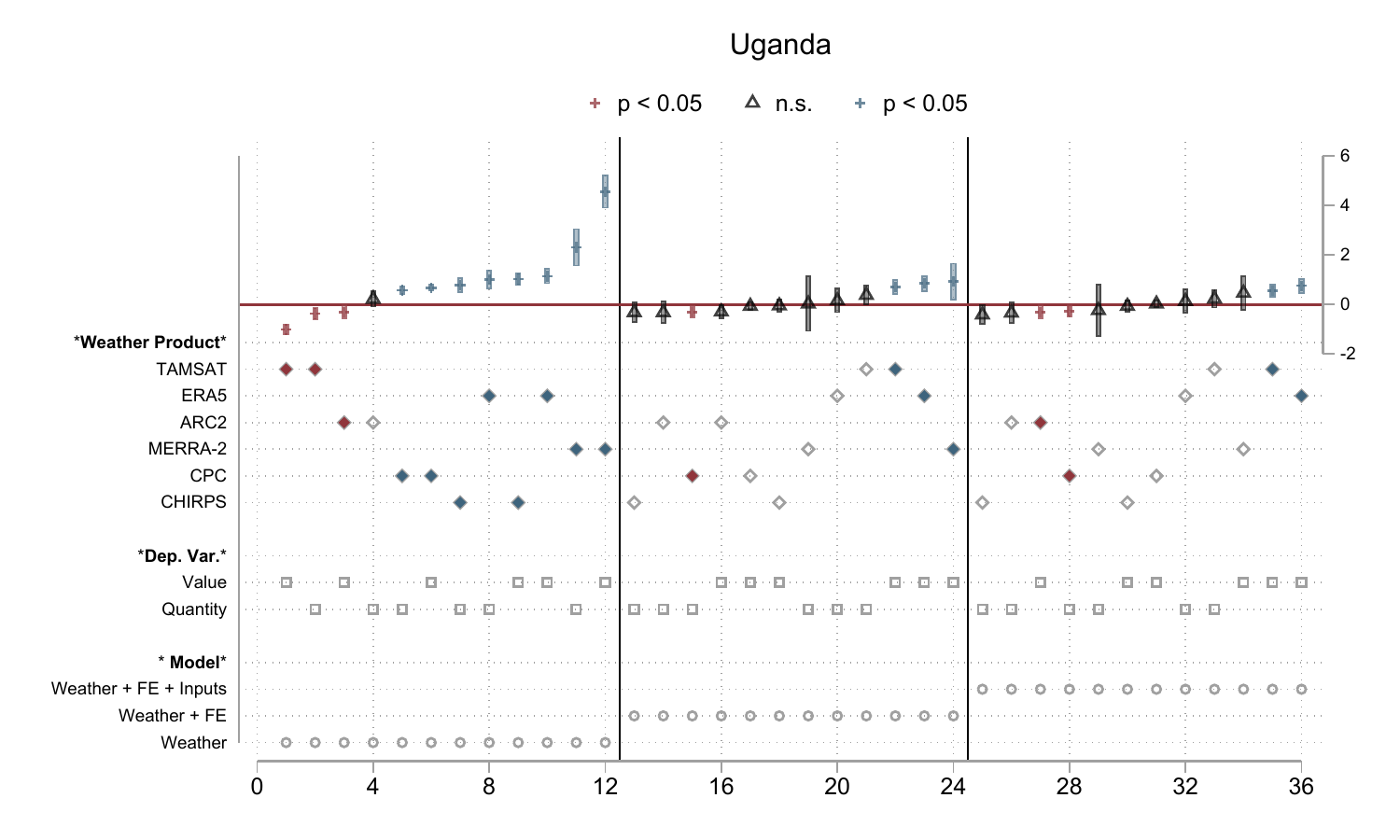}
		\end{center}
		\footnotesize  \textit{Note}: The figure presents specification curves, where each panel represents a different country, with three different models presented within each panel. Each panel includes 36 regressions, where each column represents a single regression. Significant and non-significant coefficients are designated at the top of the figure. For each Earth observation product, we also designate the significance and sign of the coefficient with color: red represents coefficients which are negative and significant; white represents insignificant coefficients, regardless of sign; and blue represents coefficients which are positive and significant.  
	\end{minipage}	
\end{figure}
\end{center}

\begin{center}
\begin{figure}[!htbp]
	\begin{minipage}{\linewidth}
		\caption{Specification Charts for Deviation in Total Rainfall}
		\label{fig:pval_v6}
		\begin{center}
			\includegraphics[width=.49\linewidth,keepaspectratio]{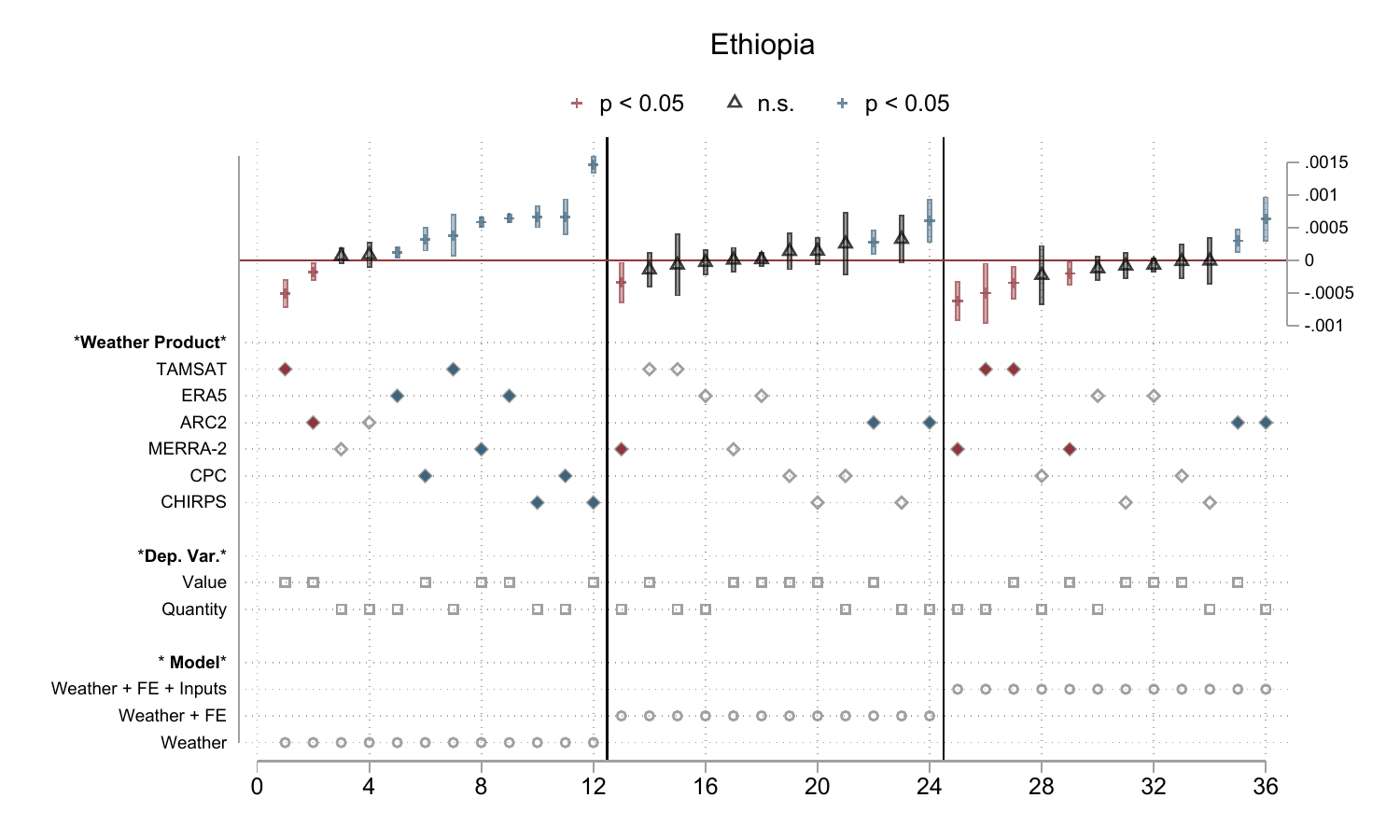}
			\includegraphics[width=.49\linewidth,keepaspectratio]{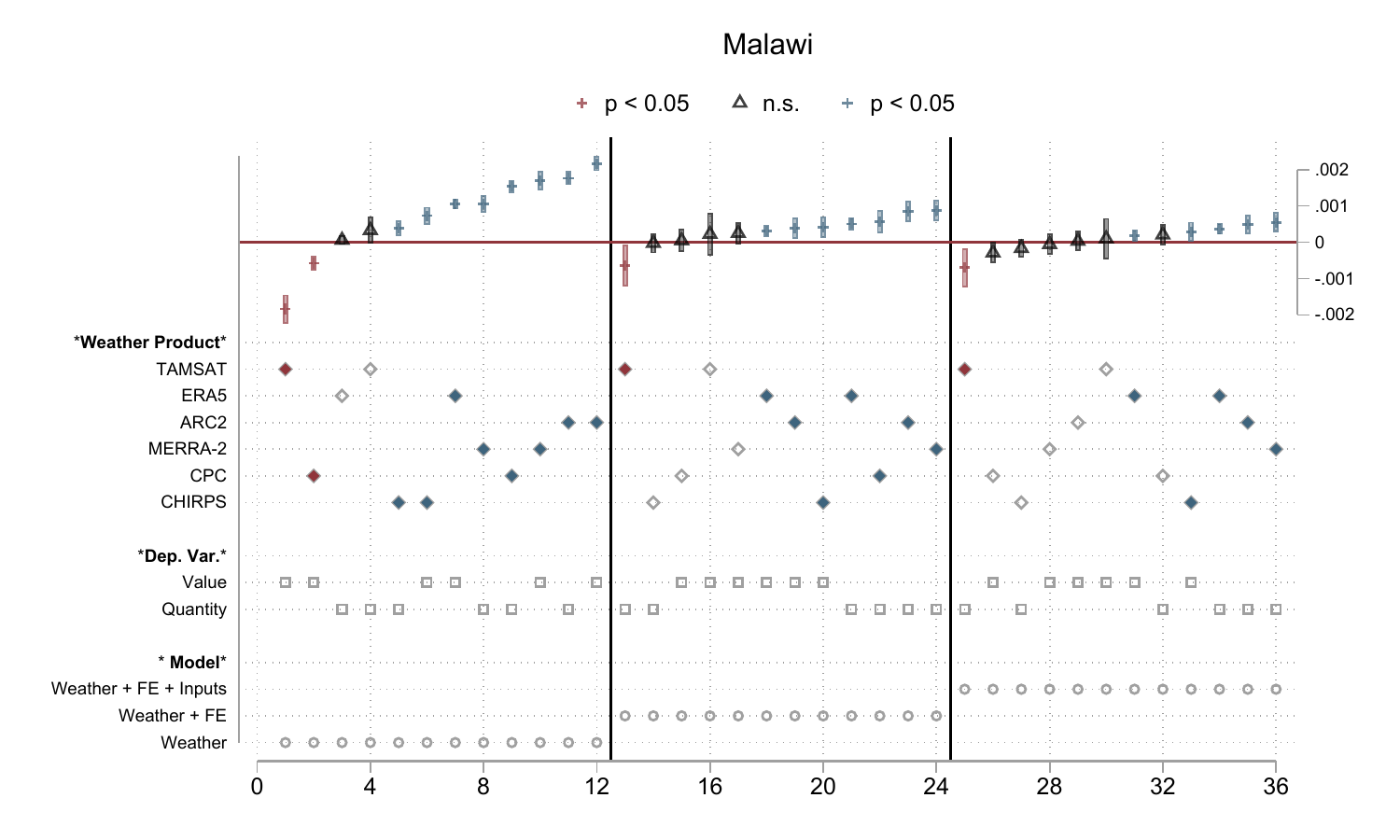}
			\includegraphics[width=.49\linewidth,keepaspectratio]{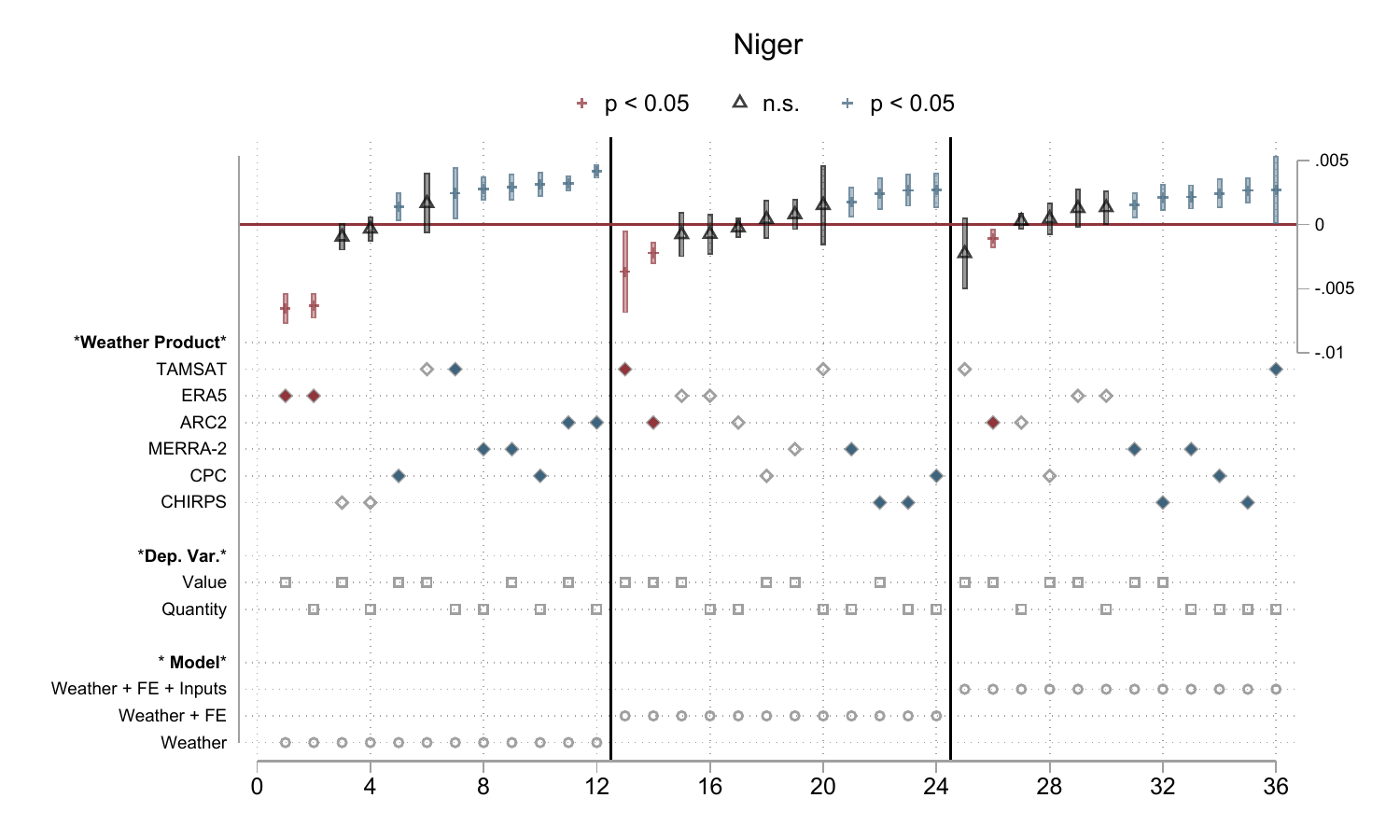}
			\includegraphics[width=.49\linewidth,keepaspectratio]{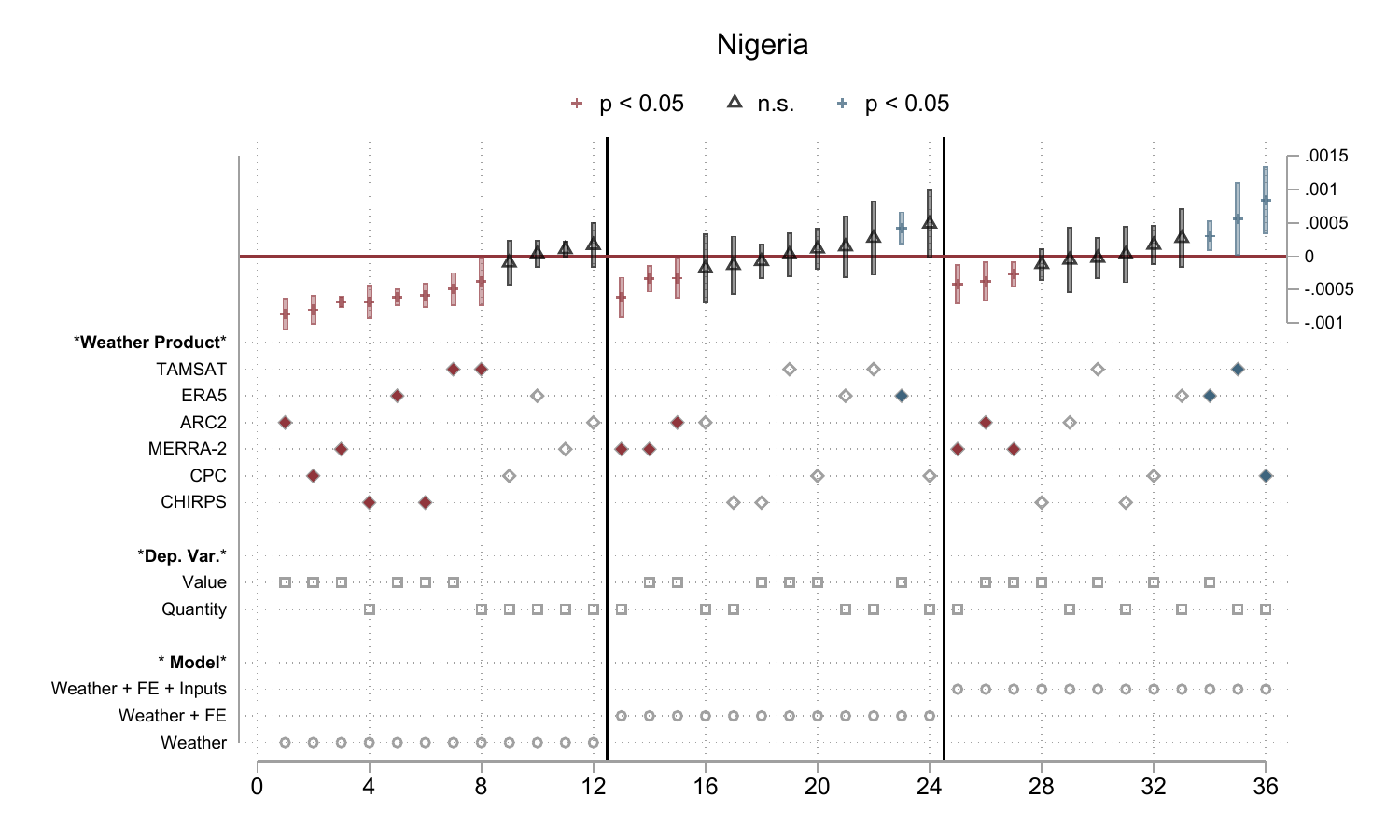}
			\includegraphics[width=.49\linewidth,keepaspectratio]{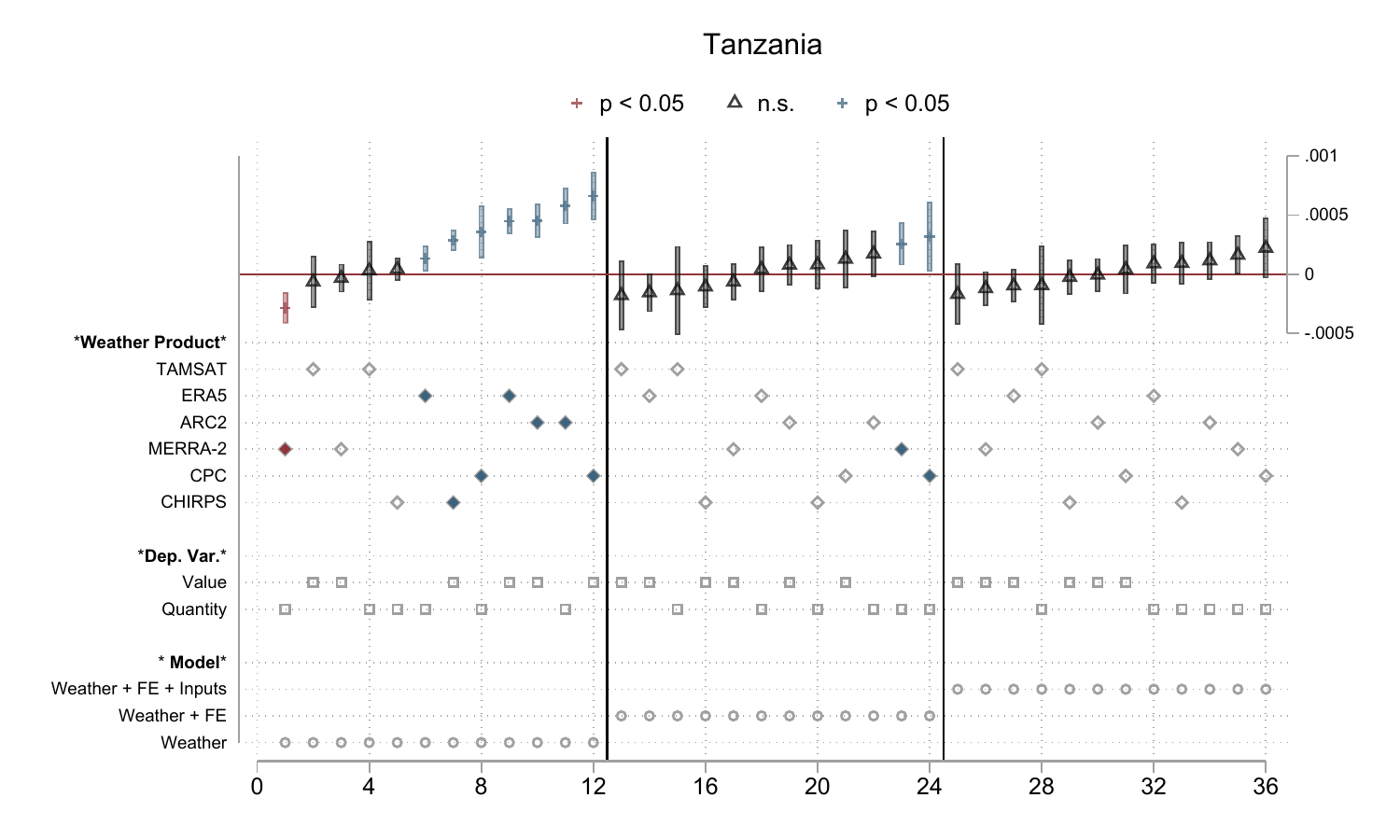}
			\includegraphics[width=.49\linewidth,keepaspectratio]{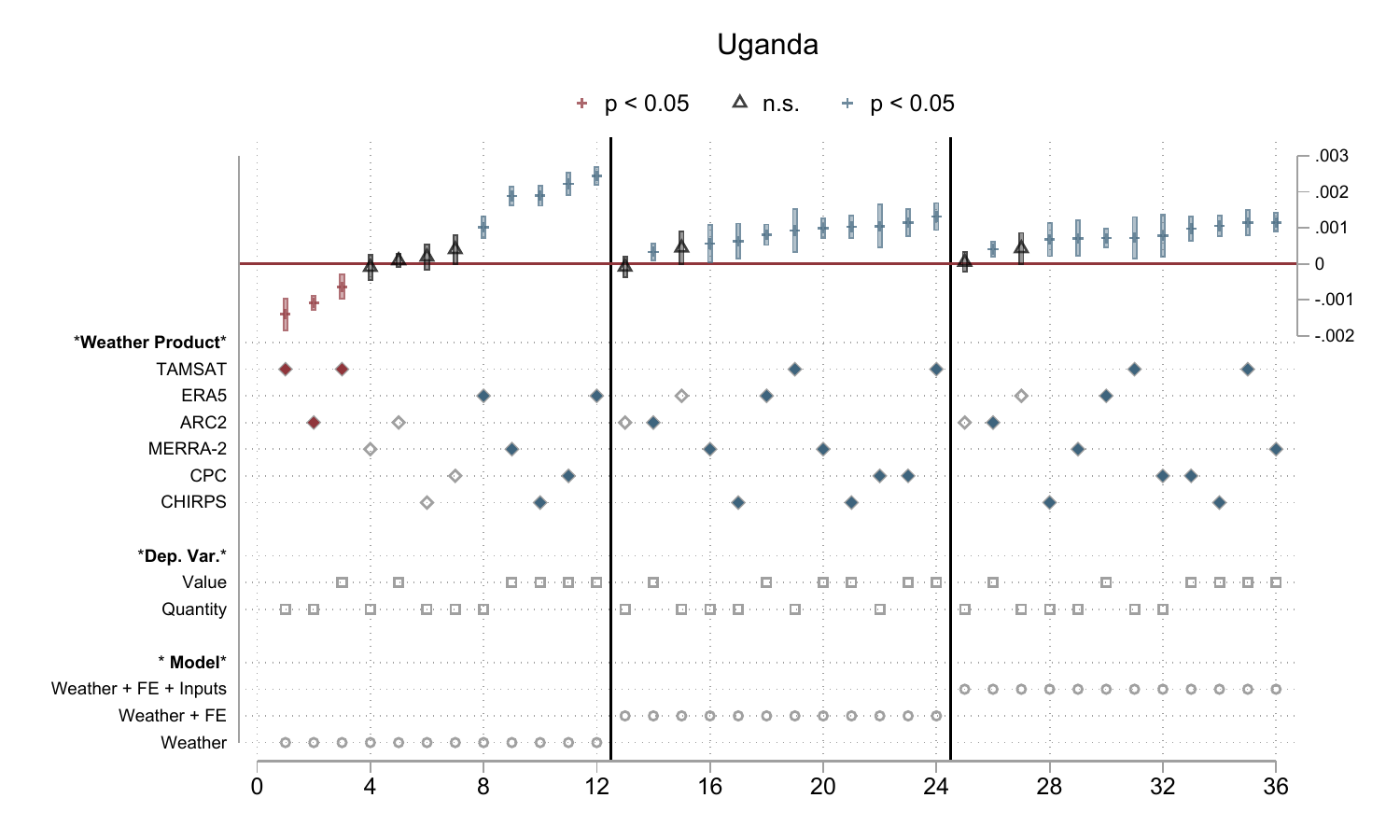}
		\end{center}
		\footnotesize  \textit{Note}: The figure presents specification curves, where each panel represents a different country, with three different models presented within each panel. Each panel includes 36 regressions, where each column represents a single regression. Significant and non-significant coefficients are designated at the top of the figure. For each Earth observation product, we also designate the significance and sign of the coefficient with color: red represents coefficients which are negative and significant; white represents insignificant coefficients, regardless of sign; and blue represents coefficients which are positive and significant.  
	\end{minipage}	
\end{figure}
\end{center}

\begin{center}
\begin{figure}[!htbp]
	\begin{minipage}{\linewidth}
		\caption{Specification Charts for z-Score of Total Rainfall}
		\label{fig:pval_v7}
		\begin{center}
			\includegraphics[width=.49\linewidth,keepaspectratio]{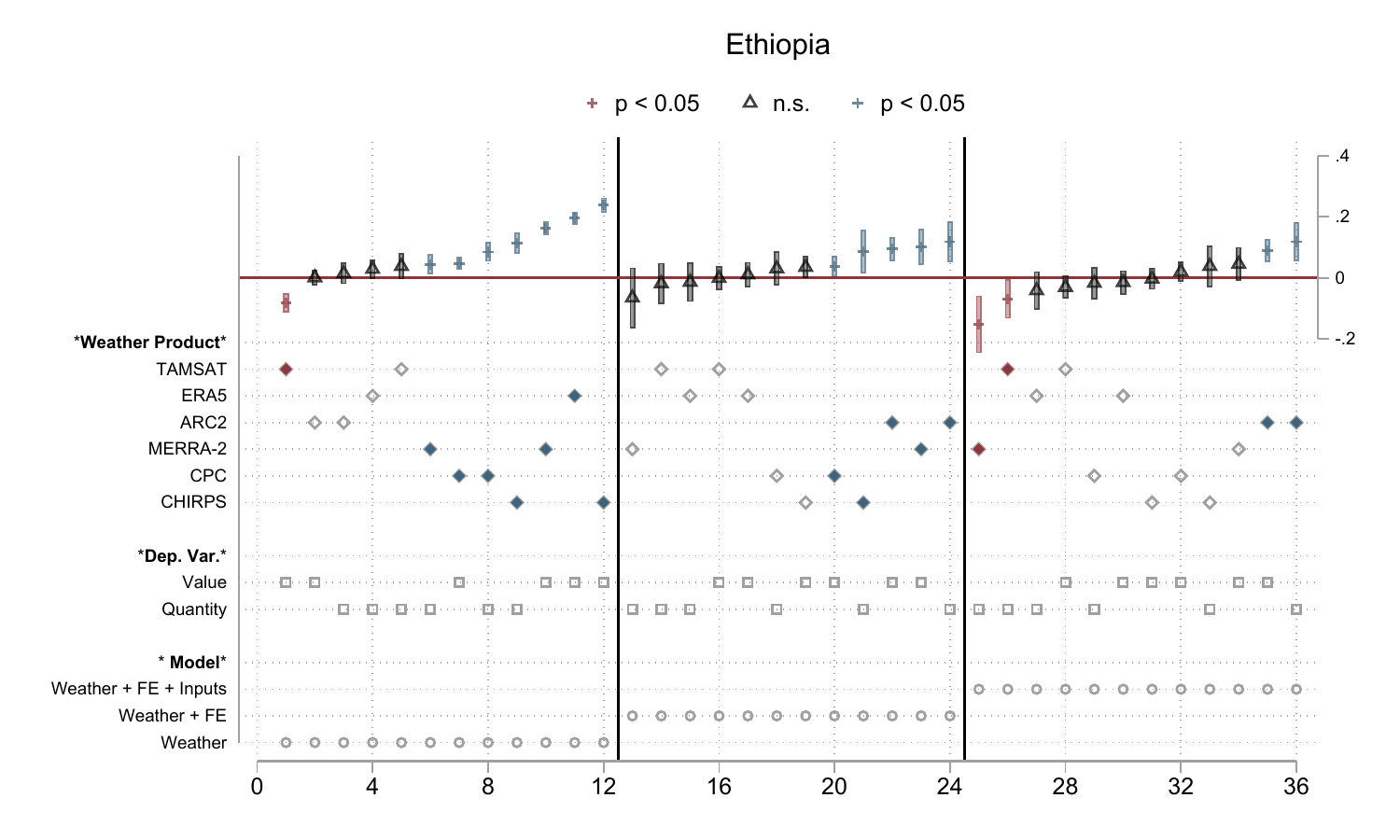}
			\includegraphics[width=.49\linewidth,keepaspectratio]{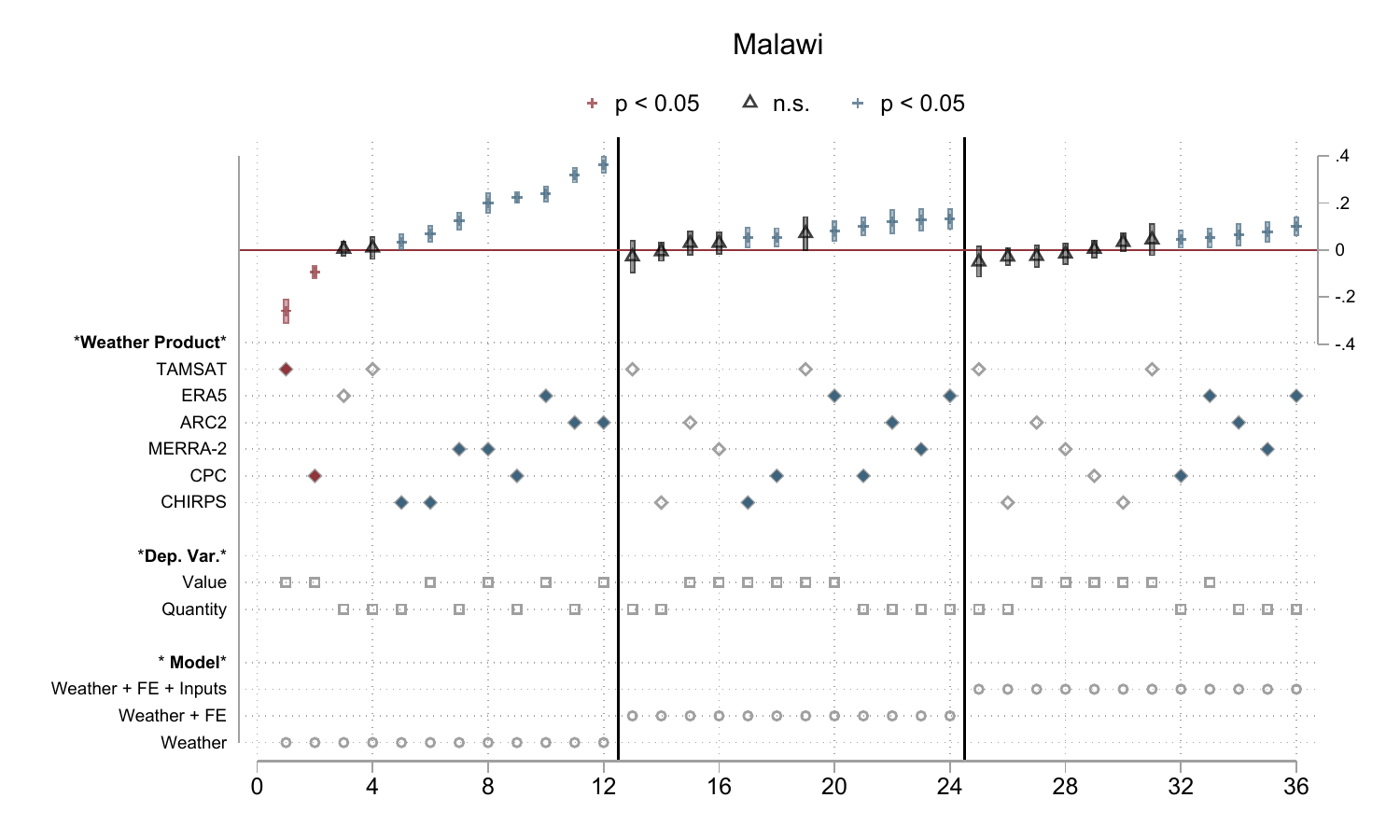}
			\includegraphics[width=.49\linewidth,keepaspectratio]{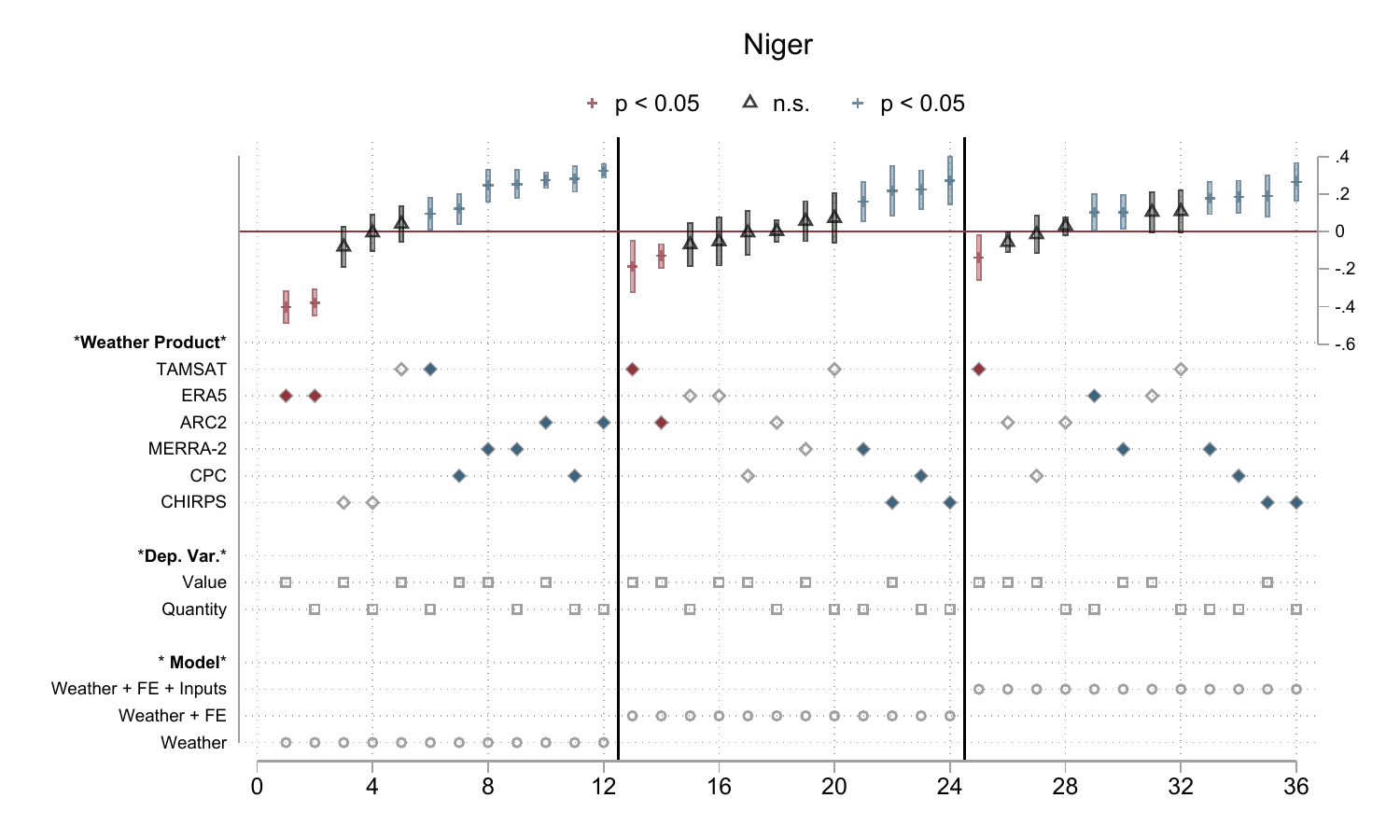}
			\includegraphics[width=.49\linewidth,keepaspectratio]{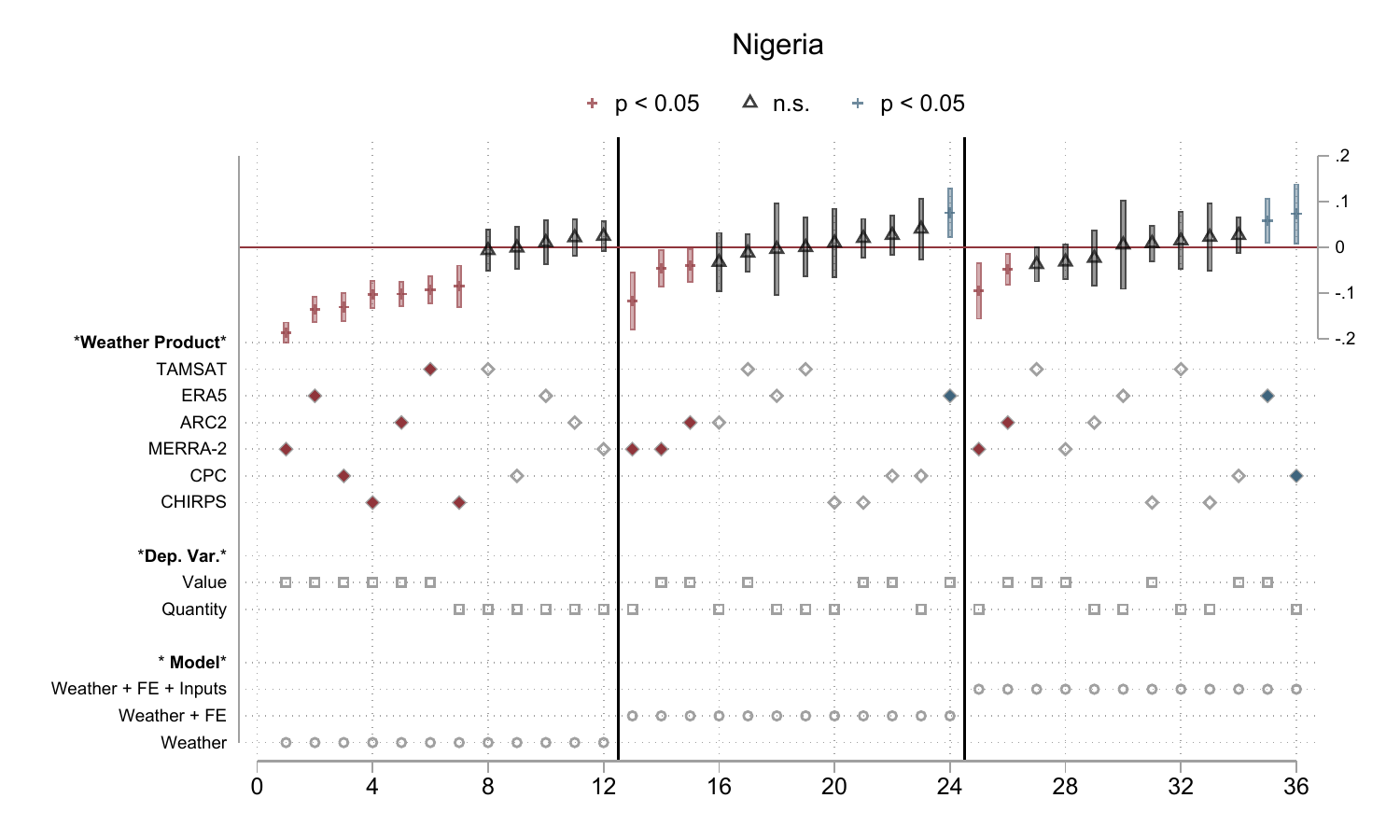}
			\includegraphics[width=.49\linewidth,keepaspectratio]{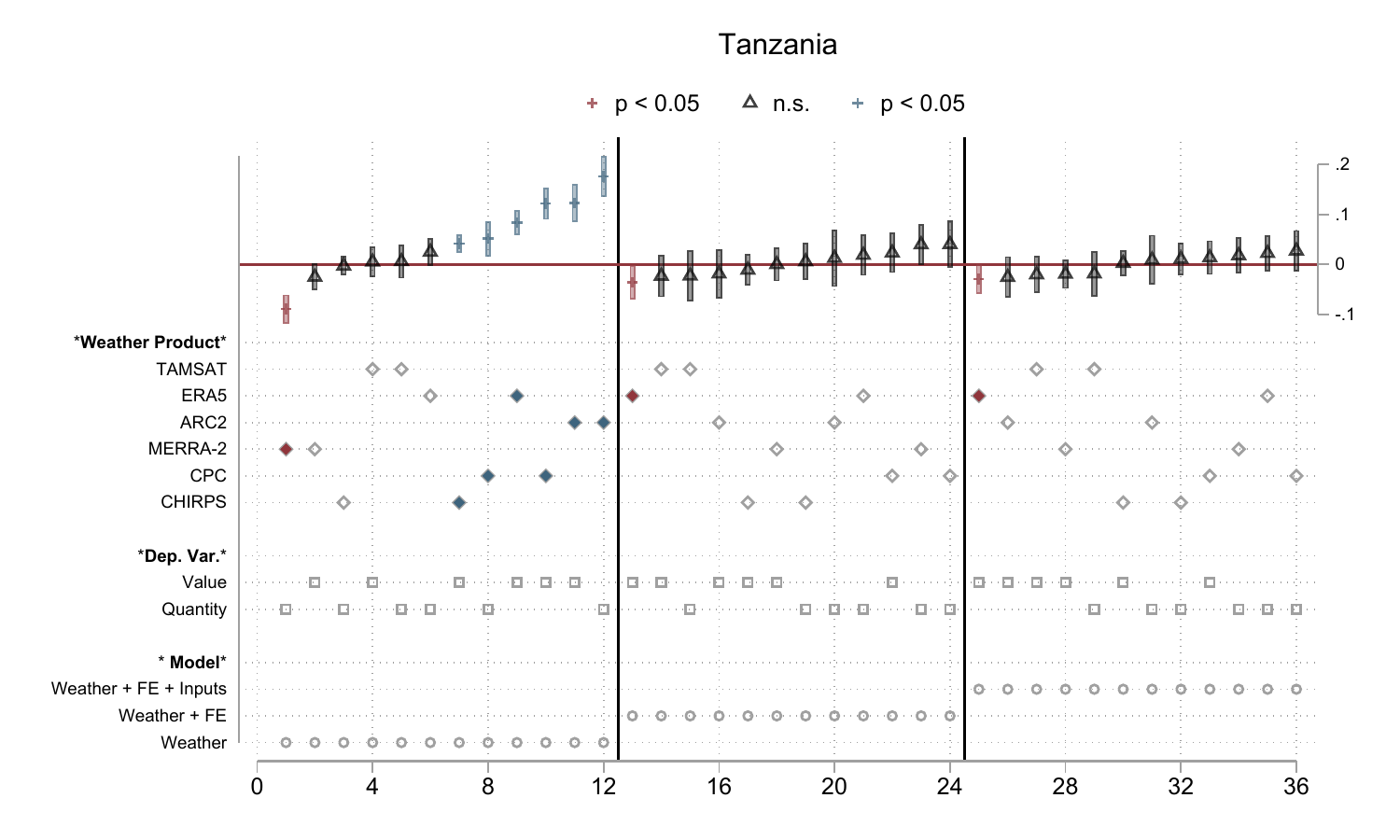}
			\includegraphics[width=.49\linewidth,keepaspectratio]{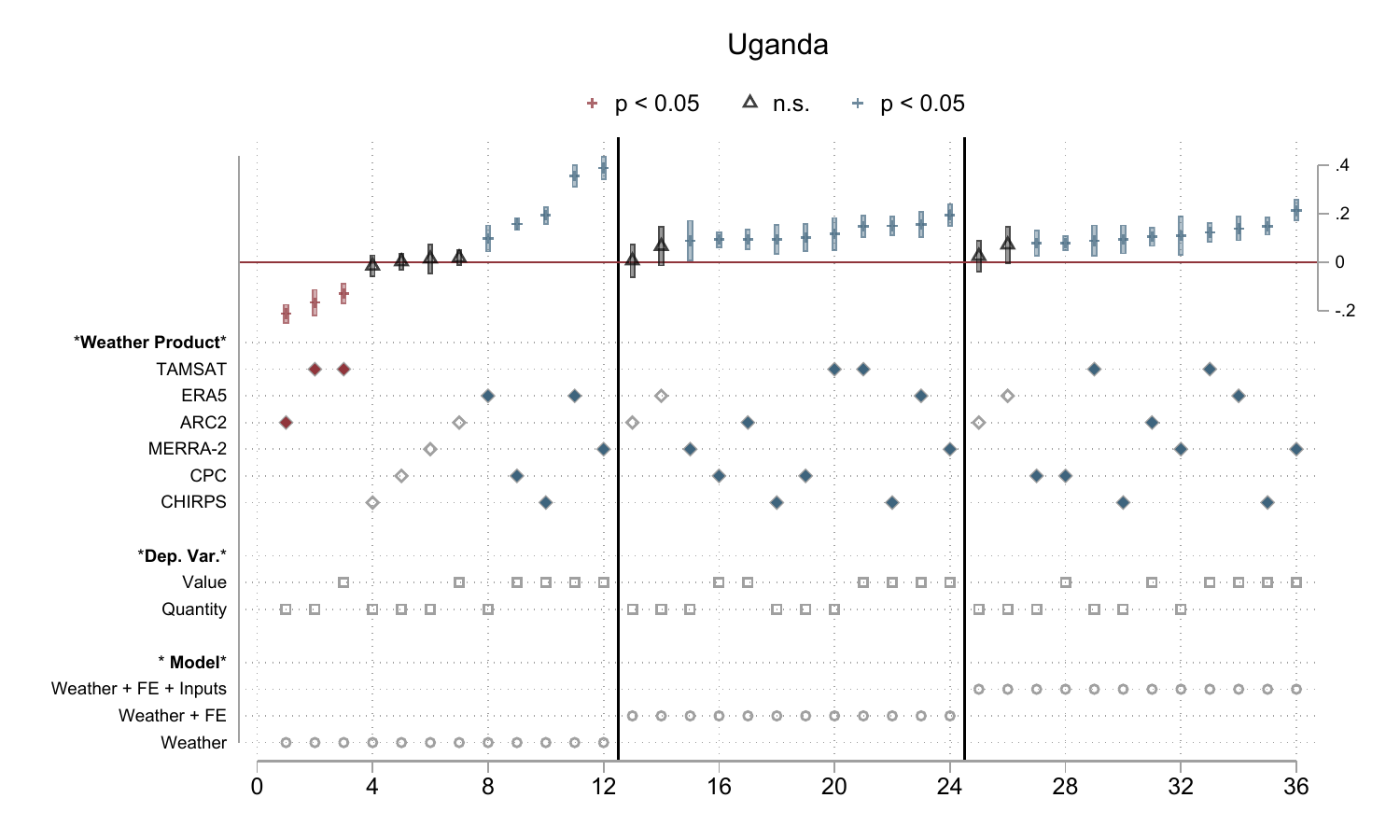}
		\end{center}
		\footnotesize  \textit{Note}: The figure presents specification curves, where each panel represents a different country, with three different models presented within each panel. Each panel includes 36 regressions, where each column represents a single regression. Significant and non-significant coefficients are designated at the top of the figure. For each Earth observation product, we also designate the significance and sign of the coefficient with color: red represents coefficients which are negative and significant; white represents insignificant coefficients, regardless of sign; and blue represents coefficients which are positive and significant.  
	\end{minipage}	
\end{figure}
\end{center}

\begin{center}
\begin{figure}[!htbp]
	\begin{minipage}{\linewidth}
		\caption{Specification Charts for Number Days with Rain}
		\label{fig:pval_v8}
		\begin{center}
			\includegraphics[width=.49\linewidth,keepaspectratio]{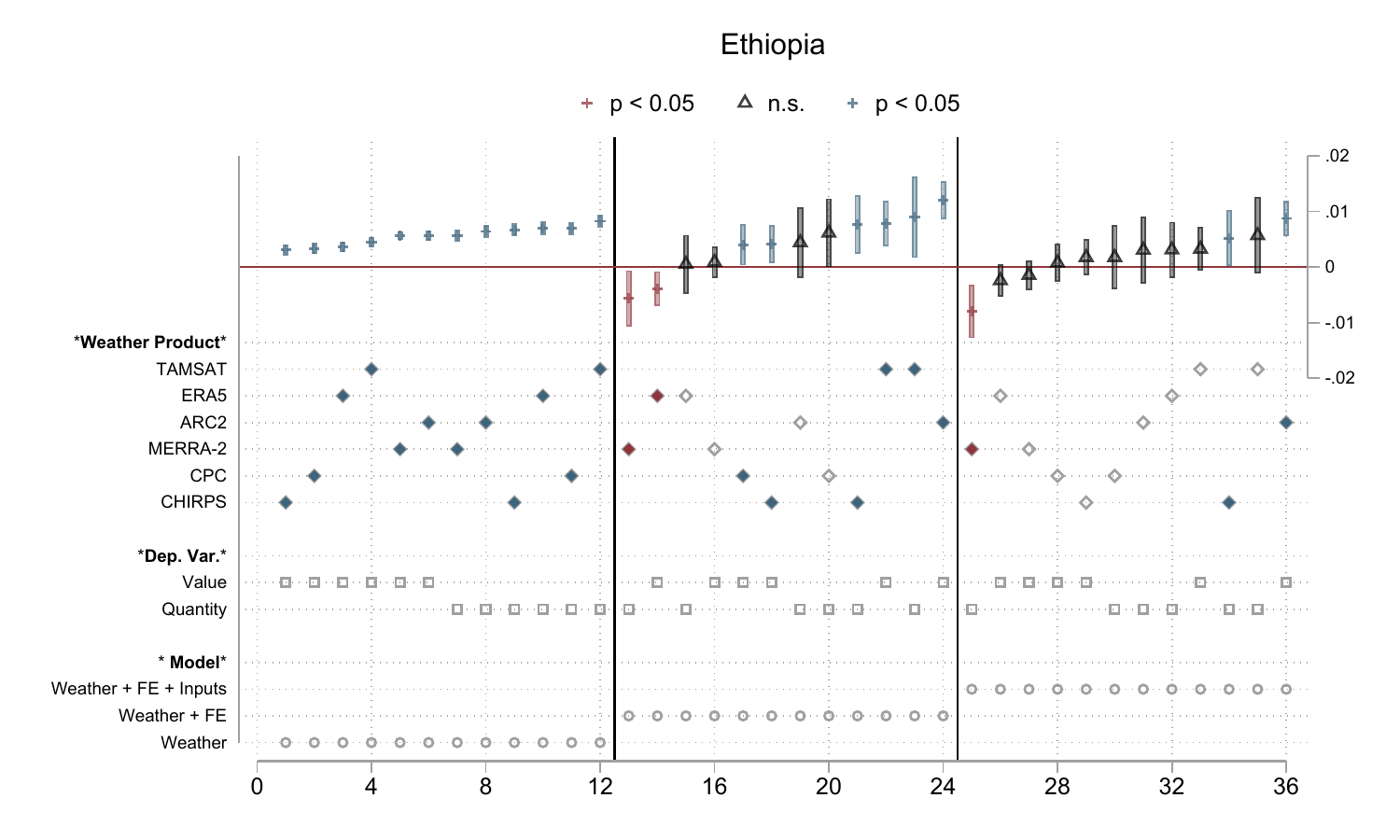}
			\includegraphics[width=.49\linewidth,keepaspectratio]{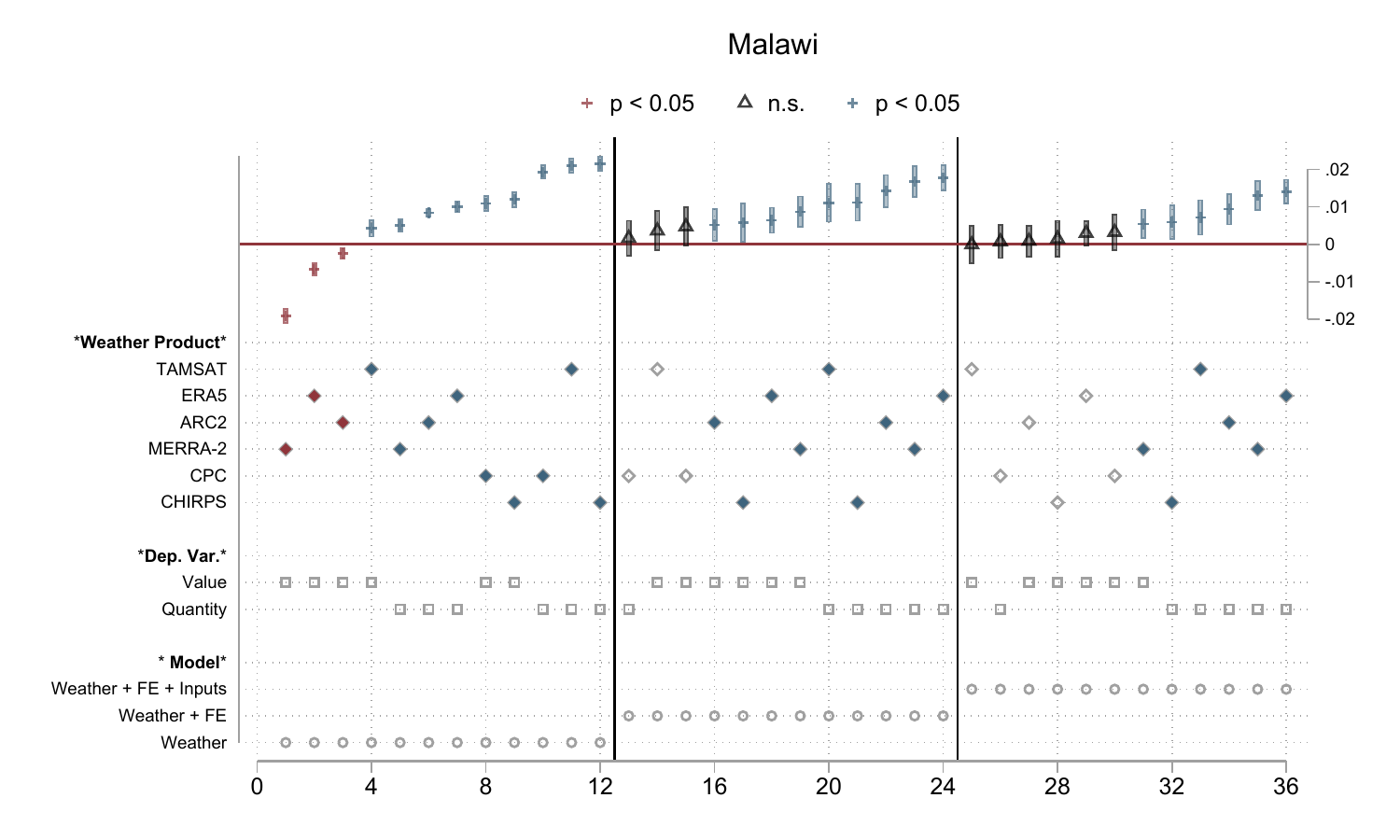}
			\includegraphics[width=.49\linewidth,keepaspectratio]{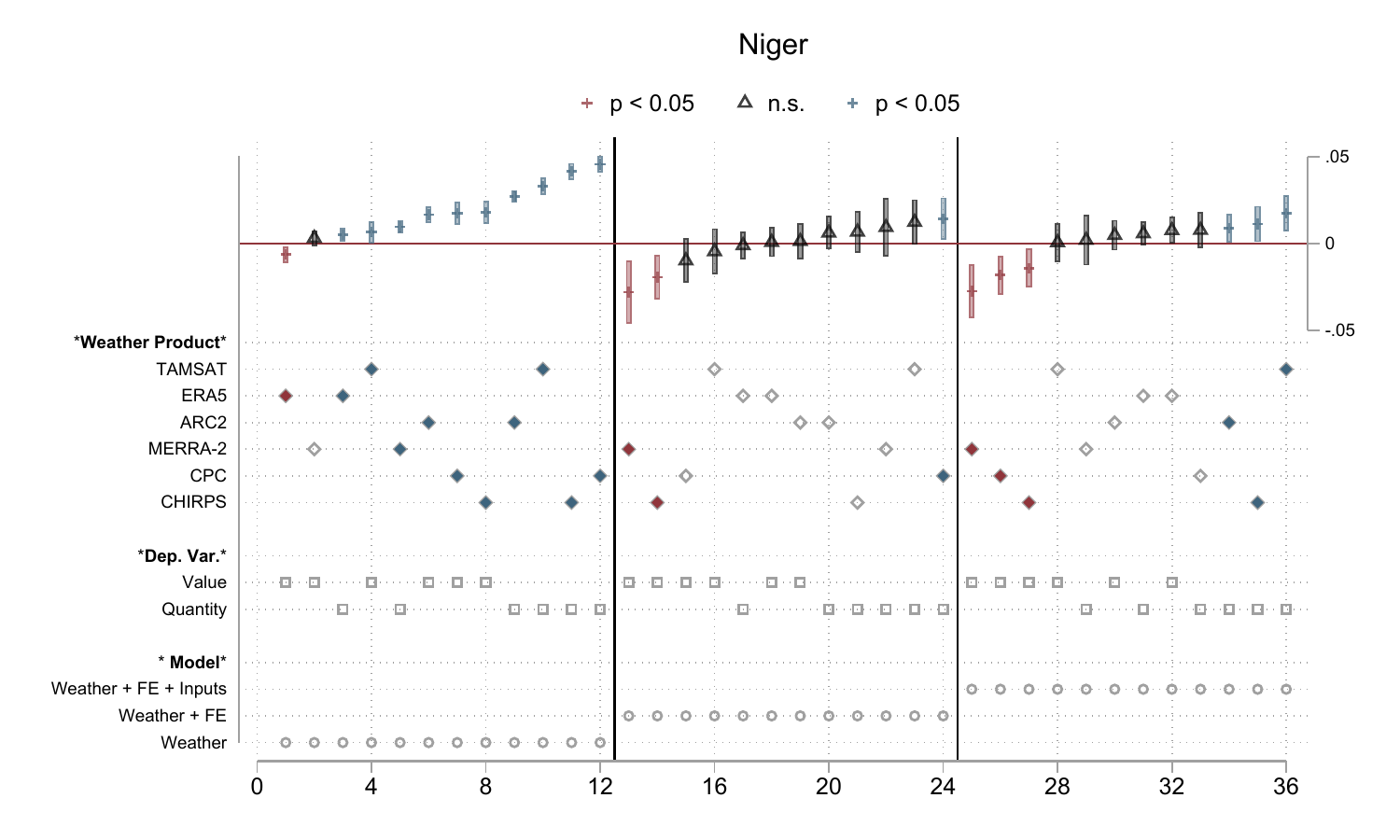}
			\includegraphics[width=.49\linewidth,keepaspectratio]{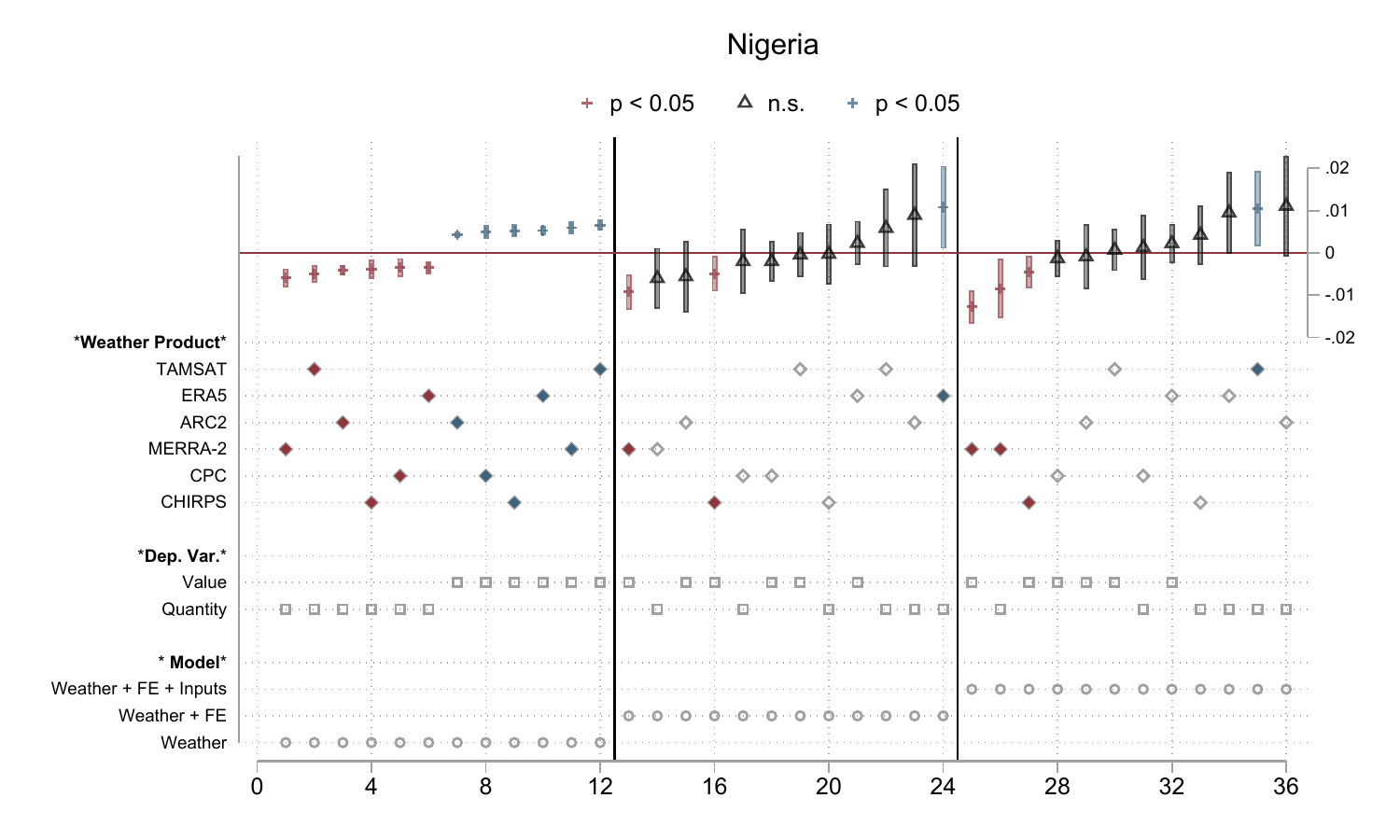}
			\includegraphics[width=.49\linewidth,keepaspectratio]{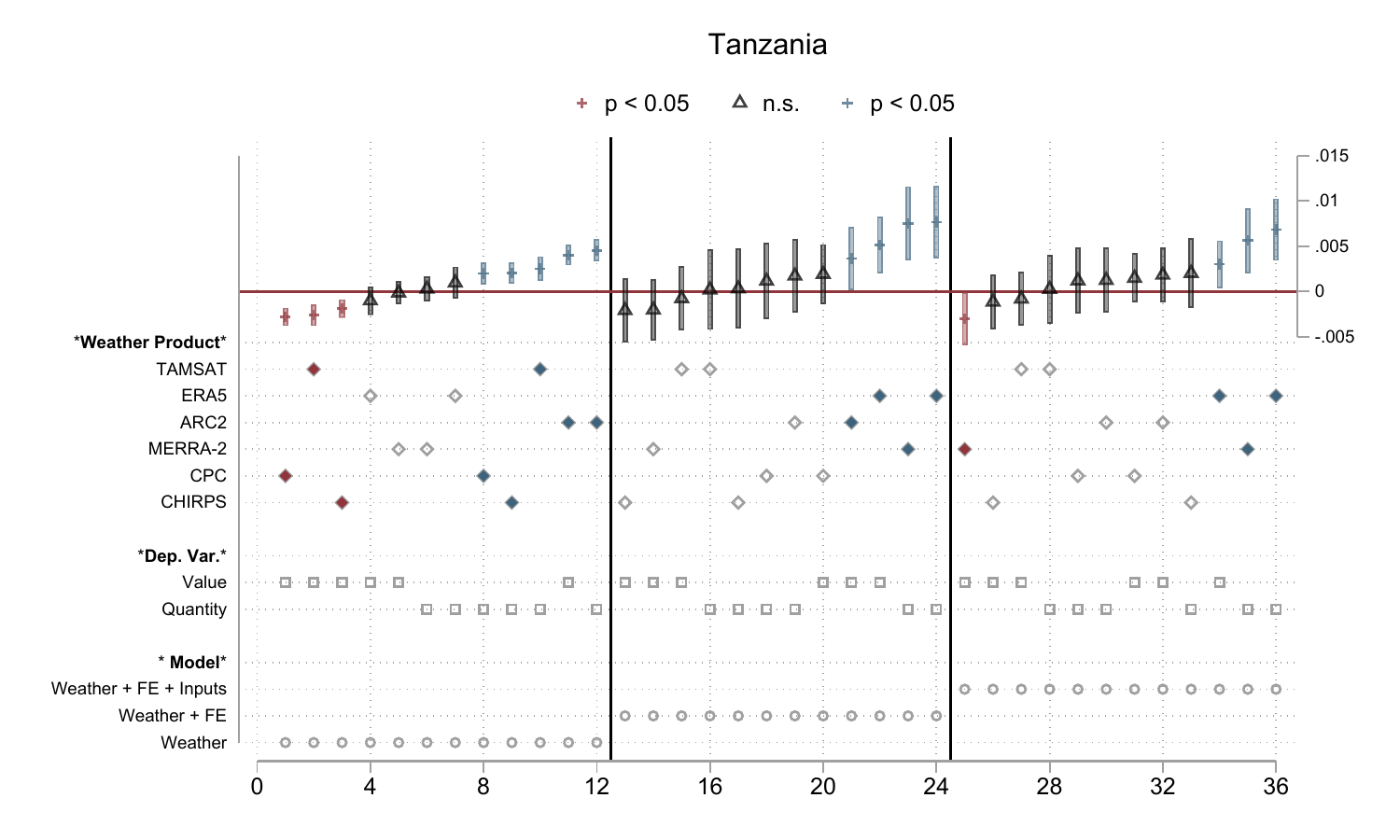}
			\includegraphics[width=.49\linewidth,keepaspectratio]{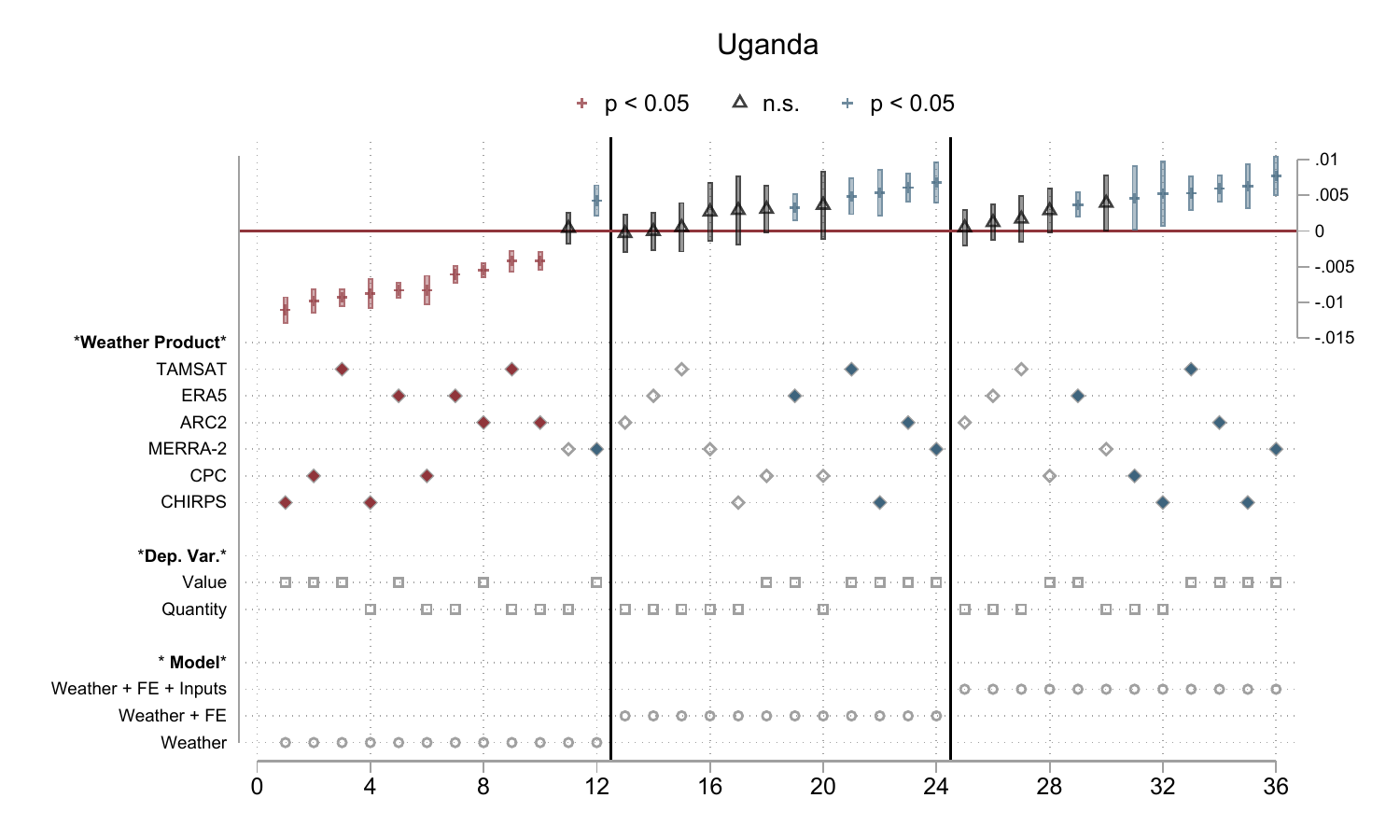}
		\end{center}
		\footnotesize  \textit{Note}: The figure presents specification curves, where each panel represents a different country, with three different models presented within each panel. Each panel includes 36 regressions, where each column represents a single regression. Significant and non-significant coefficients are designated at the top of the figure. For each Earth observation product, we also designate the significance and sign of the coefficient with color: red represents coefficients which are negative and significant; white represents insignificant coefficients, regardless of sign; and blue represents coefficients which are positive and significant.  
	\end{minipage}	
\end{figure}
\end{center}

\begin{center}
\begin{figure}[!htbp]
	\begin{minipage}{\linewidth}
		\caption{Specification Charts for Deviations in Days with Rain}
		\label{fig:pval_v9}
		\begin{center}
			\includegraphics[width=.49\linewidth,keepaspectratio]{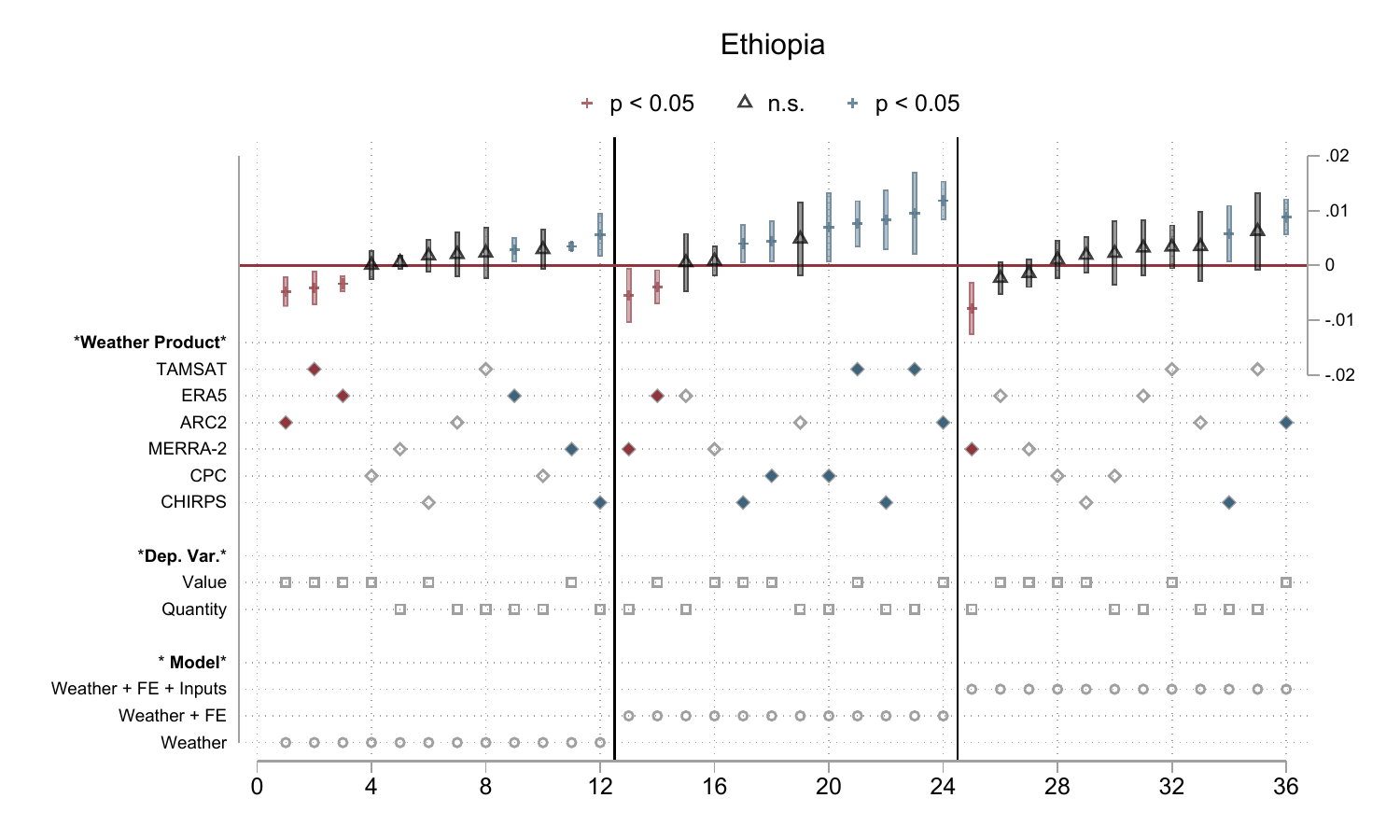}
			\includegraphics[width=.49\linewidth,keepaspectratio]{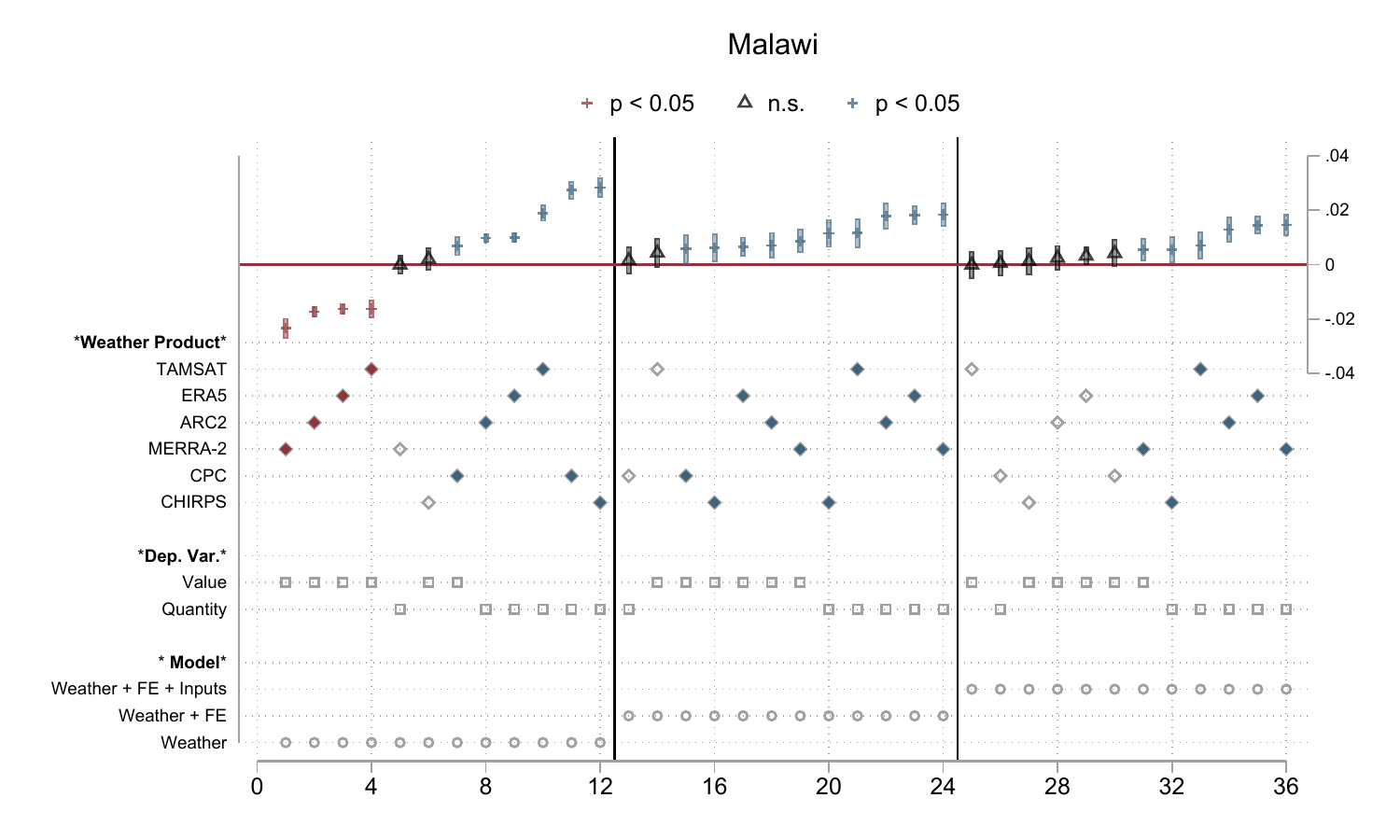}
			\includegraphics[width=.49\linewidth,keepaspectratio]{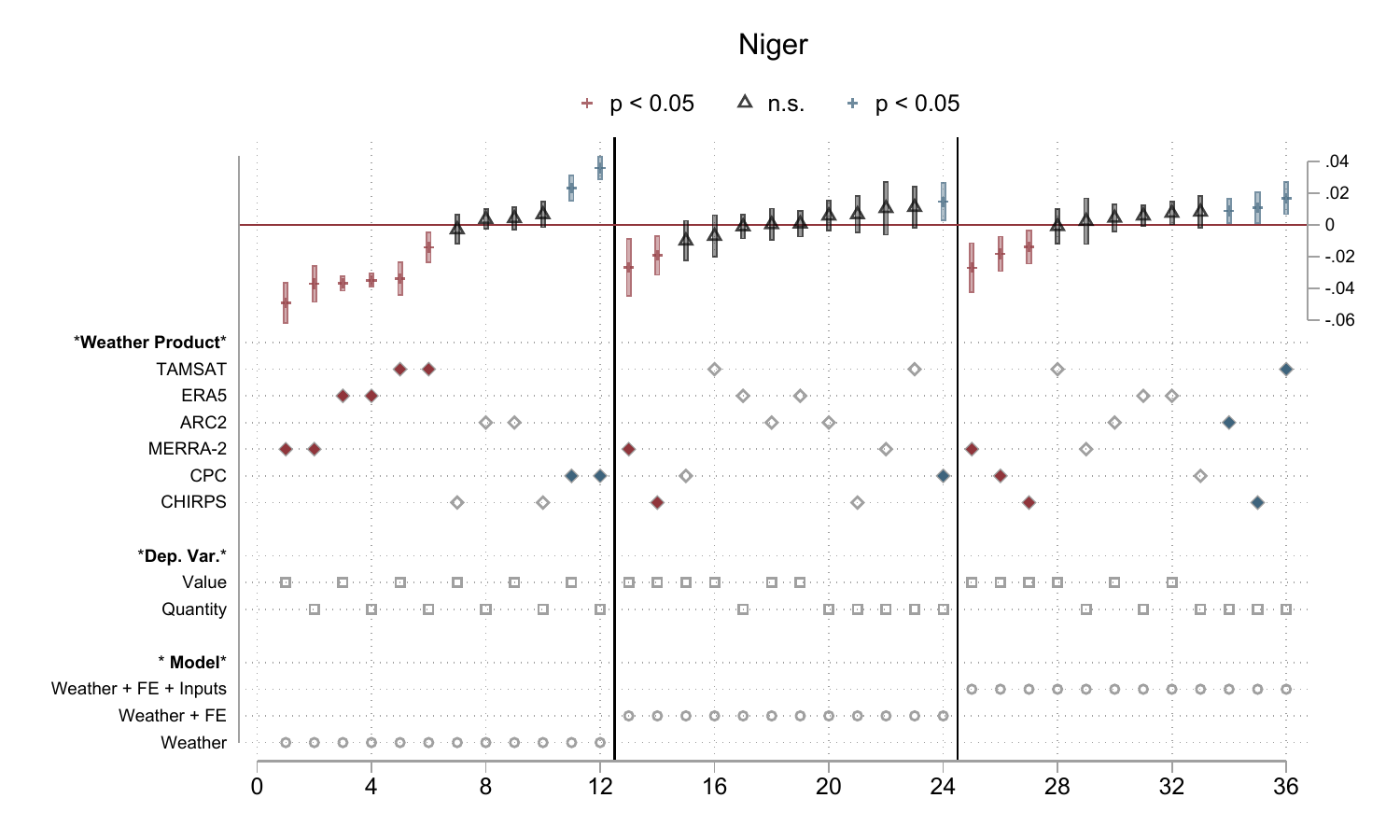}
			\includegraphics[width=.49\linewidth,keepaspectratio]{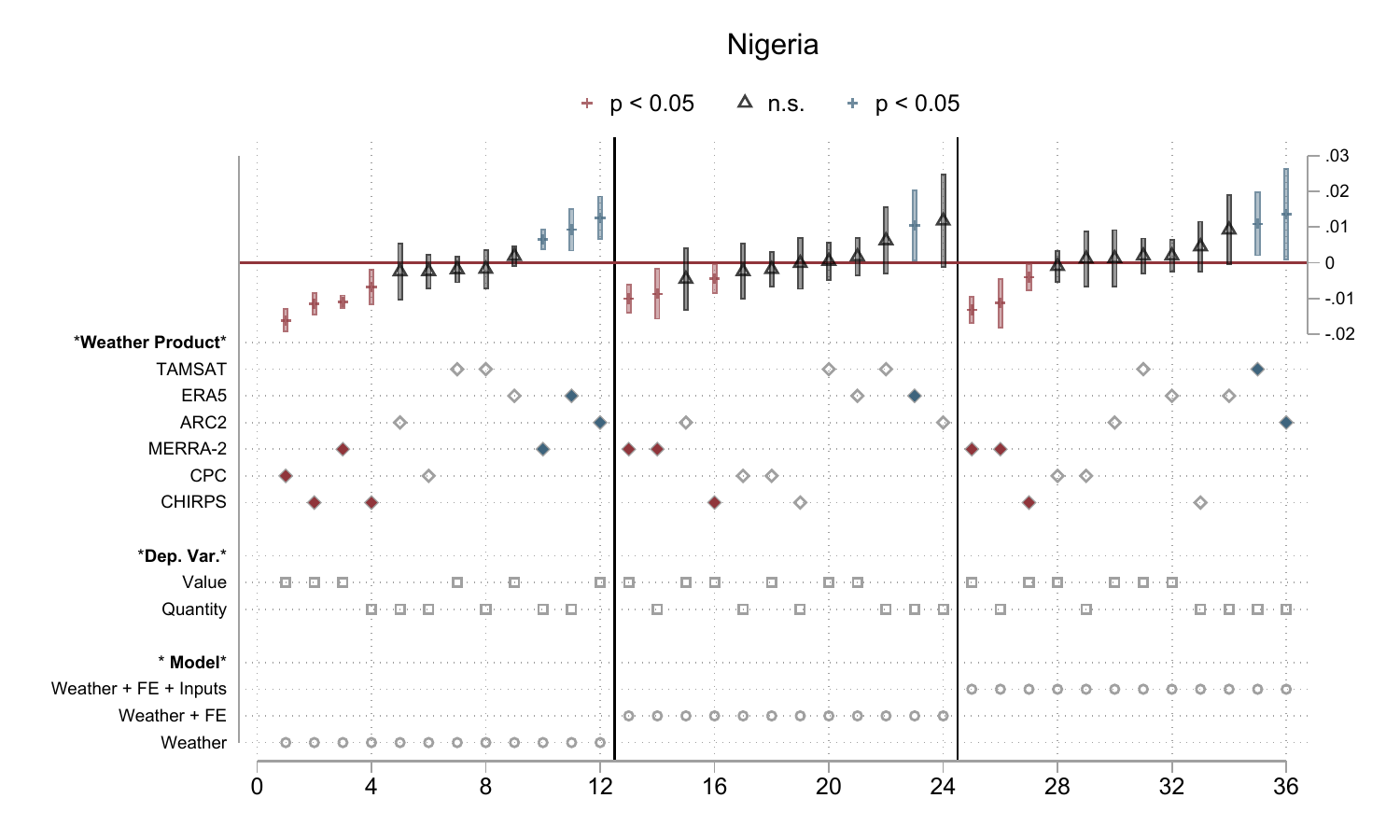}
			\includegraphics[width=.49\linewidth,keepaspectratio]{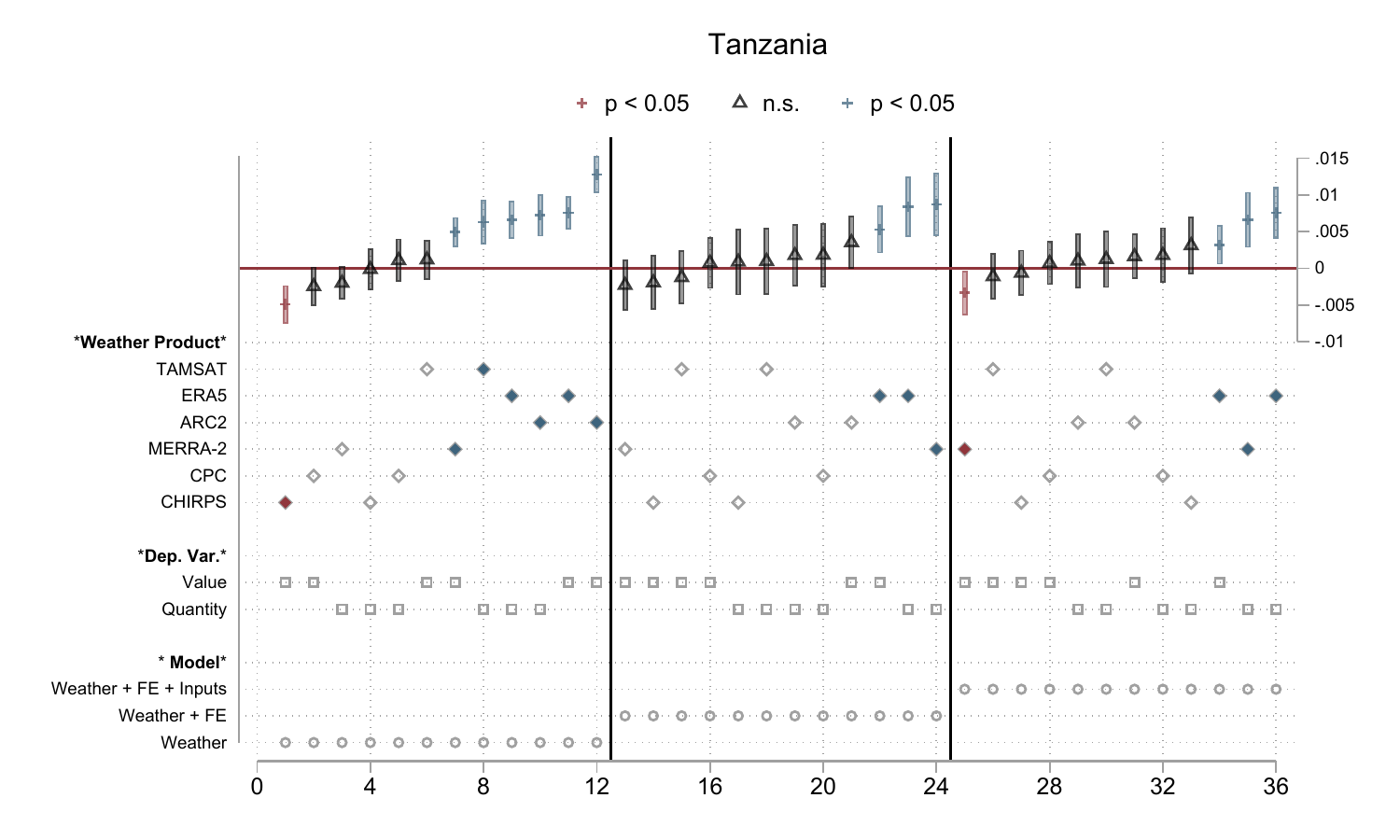}
			\includegraphics[width=.49\linewidth,keepaspectratio]{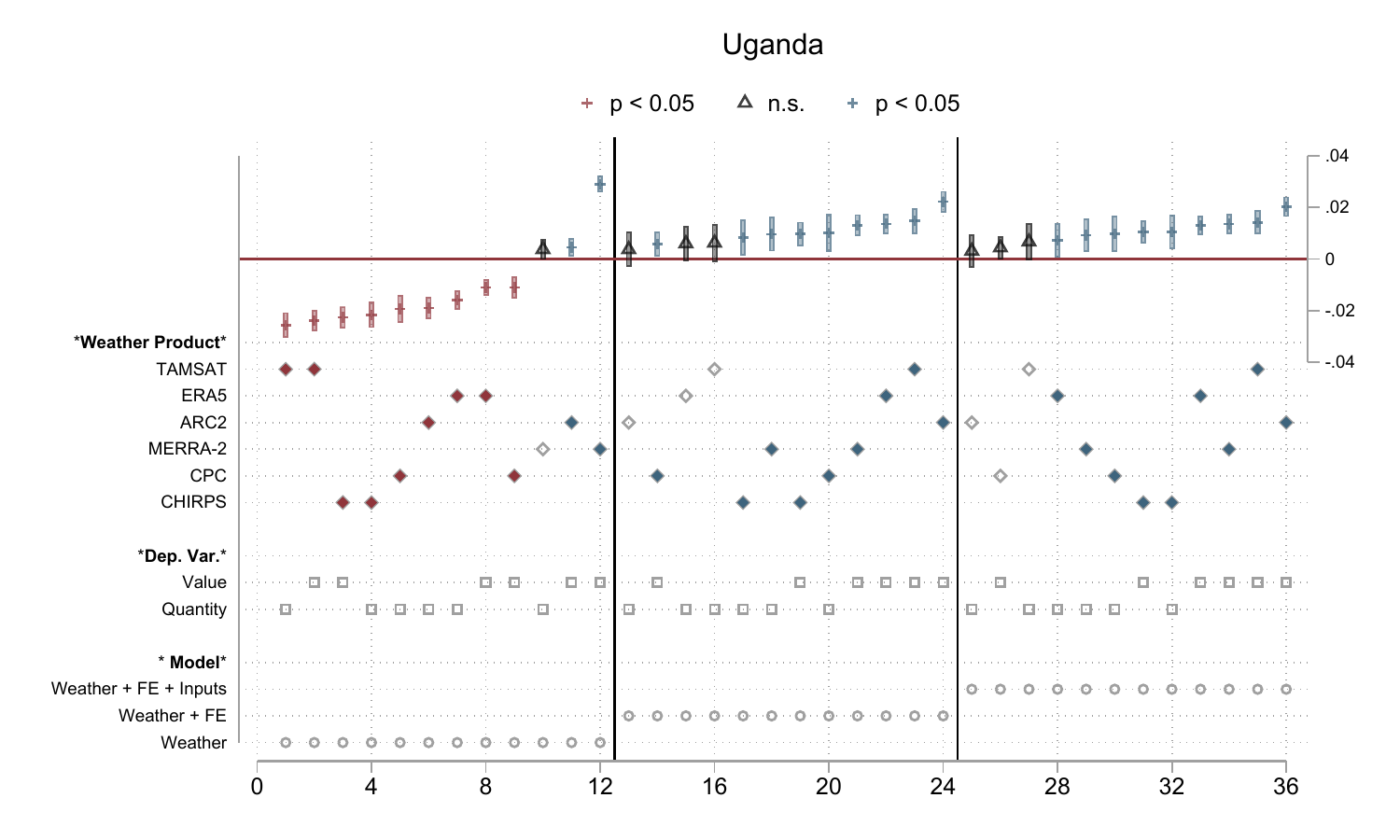}
		\end{center}
		\footnotesize  \textit{Note}: The figure presents specification curves, where each panel represents a different country, with three different models presented within each panel. Each panel includes 36 regressions, where each column represents a single regression. Significant and non-significant coefficients are designated at the top of the figure. For each Earth observation product, we also designate the significance and sign of the coefficient with color: red represents coefficients which are negative and significant; white represents insignificant coefficients, regardless of sign; and blue represents coefficients which are positive and significant.  
	\end{minipage}	
\end{figure}
\end{center}

\begin{center}
\begin{figure}[!htbp]
	\begin{minipage}{\linewidth}
		\caption{Specification Charts for Deviations in Days with No Rain}
		\label{fig:pval_v11}
		\begin{center}
			\includegraphics[width=.49\linewidth,keepaspectratio]{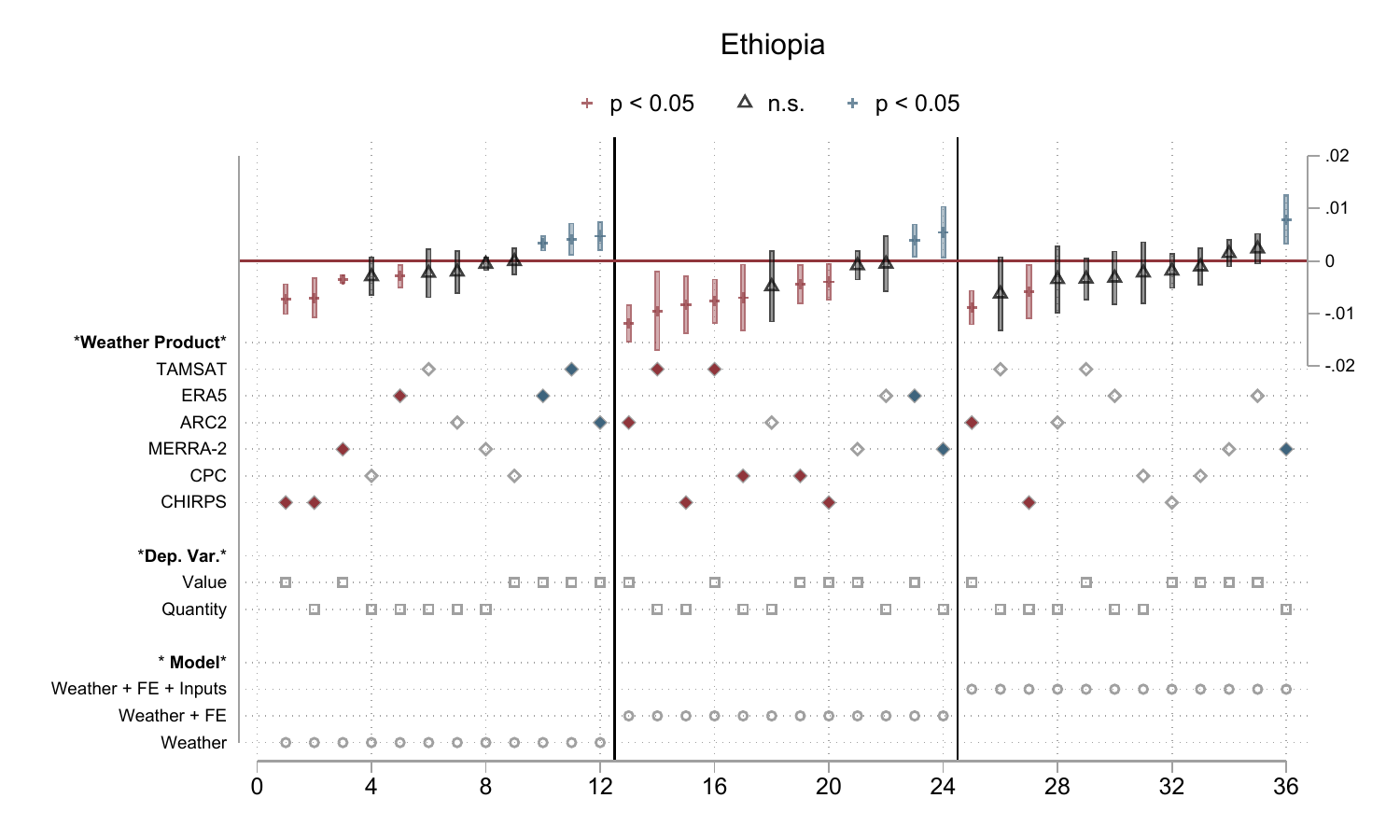}
			\includegraphics[width=.49\linewidth,keepaspectratio]{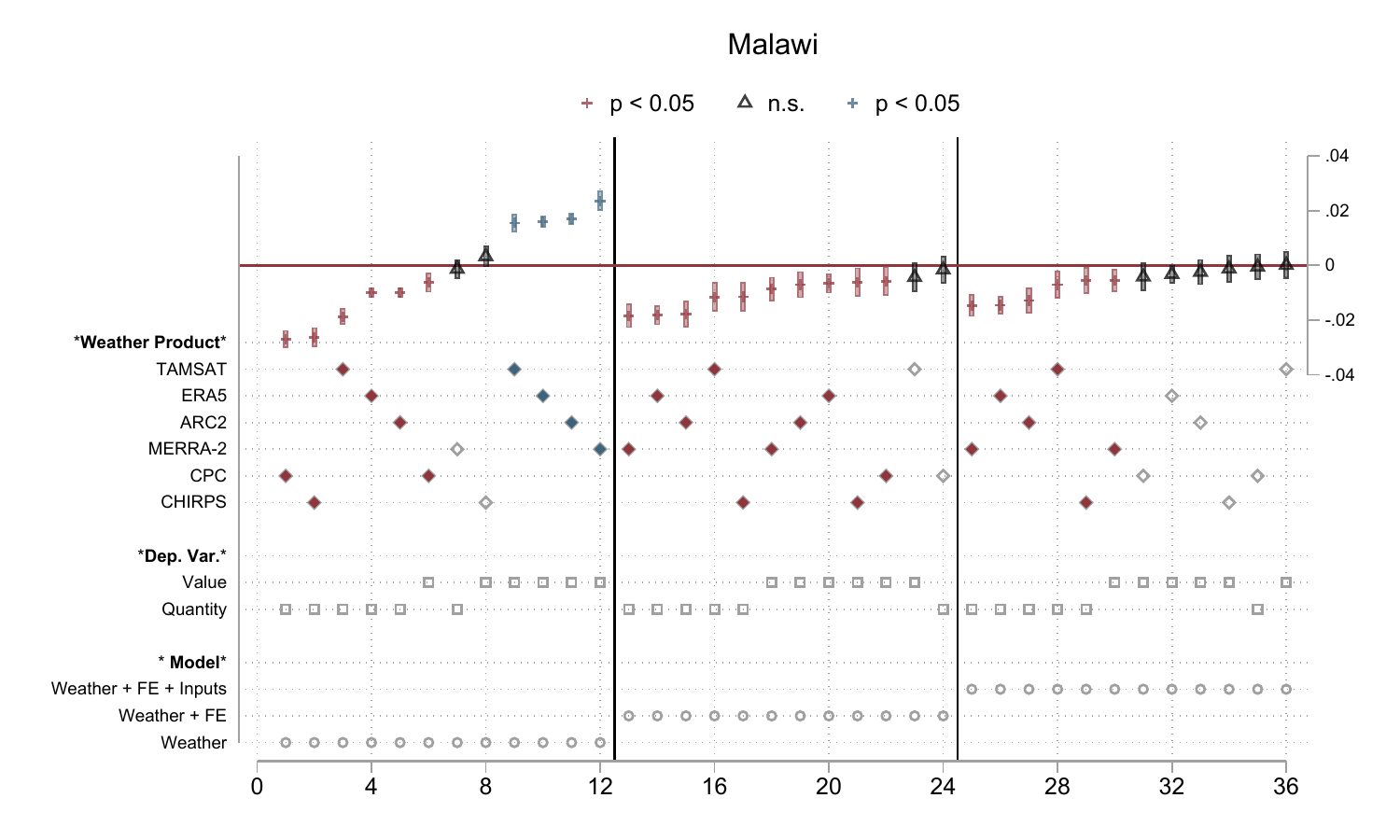}
			\includegraphics[width=.49\linewidth,keepaspectratio]{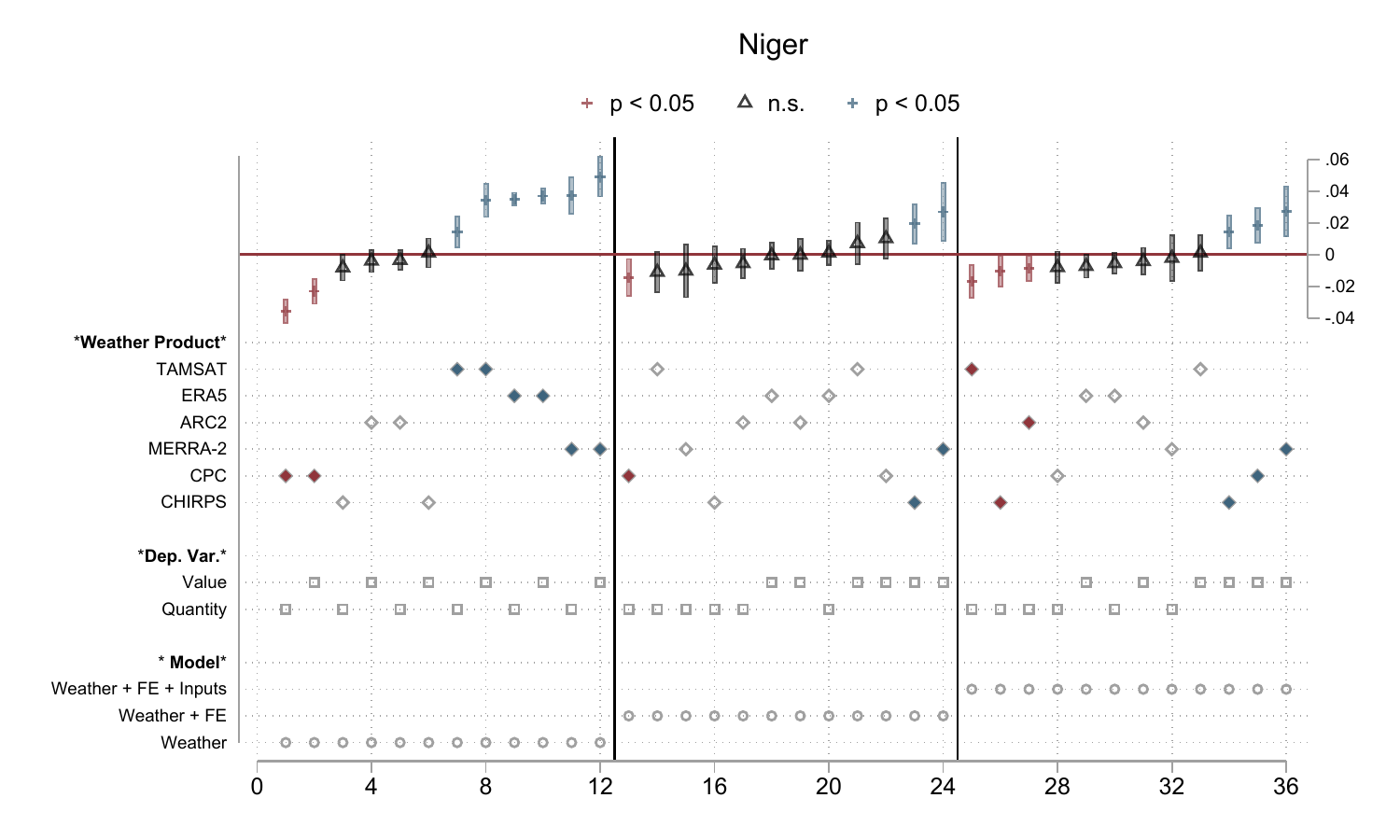}
			\includegraphics[width=.49\linewidth,keepaspectratio]{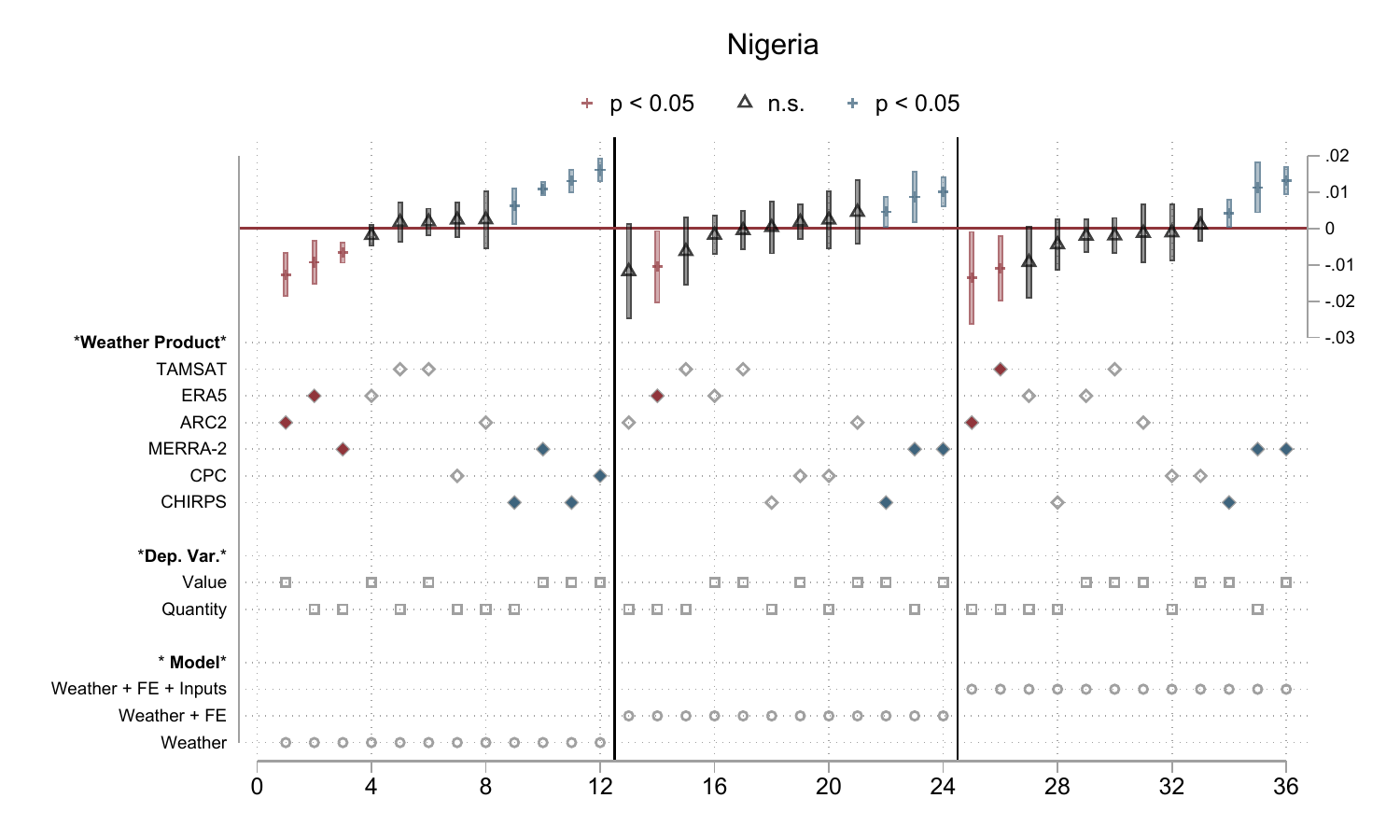}
			\includegraphics[width=.49\linewidth,keepaspectratio]{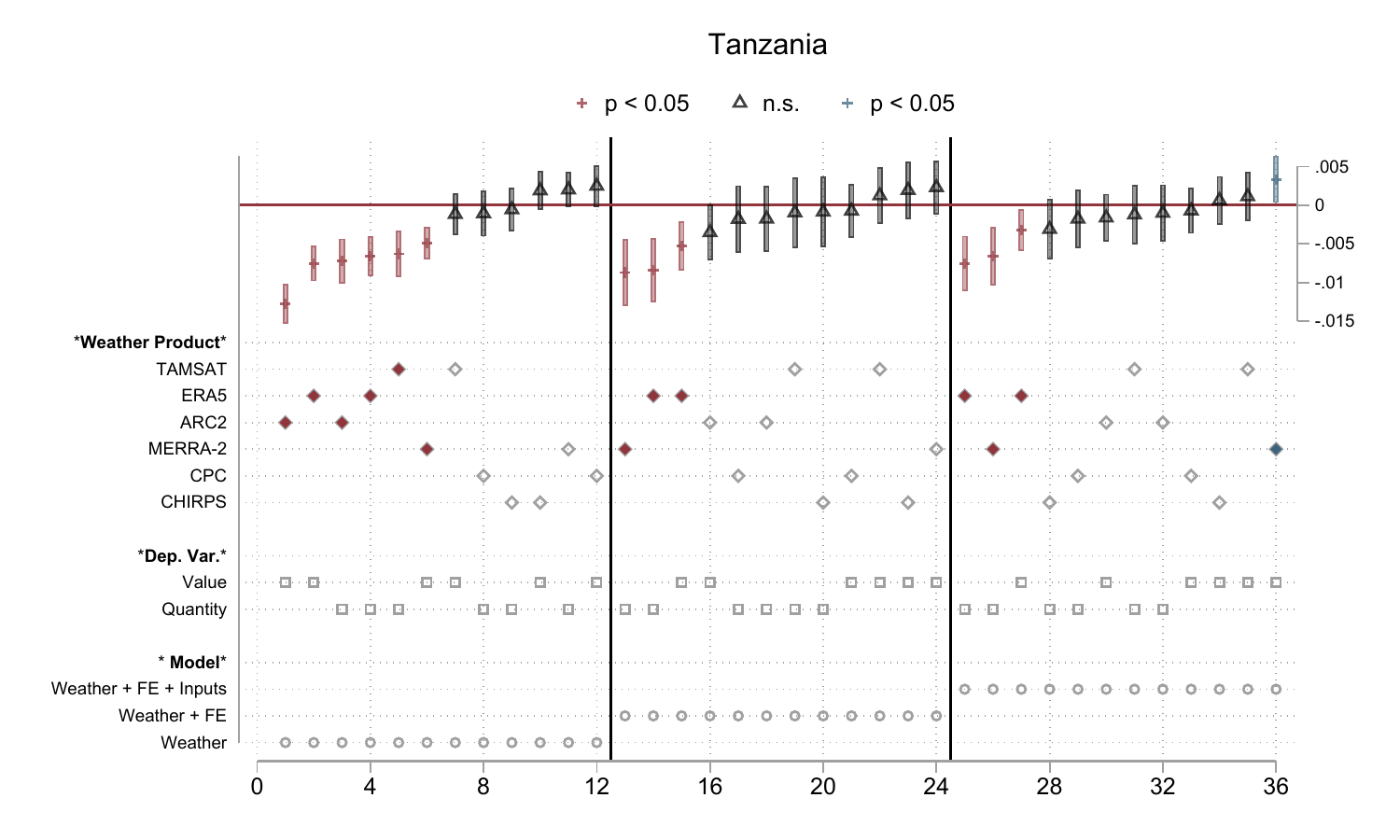}
			\includegraphics[width=.49\linewidth,keepaspectratio]{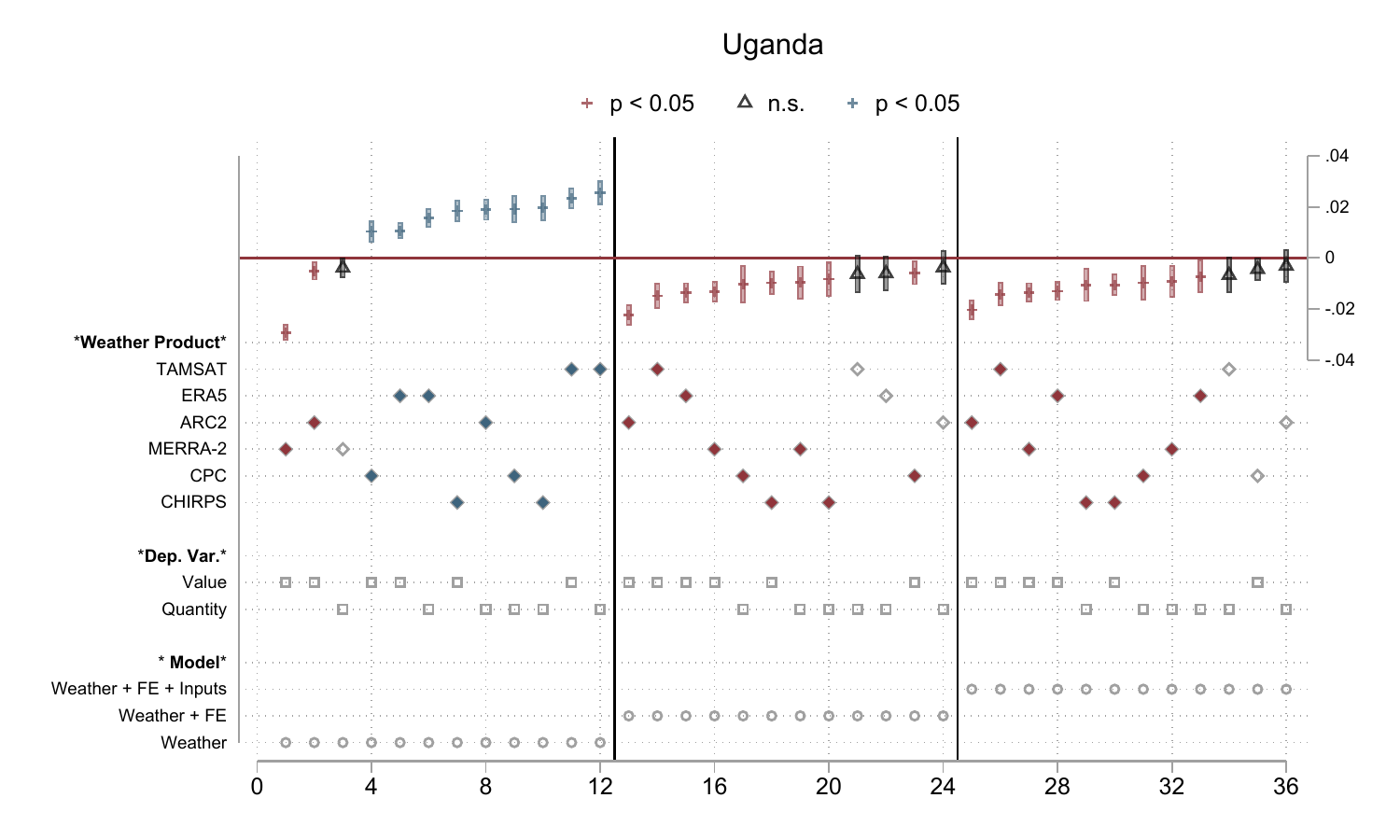}
		\end{center}
		\footnotesize  \textit{Note}: The figure presents specification curves, where each panel represents a different country, with three different models presented within each panel. Each panel includes 36 regressions, where each column represents a single regression. Significant and non-significant coefficients are designated at the top of the figure. For each Earth observation product, we also designate the significance and sign of the coefficient with color: red represents coefficients which are negative and significant; white represents insignificant coefficients, regardless of sign; and blue represents coefficients which are positive and significant.  
	\end{minipage}	
\end{figure}
\end{center}

\begin{center}
\begin{figure}[!htbp]
	\begin{minipage}{\linewidth}
		\caption{Specification Charts for Percentage of Days with Rain}
		\label{fig:pval_v12}
		\begin{center}
			\includegraphics[width=.49\linewidth,keepaspectratio]{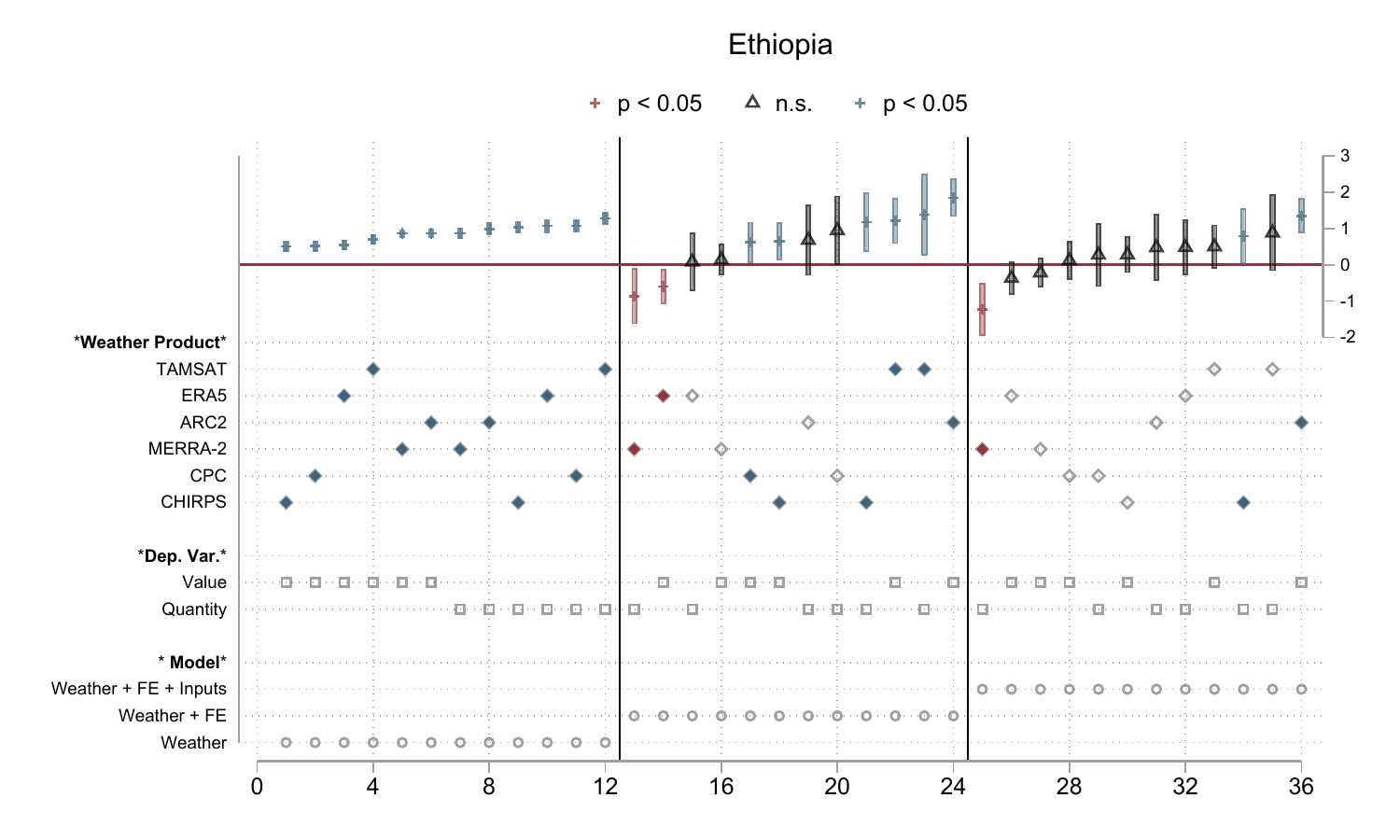}
			\includegraphics[width=.49\linewidth,keepaspectratio]{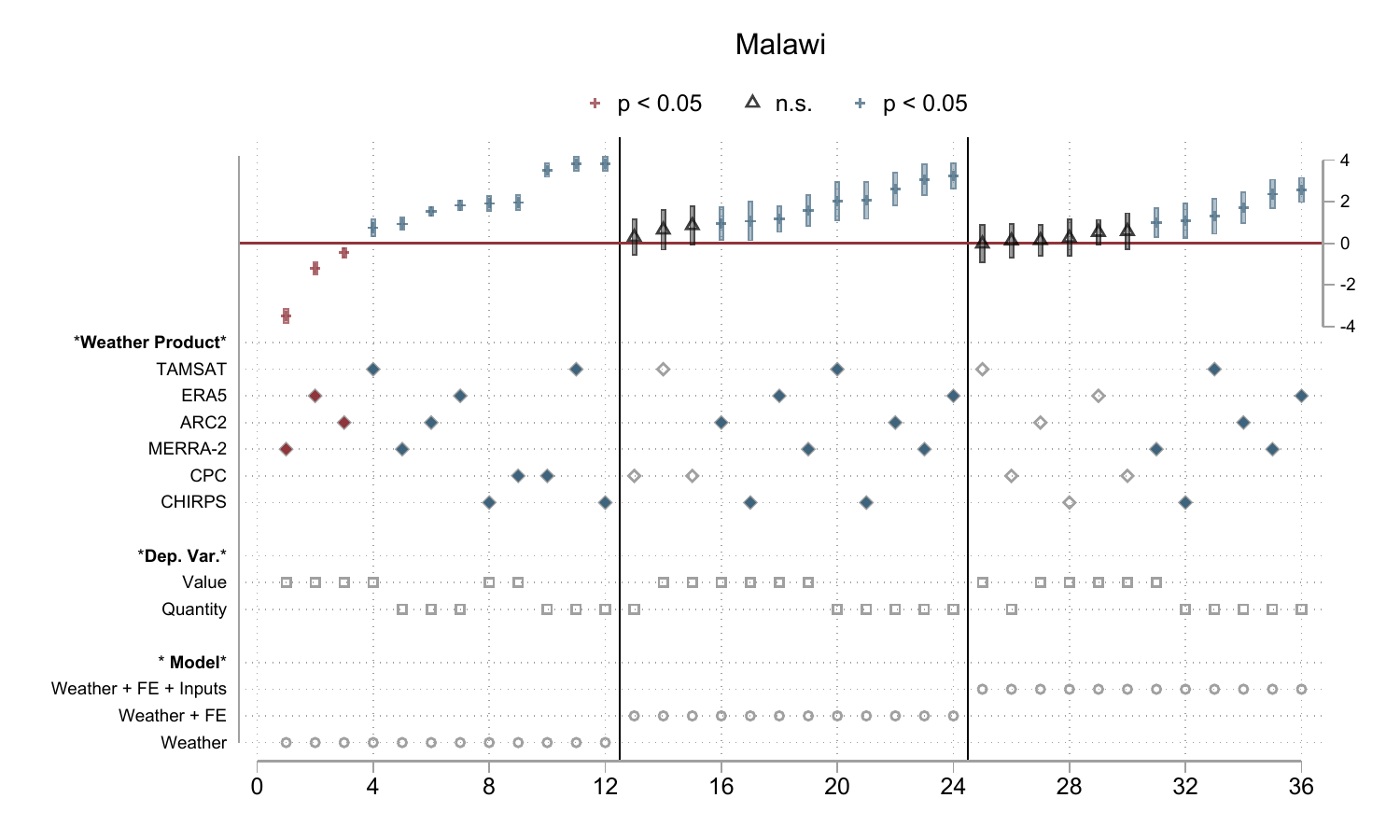}
			\includegraphics[width=.49\linewidth,keepaspectratio]{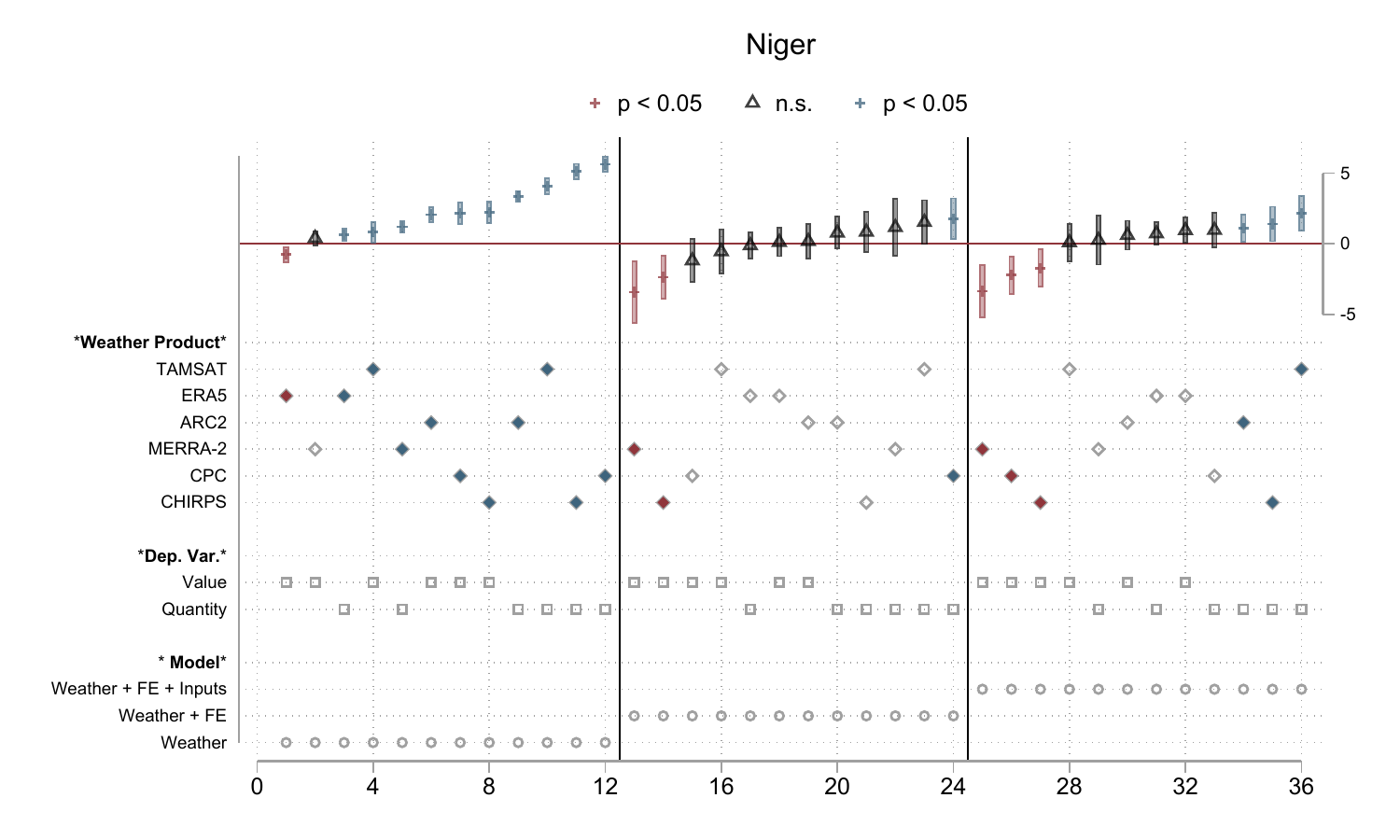}
			\includegraphics[width=.49\linewidth,keepaspectratio]{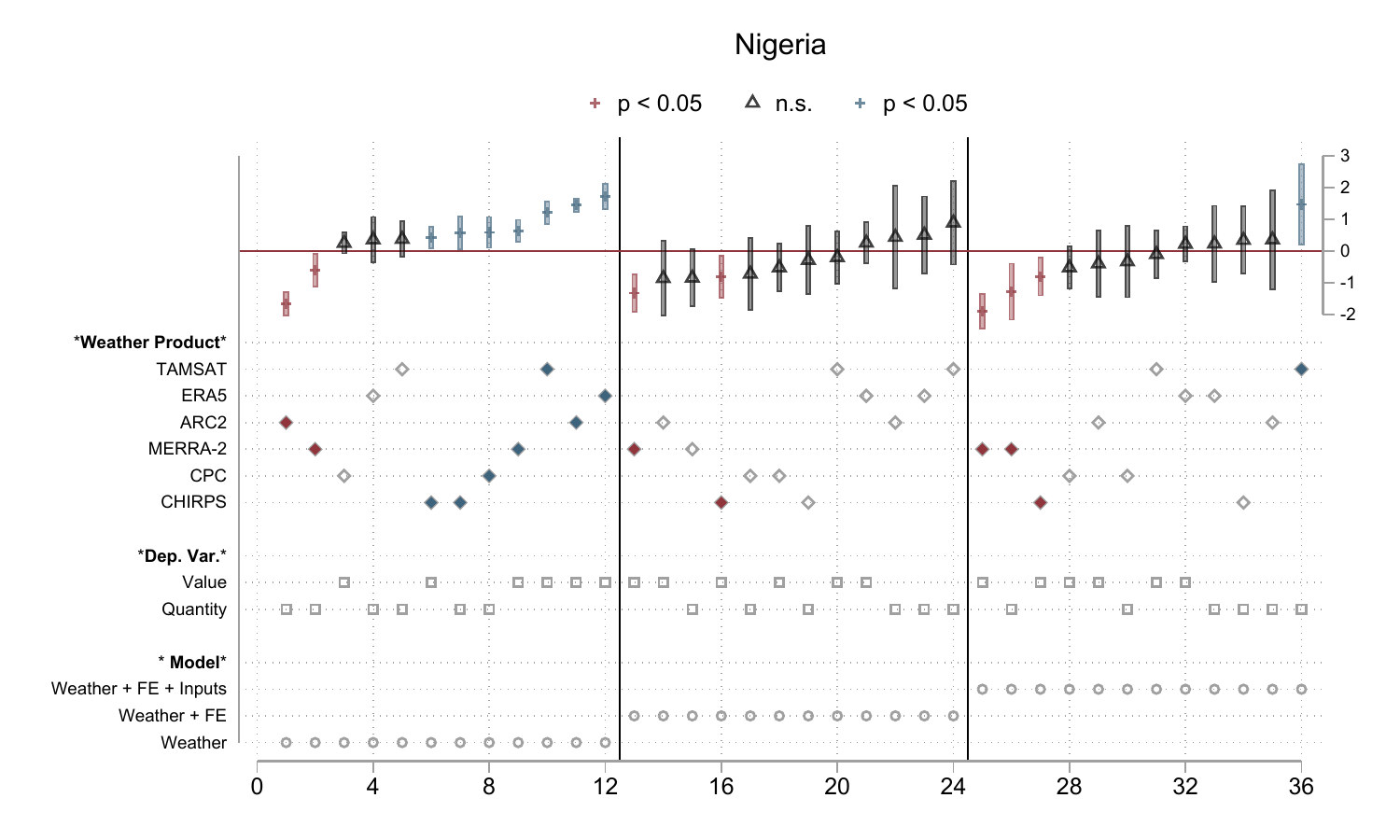}
			\includegraphics[width=.49\linewidth,keepaspectratio]{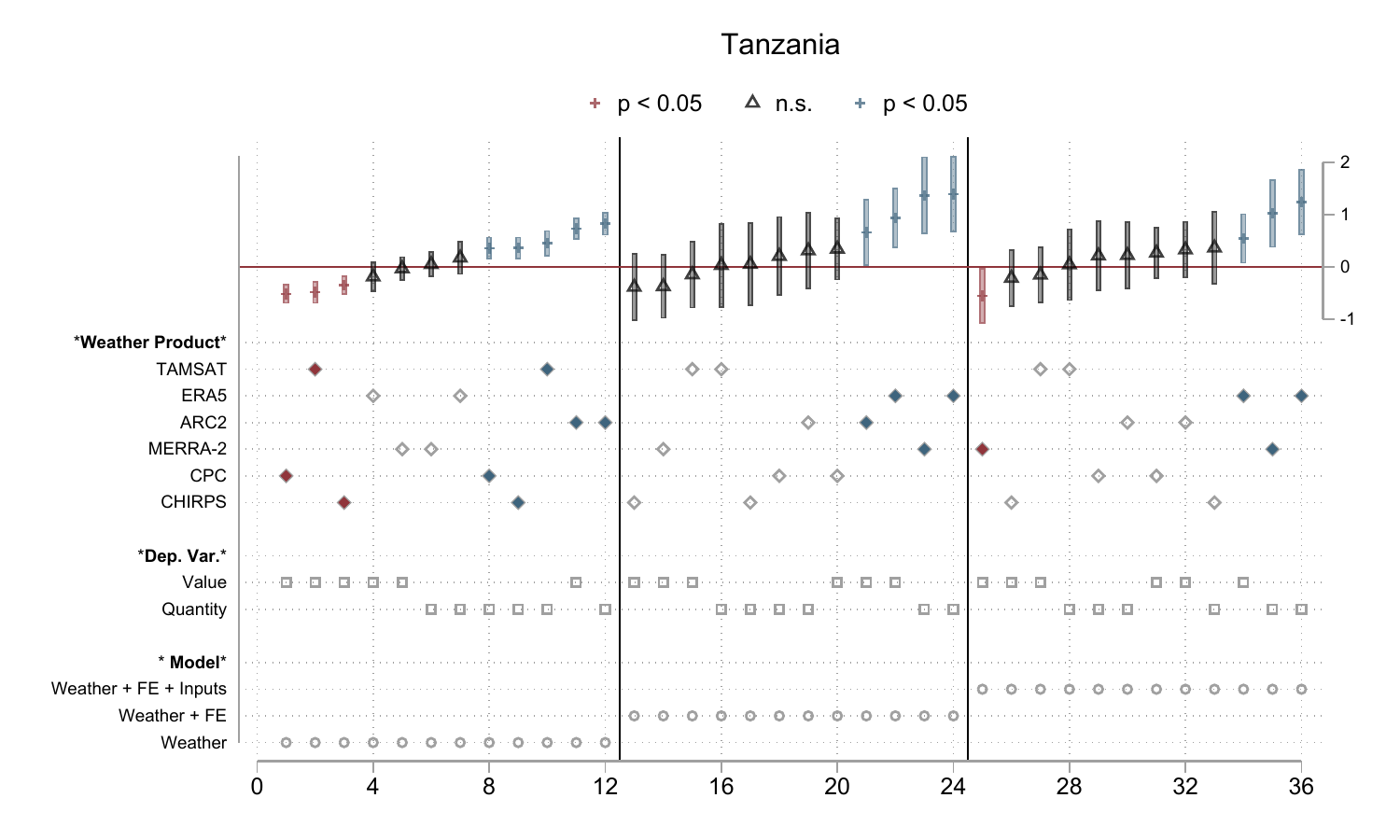}
			\includegraphics[width=.49\linewidth,keepaspectratio]{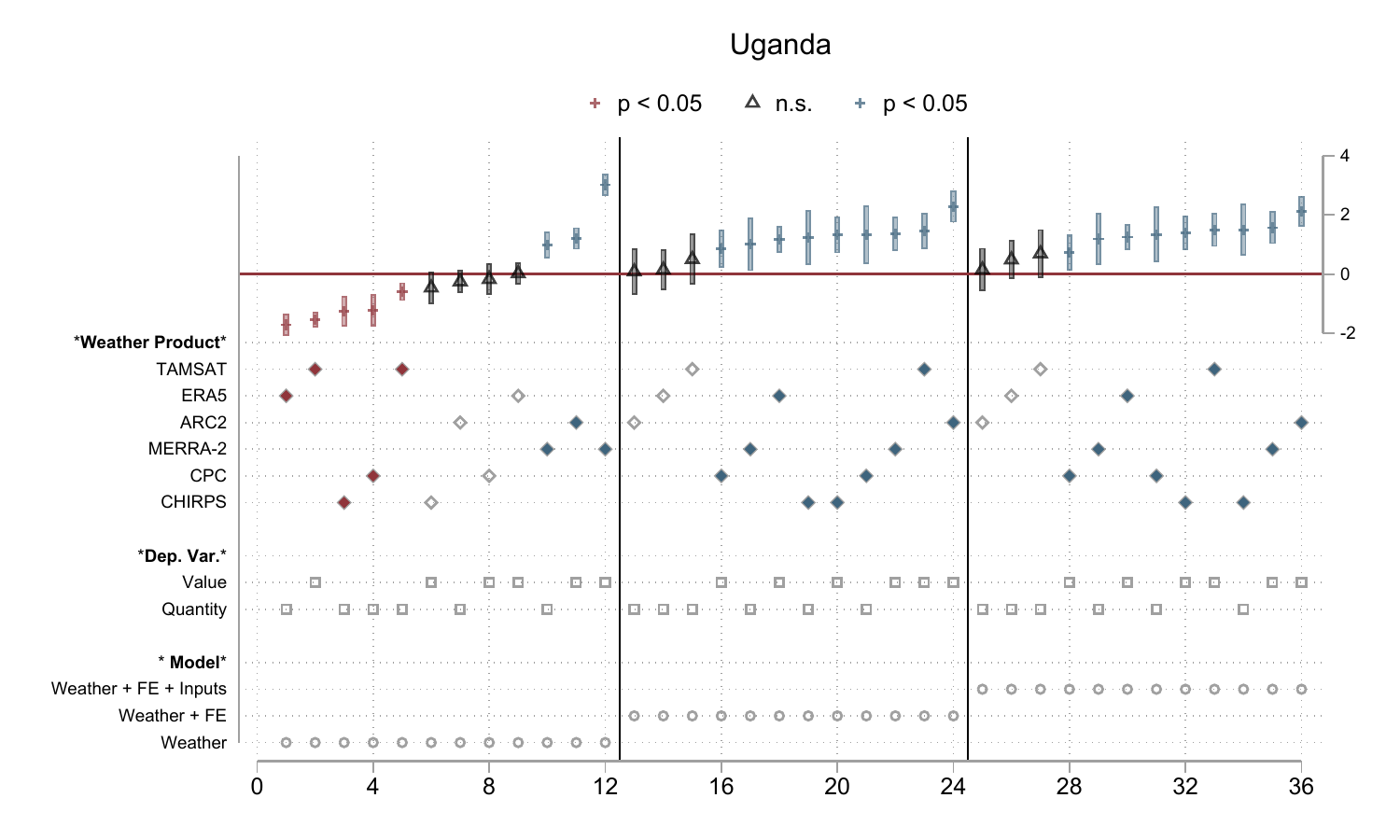}
		\end{center}
		\footnotesize \textit{Note}: The figure presents specification curves, where each panel represents a different country, with three different models presented within each panel. Each panel includes 36 regressions, where each column represents a single regression. Significant and non-significant coefficients are designated at the top of the figure. For each Earth observation product, we also designate the significance and sign of the coefficient with color: red represents coefficients which are negative and significant; white represents insignificant coefficients, regardless of sign; and blue represents coefficients which are positive and significant.  
	\end{minipage}	
\end{figure}
\end{center}

\begin{center}
\begin{figure}[!htbp]
	\begin{minipage}{\linewidth}
		\caption{Specification Charts for Deviations in Percentage of Days with Rain}
		\label{fig:pval_v13}
		\begin{center}
			\includegraphics[width=.49\linewidth,keepaspectratio]{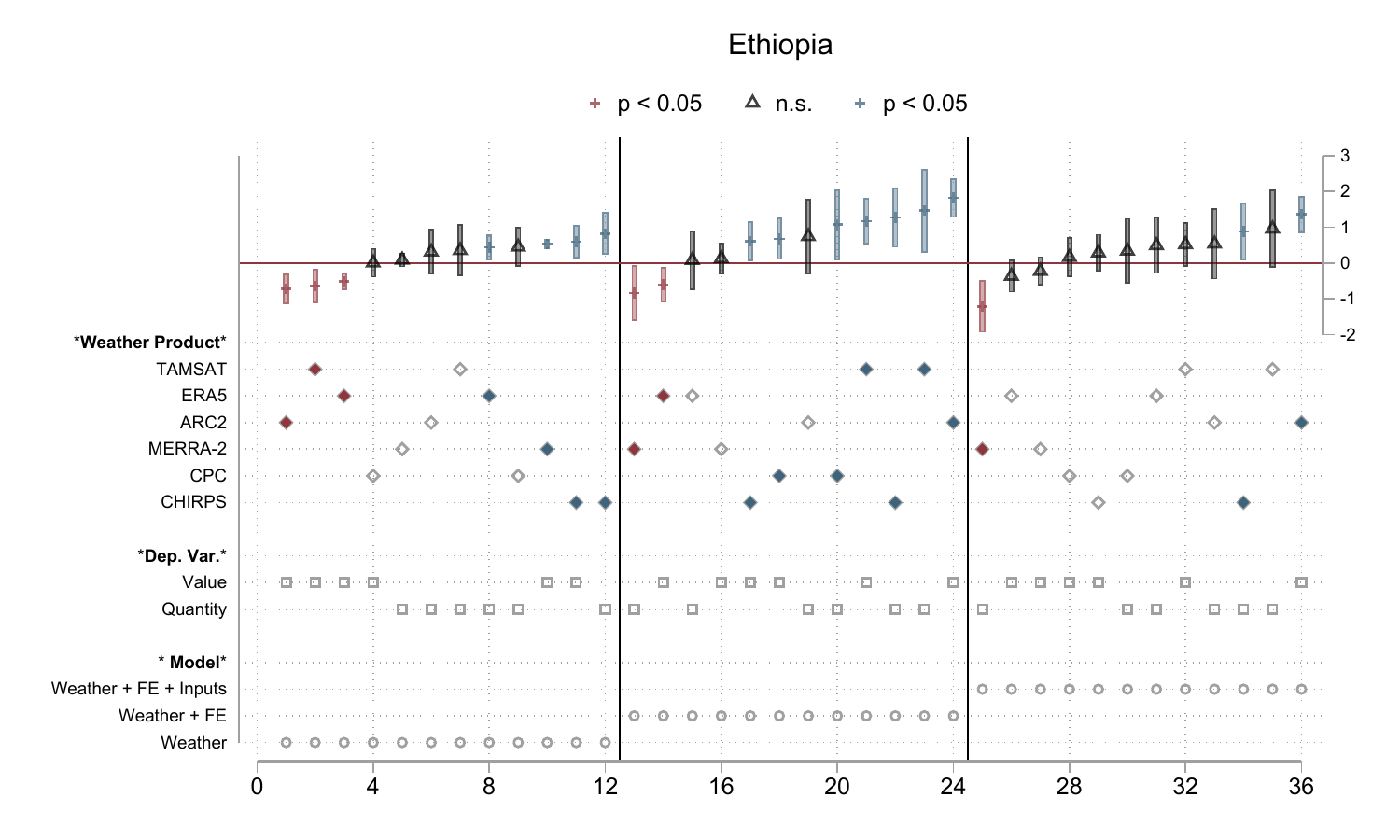}
			\includegraphics[width=.49\linewidth,keepaspectratio]{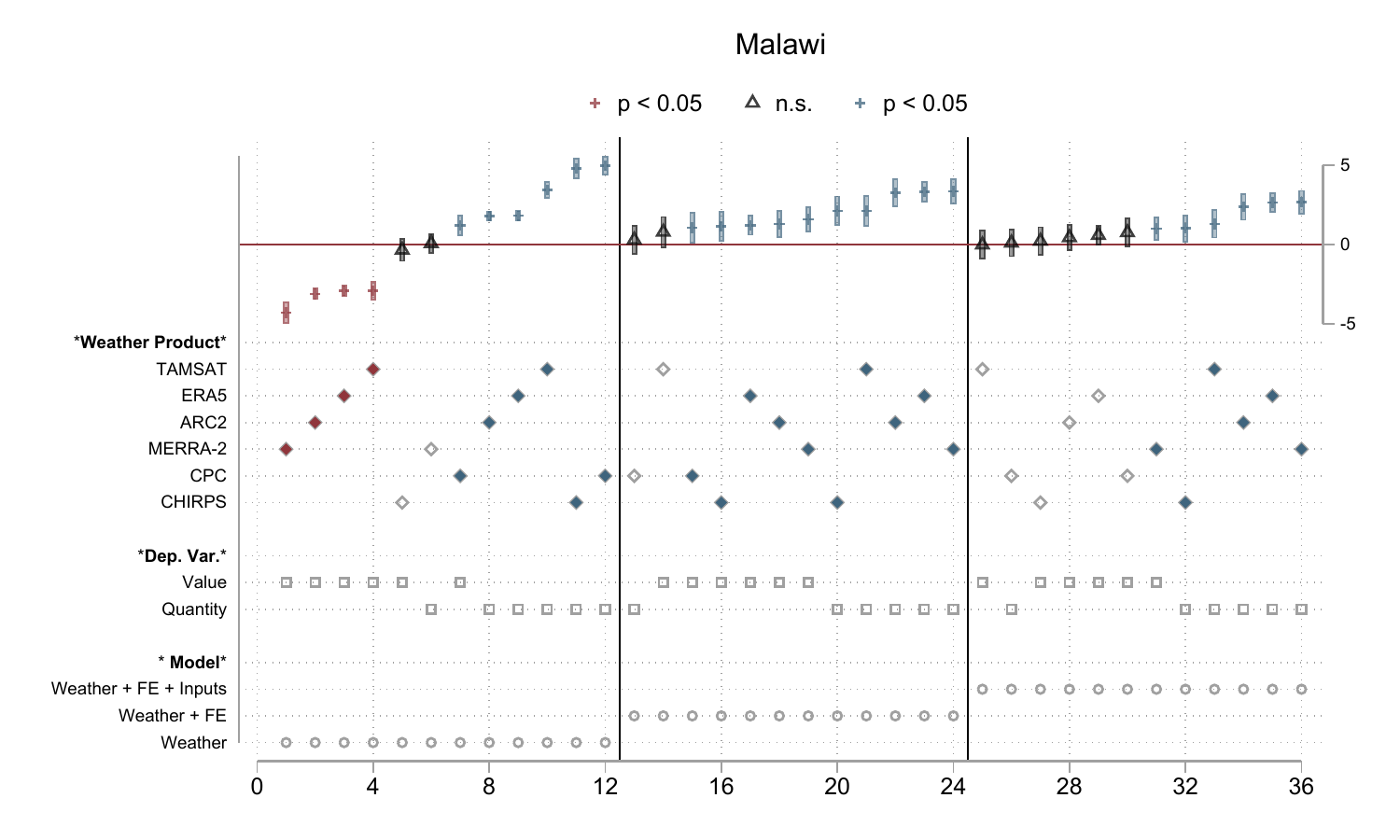}
			\includegraphics[width=.49\linewidth,keepaspectratio]{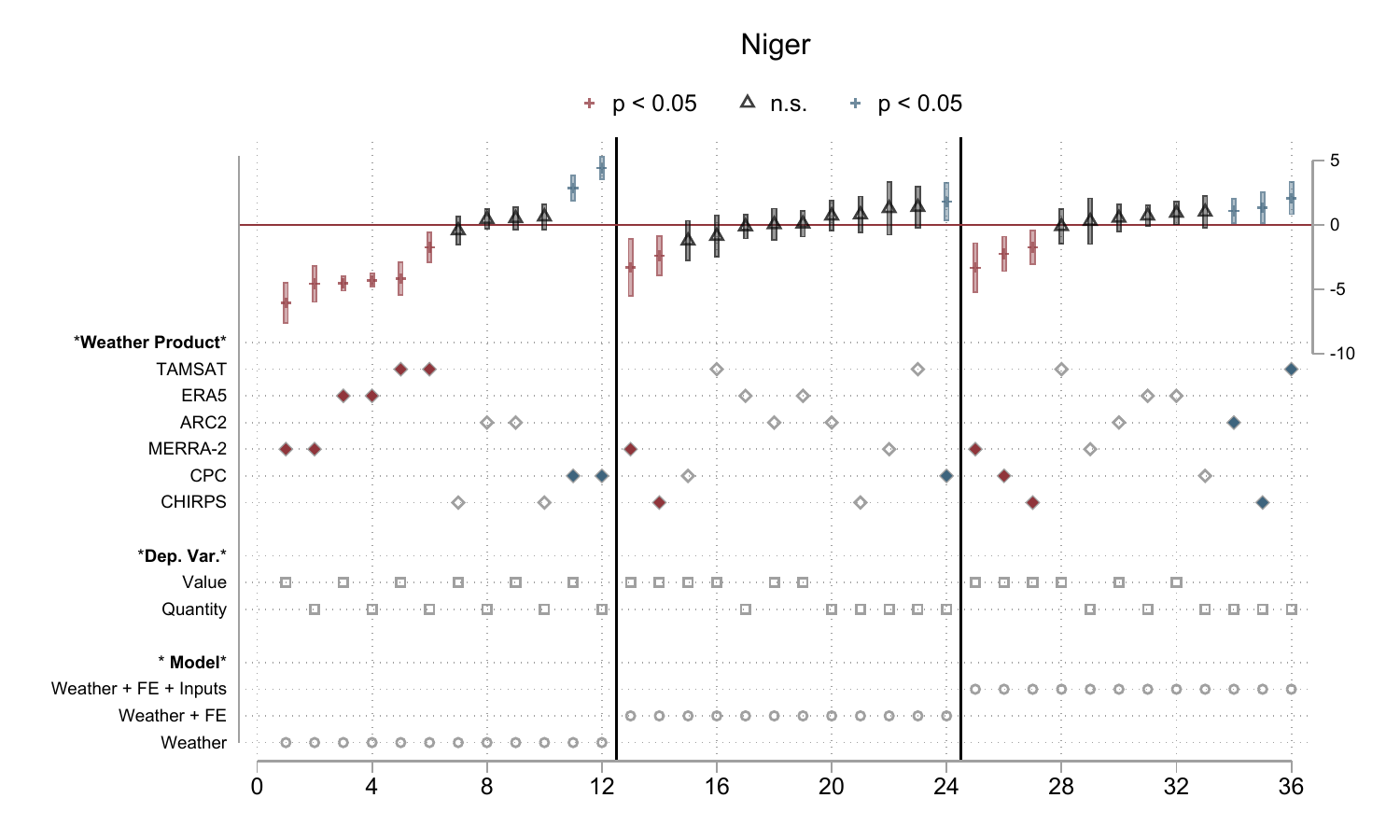}
			\includegraphics[width=.49\linewidth,keepaspectratio]{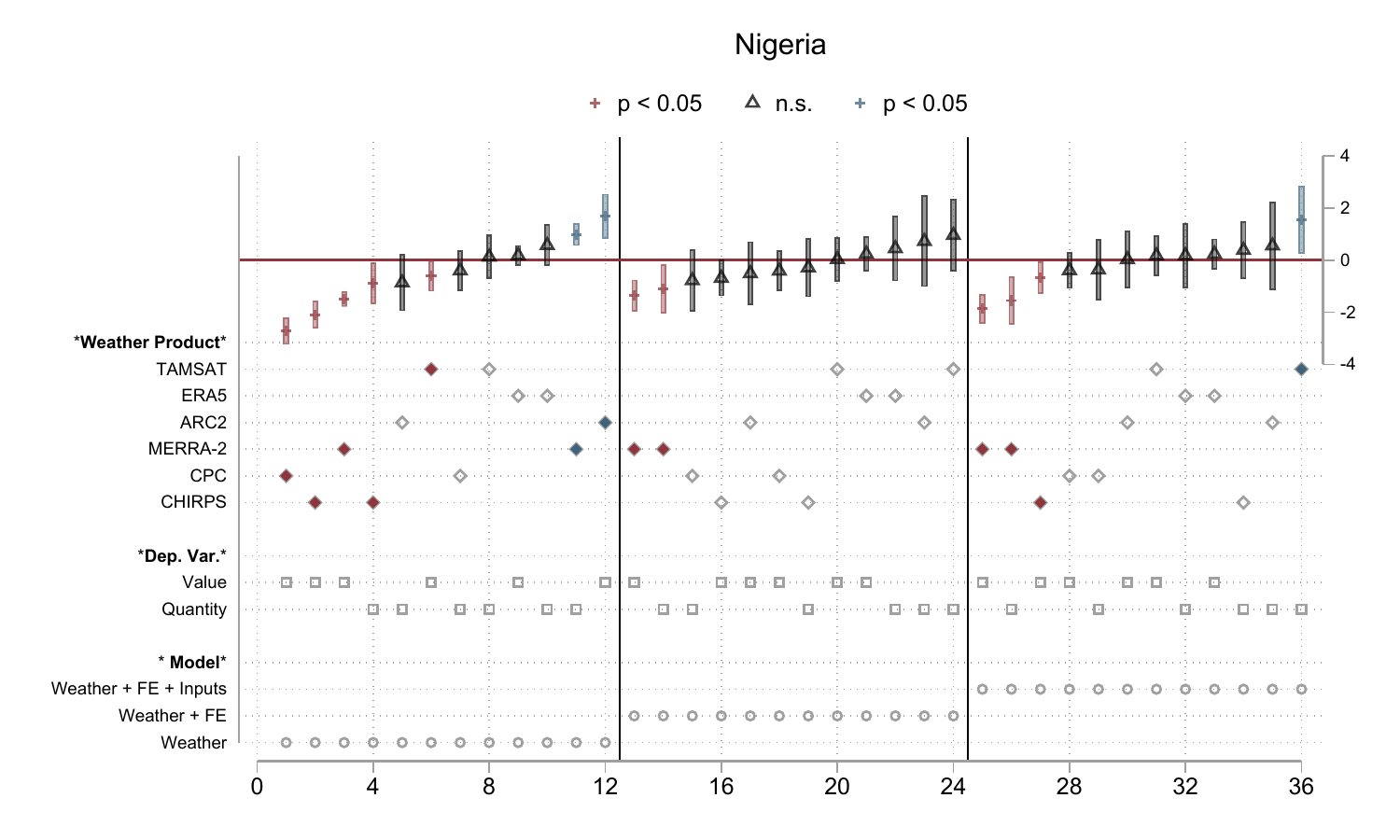}
			\includegraphics[width=.49\linewidth,keepaspectratio]{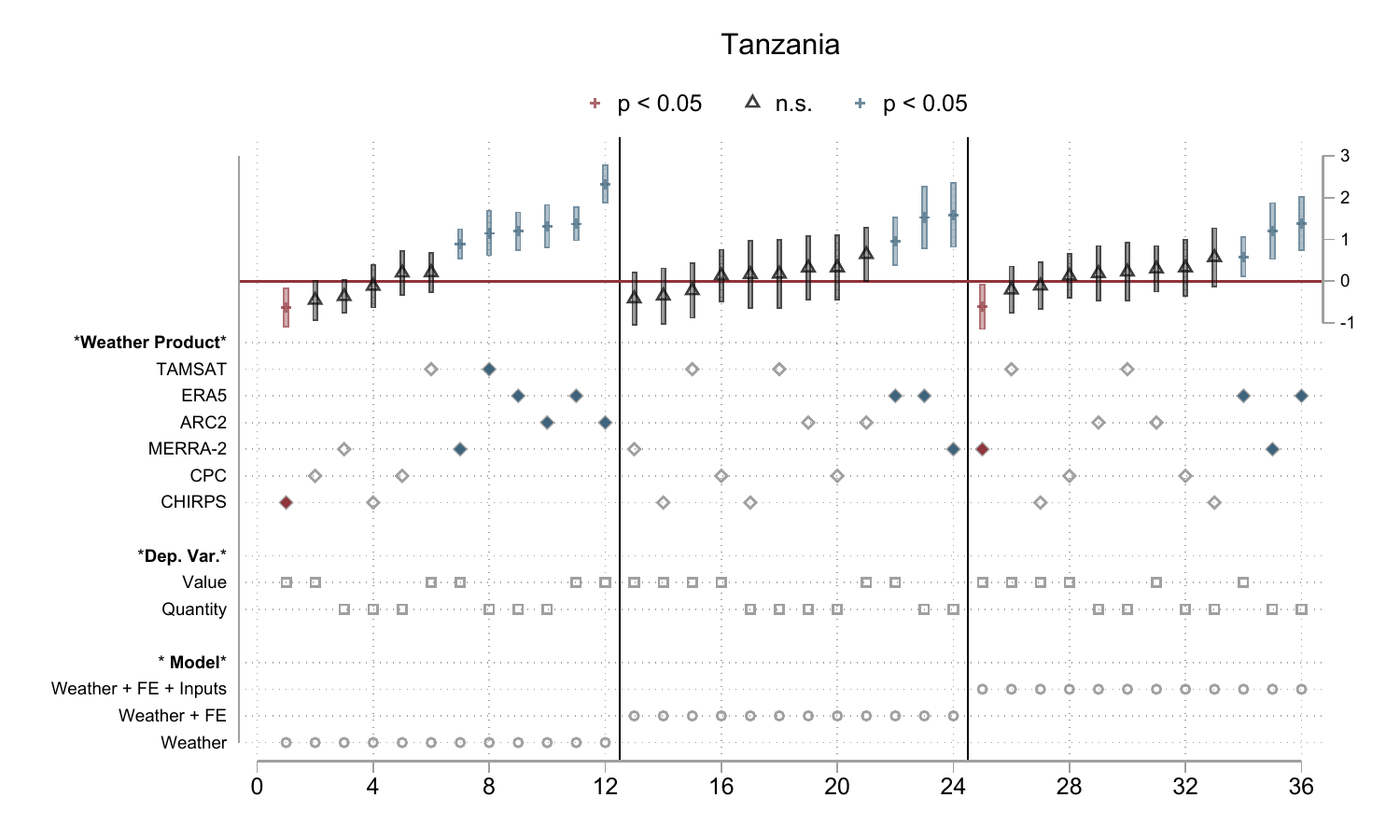}
			\includegraphics[width=.49\linewidth,keepaspectratio]{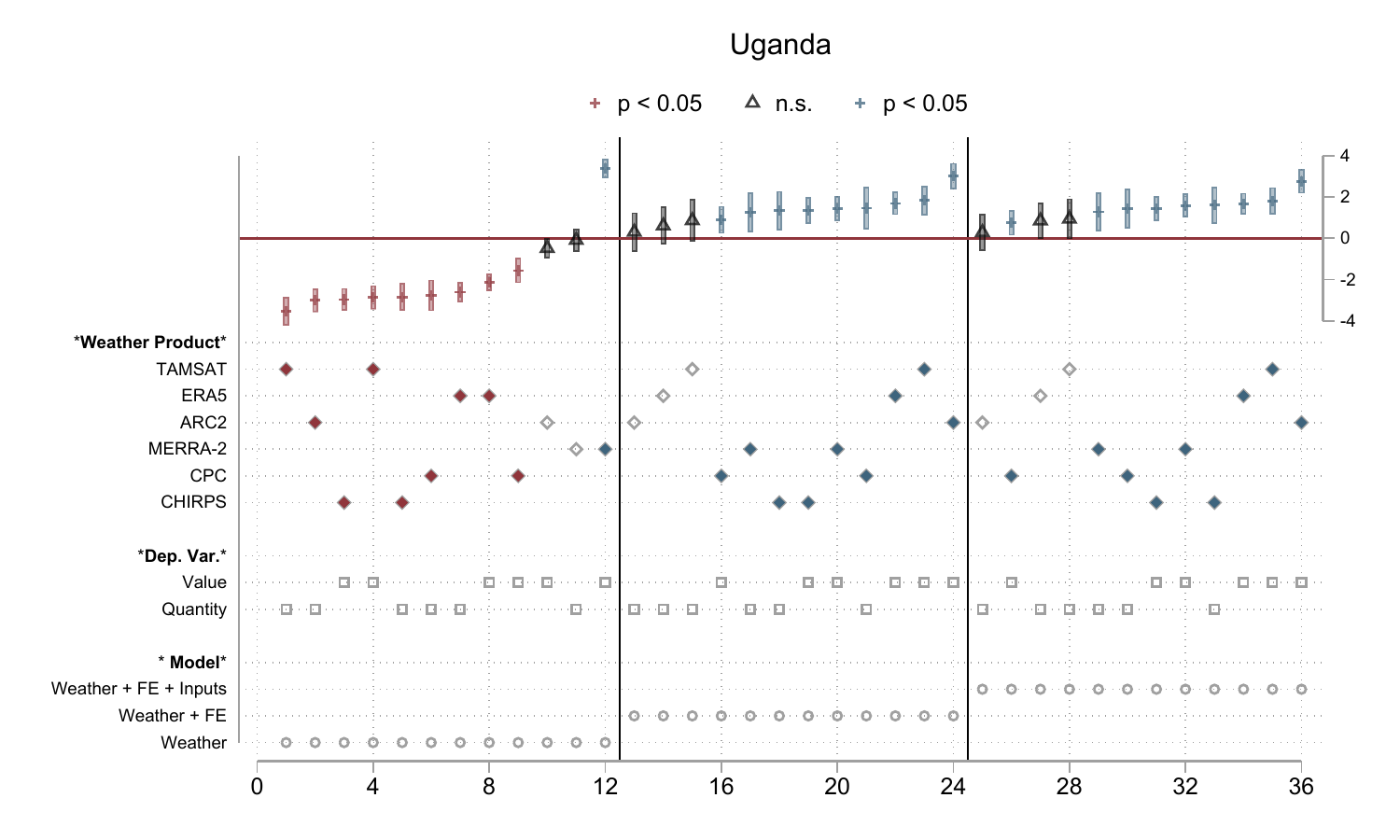}
		\end{center}
		\footnotesize  \textit{Note}: The figure presents specification curves, where each panel represents a different country, with three different models presented within each panel. Each panel includes 36 regressions, where each column represents a single regression. Significant and non-significant coefficients are designated at the top of the figure. For each Earth observation product, we also designate the significance and sign of the coefficient with color: red represents coefficients which are negative and significant; white represents insignificant coefficients, regardless of sign; and blue represents coefficients which are positive and significant.  
	\end{minipage}	
\end{figure}
\end{center}

\begin{center}
\begin{figure}[!htbp]
	\begin{minipage}{\linewidth}
		\caption{Specification Charts for Longest Dry Spell}
		\label{fig:pval_v14}
		\begin{center}
			\includegraphics[width=.49\linewidth,keepaspectratio]{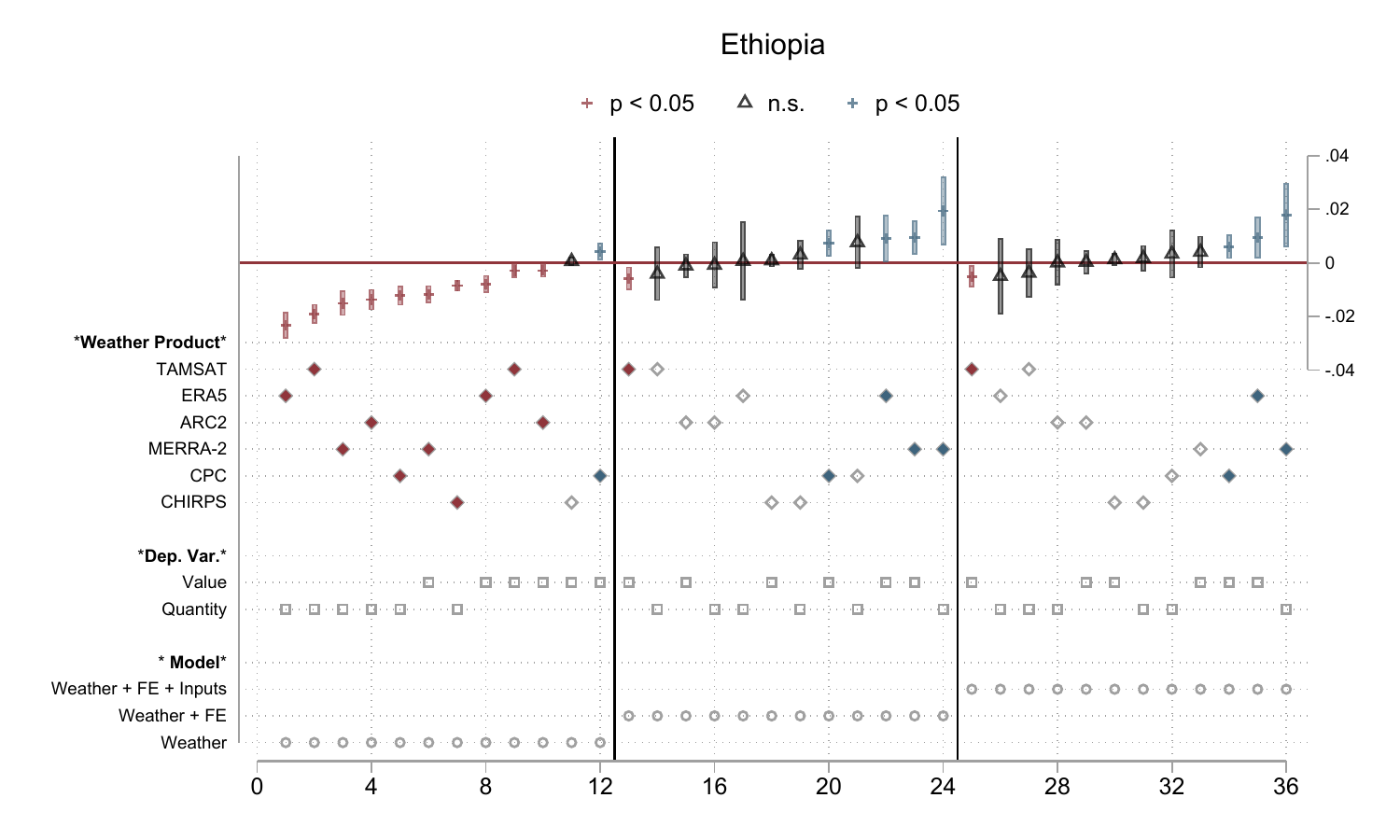}
			\includegraphics[width=.49\linewidth,keepaspectratio]{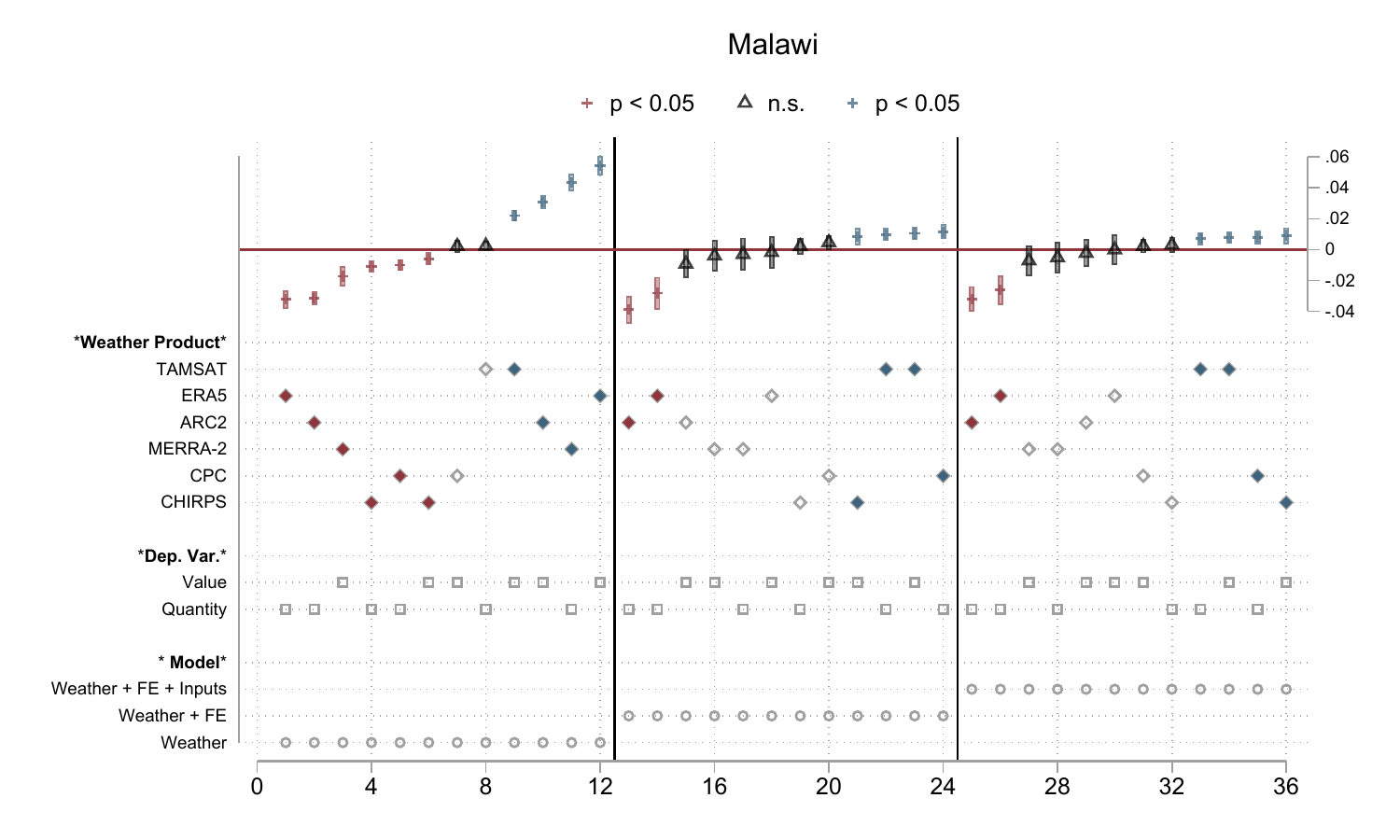}
			\includegraphics[width=.49\linewidth,keepaspectratio]{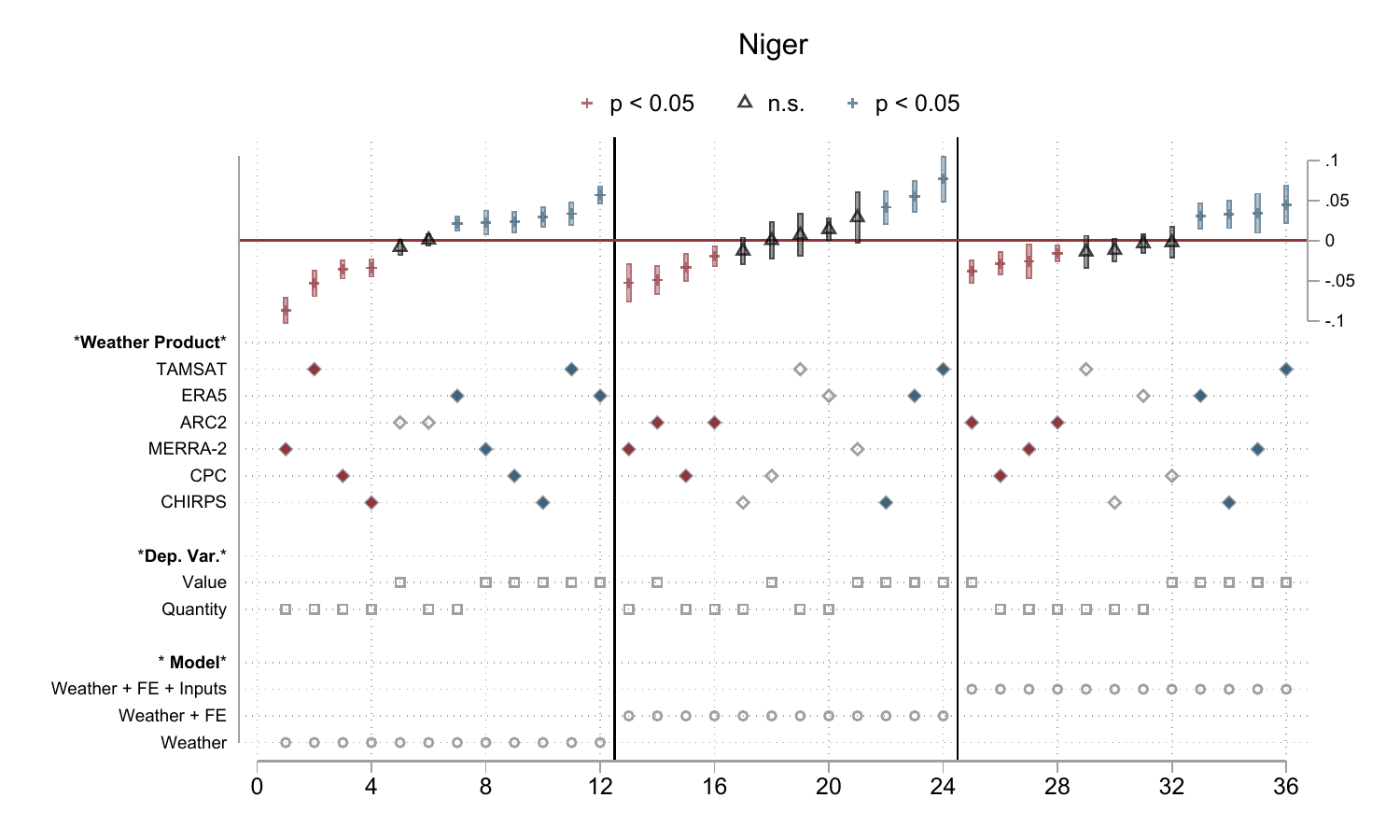}
			\includegraphics[width=.49\linewidth,keepaspectratio]{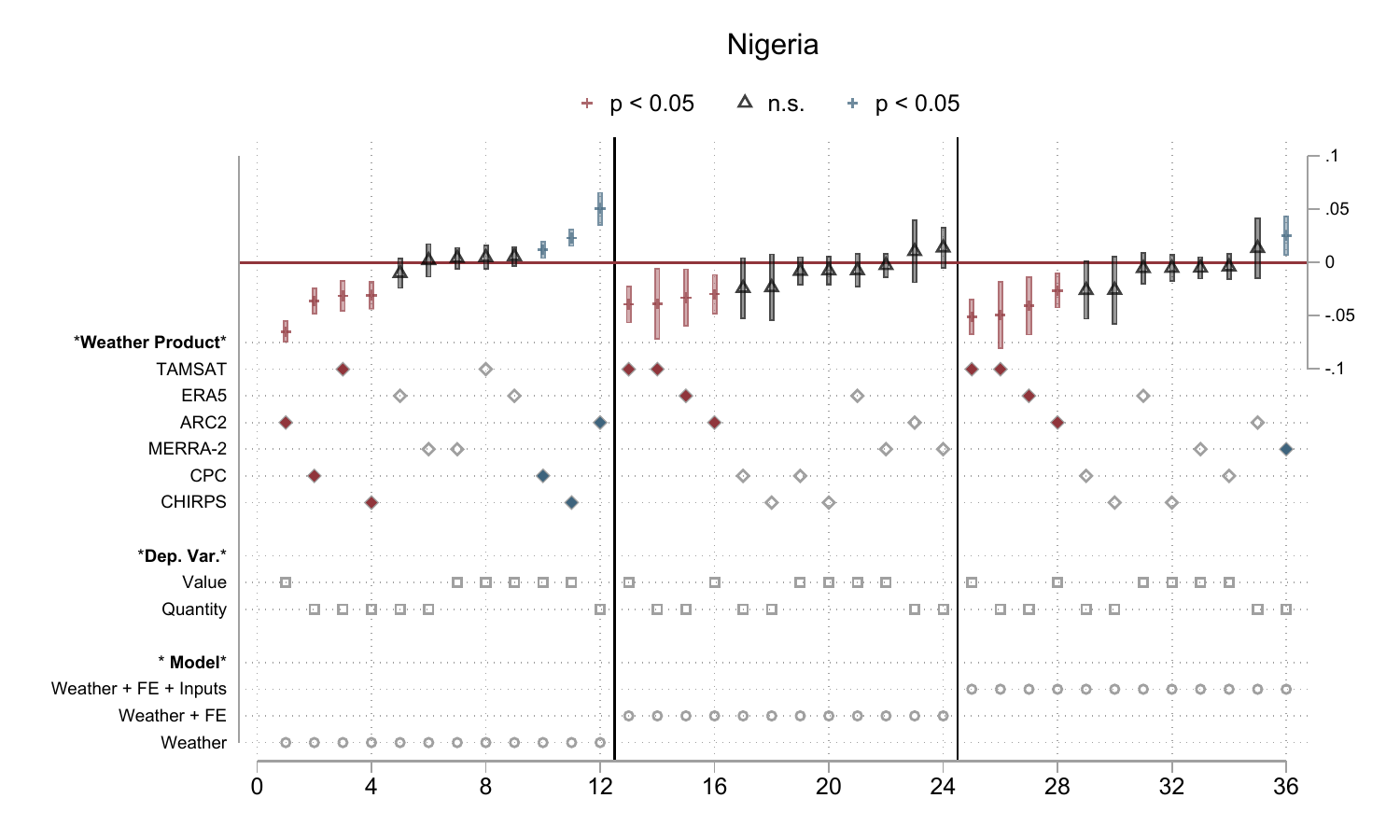}
			\includegraphics[width=.49\linewidth,keepaspectratio]{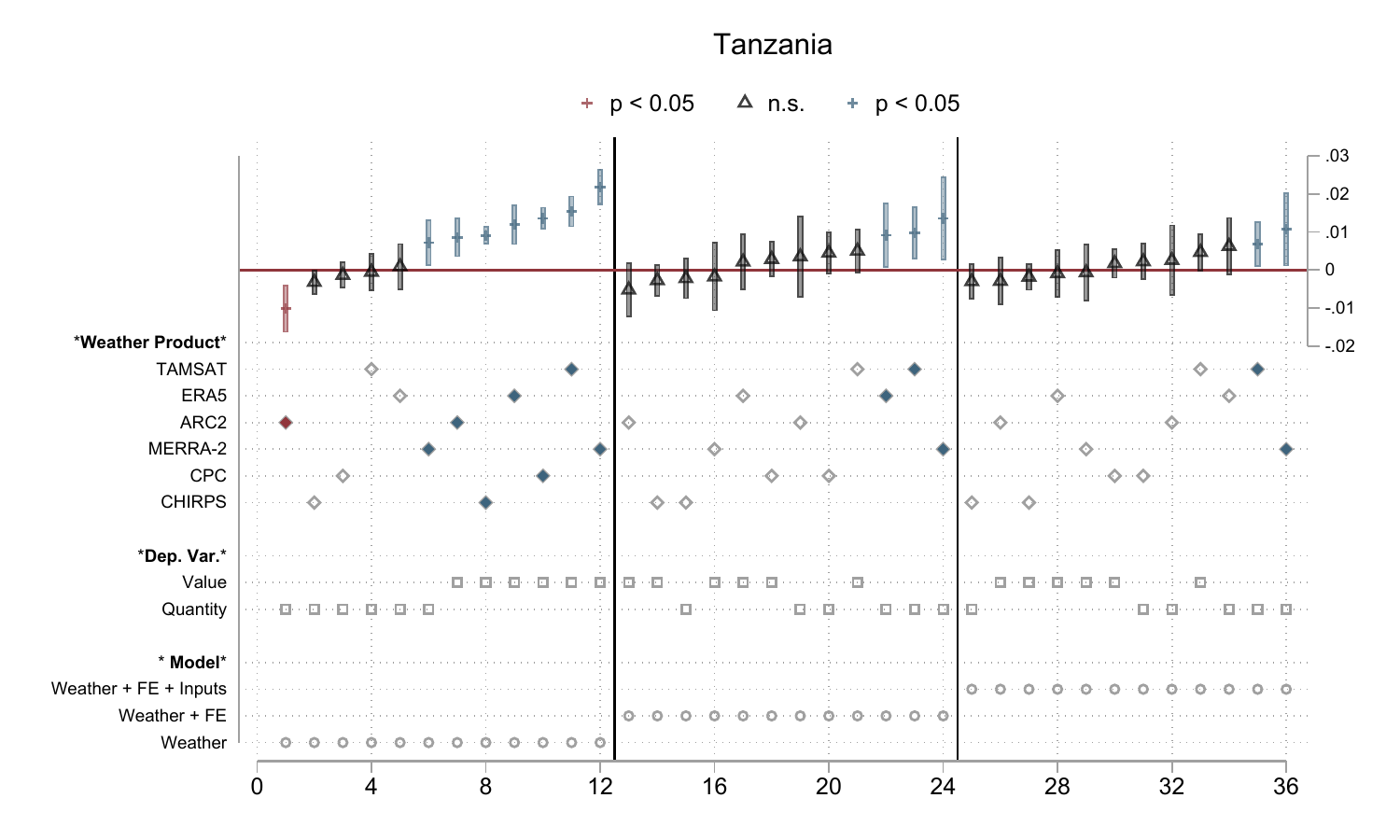}
			\includegraphics[width=.49\linewidth,keepaspectratio]{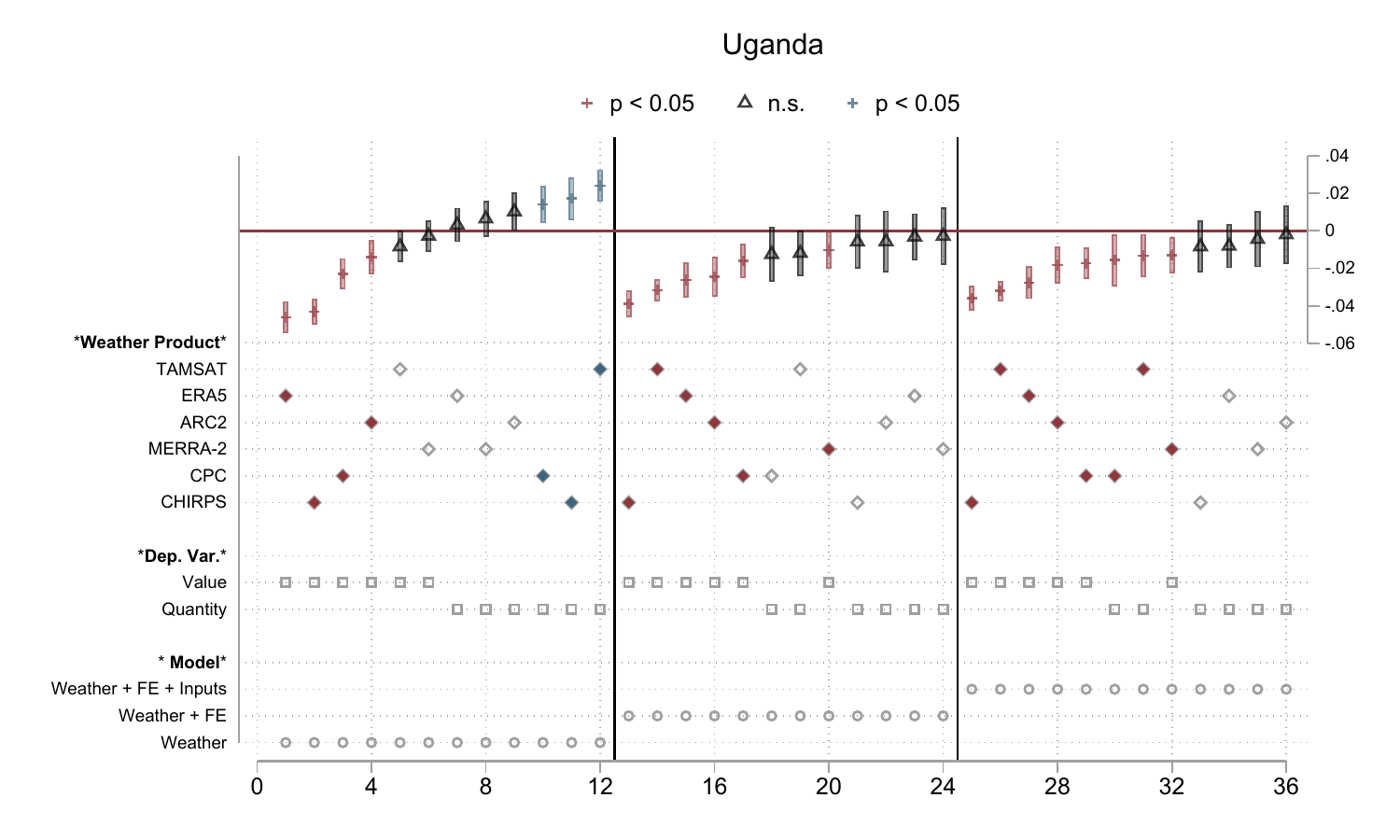}
		\end{center}
		\footnotesize  \textit{Note}: The figure presents specification curves, where each panel represents a different country, with three different models presented within each panel. Each panel includes 36 regressions, where each column represents a single regression. Significant and non-significant coefficients are designated at the top of the figure. For each Earth observation product, we also designate the significance and sign of the coefficient with color: red represents coefficients which are negative and significant; white represents insignificant coefficients, regardless of sign; and blue represents coefficients which are positive and significant.  
	\end{minipage}	
\end{figure}
\end{center}

\begin{center}
\begin{figure}[!htbp]
	\begin{minipage}{\linewidth}
		\caption{Specification Charts for Median Daily Temperature}
		\label{fig:pval_v16}
		\begin{center}
			\includegraphics[width=.49\linewidth,keepaspectratio]{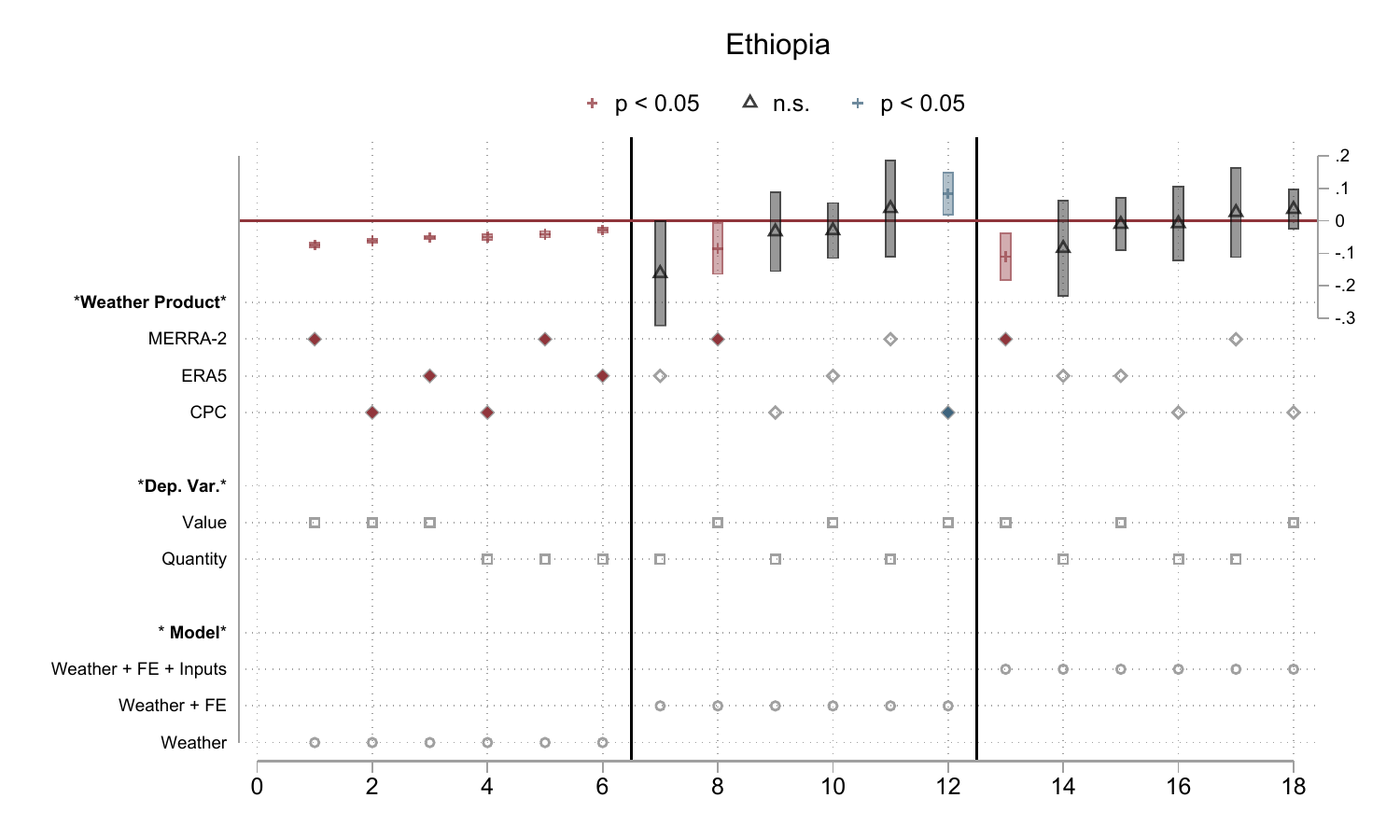}
			\includegraphics[width=.49\linewidth,keepaspectratio]{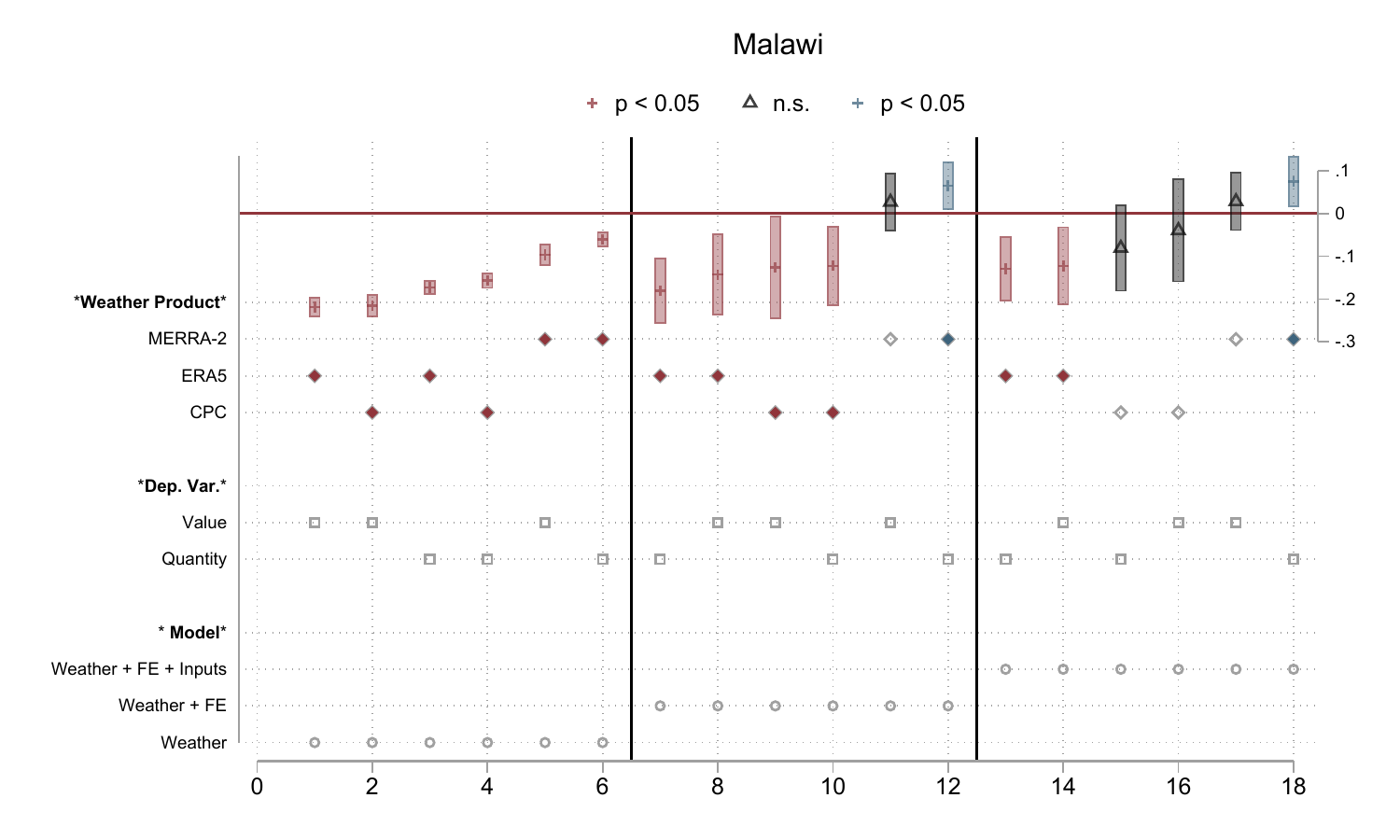}
			\includegraphics[width=.49\linewidth,keepaspectratio]{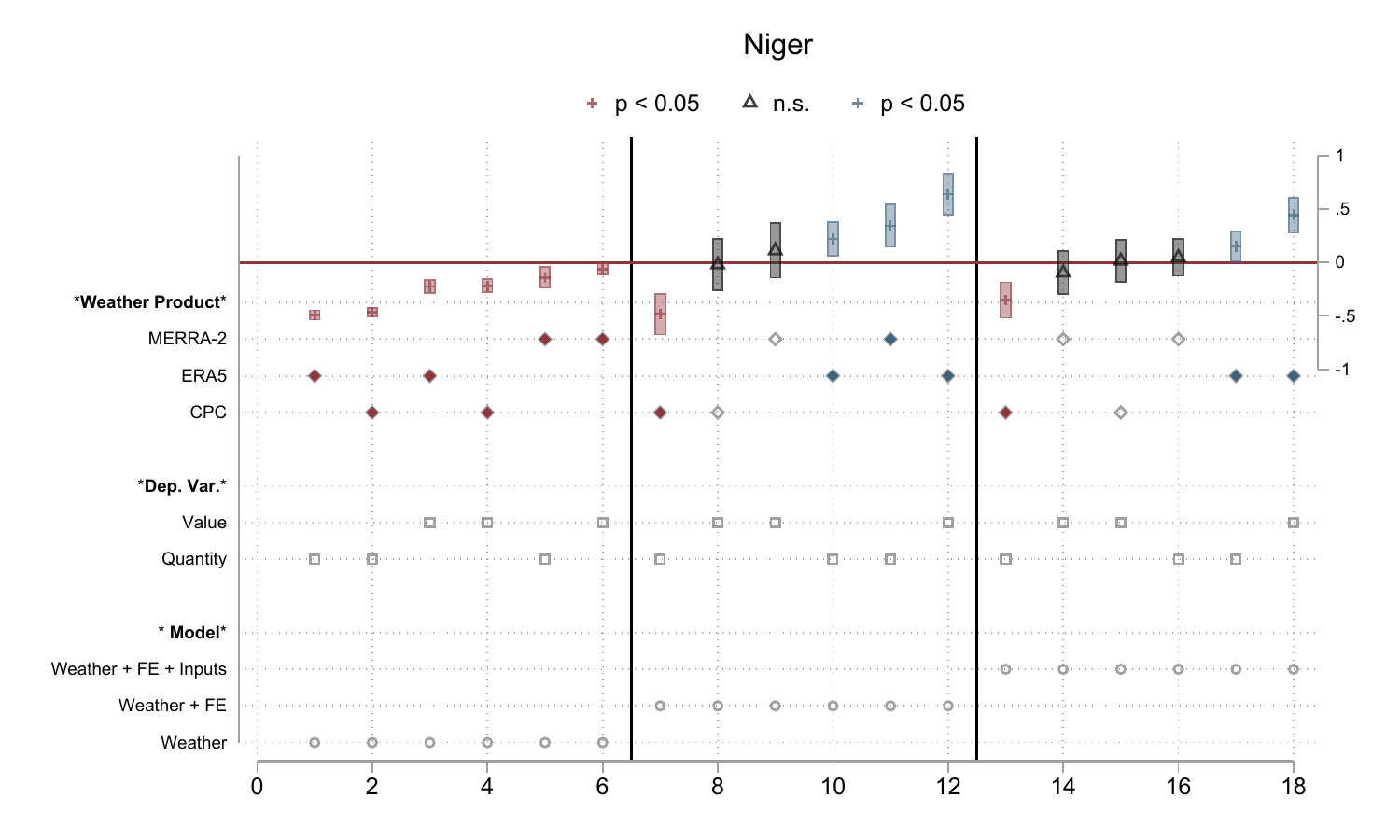}
			\includegraphics[width=.49\linewidth,keepaspectratio]{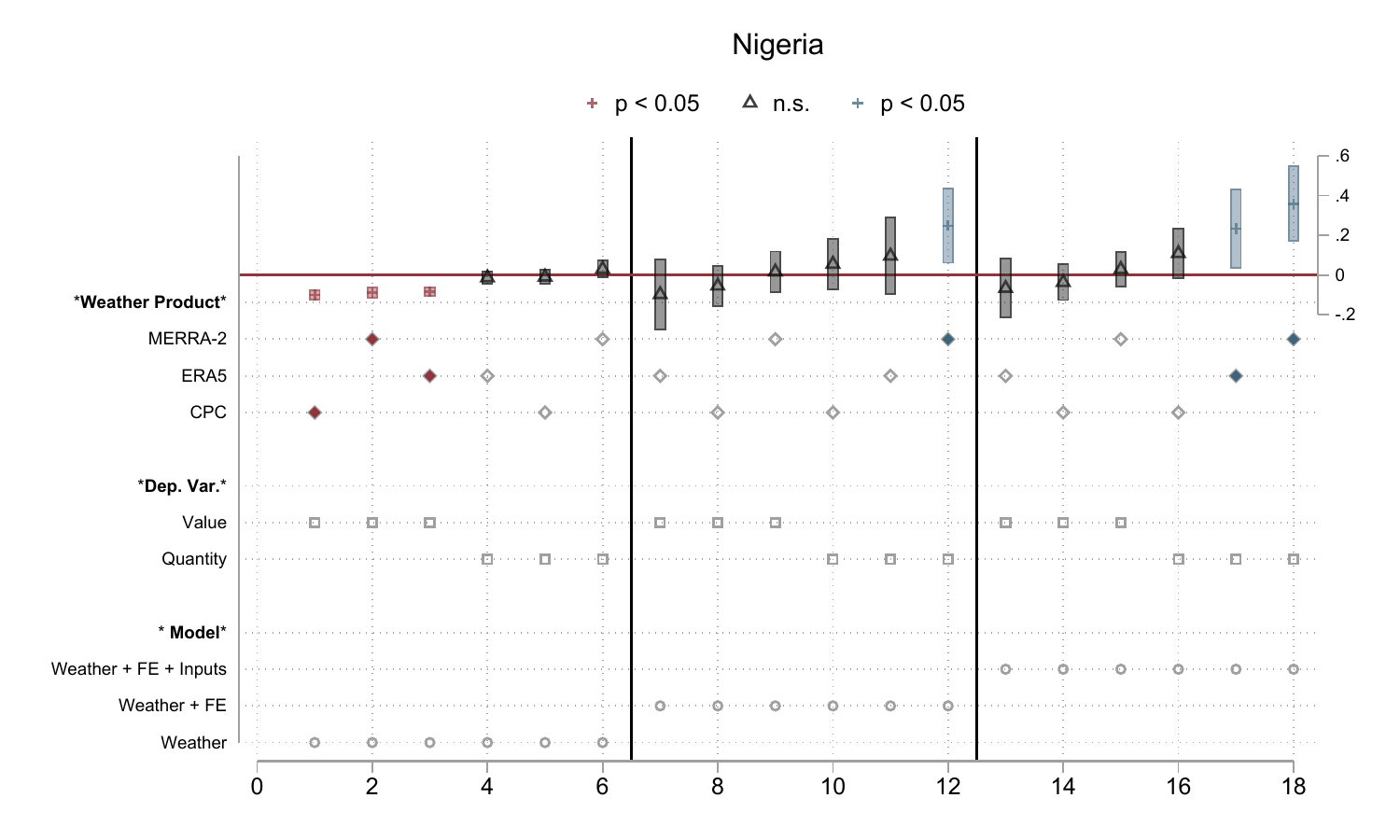}
			\includegraphics[width=.49\linewidth,keepaspectratio]{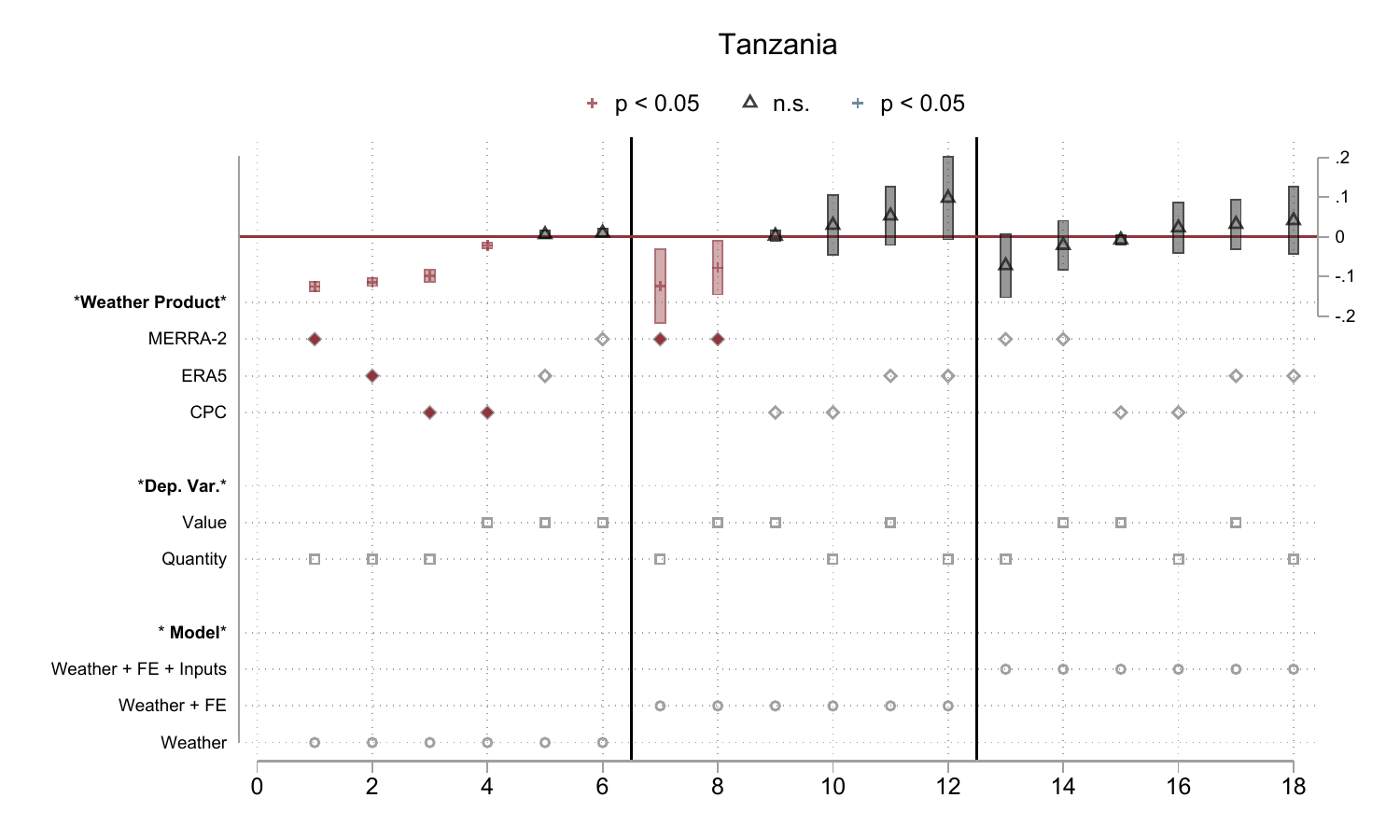}
			\includegraphics[width=.49\linewidth,keepaspectratio]{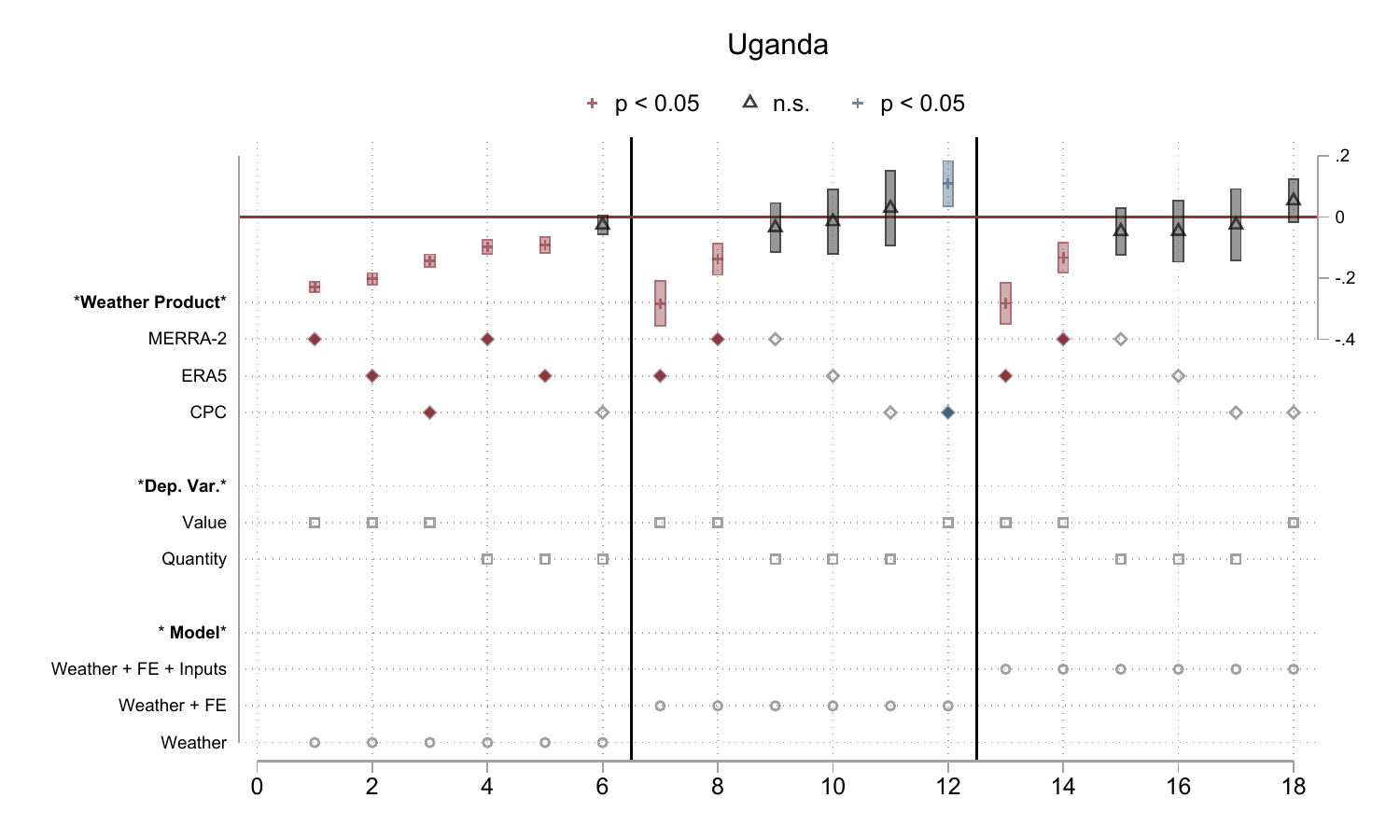}
		\end{center}
		\footnotesize  \textit{Note}: The figure presents specification curves, where each panel represents a different country, with three different models presented within each panel. Each panel includes 18 regressions, where each column represents a single regression. Significant and non-significant coefficients are designated at the top of the figure. For each Earth observation product, we also designate the significance and sign of the coefficient with color: red represents coefficients which are negative and significant; white represents insignificant coefficients, regardless of sign; and blue represents coefficients which are positive and significant.  
	\end{minipage}	
\end{figure}
\end{center}

\begin{center}
\begin{figure}[!htbp]
	\begin{minipage}{\linewidth}
		\caption{Specification Charts for Variance of Daily Temperature}
		\label{fig:pval_v17}
		\begin{center}
			\includegraphics[width=.49\linewidth,keepaspectratio]{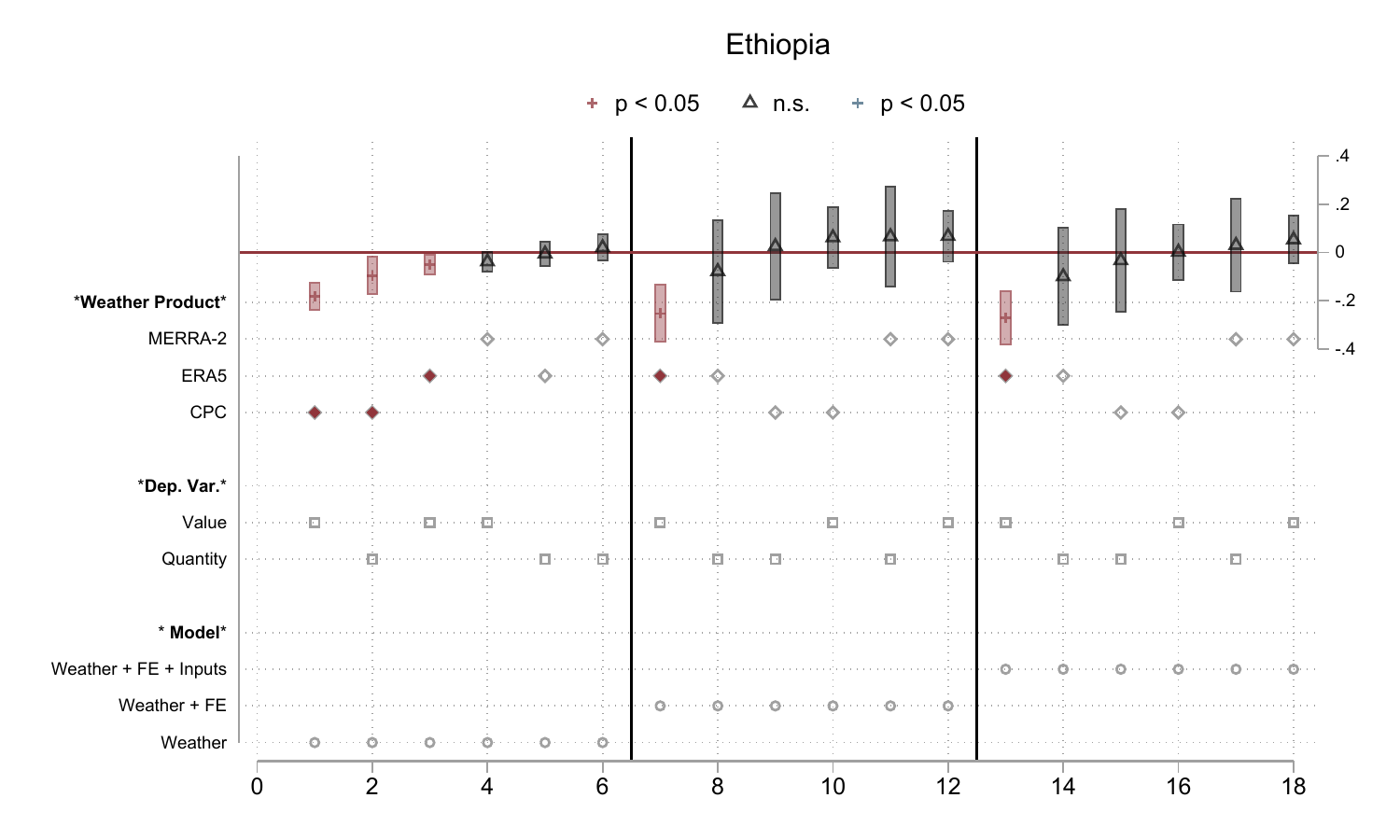}
			\includegraphics[width=.49\linewidth,keepaspectratio]{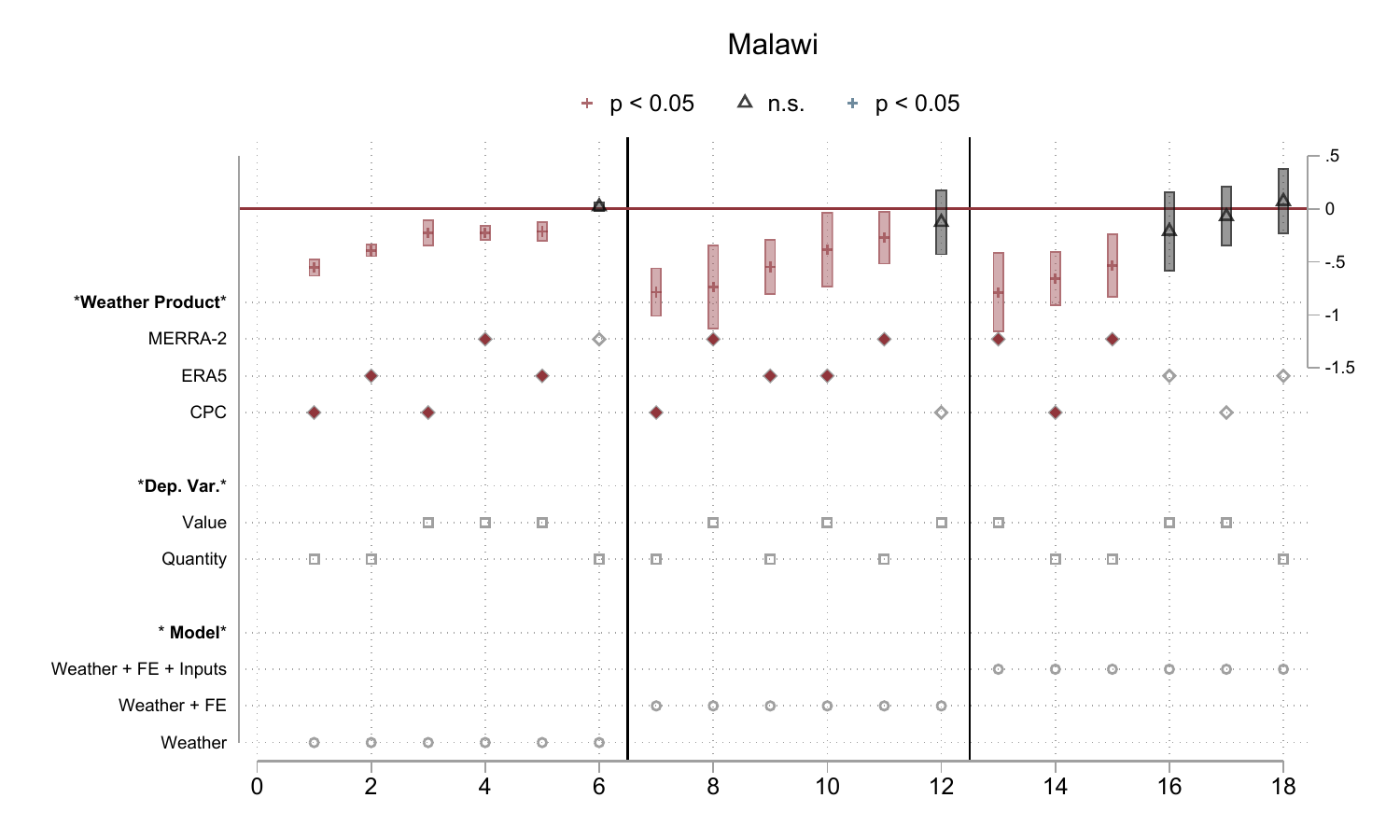}
			\includegraphics[width=.49\linewidth,keepaspectratio]{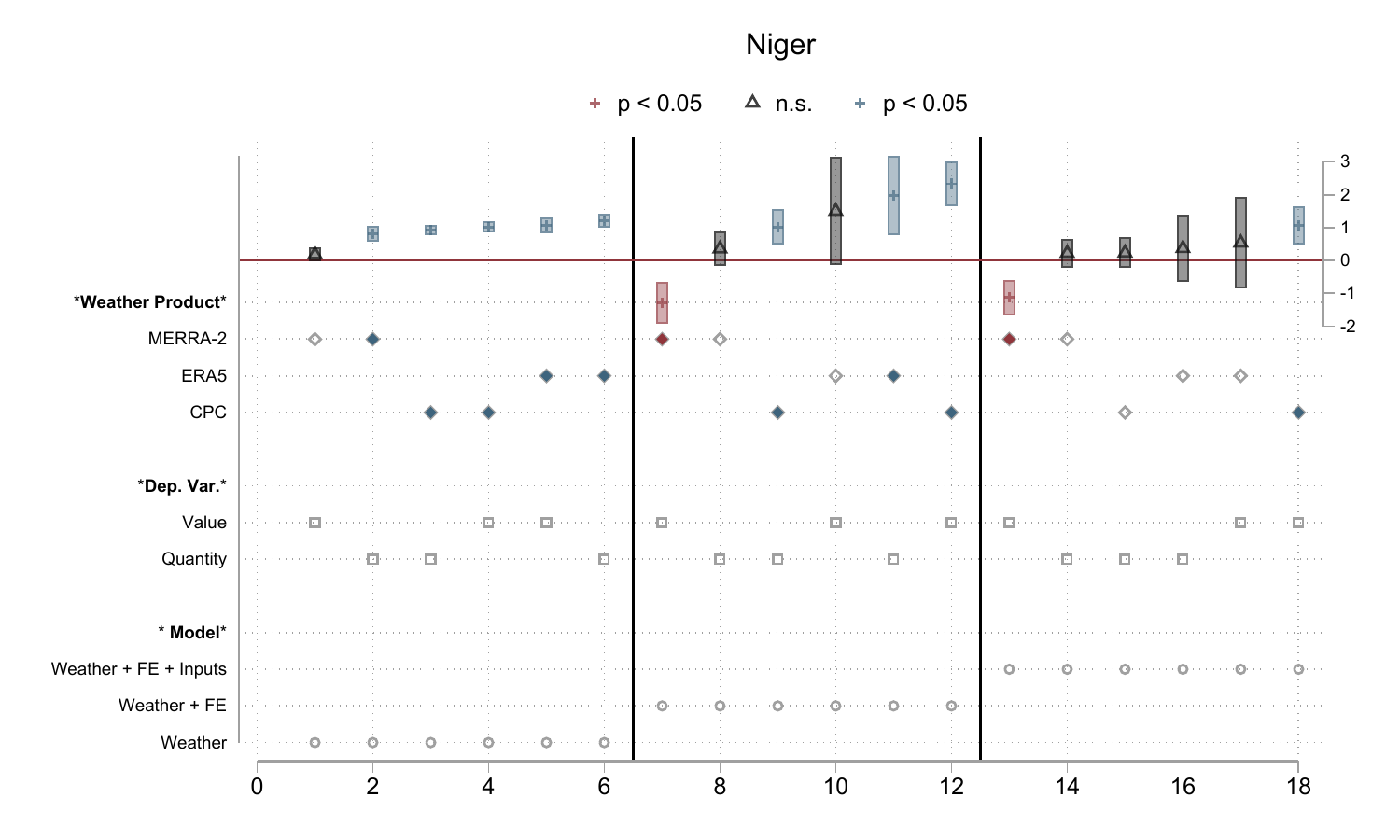}
			\includegraphics[width=.49\linewidth,keepaspectratio]{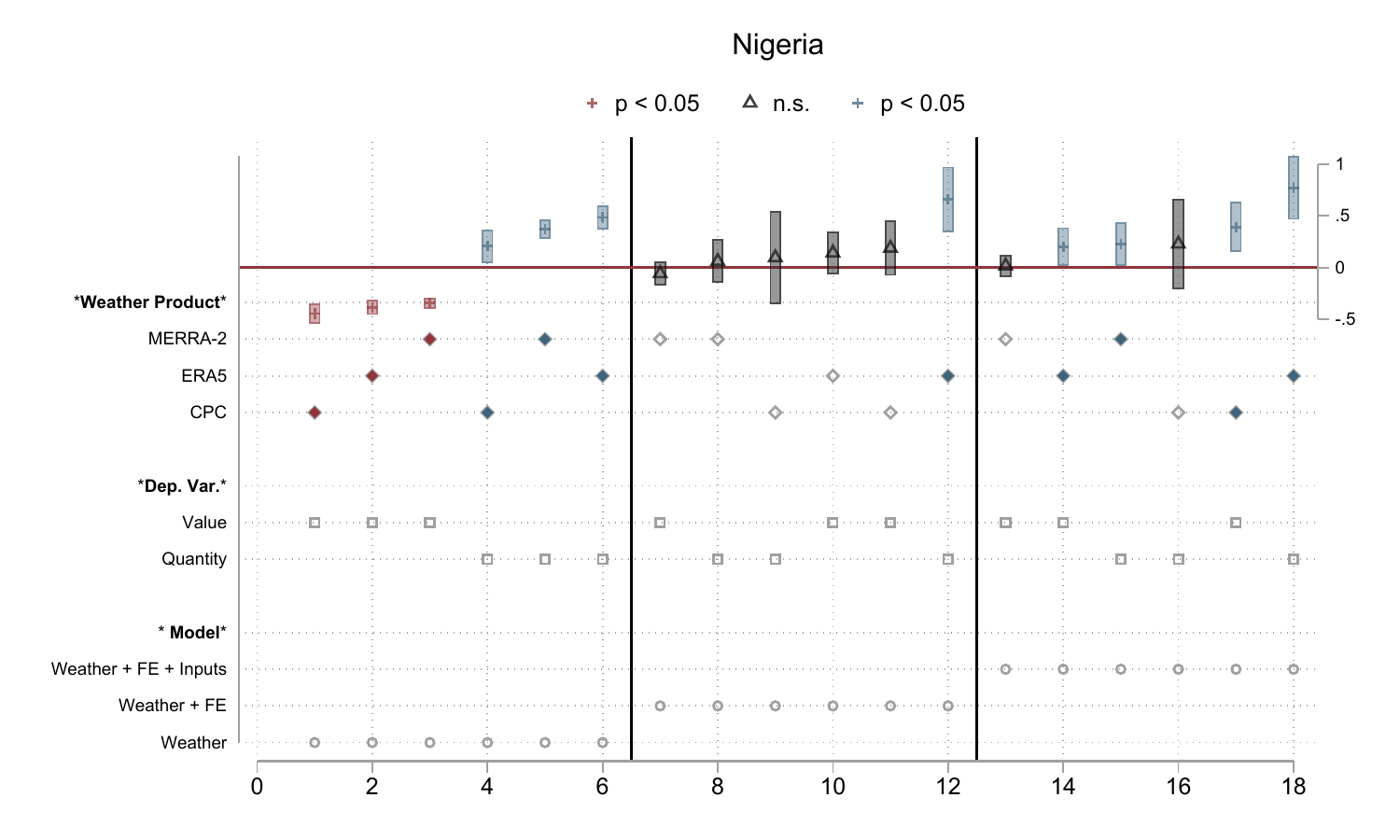}
			\includegraphics[width=.49\linewidth,keepaspectratio]{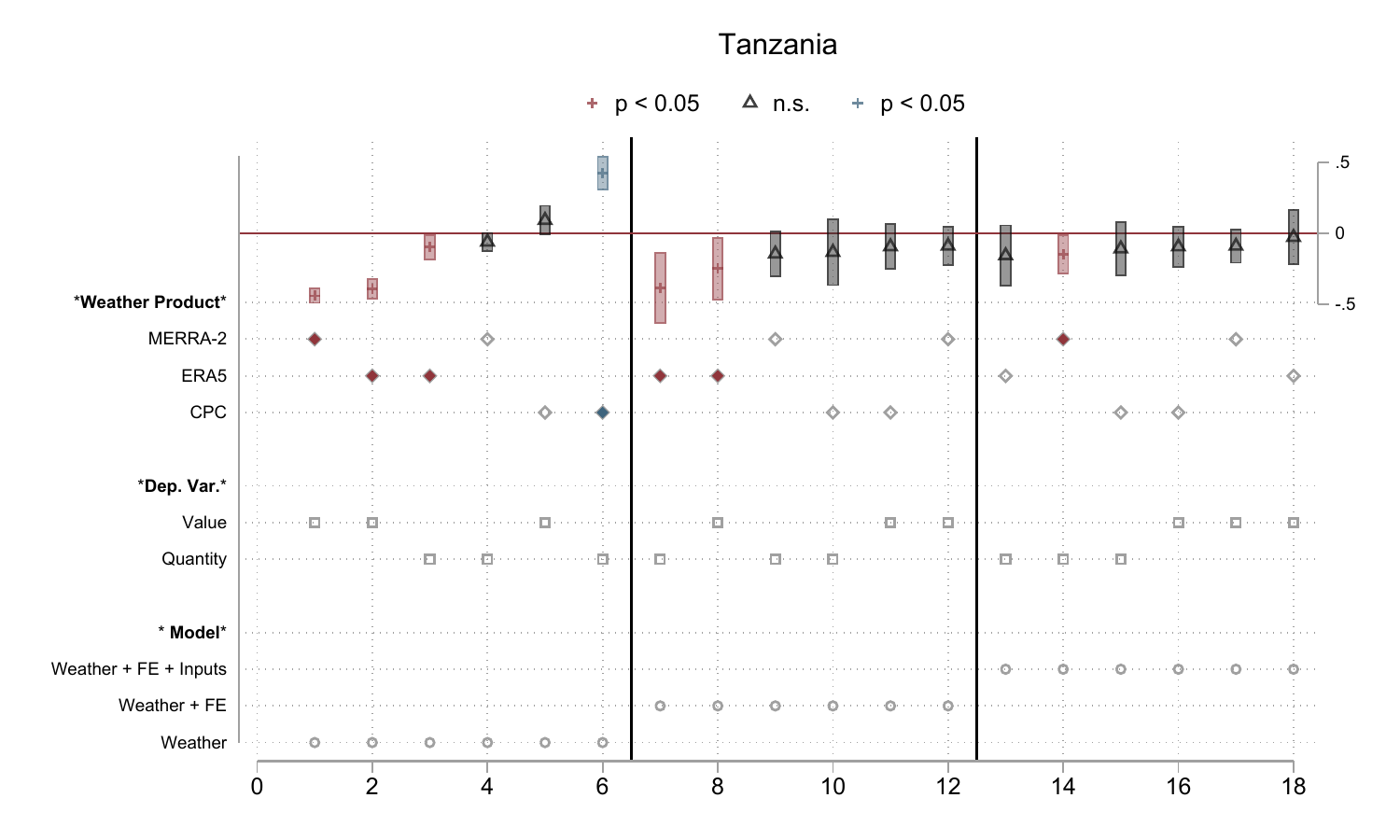}
			\includegraphics[width=.49\linewidth,keepaspectratio]{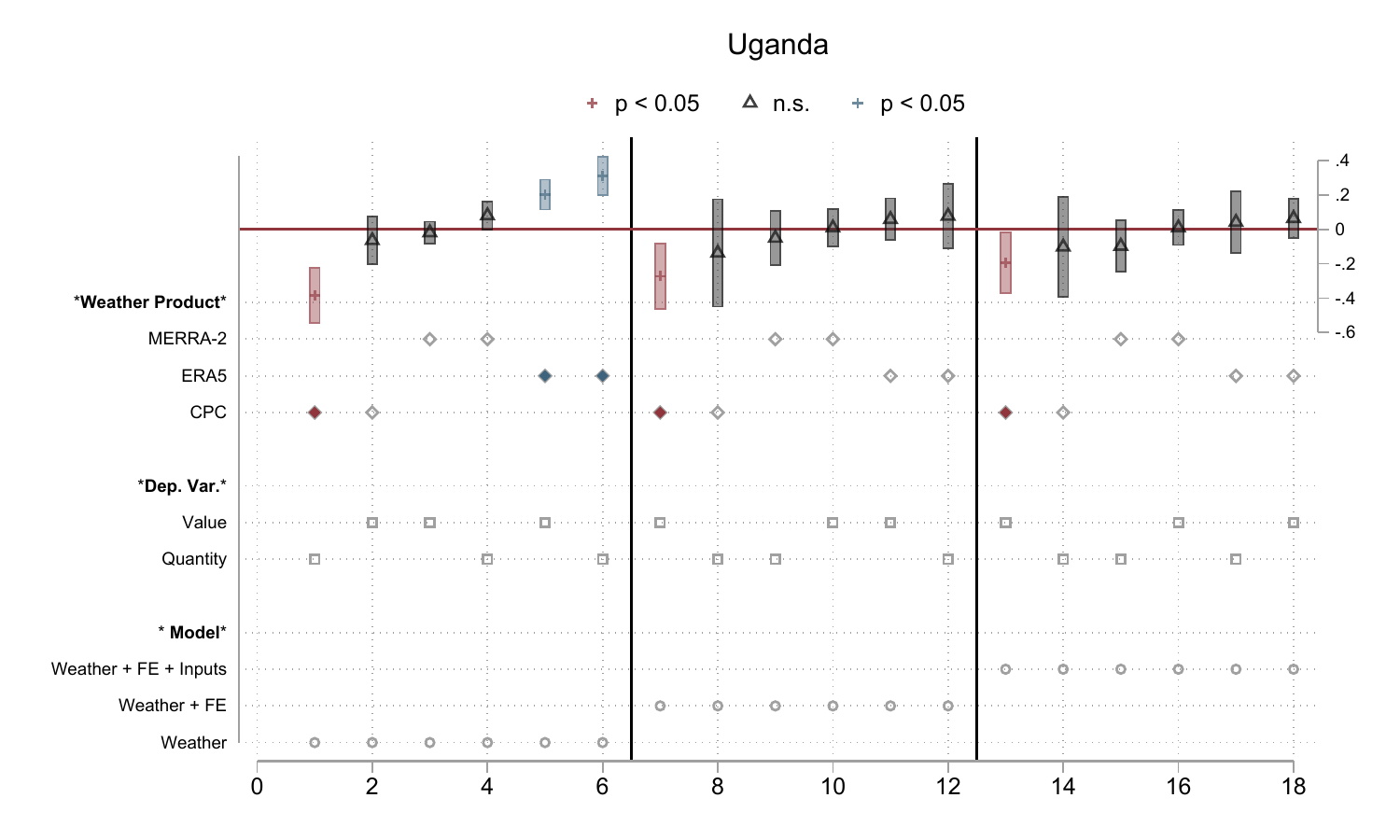}
		\end{center}
		\footnotesize  \textit{Note}: The figure presents specification curves, where each panel represents a different country, with three different models presented within each panel. Each panel includes 18 regressions, where each column represents a single regression. Significant and non-significant coefficients are designated at the top of the figure. For each Earth observation product, we also designate the significance and sign of the coefficient with color: red represents coefficients which are negative and significant; white represents insignificant coefficients, regardless of sign; and blue represents coefficients which are positive and significant.  
	\end{minipage}	
\end{figure}
\end{center}

\begin{center}
\begin{figure}[!htbp]
	\begin{minipage}{\linewidth}
		\caption{Specification Charts for Skew of Daily Temperature}
		\label{fig:pval_v18}
		\begin{center}
			\includegraphics[width=.49\linewidth,keepaspectratio]{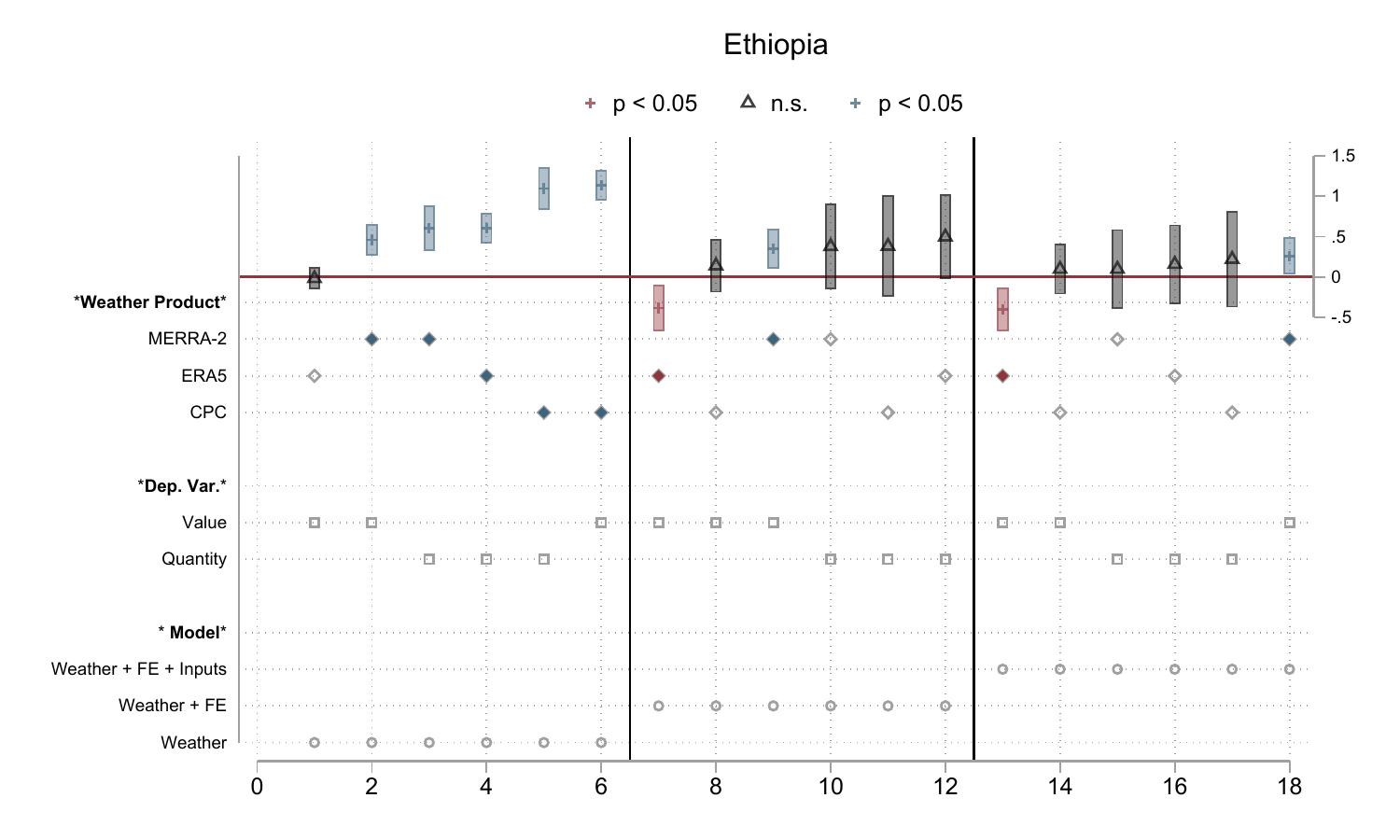}
			\includegraphics[width=.49\linewidth,keepaspectratio]{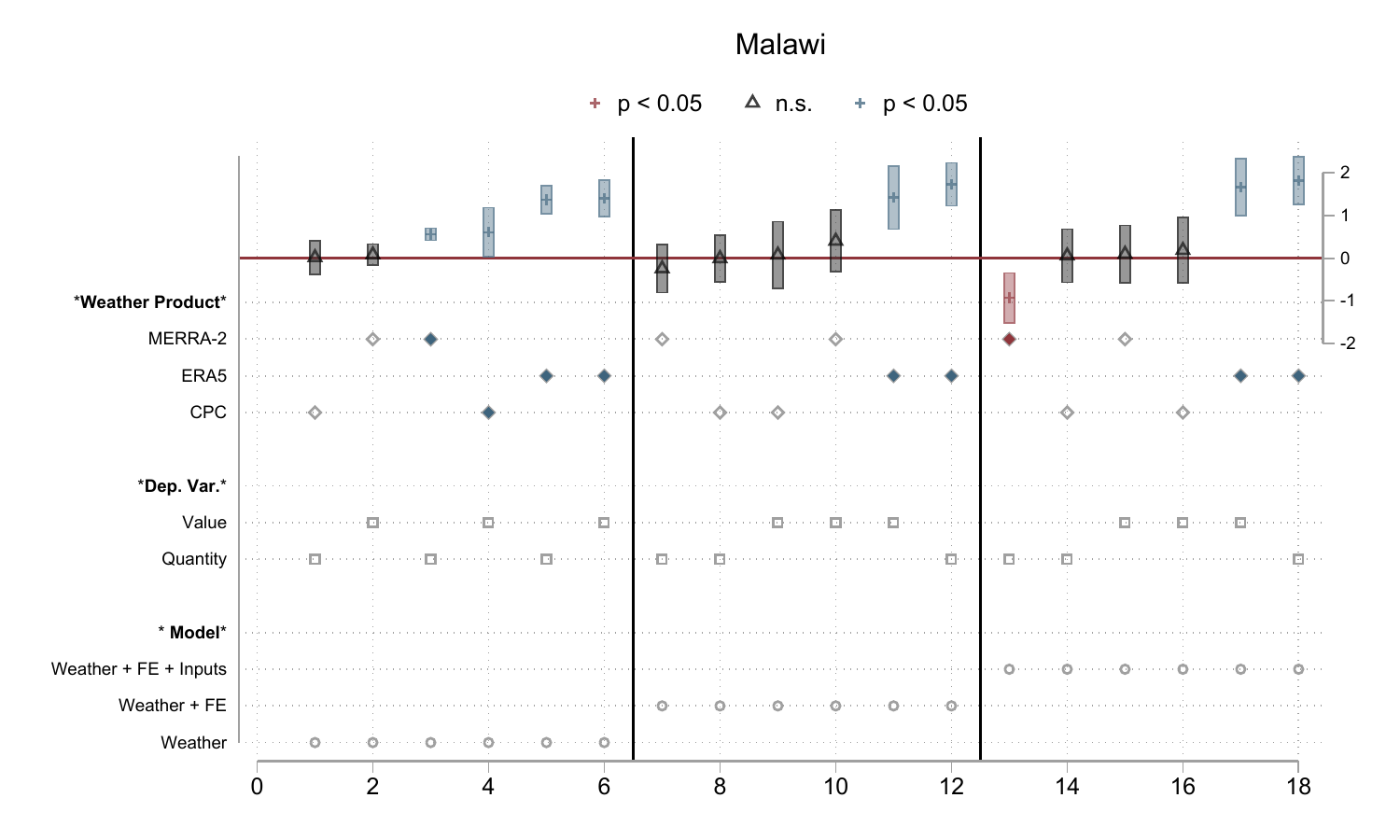}
			\includegraphics[width=.49\linewidth,keepaspectratio]{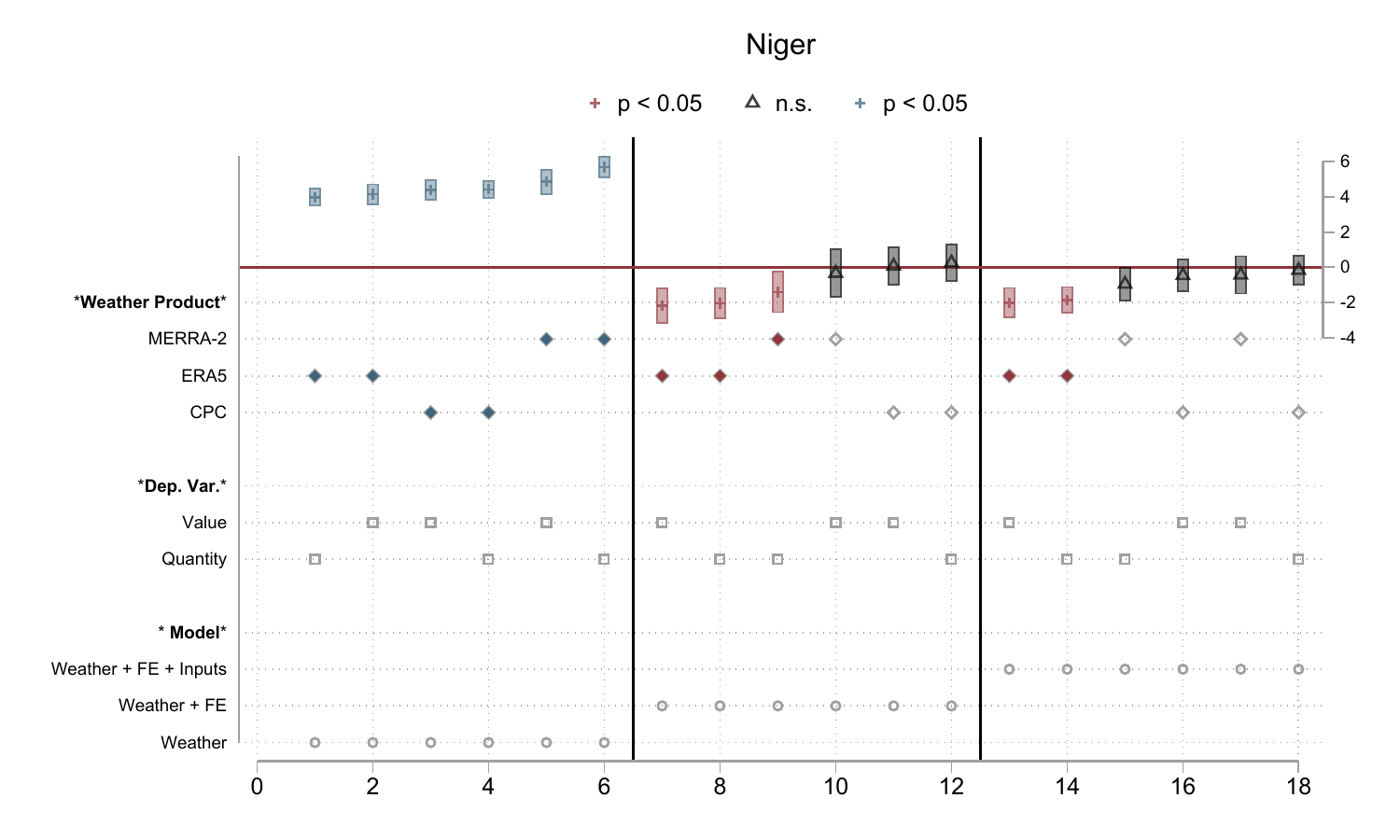}
			\includegraphics[width=.49\linewidth,keepaspectratio]{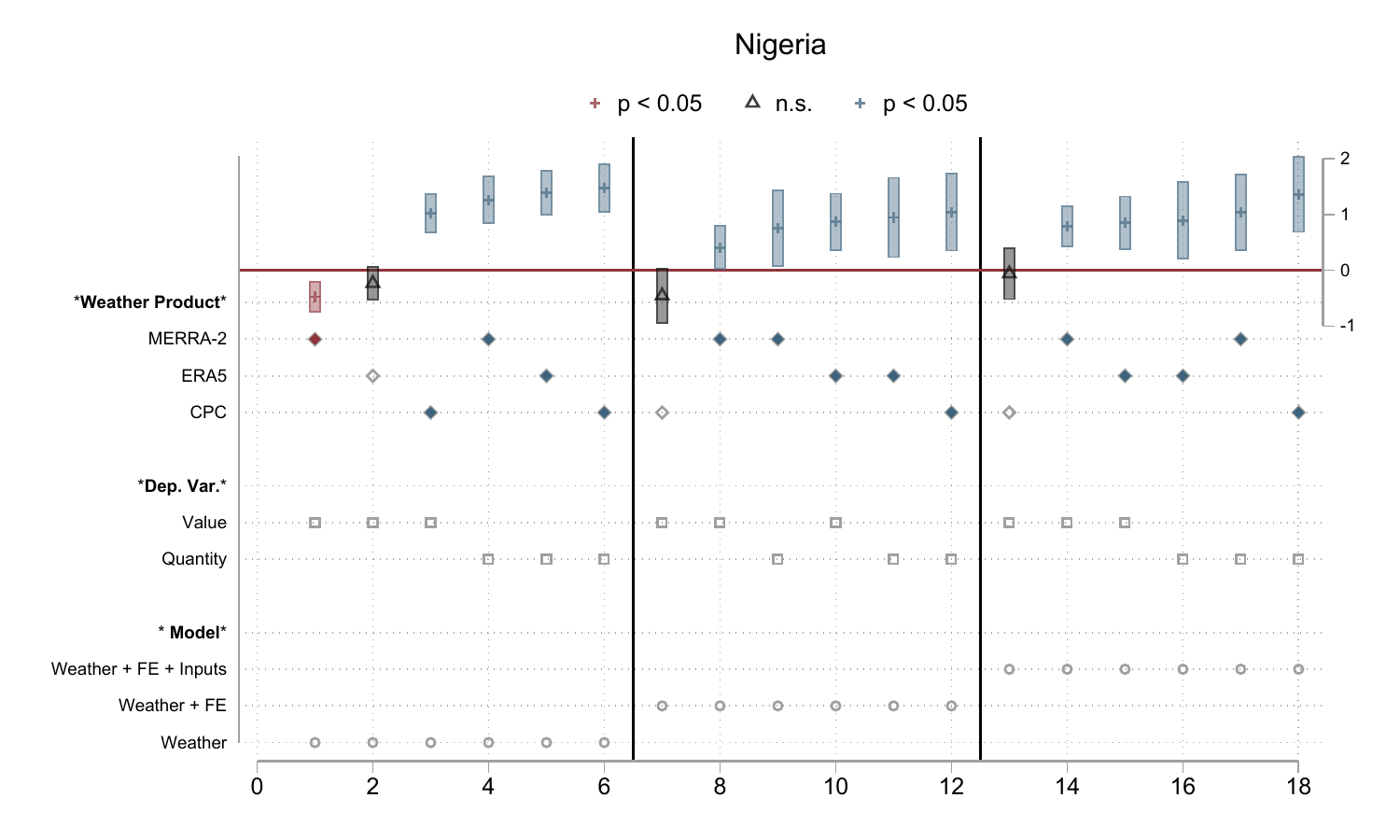}
			\includegraphics[width=.49\linewidth,keepaspectratio]{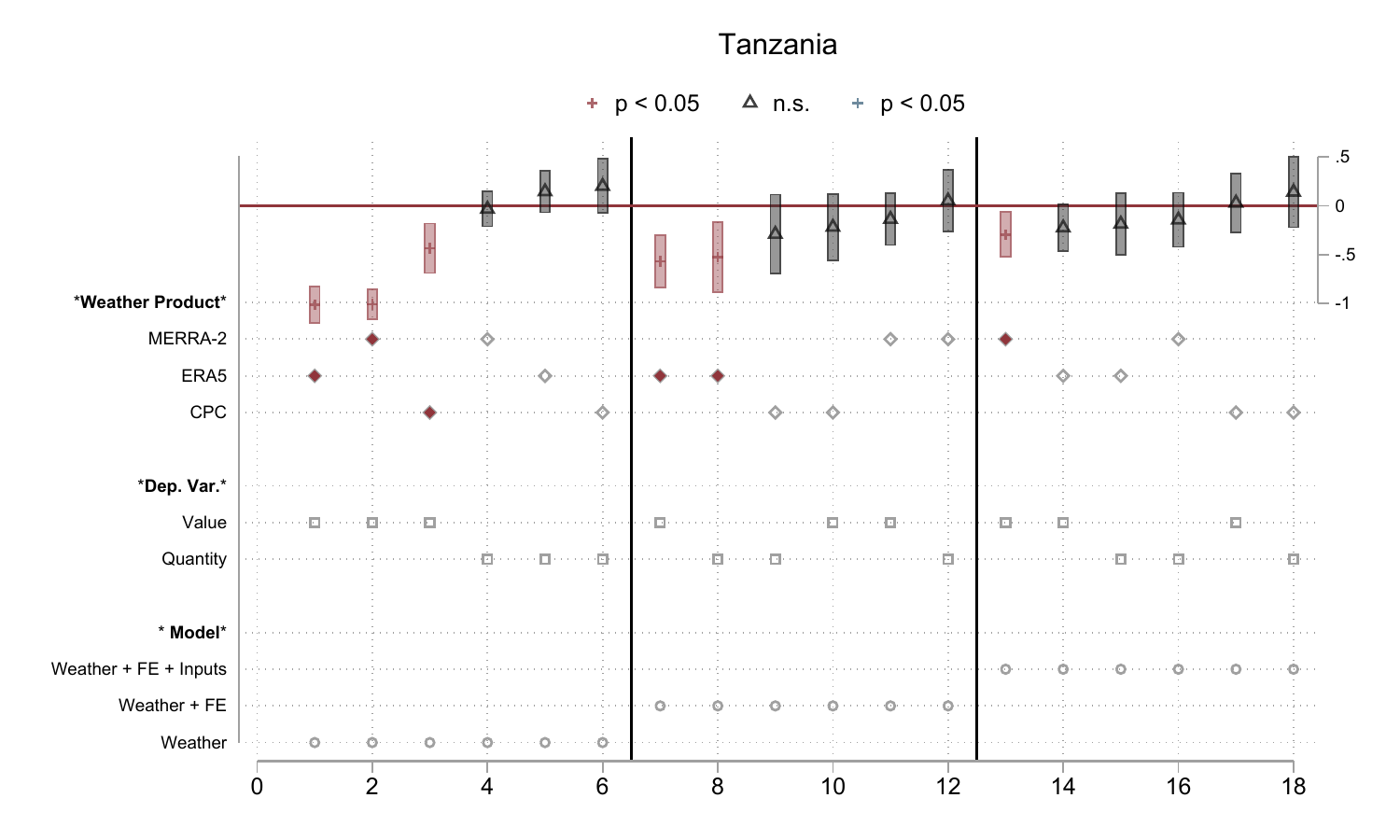}
			\includegraphics[width=.49\linewidth,keepaspectratio]{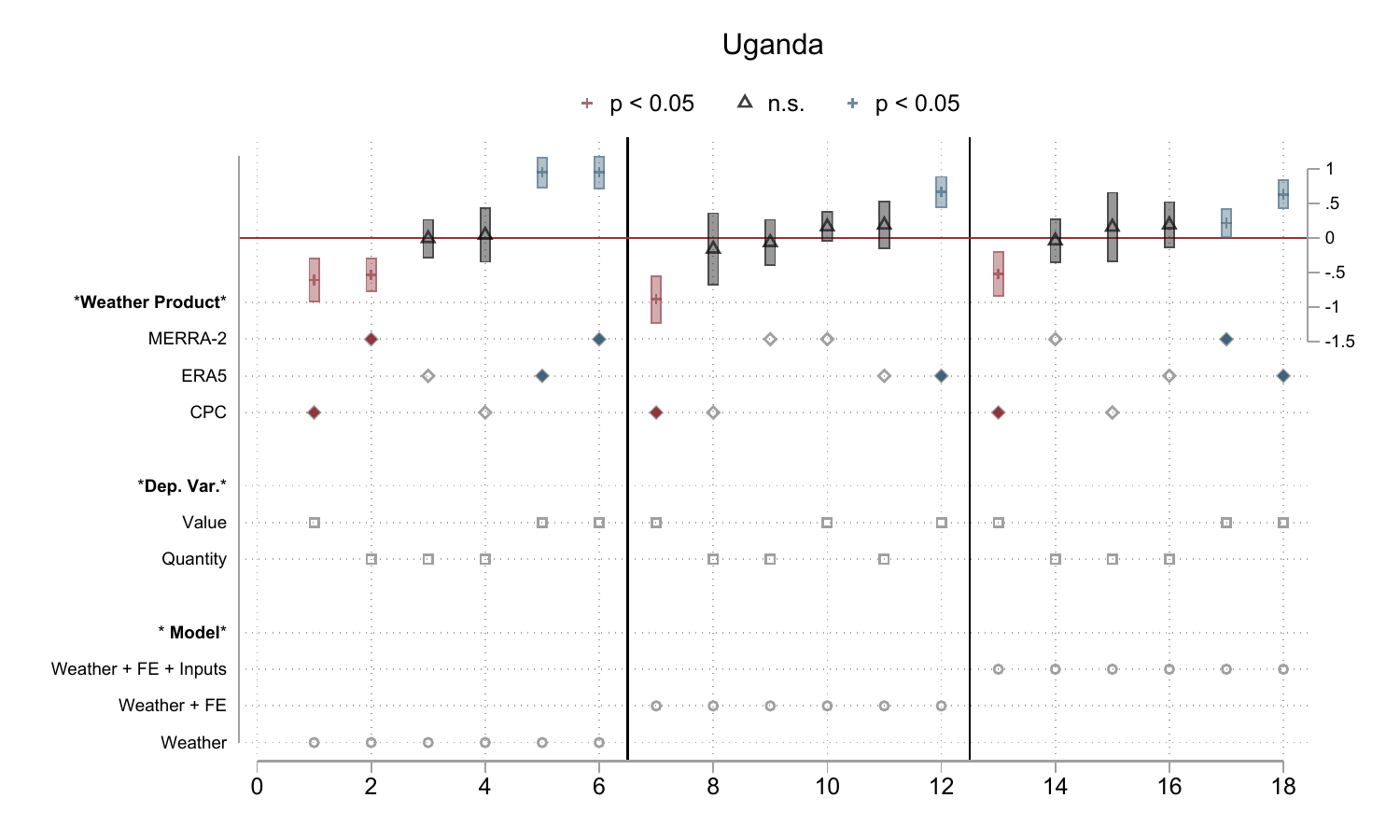}
		\end{center}
		\footnotesize  \textit{Note}: The figure presents specification curves, where each panel represents a different country, with three different models presented within each panel. Each panel includes 18 regressions, where each column represents a single regression. Significant and non-significant coefficients are designated at the top of the figure. For each Earth observation product, we also designate the significance and sign of the coefficient with color: red represents coefficients which are negative and significant; white represents insignificant coefficients, regardless of sign; and blue represents coefficients which are positive and significant.  
	\end{minipage}	
\end{figure}
\end{center}

\begin{center}
\begin{figure}[!htbp]
	\begin{minipage}{\linewidth}
		\caption{Specification Charts for Deviations in GDD}
		\label{fig:pval_v20}
		\begin{center}
			\includegraphics[width=.49\linewidth,keepaspectratio]{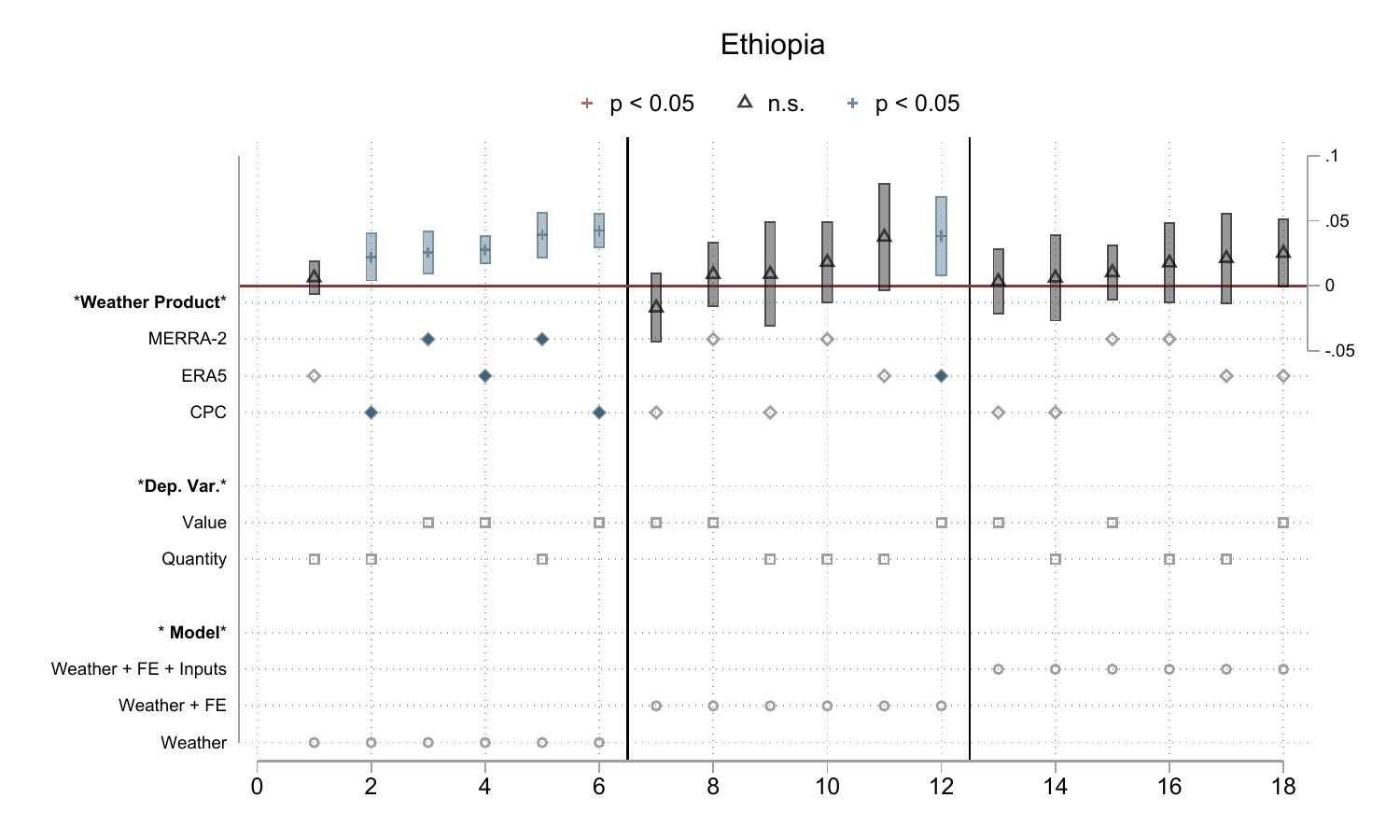}
			\includegraphics[width=.49\linewidth,keepaspectratio]{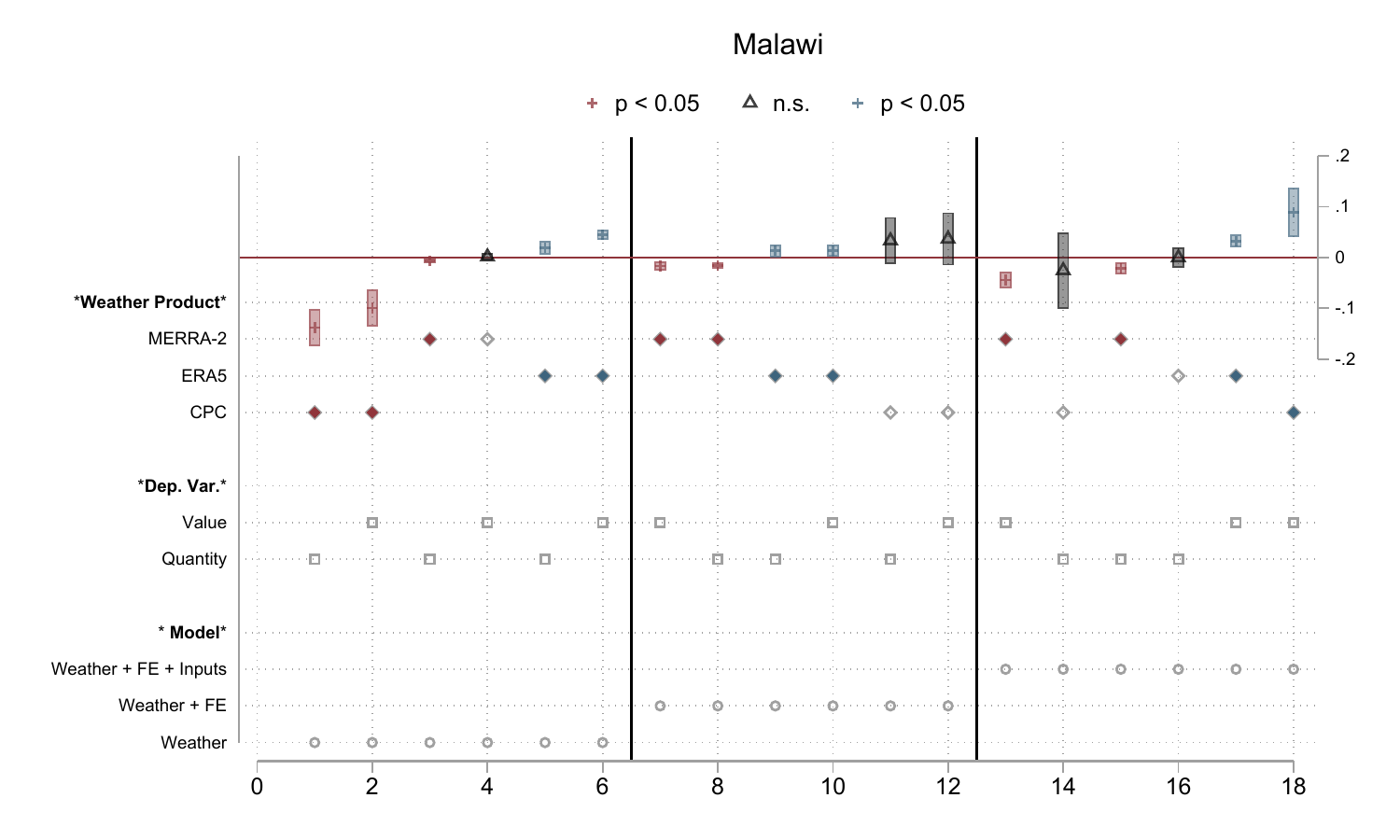}
			\includegraphics[width=.49\linewidth,keepaspectratio]{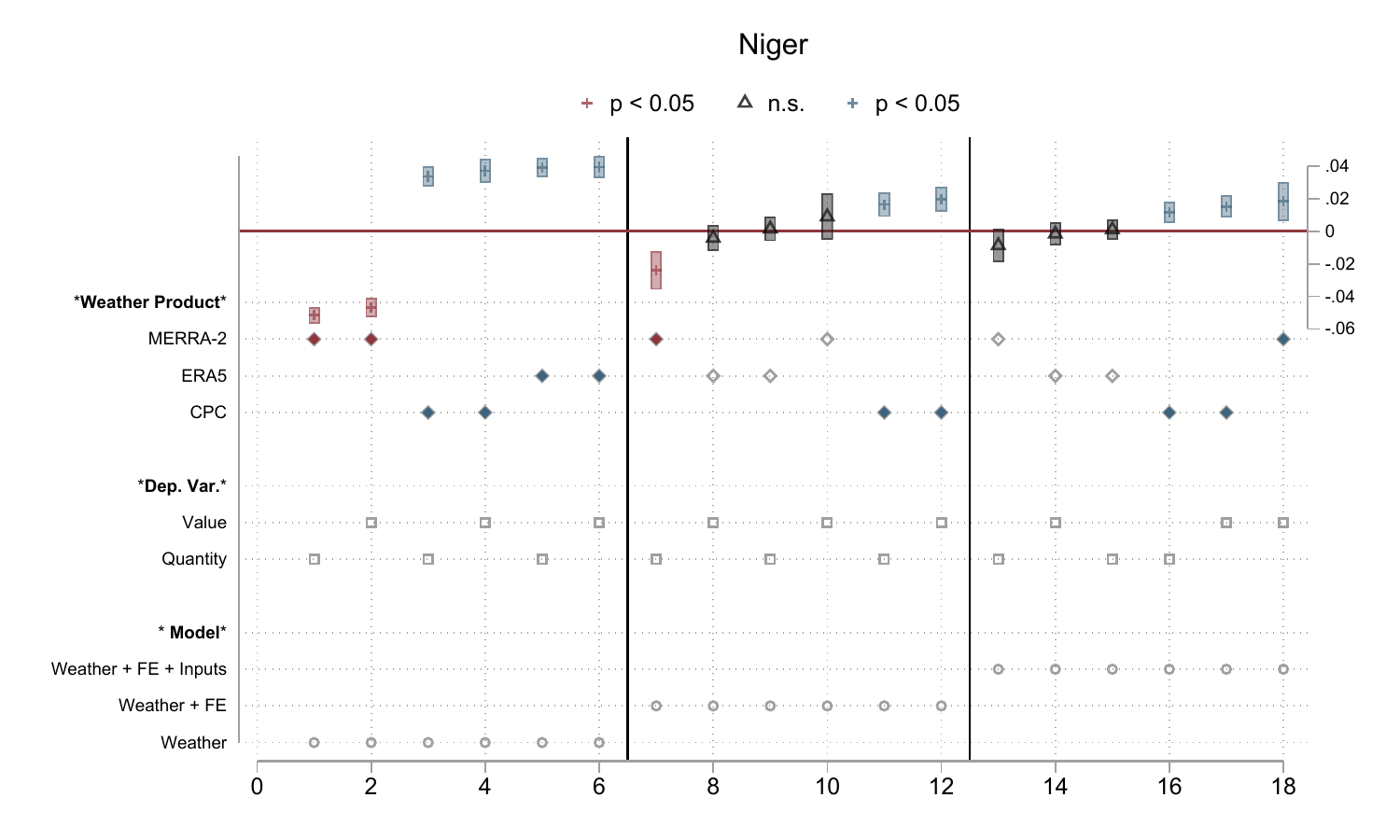}
			\includegraphics[width=.49\linewidth,keepaspectratio]{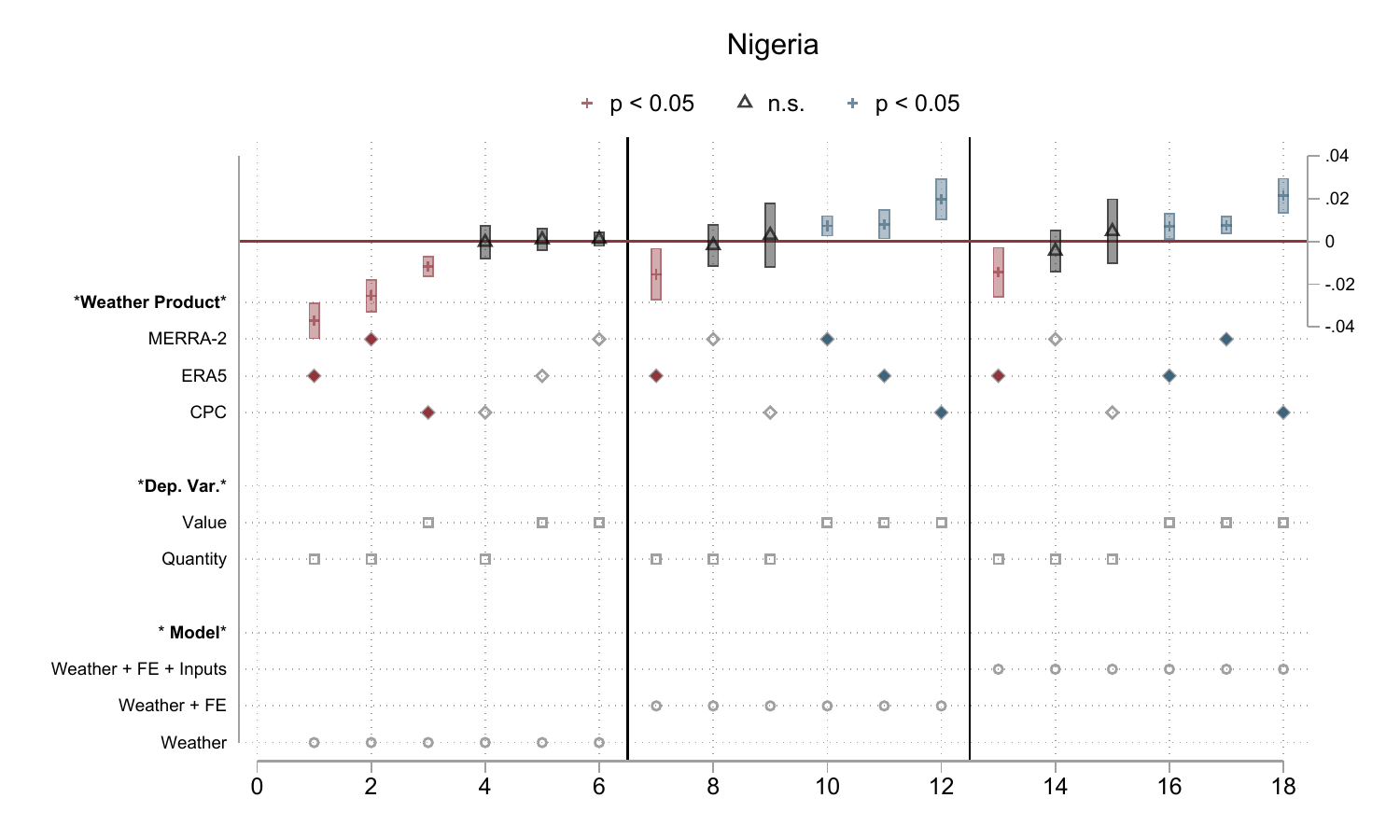}
			\includegraphics[width=.49\linewidth,keepaspectratio]{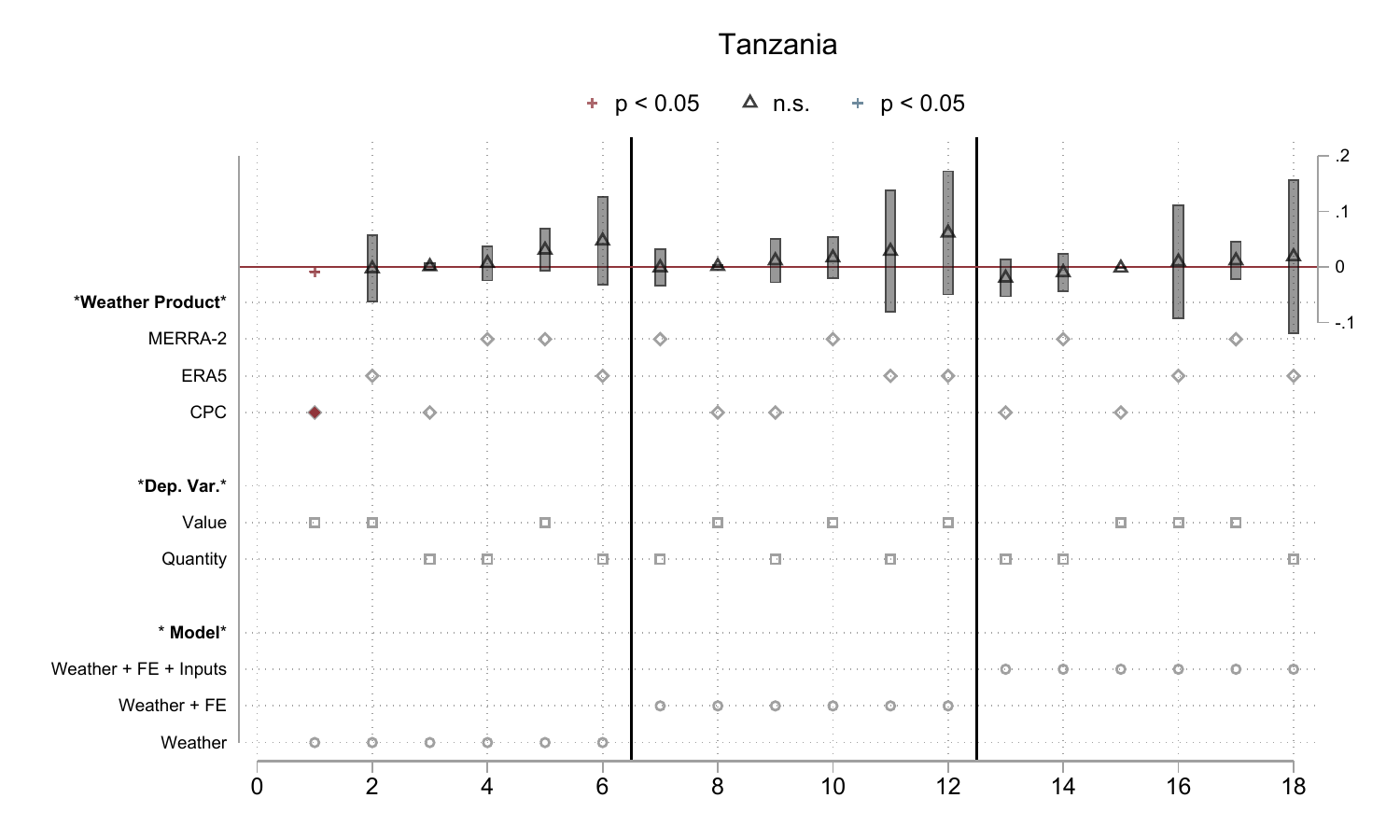}
			\includegraphics[width=.49\linewidth,keepaspectratio]{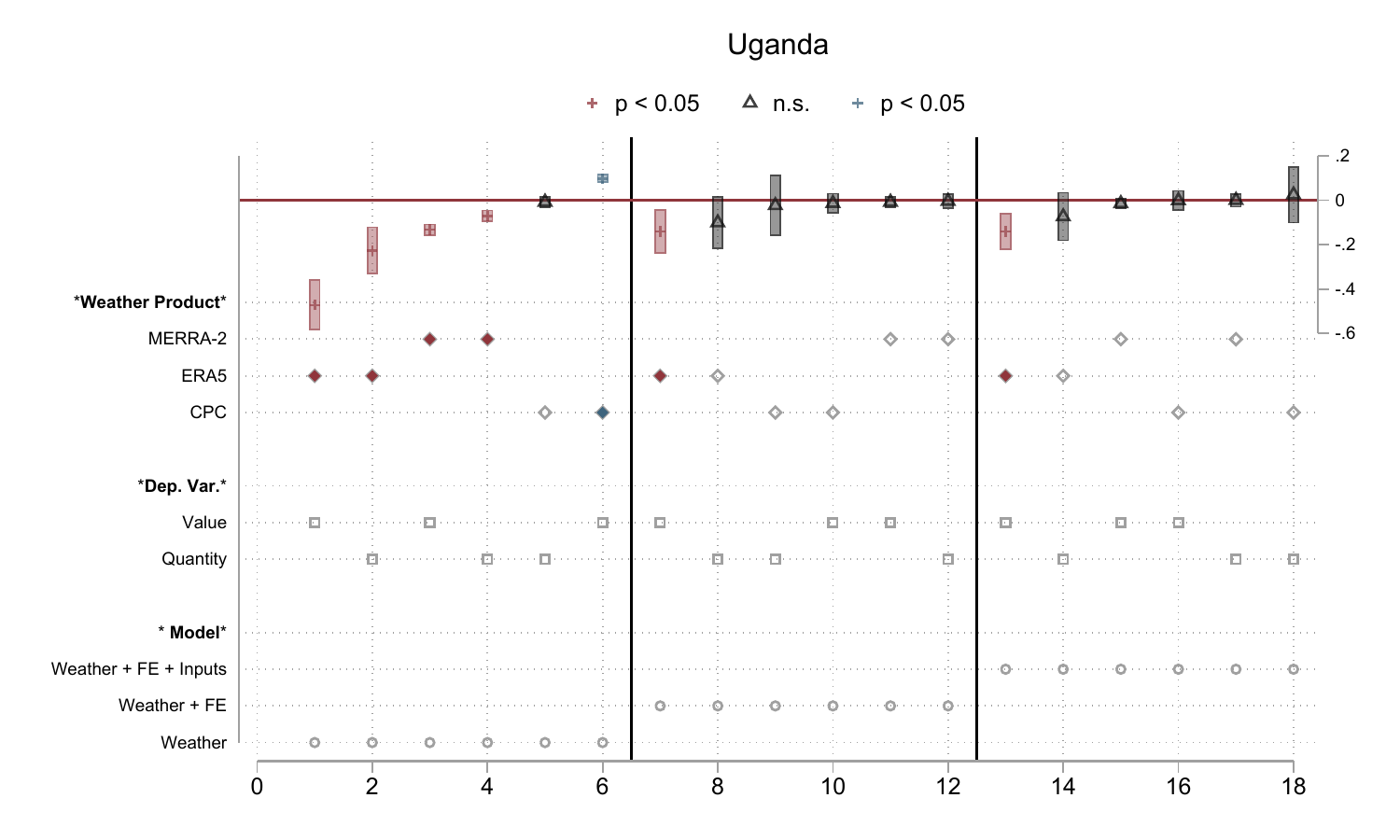}
		\end{center}
		\footnotesize  \textit{Note}: The figure presents specification curves, where each panel represents a different country, with three different models presented within each panel. Each panel includes 18 regressions, where each column represents a single regression. Significant and non-significant coefficients are designated at the top of the figure. For each Earth observation product, we also designate the significance and sign of the coefficient with color: red represents coefficients which are negative and significant; white represents insignificant coefficients, regardless of sign; and blue represents coefficients which are positive and significant.  
	\end{minipage}	
\end{figure}
\end{center}

\begin{center}
\begin{figure}[!htbp]
	\begin{minipage}{\linewidth}
		\caption{Specification Charts for z-Score of GDD}
		\label{fig:pval_v21}
		\begin{center}
			\includegraphics[width=.49\linewidth,keepaspectratio]{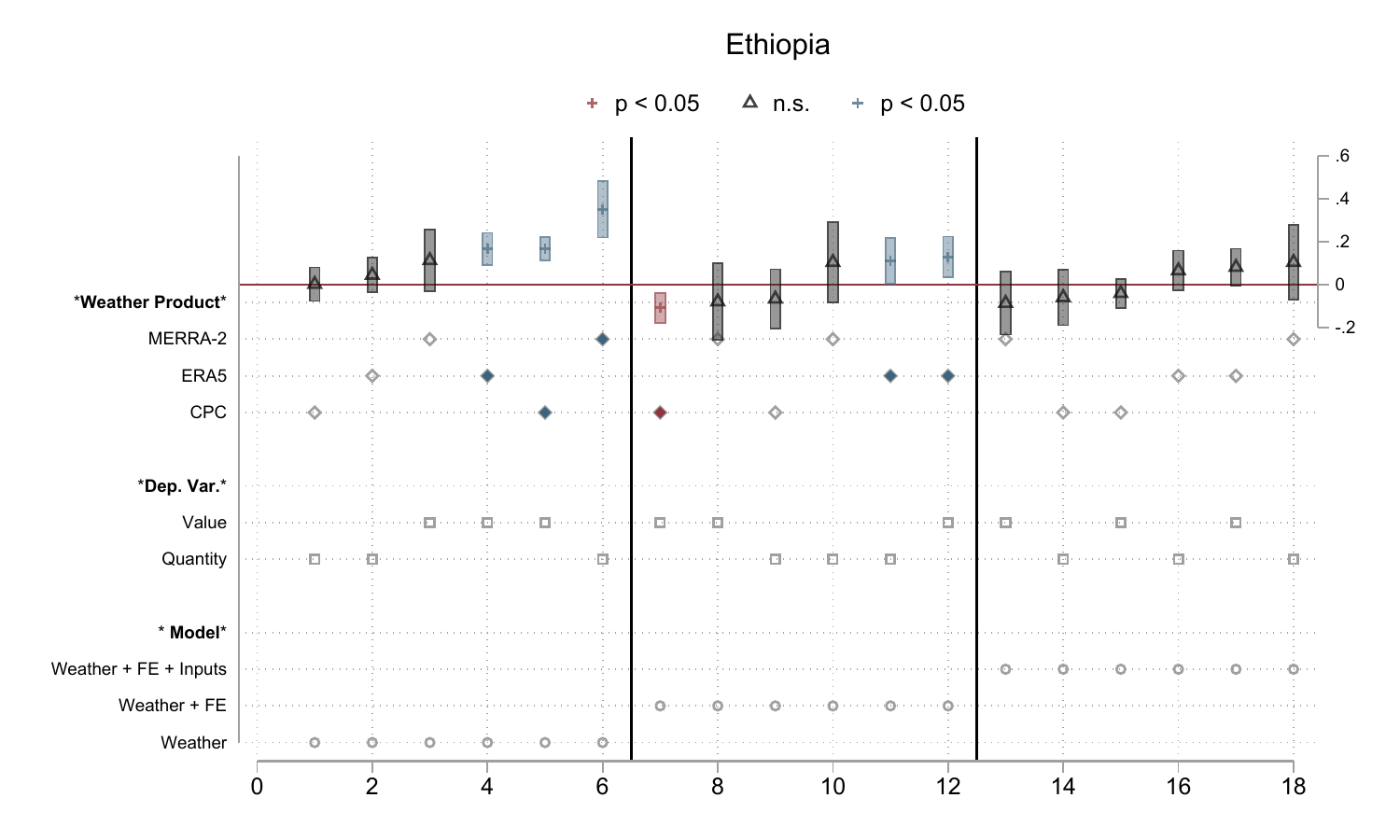}
			\includegraphics[width=.49\linewidth,keepaspectratio]{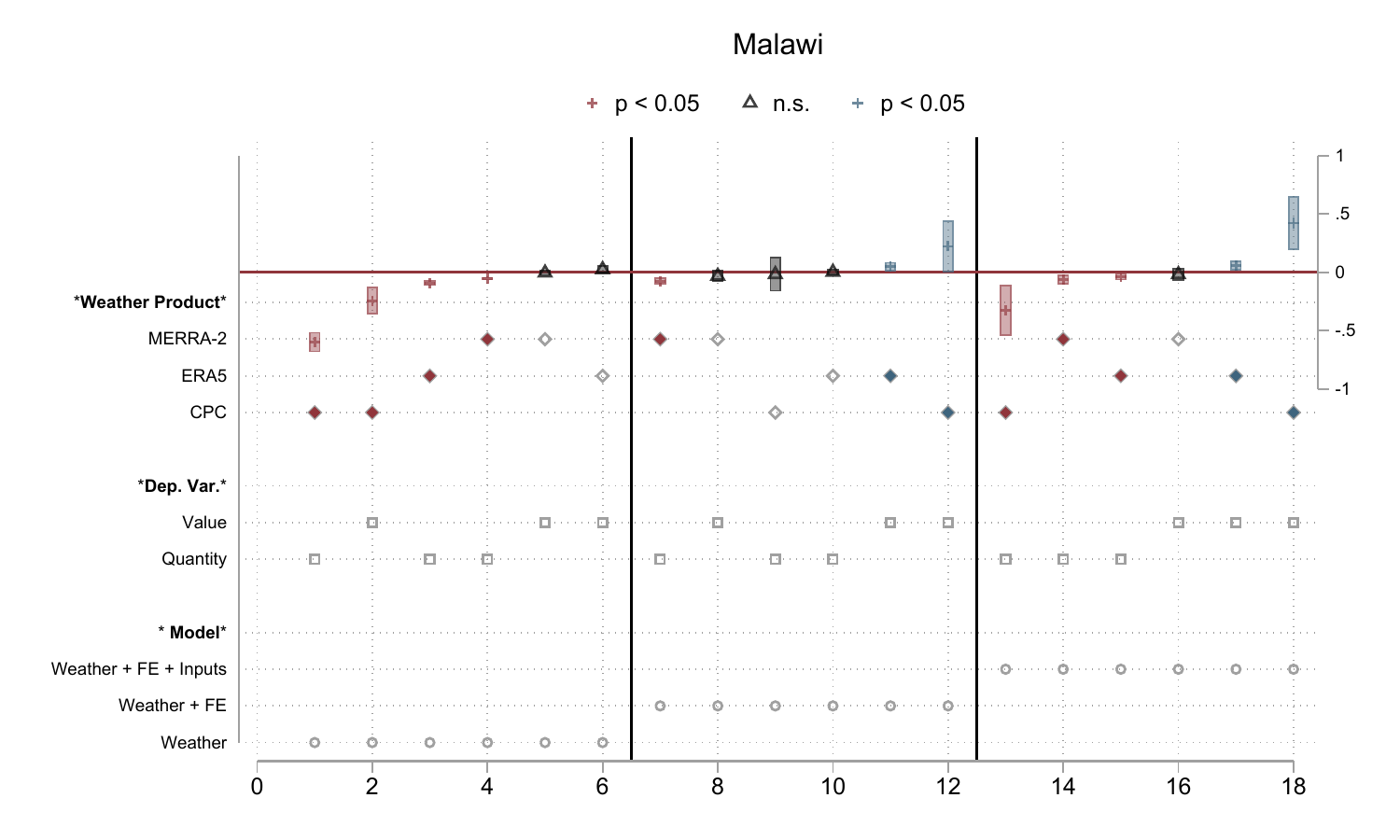}
			\includegraphics[width=.49\linewidth,keepaspectratio]{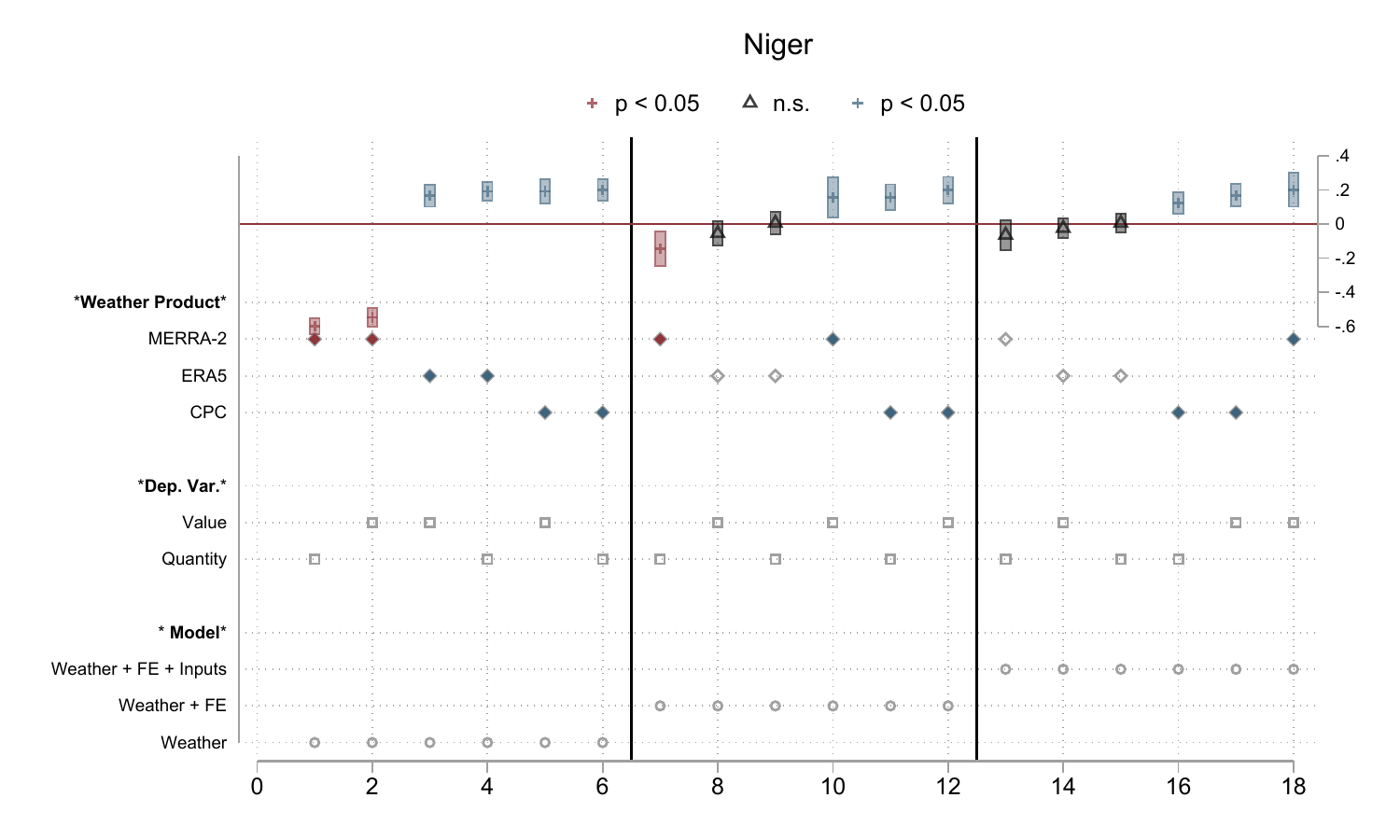}
			\includegraphics[width=.49\linewidth,keepaspectratio]{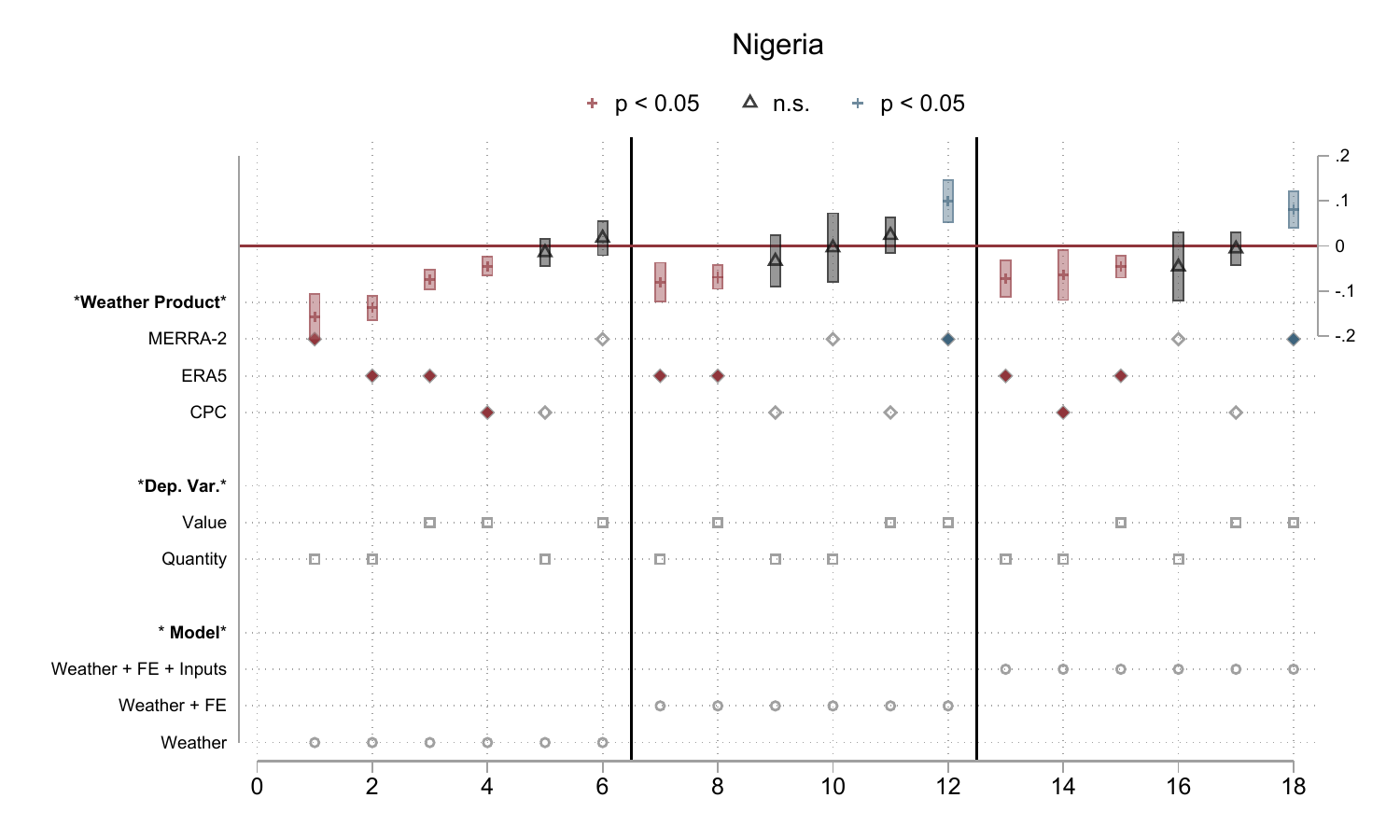}
			\includegraphics[width=.49\linewidth,keepaspectratio]{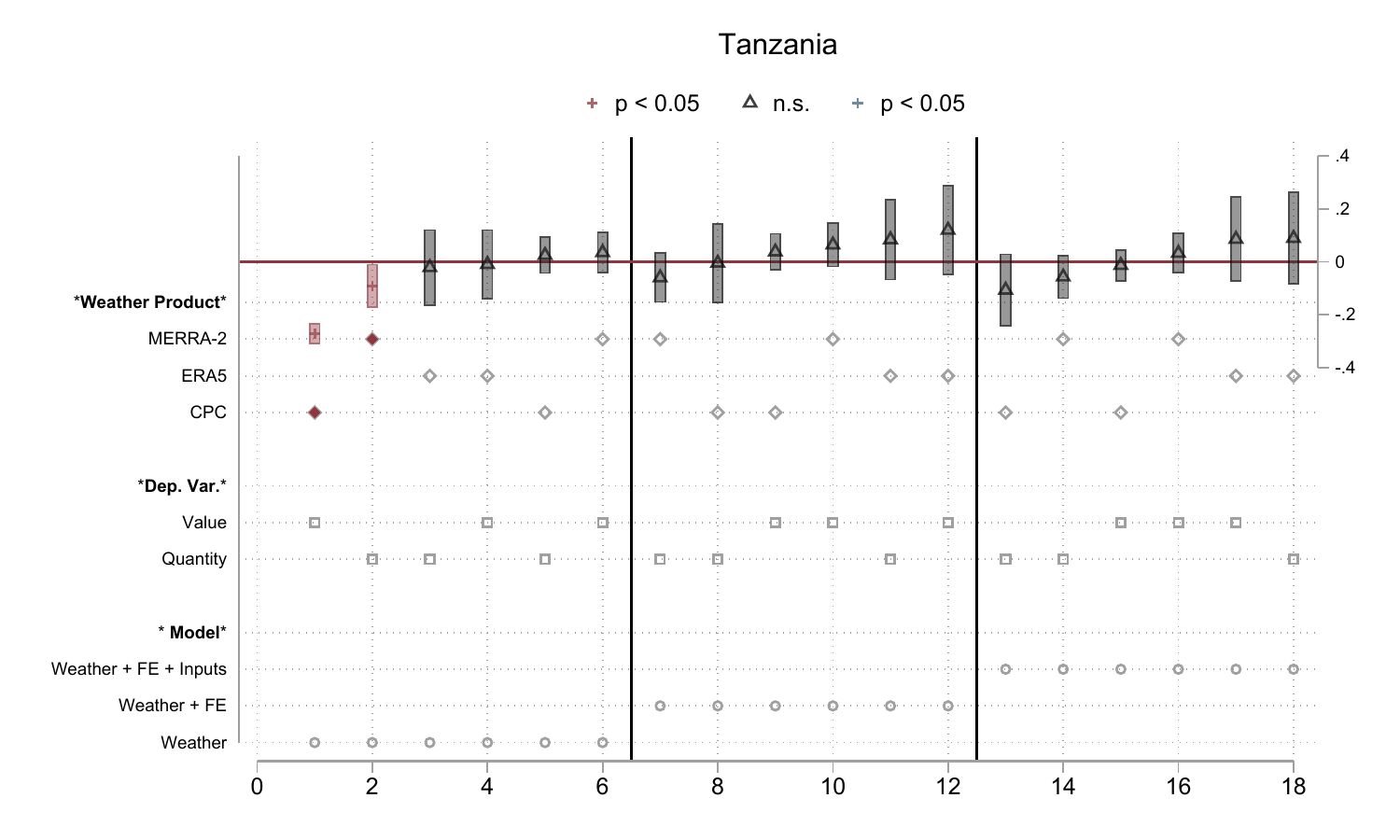}
			\includegraphics[width=.49\linewidth,keepaspectratio]{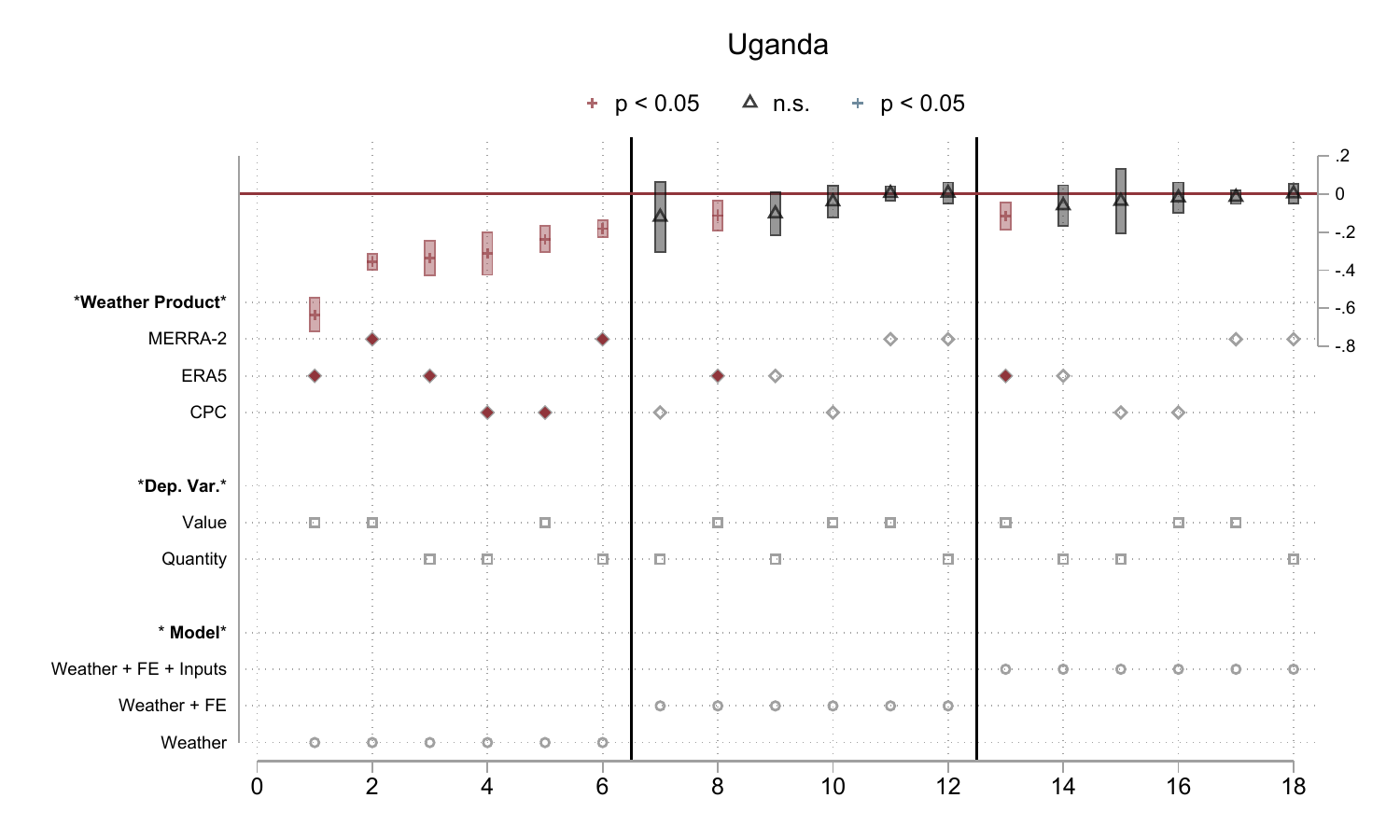}
		\end{center}
		\footnotesize  \textit{Note}: The figure presents specification curves, where each panel represents a different country, with three different models presented within each panel. Each panel includes 18 regressions, where each column represents a single regression. Significant and non-significant coefficients are designated at the top of the figure. For each Earth observation product, we also designate the significance and sign of the coefficient with color: red represents coefficients which are negative and significant; white represents insignificant coefficients, regardless of sign; and blue represents coefficients which are positive and significant.  
	\end{minipage}	
\end{figure}
\end{center}

\begin{center}
\begin{figure}[!htbp]
	\begin{minipage}{\linewidth}
		\caption{Specification Charts for Maximum Daily Temperature}
		\label{fig:pval_v22}
		\begin{center}
			\includegraphics[width=.49\linewidth,keepaspectratio]{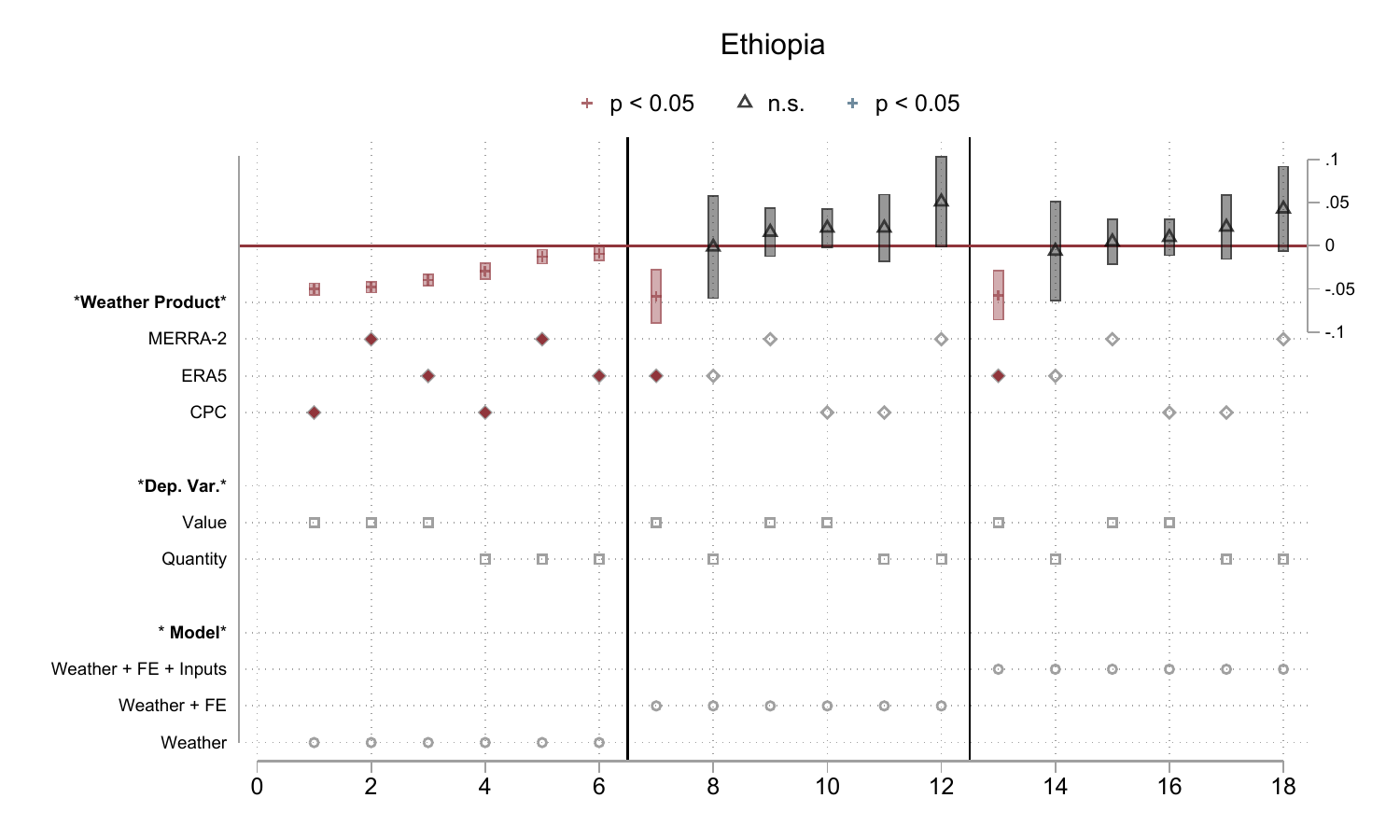}
			\includegraphics[width=.49\linewidth,keepaspectratio]{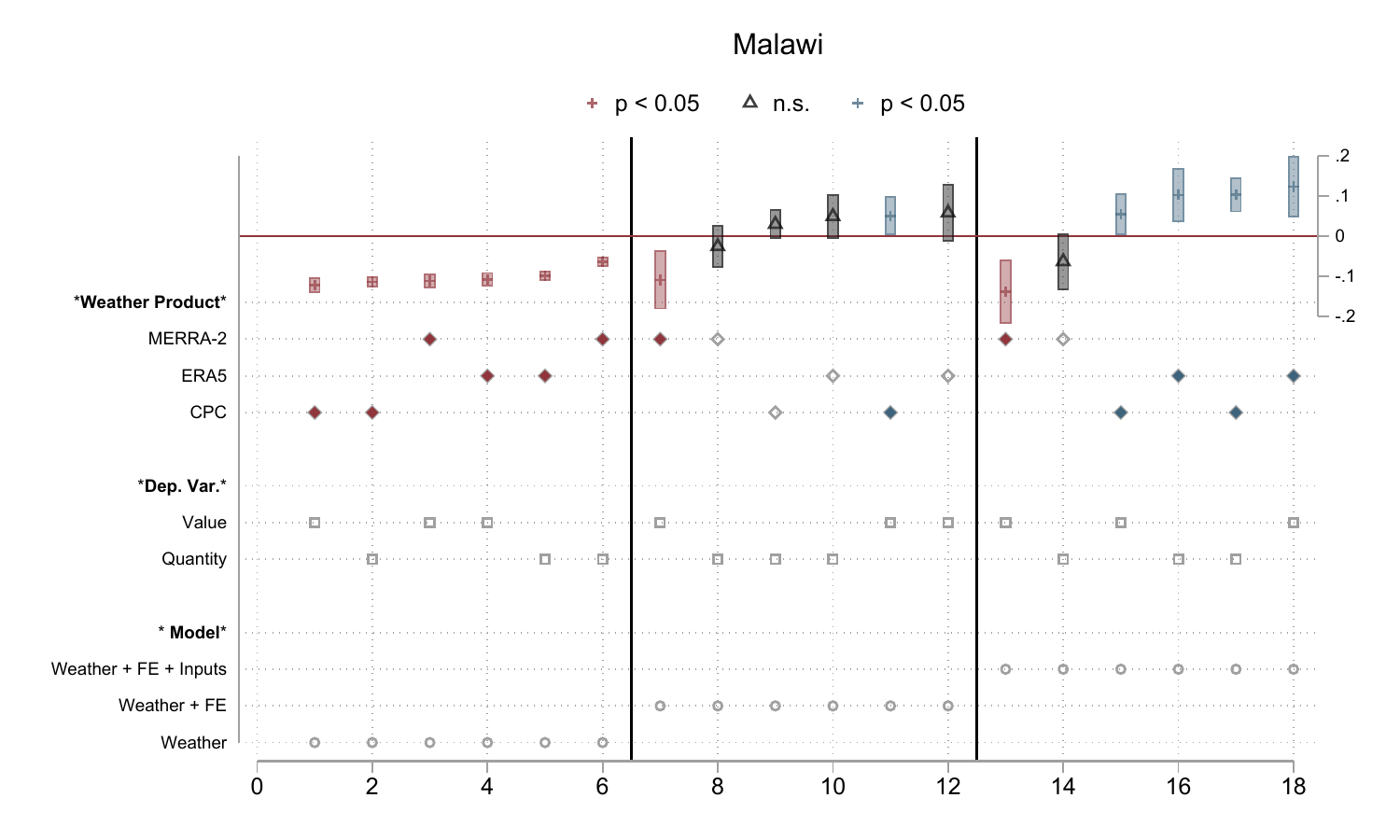}
			\includegraphics[width=.49\linewidth,keepaspectratio]{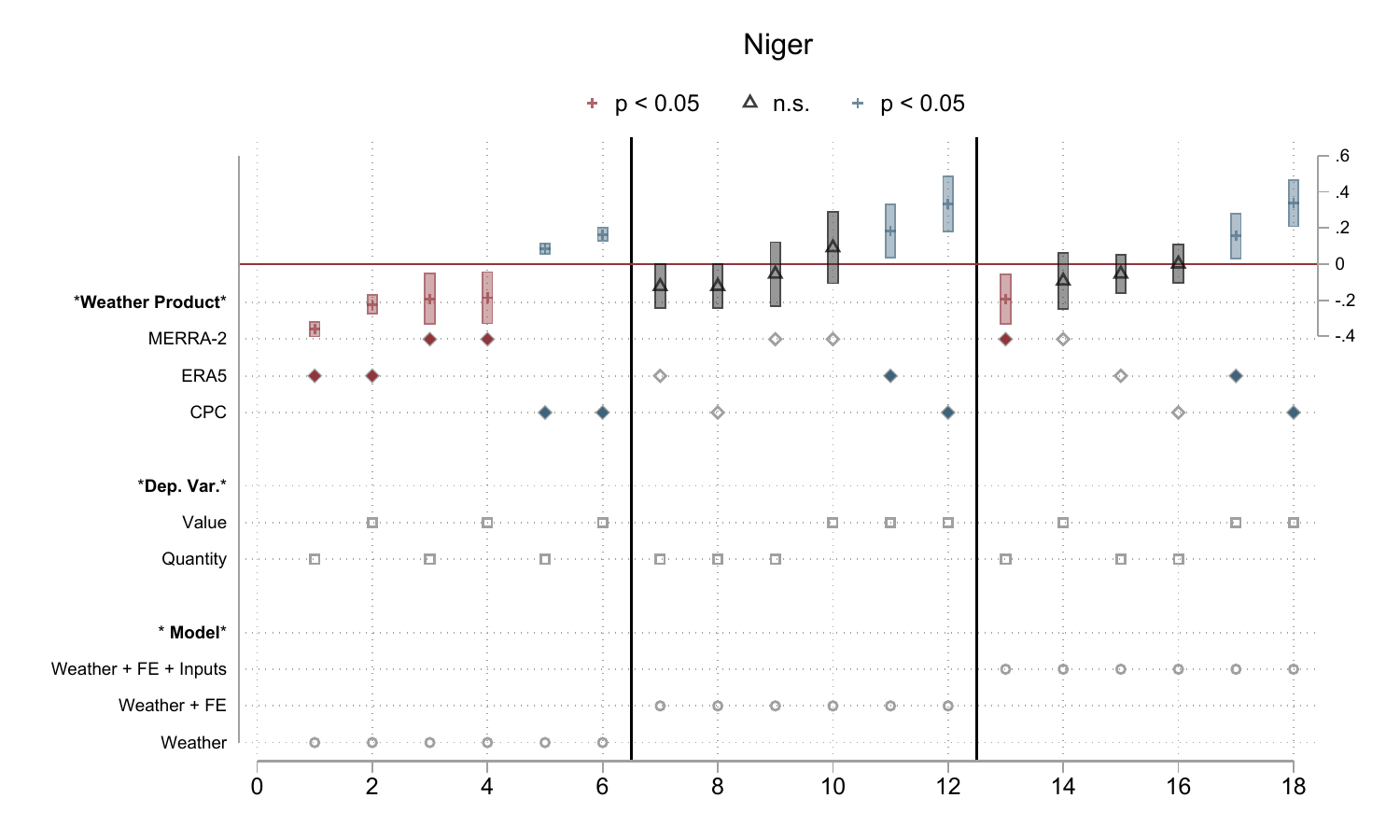}
			\includegraphics[width=.49\linewidth,keepaspectratio]{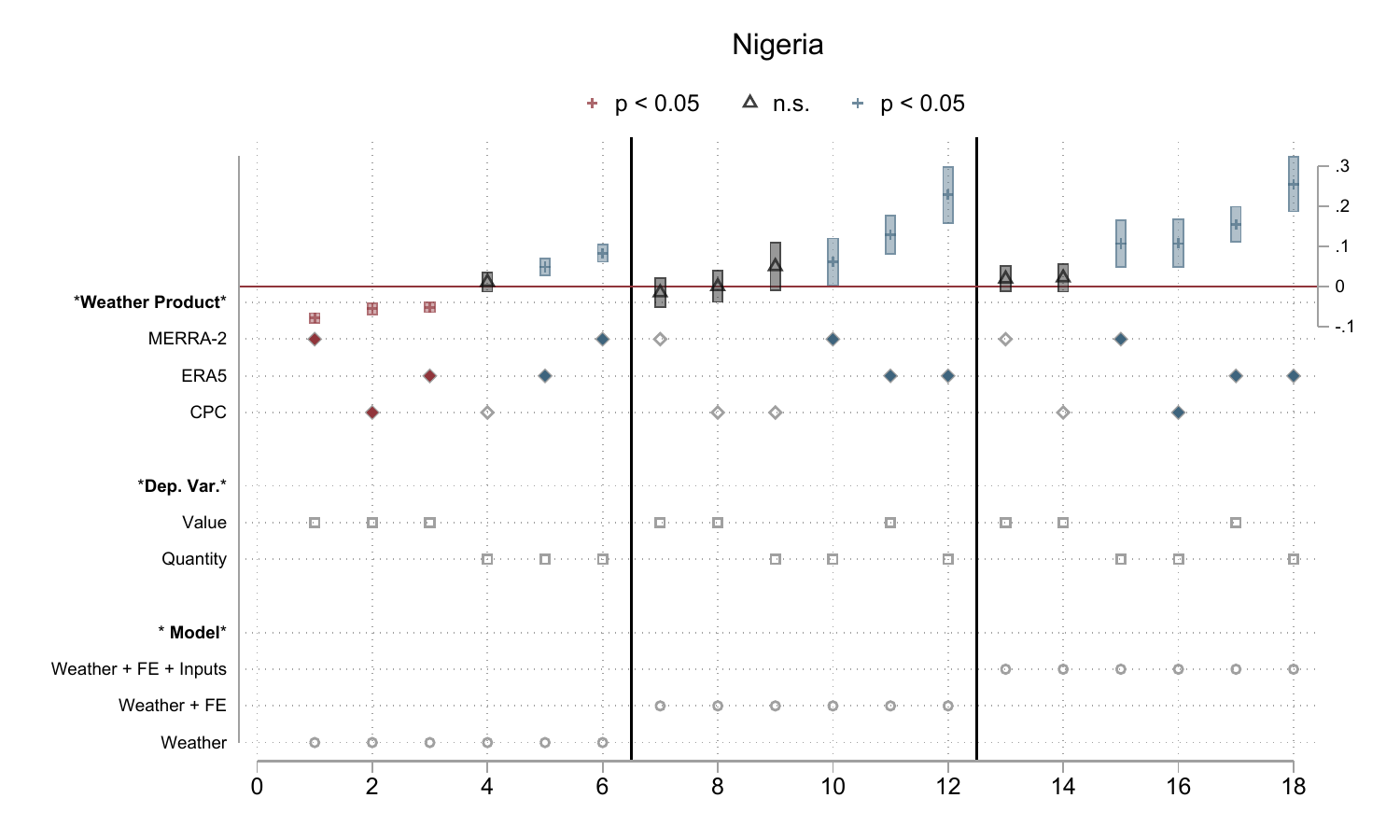}
			\includegraphics[width=.49\linewidth,keepaspectratio]{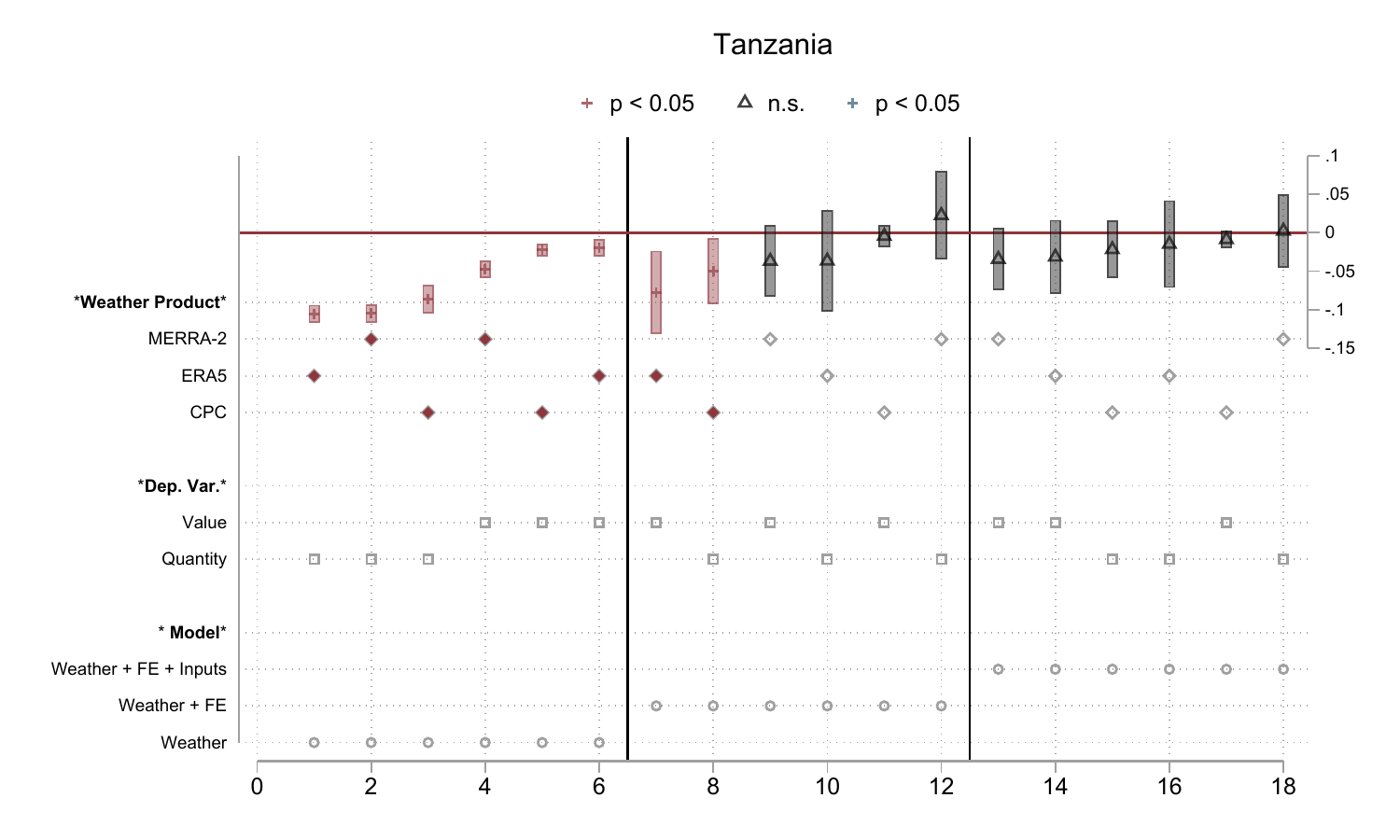}
			\includegraphics[width=.49\linewidth,keepaspectratio]{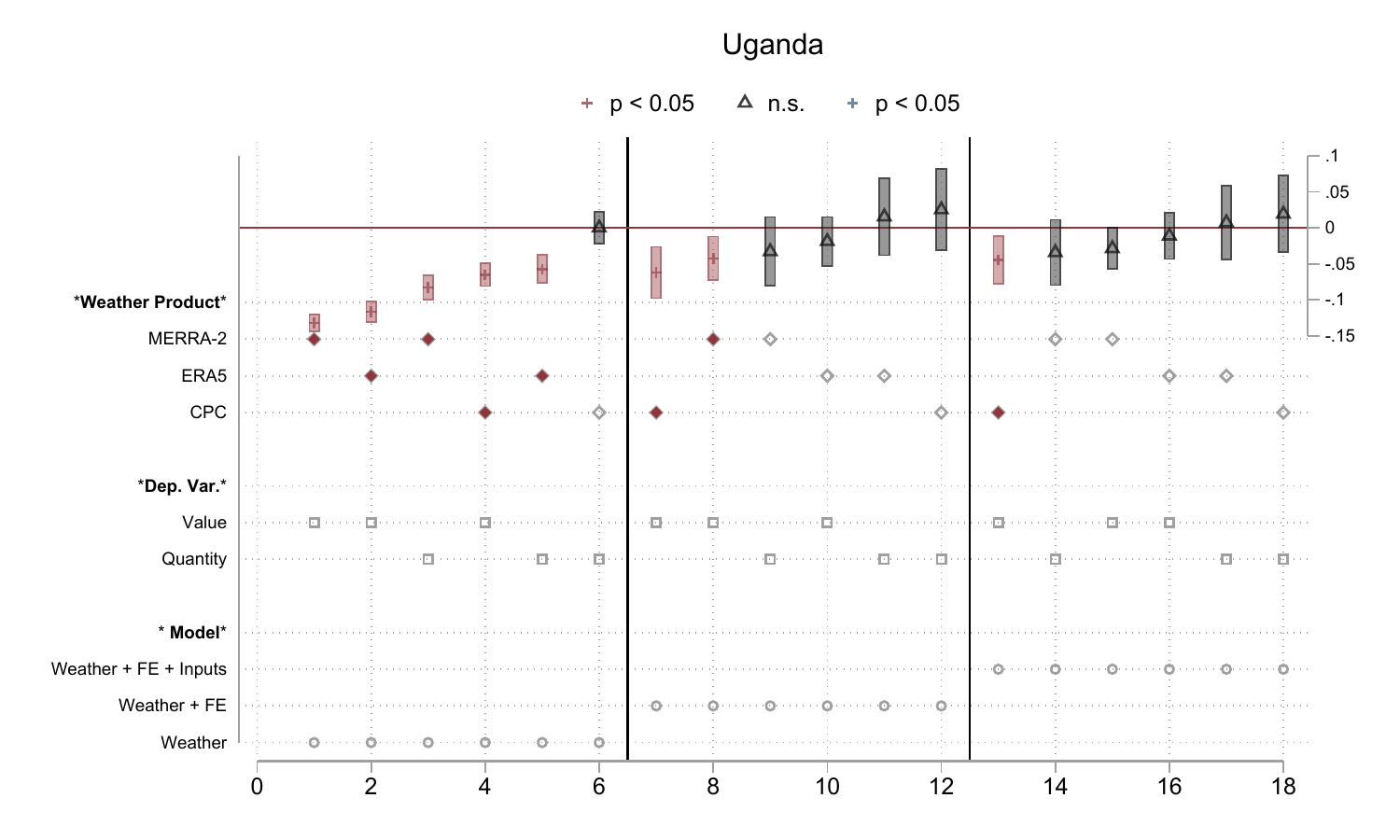}
		\end{center}
		\footnotesize  \textit{Note}: The figure presents specification curves, where each panel represents a different country, with three different models presented within each panel. Each panel includes 18 regressions, where each column represents a single regression. Significant and non-significant coefficients are designated at the top of the figure. For each Earth observation product, we also designate the significance and sign of the coefficient with color: red represents coefficients which are negative and significant; white represents insignificant coefficients, regardless of sign; and blue represents coefficients which are positive and significant.  
	\end{minipage}	
\end{figure}
\end{center}

\newpage 
\begin{landscape}
\begin{center}
\begin{figure}[!htbp]
	\begin{minipage}{\linewidth}
		\caption{Bumpline: Days Without Rain}
		\label{fig:bump_norain}
		\begin{center}
			\includegraphics[width=\linewidth,keepaspectratio]{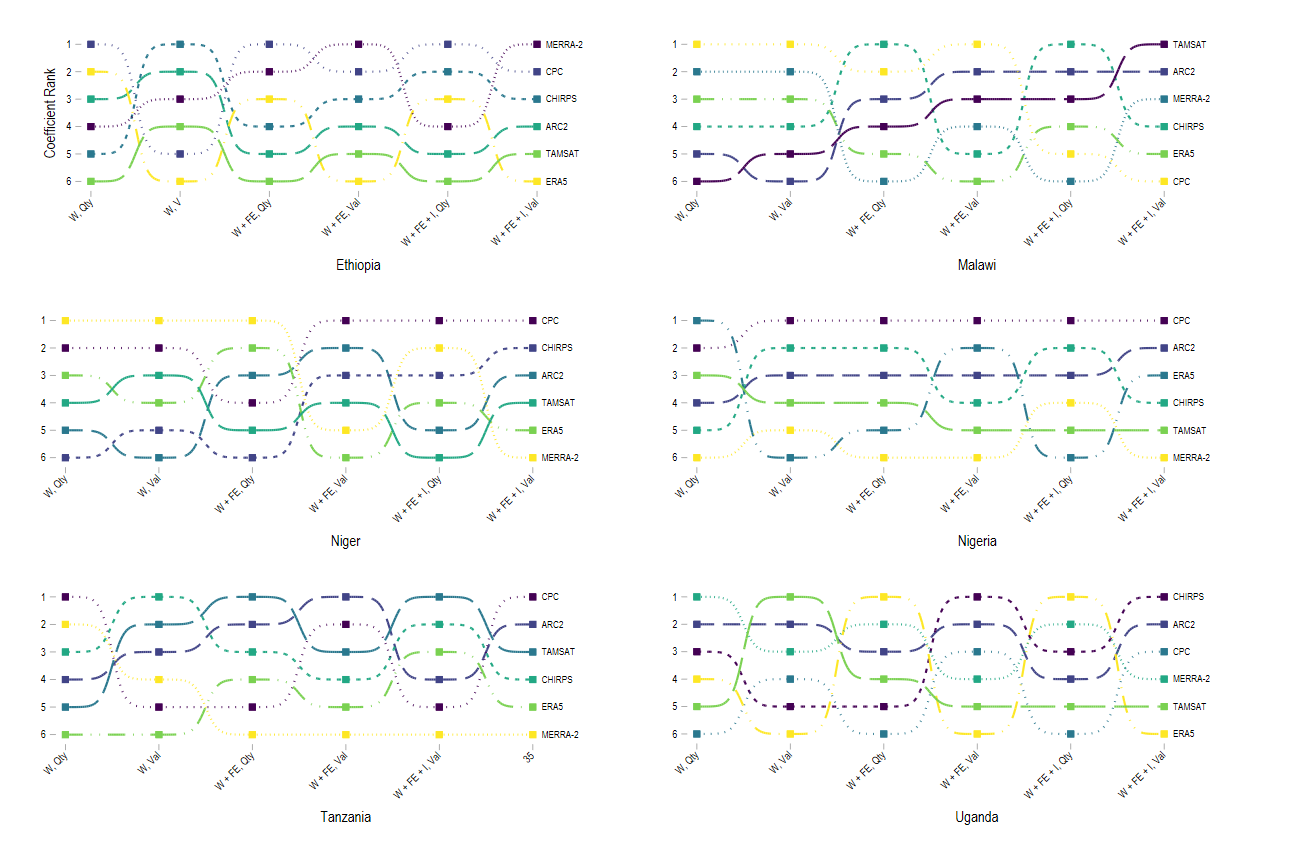}
		\end{center}
		\footnotesize  \textit{Note}: The figure presents a bumpline plot, where each panel represents a different country, where each column represents a different models and a different outcome variable for each model; each row represents the rank of each coefficient (one to six, with one representing the largest number and six representing the smallest number) within those regressions. Each panel includes 36 regressions. The bumpline package was developed by \cite{bumpline}.
	\end{minipage}	
\end{figure}
\end{center}    
\end{landscape}

\newpage 
\begin{landscape}
\begin{center}
\begin{figure}[!htbp]
	\begin{minipage}{\linewidth}
		\caption{Bumpline: Growing Degree Days}
		\label{fig:bump_gdd}
		\begin{center}
			\includegraphics[width=\linewidth,keepaspectratio]{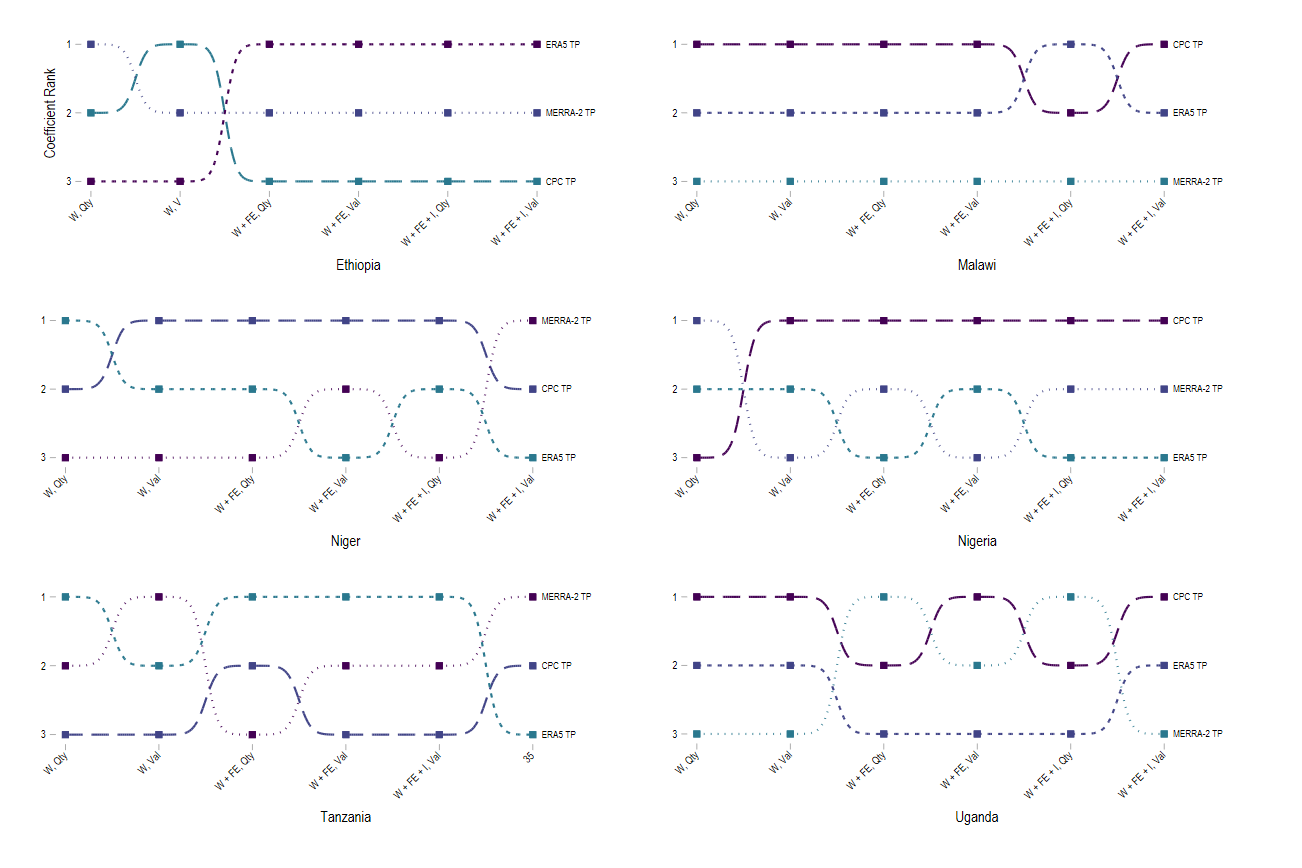}
		\end{center}
		\footnotesize  \textit{Note}: The figure presents a bumpline plot, where each panel represents a different country, where each column represents a different models and a different outcome variable for each model; each row represents the rank of each coefficient (one to six, with one representing the largest number and six representing the smallest number) within those regressions. Each panel includes 18 regressions. The bumpline package was developed by \cite{bumpline}.
	\end{minipage}	
\end{figure}
\end{center}    
\end{landscape}

\end{document}

%% file: tables/summary_stats.tex
\begin{tabular}{l*{6}{c}} \\ [-1.8ex]\hline \hline \\[-1.8ex] 
& \multicolumn{1}{c}{Ethiopia} & \multicolumn{1}{c}{Malawi} &  
\multicolumn{1}{c}{Niger} & \multicolumn{1}{c}{Nigeria} & 
\multicolumn{1}{c}{Tanzania} & \multicolumn{1}{c}{Uganda} \\ 
\midrule &&&&&& \\ \multicolumn{7}{l}{\emph{\textbf{Panel A}: 
Total Farm Production}} \\ 
\midrule
Total farm production (2015 USD)&       340.3&       395.0&       188.5&       639.3&       223.3&       145.5\\
                    &     (538.8)&    (3913)&     (201.7)&     (689.3)&     (369.2)&     (195.8)\\
Total farmed area (ha)&       0.959&       0.852&      11.14&       1.999&       1.748&       3.872\\
                    &     (4.709)&     (0.858)&    (15.59)&     (3.092)&     (3.691)&     (8.939)\\
Total farm yield (2015 USD/ha)&      605.0&      367.3&       60.93&      728.9&      233.1&       98.90\\
                    &     (909.7)&     (393.9)&     (169.0)&     (929.3)&     (306.0)&     (136.4)\\
Total farm labor rate (days/ha)&      480.1&      241.4&       94.82&      161.9&      299.2&      286.2\\
                    &     (987.5)&     (187.6)&     (219.1)&     (218.4)&     (359.6)&     (363.8)\\
Total farm fertilizer rate (kg/ha)&      67.02&     126.8&       5.120&     109.2&      23.87&       0.376\\
                    &    (108.35)&    (279.7)&     (23.83)&    (240.8)&     (79.05)&      (2.601)\\
Total farm pesticide use (\%)&       0.072&       0.058&       0.080&       0.211&       0.087&       0.069\\
                    &     (0.259)&     (0.234)&     (0.271)&     (0.408)&     (0.282)&     (0.253)\\
Total farm herbicide use (\%)&       0.229&       0.016&       0.037&       0.307&       0.085&       0.036\\
                    &     (0.420)&     (0.125)&     (0.190)&     (0.461)&     (0.279)&     (0.185)\\
Total farm irrigation use (\%)&       0.070&       0.016&       0.000&       0.028&       0.036&       0.015\\
                    &     (0.255)&     (0.126)&     (0.000)&     (0.165)&     (0.187)&     (0.123)\\
 \midrule Observations & 10674 & 8897 & 3913 & 9145 & 9916 & 11692 \\ 
Households & 5333 & 3833 & 2320 & 3412 & 4804 & 4003 \\ 
\midrule &&&&&& \\ \multicolumn{7}{l}{\emph{\textbf{Panel B}: 
Primary Crop Production}} \\  
Primary crop production (kg)&      330.1&     1043&      471.5&      607.8&      660.9&       44.38\\
                    &    (492.2)&   (3034.8)&    (488.5)&    (962.2)&    (881.0)&     (75.11)\\
Primary crop farmed area (ha)&       0.295&       0.712&       4.930&       0.677&       0.984&       1.121\\
                    &     (1.083)&     (0.760)&     (6.451)&     (0.999)&     (1.691)&     (2.002)\\
Primary crop yield (kg/ha)&     1711&     1491&      219.8&     1840&     1160&       82.06\\
                    &    (1893)&    (1379)&     (363.5)&    (1992)&    (1322)&     (115.6)\\
Primary crop labor rate (days/ha)&      493.3&      237.8&       70.71&      167.2&      259.2&      328.2\\
                    &     (712.5)&     (193.8)&     (121.1)&     (232.2)&     (302.6)&     (435.1)\\
Primary crop fertilizer rate (kg/ha)&      91.08&     118.7&       1.945&     150.0&      28.23&       0.471\\
                    &   (159.9)&   (250.6)&     (7.754)&   (296.5)&    (89.41)&     (3.450)\\
Primary crop pesticide use (\%)&       0.044&       0.027&       0.055&       0.185&       0.077&       0.037\\
                    &     (0.204)&     (0.163)&     (0.228)&     (0.388)&     (0.266)&     (0.189)\\
Primary crop herbicide use (\%)&       0.044&       0.009&       0.028&       0.390&       0.072&       0.033\\
                    &     (0.206)&     (0.094)&     (0.164)&     (0.488)&     (0.259)&     (0.178)\\
Primary crop irrigation use (\%)&       0.050&       0.013&       0.000&       0.022&       0.020&       0.012\\
                    &     (0.219)&     (0.115)&     (0.000)&     (0.146)&     (0.141)&     (0.110)\\
 \midrule Observations & 5904 & 7866 & 3492 & 5349 & 7341 & 6156 \\ 
Households & 3490 & 3594 & 2079 & 3055 & 3837 & 2872 \\ 
\hline \hline \\[-1.8ex] \multicolumn{7}{p{\linewidth}}{\small 
\noindent \textit{Note}: The table presents the mean  and 
(standard deviation) of total farm production and primary crop 
production. Statistics are calculated for each country, 
aggregated across all waves used.}  \end{tabular}